\documentclass[10pt,a4paper]{article}
\pdfoutput=1
\usepackage{jstyle}
\usepackage{amsmath, amsfonts, amssymb,subcaption}
\usepackage{graphicx, hyperref, color, verbatim}
\usepackage{diagbox,physics}

\usepackage{xcolor}
\usepackage{tikz}
\usepackage{amsmath}

\usepackage{tikz-cd}
\usetikzlibrary{matrix, calc, arrows}

\newcommand{\bea}{\begin{eqnarray}}
\newcommand{\eea}{\end{eqnarray}}
\newcommand\restr[2]{{\left.\kern-\nulldelimiterspace#1\vphantom{\big|}\right|_{#2}}}

\newcommand{\QforOtilde}{\mathsf{Q}}
\newcommand{\SforOtilde}{\mathsf{S}}

\newenvironment{psmallmatrix}
  {\left(\begin{smallmatrix}}
  {\end{smallmatrix}\right)}

\def \mapright#1{\smash{\mathop{\longrightarrow}\limits^{#1}}}

\newcommand{\weight}{\omega}
\newcommand{\myzeta}{\zeta}

\newcommand{\ffcorrshort}{f}

\newcommand{\sX}{\mathfrak{X}}

\newcommand{\D}{\Delta}
\newcommand{\h}{h}

\newcommand{\DeltaExt}{\h_{\mathtt{ext}}}

\newcommand{\len}{l}
\newcommand{\vv}{v}
\newcommand{\ww}{w}

\title{Unmixing the Wilson line defect CFT. \\
Part I: spectrum and kinematics}

\author[\varphi]{Pietro Ferrero}
\author[\Psi,f]{Carlo Meneghelli} 
\affiliation[\varphi]{Simons Center for Geometry and Physics, SUNY, Stony Brook, NY 11794, USA}
\affiliation[\Psi]{Dipartimento SMFI, Universit\`{a} di Parma, Viale G.P.
Usberti 7/A, 43100, Parma, Italy}
\affiliation[f]{INFN Gruppo Collegato di Parma}

\emailAdd{pferrero@scgp.stonybrook.edu, \\ \hskip 38pt carlo.meneghelli@unipr.it}
\abstract{This is the first of a series of two papers in which we study the one-dimensional defect CFT defined by insertions of local operators along a $\tfrac{1}{2}$-BPS Wilson line in $\mathcal{N}=4$ super Yang-Mills. In this first paper we focus on the kinematical implications of invariance under the $\mathfrak{osp}(4^*|4)$ superconformal algebra preserved by the line. We study correlation functions involving both protected and unprotected supermultiplets and derive the associated superconformal blocks, using two types of superspace for short and long representations. We also discuss the spectrum of defect theories defined by the Wilson line, focusing in particular on fundamental lines in the planar limit: in this case we provide a detailed analysis of the type and number of states both at weak 't~Hooft coupling, via the free gauge theory description of the defect CFT, and at strong coupling, where there is a dual description via AdS/CFT. Focusing on the strongly-coupled regime, which will be subject to a detailed analysis using analytic bootstrap techniques in \cite{Ferrero:2023gnu}, we also develop a strategy that allows to explicitly build superconformal primary operators and their superconformal descendants in terms of the elementary fields in the AdS Lagrangian description. The explicit results will be used in \cite{Ferrero:2023gnu} to address the problem of operators mixing at strong coupling. This paper and the companion \cite{Ferrero:2023gnu} provide an extended version of the results presented in \cite{Ferrero:2021bsb}.}

\begin{document}
\maketitle
\tableofcontents

\newpage

\section{Introduction}\label{sec:intro}

Conformal field theories (CFTs) are a cornerstone of our understanding of the space of quantum field theories (QFTs) and their study has seen a revival with the modern conformal bootstrap program \cite{Rattazzi:2008pe}. In the bootstrap approach a crucial role is played by symmetries and in particular the implications of conformal invariance are exploited to obtain non-perturbative constraints on CFTs, without relying on a weakly-coupled description. The bootstrap constraints become even more powerful when combined with supersymmetry, although in that case most of the explorations have been so far limited to the sector of absolutely protected multiplets. Recently some attention has been devoted to the study of CFTs in the presence of {\it defects}: these can be realized, for instance, by boundary conditions or extended operators and they are interesting for both low- and high-energy physics applications. The study of conformal defects dates back to \cite{Cardy:1984bb,Cardy:1989ir} and they were revived more recently from a bootstrap perspective in \cite{Liendo:2012hy,Gaiotto:2013nva,Gliozzi:2015qsa,Gliozzi:2016cmg,Billo:2016cpy}. 

We will be particularly interested in the case of line defects preserving some amount of supersymmetry, for which all possible superconformal algebras and their representations were classified in \cite{Agmon:2020pde}. Moreover, while this is surely an interesting possibility we shall not consider bulk correlators in the presence of a defect\footnote{For the bootstrap of bulk correlators in presence of a half-BPS Wilson line in 4d $\mathcal{N}=4$ super Yang-Mills, see {\it e.g.} \cite{Barrat:2020vch,Barrat:2021yvp,Gimenez-Grau:2023fcy} and for half-BPS surface defects in the 6d $(2,0)$ theory see \cite{Meneghelli:2022gps,Chen:2023yvw}.}, but rather focus on correlation functions between excitations localized on the defect, so that our setup will be purely one-dimensional. More specifically, here and in the companion work \cite{Ferrero:2023gnu} we consider one-half BPS Wilson lines in 4d $\mathcal{N}=4$ super Yang-Mills (SYM) theory \cite{Maldacena:1998im}. The insertion of local operators along such a supersymmetric Wilson line defines a one-dimensional defect CFT preserving an $\mathfrak{osp}(4^*|4)$ superconformal algebra and therefore yields a one-dimensional CFT with sixteen supercharges.

One reason for the interest in such defect CFT is the fact that the expectation value of supersymmetric Wilson loops of arbitrary shape can be computed by perturbing away from the straight (or circular) one-half BPS contour with insertions of the displacement operator \cite{Drukker:2006xg}. This makes it particularly interesting to be able to compute correlation functions of such operator along the line. Moreover, as we shall stress various times, the high amount of supersymmetry and the simple one-dimensional kinematics make this model a simple playground where various ideas and techniques can be tested. This is particularly profitable since the model preserves most of the appealing features of the bulk 4d $\mathcal{N}=4$ SYM theory, so that one can study the half-BPS defect CFT using holography \cite{Drukker:2005kx,Gomis:2006sb,Gomis:2006im,Giombi:2017cqn,Giombi:2020kvo}, integrability \cite{Kiryu:2018phb,Grabner:2020nis,Cavaglia:2021bnz,Cavaglia:2022qpg}, supersymmetric localization \cite{Giombi:2018qox,Giombi:2018hsx,Giombi:2020amn}, large charge expansions \cite{Giombi:2021zfb,Giombi:2022anm} and the conformal bootstrap \cite{Liendo:2016ymz,Liendo:2018ukf,Ferrero:2021bsb}. Interestingly, the half-BPS Wilson loop also arises as the IR fixed point of a defect RG flow from a non-supersymmetric loop, as first observed in \cite{Polchinski:2011im} and then studied in a series of papers \cite{Beccaria:2017rbe,Beccaria:2019dws,Beccaria:2021rmj,Beccaria:2022bcr} (see also \cite{Cuomo:2021rkm} for a discussion of general RG flows on line defects). Recently, a study of out-of-time-order correlation functions has also been initiated in \cite{Giombi:2022pas}.

In this and the companion paper \cite{Ferrero:2023gnu} we expand on the results sketched in \cite{Ferrero:2021bsb}, while also providing several novel results for the half-BPS Wilson line in the fundamental representation, in planar 4d $\mathcal{N}=4$ SYM. Note, however, that most of the developments related to superconformal kinematics apply more generally to arbitrary one-dimensional theories with $\mathfrak{osp}(4^*|4)$ invariance, regardless of their microscopic definition. In particular, while \cite{Ferrero:2023gnu} has a bootstrap flavor and is devoted to the study of correlation functions for large 't Hooft coupling, here we shall focus on two non-dynamical aspects: the superconformal kinematics for various types of correlation functions involving short and long multiplets (which only relies on $\mathfrak{osp}(4^*|4)$ invariance), as well as the counting of states for the fundamental Wilson line in the planar limit, both at weak and at strong coupling.

In terms of kinematics, we consider three- and four-point functions (as well as the associated superconformal blocks in the latter case) that are relevant for the bootstrap computations of \cite{Ferrero:2023gnu}: for instance, we study the superconformal blocks for correlators of the type $\langle \mathcal{D}\mathcal{D}\mathcal{D}\mathcal{D}\rangle$ (already discussed in \cite{Liendo:2018ukf}), $\langle \mathcal{D}\mathcal{D}\mathcal{D}\mathcal{L}\rangle$ and $\langle \mathcal{D}\mathcal{D}\mathcal{L}\mathcal{L}\rangle$, where $\mathcal{D}$ are one-half BPS multiplets and $\mathcal{L}$ are long multiplets. We only consider correlators that are relevant for \cite{Ferrero:2023gnu}, but using the same techniques additional results could be derived for other correlators. This is particularly interesting given the fact that, despite the great attention devoted to supersymmetric CFTs (SCFTs), in higher dimensions not much is known about conformal blocks beyond four-point functions of one-half BPS operators while the simplified kinematics of one-dimensional theories offers the chance to develop new tools for the study of correlation functions involving long multiplets (see \cite{Cornagliotto:2017dup,Kos:2018glc,Buric:2020buk,Buric:2020qzp} for some results in this direction).

The other main theme of the paper is the study of the spectrum of the half-BPS line defect in the fundamental representation, in planar $\mathcal{N}=4$ SYM. While it is an extremely interesting and challenging problem to compute the spectrum for arbitrary values of the 't Hooft coupling $\lambda$ (see \cite{Grabner:2020nis,Cavaglia:2021bnz,Cavaglia:2022qpg} for progress in this direction using integrability), here we shall address a much simpler question and study the spectrum at weak coupling ($\lambda=0$), where the model admits is a free gauge theory description, and at strong coupling ($\lambda=\infty$), where one has a free theory description in terms of semiclassical string theory in AdS$_2$ \cite{Giombi:2017cqn}. We shall discuss the techniques that are necessary to perform counting of operators and discuss some results that provide an important input for integrability-based computations. Besides counting, we also consider a different problem related to the spectrum at strong coupling: the explicit construction of superconformal primaries and their descendants in terms of the fundamental fields in the AdS$_2$ Lagrangian description. While this is purely a technical issue, its resolution is crucial for us in order to address the problem of operators degeneracy, that as discussed in \cite{Ferrero:2021bsb} (see \cite{Ferrero:2023gnu} for more details) is the main obstacle to performing analytic bootstrap computations at all orders in the expansion for large $\lambda$.

The structure of the paper is as follows. In Section \ref{sec:superalgebras} we present the $\mathfrak{osp}(4^*|4)$ superconformal algebra and provide a brief description of its irreducible representations, focusing in particular on absolutely protected $\tfrac{1}{2}$-BPS multiplets and generic long supermultiplets, introducing for each a suitable superspace representation of the algebra. In Section \ref{sec:kinematics} we use such superspace description to build two-, three- and four-point functions involving short and long multiplets, exploring their superconformal kinematics and the associated superconformal blocks. In Section \ref{sec:spectrum} we discuss how one can study the spectrum of local operators in Wilson line theories and focus in particular on the $\tfrac{1}{2}$-BPS defect in the fundamental representation in the planar limit, where using characters we give a precise determination of the spectrum for both weak and strong coupling. While counting is often enough, it is sometimes useful to have an explicit characterization of local operators in terms of elementary fields, and this is the subject of Section \ref{sec:operators}, where the focus is on the strongly-coupled regime. We conclude with a discussion and some outlook in Section \ref{sec:discussion}. Finally, several technical points and some of our results are relegated to the appendices: in Appendix \ref{app:notation} we summarize our choices in terms of notation; in Appendix \ref{app:supermultiplets} we spell out in detail the content of superconformal descendants for all supermultiplets of $\mathfrak{osp}(4^*|4)$; in Appendix \ref{app:counting} we provide details on the construction of characters as well as how the counting of states is performed at weak and strong coupling, including tables with the degeneracies of the lowest-lying states; in Appendix \ref{app:forOperatorConstuction} we collect additional information relevant for the explicit construction of local operators and their correlators in the free theory at strong coupling. Appendix \ref{app:blocks} contains explicit results related to various types of superconformal blocks that we chose not to include in the main text, and finally Appendix \ref{app:operules} contains a summary and justification of various OPE rules that will be useful for \cite{Ferrero:2023gnu}.

\section{$\mathfrak{osp}(4^*|4)$: algebra, representations and superspace}\label{sec:superalgebra}
\label{sec:superalgebras}

\subsection{Superalgebra and unitary multiplets}

Let us start by describing the relevant 1d superconformal algebra $\mathfrak{osp}(4^*|4)$. 
Its bosonic subalgebra is $\mathfrak{g}_0=\mathfrak{so}(4^*)\oplus \mathfrak{sp}(4)\simeq \mathfrak{sl}(2)\oplus \mathfrak{su}(2)\oplus \mathfrak{sp}(4)$, where the first term corresponds to the 1d conformal symmetry $\mathfrak{sl}(2)\simeq \mathfrak{so}(2,1)$, generated by translation $\mathfrak{P}$, dilatations $\mathfrak{D}$ and special conformal transformations $\mathfrak{K}$. On the other hand, we denote with $\mathfrak{L}_{\alpha}^{\,\,\,\beta}$ the generators of transverse rotations $\mathfrak{su}(2)$, with Greek indices $\alpha,\beta,\ldots\in \{1,2\}$, and with $\mathfrak{R}_{A}^{\,\,\,B}$ the generators of the R symmetry algebra $\mathfrak{sp}(4)$, with capital Latin indices $A,B,\ldots\in \{1,4\}$. We raise and lower $\mathfrak{su}(2)$ indices with the two-dimensional Levi-Civita tensor $\epsilon_{\alpha\beta}$ and similarly for $\mathfrak{sp}(4)$ indices with the symplectic form $\Omega_{AB}$ -- see Appendix \ref{app:notation} for our conventions. The sixteen fermionic generators transform in the $(\mathbf{2},\mathbf{2},\mathbf{4})$ representation\footnote{Here we identify representations with their dimensions, in the following we will also use Dynkin labels.} of $\mathfrak{g}_0$ and can be divided in eight Poincar\'e supercharges $\mathfrak{Q}_{\alpha A}$ and their eight superconformal partners $\mathfrak{S}_{\alpha A}$. When the 1d SCFT is realized as a defect in a 4d $\mathcal{N}=4$ SCFT, the symmetry $\mathfrak{osp}(4^*|4)$ arises as a subalgebra of the 4d $\mathcal{N}=4$ superconformal algebra $\mathfrak{psu}(2,2|4)$, where one breaks $\mathfrak{su}(2,2)\to \mathfrak{sl}(2)\oplus \mathfrak{su}(2)$, $\mathfrak{su}(4)\to \mathfrak{sp}(4)$ as well as sixteen of the original thirty-two supercharges.

In our conventions, the bosonic subalgebra $\mathfrak{g}_0$ is 
\begin{equation}\label{ospalgebra_BB}
\begin{gathered}
\text{[}\mathfrak{D},\mathfrak{P}]=\mathfrak{P}\, ,\qquad
[\mathfrak{D},\mathfrak{K}]=-\mathfrak{K}\, ,\qquad
[\mathfrak{K},\mathfrak{P}]=2\,\mathfrak{D}\,,\qquad [\mathfrak{L}_{\alpha}^{\,\,\,\beta},\mathfrak{L}_{\gamma}^{\,\,\,\delta}]=\delta_{\alpha}^{\,\,\,\delta}\,\mathfrak{L}_{\gamma}^{\,\,\,\beta}-\delta_{\gamma}^{\,\,\,\beta}\,\mathfrak{L}_{\alpha}^{\,\,\,\delta}\,, \\[5pt]
[\mathfrak{R}_A^{\,\,\,B},\mathfrak{R}_C^{\,\,\,D}]=\delta_A^{\,\,\,D}\,\mathfrak{R}_C^{\,\,\,B}-\delta_C^{\,\,\,B}\,\Omega_{AF}\,\Omega^{DE}\,\mathfrak{R}_E^{\,\,\,F}+\Omega^{BD}\,\Omega_{AE}\,\mathfrak{R}_C^{\,\,\,E}-\Omega_{AC}\,\Omega^{DE}\,\mathfrak{R}_E^{\,\,\,B}\,.
\end{gathered}
\end{equation}
The anti-commutation relations between the fermionic generators are given by
\begin{equation}\label{ospalgebra_FF}
\begin{gathered}
 \{\mathfrak{Q}_{\alpha A},\,\mathfrak{S}^{\beta B}\}\,=\,
\delta_{\alpha}^{\,\,\,\beta}\,\mathfrak{R}_A^{\,\,\,B}-\delta_A^{\,\,\,B}\,\mathfrak{L}_{\alpha}^{\,\,\,\beta}+\delta_{\alpha}^{\,\,\,\beta}\,\delta_A^{\,\,\,B}\,\mathfrak{D}\,,
\qquad\,\,
\begin{aligned}
 \{\mathfrak{Q}_{\alpha A},\,\mathfrak{Q}_{\beta B}\}\,=\,\epsilon_{\alpha \beta}\,\Omega_{AB}\,
 \mathfrak{P}\,,\\
 \{\mathfrak{S}^{\alpha A},\,\mathfrak{S}^{\beta B}\}\,=\,\epsilon^{\alpha \beta}\,\Omega^{AB}\,
 \mathfrak{K}\,,
\end{aligned}
\end{gathered}
\end{equation}
while their commutators with the bosonic part of the algebra read
\begin{equation}\label{ospalgebra_BF}
\begin{gathered}
\text{[}\mathfrak{P},\mathfrak{Q}_{\alpha A}\text{]}=0\,,\qquad
[\mathfrak{D},\mathfrak{Q}_{\alpha A}]=\tfrac{1}{2}\mathfrak{Q}_{\alpha A}\,,\qquad
[\mathfrak{K},\mathfrak{Q}_{\alpha A}]=\epsilon_{\alpha\beta}\,\Omega_{AB}\,\mathfrak{S}^{\beta B}\,, \\[5pt]
[\mathfrak{P},\mathfrak{S}^{\alpha A}]=-\epsilon^{\alpha\beta}\,\Omega^{AB}\,\mathfrak{Q}_{\beta B}\,,\qquad
[\mathfrak{D},\mathfrak{S}^{\alpha A}]=-\tfrac{1}{2}\mathfrak{S}^{\alpha A}\,,\qquad
[\mathfrak{K},\mathfrak{S}^{\alpha A}]=0\,,\\[5pt]
[\mathfrak{L}_{\alpha}^{\,\,\,\beta},\mathfrak{Q}_{\gamma C}]=-\delta_{\gamma}^{\,\,\,\beta}\,\mathfrak{Q}_{\alpha C}+\tfrac{1}{2}\,\delta_{\alpha}^{\,\,\,\beta}\,\mathfrak{Q}_{\gamma C}\,, \qquad
[\mathfrak{L}_{\alpha}^{\,\,\,\beta},\mathfrak{S}^{\gamma C}]=\delta_{\alpha}^{\,\,\,\gamma}\,\mathfrak{S}^{\beta C}-\tfrac{1}{2}\,\delta_{\alpha}^{\,\,\,\beta}\,\mathfrak{S}^{\gamma C}\,,\\[5pt]
[\mathfrak{R}_A^{\,\,\,B},\mathfrak{Q}_{\gamma C}]=-\delta_C^{\,\,\,B}\,\mathfrak{Q}_{\gamma A}-\Omega_{AC}\,\Omega^{BD}\,\mathfrak{Q}_{\gamma D}\,, \qquad 
[\mathfrak{R}_A^{\,\,\,B},\mathfrak{S}^{\gamma C}]=\delta_A^{\,\,\,C}\,\mathfrak{S}^{\gamma B}+\Omega^{BC}\,\Omega_{AD}\,\mathfrak{S}^{\gamma D}\,. 
\end{gathered}
\end{equation}
The algebra admits a quadratic Casimir operator, which in our conventions reads
\begin{align}\label{ospCasimir}
\mathfrak{C}_{\mathfrak{osp}(4^*|4)}=\mathfrak{D}^2-\frac{1}{2}\{\mathfrak{P},\mathfrak{K}\}+\frac{1}{2}\,\mathfrak{L}_{\alpha}^{\,\,\,\beta}\,\mathfrak{L}_{\beta}^{\,\,\,\alpha}-\frac{1}{4}\mathfrak{R}_{A}^{\,\,\,B}\,\mathfrak{R}_{B}^{\,\,\,A}-\frac{1}{2}\,[\mathfrak{Q}_{\alpha A},\mathfrak{S}^{\alpha A}]\,.
\end{align}

Unitary highest weight representations of $\mathfrak{osp}(4^*|4)$ are uniquely specified by the scaling dimensions and R-symmetry representation\footnote{
From the 1d perspective the R-symmetry is $\mathfrak{su}(2)\oplus \mathfrak{sp}(4)$ even though the $\mathfrak{su}(2)$ factor is part of the four dimensional space-time symmetry when the theory is realized as a defect CFT.
} $\weight=\{\h,s,[a,b]\}$ of their superconformal primary. Here $\h$ is the scaling dimension, $s\in \mathbb{N}$ corresponds to the $s+1$-dimensional representation of $\mathfrak{su}(2)$  while $[a,b]$ are $ \mathfrak{sp}(4)$ Dynkin labels, chosen in such a way that $[1,0]=\mathbf{4}$ and $[0,1]=\mathbf{5}$. We make here an important remark regarding notation. In the rest of this paper as well as in \cite{Ferrero:2023gnu}, we shall denote the {\it non-perturbative} dimension of operators with the symbol $h$, while conformal dimensions in free theories will be denoted by $\Delta$, with $h=\Delta+\ldots$ when expanding around a free theory point (as it will be the case in \cite{Ferrero:2023gnu}).

We record here the eigenvalue of the quadratic Casimir operator \eqref{ospCasimir} on such representations, which in our conventions reads
\begin{align}\label{ospCasimir_eigenvalue}
\mathfrak{c}_{\mathfrak{osp}(4^*|4)}=\h(\h+3)+\tfrac{1}{4}s(s+2)-\tfrac{1}{2}a^2-a(b+2)-b(b+3)\,.
\end{align}
The unitary superconformal multiplets of $\mathfrak{osp}(4^*|4)$ were classified, for example, in \cite{Gunaydin:1990ag,Dorey:2018klg,phdthesis,Agmon:2020pde} and we review them in detail in Appendix \ref{app:supermultiplets}, where we also provide explicitly their operators content. The complete classification includes absolutely protected short multiplets, semi-short multiplets and generic long ones. However, in this work we will mostly deal with two type of representations: 
\begin{itemize}
\item[(i)] Short ($\tfrac{1}{2}$-BPS) multiplets  $\mathcal{D}_k$, with $\weight=\{k,0,[0,k]\}$, whose dimension is absolutely protected.
\item[(ii)] Long multiplets $\mathcal{L}_{s,[a,b]}^{\h}$, where the scaling dimension is subject to the unitarity bound 
\begin{equation}
\h \geq a+b+1 +\tfrac{1}{2}s\,.
\end{equation}
\end{itemize}
All the other unitary irreducible representations can be obtained by taking subrepresentations and quotients of the long representations at the unitarity bound, see Appendix \ref{app:supermultiplets} for more details\footnote{More results on the representation theory of $\mathfrak{osp}(4^*|4)$ can be found in \cite{Gunaydin:1990ag,Liendo:2016ymz,Liendo:2018ukf,Dorey:2018klg,phdthesis,Agmon:2020pde}.}. The reason why we choose to focus on these two types of supermultiplets is that we are ultimately interested in studying the Maldacena Wilson line defect in 4d $\mathcal{N}=4$ SYM at strong coupling in the planar limit, where all other representations are absent, as argued in \cite{Liendo:2018ukf,Ferrero:2021bsb} and reviewed in Appendix \ref{app:counting}.

While the decomposition of all supermultiplets in conformal primaries of $\mathfrak{sl}(2)$ is given in Appendix \ref{app:supermultiplets}, we would like to review here some key examples in order to highlight certain features. For this purpose, we introduce the notation $[a,b]_s^{\h}$ to denote conformal primaries with dimension $\h$, $\mathfrak{su}(2)$ spin $s$ and transforming in the representation with Dynkin labels $[a,b]$ of $\mathfrak{sp}(4)$. Let us start from the short multiplet $\mathcal{D}_1$, which is always present in a 1d SCFT with $\mathfrak{osp}(4^*|4)$ symmetry that is realized as a defect in a 4d $\mathcal{N}=4$ SCFT. Its structure is given by
\begin{equation}
\label{D1Structure}
\begin{tikzpicture}
    \node (p1) at ( 0, 0) {$\varphi:\,\,\,[0,1]^{\h=1}_{s=0}$}; 
    \node (p2) at (2.7, 0) {
     $\Psi:\,\,\,[1,0]^{\h=\frac{3}{2}}_{s=1}$
    };
    \node (p3) at (5.4,0) {$f:\,\,\,[0,0]^{\h=2}_{s=2}$};
       \draw [-{Stealth}] (p1) -- (p2);
        \draw [-{Stealth}] (p2) -- (p3);
       \end{tikzpicture}
\end{equation}
where the arrows denote the action of Poincar\'e supercharges. Its components are associated to the breaking of the symmetries of the bulk theory due to the presence of the defect: the $5$ scalars $\phi$ with $\h=1$ are related to the breaking $\mathfrak{su}(4)\rightarrow \mathfrak{sp}(4)$, the fermions $\psi$ with the breaking of half of the fermionic symmetry and the displacement operator $f$ transforming as $[0,0]^{\h=2}_{s=2}$ with the breaking of the four-dimensional conformal algebra $\mathfrak{su}(2,2)$ -- see, {\it e.g.}, \cite{Billo:2016cpy}. For this reason we refer to $\mathcal{D}_1$ as the displacement supermultiplet, or superdisplacement operator. Note also that, having non-trivial components only at level one and two, this multiplet is shorter than generic $\tfrac{1}{2}$-BPS multiplets: we will call it {\it ultra-short}. The structure of $\mathcal{D}_k$ multiplets for $k>1$ reads
\begin{equation} 
\label{DkStructure}
\begin{tikzpicture}
    \node (p1) at ( 0, 0) {$[0,k]^{\h=k}_{s=0}$}; 
    \node (p2) at (2.9, 0) {
     $[1,k-1]^{\h=k+\frac{1}{2}}_{s=1}$
    };
    \node (p3) at ( 6.5,0) {$\,\,\,\,\,\,\,\,\,\,\,\,\,\,\,\oplus\,\,\,\,\,\,\,\,\,\,\,\,\,\,\,\,\,\,\,\,$
   % $\left([0,k-1]^{\h=k+1}_{s=2}\oplus[2,k-2]^{\h=k+1}_{s=0}\right)$
    };
       \node (p3up) at ( 6.5,0.6) {  $[0,k-1]^{\h=k+1}_{s=2}$
    };
       \node (p3down) at ( 6.5,-0.6) {$[2,k-2]^{\h=k+1}_{s=0}$
    };
    \node (p4) at ( 9.9,0) {$[1,k-2]^{\h=k+\frac{3}{2}}_{s=1}$};
    \node (p5) at ( 13.2,0) {$[0,k-2]^{\h=k+2}_{s=0}$};

  %  \begin{scope}[every path/.style={->}]
       \draw [-{Stealth}] (p1) -- (p2);
       
        \draw [-{Stealth}] (p2) -- (p3);
             \draw [-{Stealth}] (p3) -- (p4);
                  \draw [-{Stealth}] (p4) -- (p5);
  %  \end{scope}  
  % to zero which is  ( 5.3,-2.3)
       
       \end{tikzpicture}
\end{equation}
and contains states up to level four for all $k>1$. Finally, out of all long supermultiplets, a crucial role in \cite{Ferrero:2023gnu} will be played by those that are singlets of $\mathfrak{su}(2)\oplus \mathfrak{sp}(4)$: $\mathcal{L}_{0,[0,0]}^{\h}$. Their structure is significantly simpler than that of generic long multiplets described in Appendix \ref{app:supermultiplets} and reads
\begin{equation}
\label{L00Structure}
\begin{tikzpicture}
    \node (p1) at ( 0, 0) {$[0,0]^{\h}_{0}$}; 
    \node (p2) at (0, -1.2) {
     $[1,0]^{\h+\frac{1}{2}}_{1}$
    };
    \node (p3) at ( 0,-2.4) {$ [0,1]^{\h+1}_{2} \oplus [0,0]^{\h+1}_{2} \oplus { \color{brown} [2,0]^{\h+1}_{0}}
    $
    };
       \node (p4) at ( 0,-3.6) {$ [1,0]^{\h+\frac{3}{2}}_{3} \oplus
       { \color{brown} [1,1]^{\h+\frac{3}{2}}_{1}}
       \oplus { \color{brown} [1,0]^{\h+\frac{3}{2}}_{1}}
    $
    };
      \node (p5) at ( 0,-4.8) {$ [0,0]^{\h+2}_{4} \oplus
       { \color{brown} [2,0]^{\h+2}_{2} }
       \oplus 
       { \color{brown} [0,1]^{\h+2}_{2}}
       \oplus 
         { \color{brown} [0,2]^{\h+2}_{0}}
       \oplus 
   { \color{brown} [0,1]^{\h+2}_{0}}
       \oplus 
        { \color{brown} [0,0]^{\h+2}_{0}}
    $
    };
 \node (p6) at ( 0,-6) {$   { \color{brown} [1,0]^{\h+\frac{5}{2}}_{3}} \oplus
       { \color{brown} [1,1]^{\h+\frac{5}{2}}_{1}}
       \oplus { \color{brown} [1,0]^{\h+\frac{5}{2}}_{1}}
    $
    };
     \node (p7) at ( 0,-7.2) {${ \color{brown} [0,1]^{\h+3}_{2}} \oplus { \color{brown}[0,0]^{\h+3}_{2}} \oplus { \color{brown} [2,0]^{\h+3}_{0}}
    $
    };
      \node (p8) at (0, -8.4) {
     ${ \color{brown}[1,0]^{\h+\frac{7}{2}}_{1}}$
    };
          \node (p9) at (0, -9.6) {
     ${ \color{brown}[0,0]^{\h+4}_{0}}$
    };
  %  \begin{scope}[every path/.style={->}]
       \draw [-{Stealth}] (p1) -- (p2);
       
        \draw [-{Stealth}] (p2) -- (p3);
         \draw [-{Stealth}] (p3) -- (p4);
     \draw [-{Stealth}] (p4) -- (p5);
        \draw [-{Stealth}] (p5) -- (p6);

   \draw [-{Stealth}] (p6) -- (p7);

   \draw [-{Stealth}] (p7) -- (p8);
    \draw [-{Stealth}] (p8) -- (p9);
  %  \end{scope}  
  % to zero which is  ( 5.3,-2.3)
       \end{tikzpicture}
\end{equation}
At the unitarity bound $\h=1$, the parts in black forms a sub-representation and the one in \textcolor{brown}{brown} forms the quotient representation: in the language of \cite{Agmon:2020pde} these are $A_2[0,0]_0^{\h=1}$ and $B_1[2,0]_0^{\h=2}$, respectively, where the first is $\tfrac{1}{2}$-BPS and the second is $\tfrac{1}{4}$-BPS -- see Appendix \ref{app:supermultiplets} for more details. Let us also point out the representation of the type $[0,2]_{0}^{\h}$ appears only once in this supermultiplet: this fact will be relevant in Section \ref{sec:longcorrelators} for the study of the correlators $\langle \mathcal{D}_2\, \mathcal{L}^{\h_2}_{0,[0,0]}\, \mathcal{L}^{\h_3}_{0,[0,0]}\rangle$ and $\langle \mathcal{D}_1\mathcal{D}_1 \mathcal{D}_2\, \mathcal{L}^\h_{0,[0,0]}\rangle$, which in turn play a crucial role for the bootstrap analysis of \cite{Ferrero:2023gnu}.

\subsection{Realization of short multiplets in superspace}

A convenient superspace to describe the short supermultiplets of type $\mathcal{D}_k$ has been introduced in \cite{Liendo:2016ymz,Liendo:2018ukf} and we review it here. The $(4|4)$ superspace coordinates consist of a graded symmetric matrix\footnote{In our conventions this condition can be rewritten as $X^{\text{st}}=\Sigma \,X$ with $\Sigma=\text{diag}(-1,-1,+1,+1)$ and supertransposition is defined as
$\left(\begin{smallmatrix}P & Q\\R &S \end{smallmatrix}\right)^{\text{st}}=\left(\begin{smallmatrix}P^{\text{t}} & R^{\text{t}}\\- Q^{\text{t}} &S^{\text{t}} \end{smallmatrix}\right)$ where $\text{t}$ denotes ordinary transposition.
} 
\begin{equation}
\label{XmatrixShort}
X=
\begin{pmatrix}
t\,\epsilon^{\alpha \beta} & \theta^{\alpha b} \\
-\theta^{\beta a}  & y^{(ab)}
\end{pmatrix}\,,
\end{equation}
where $a,b=1,2$, $\alpha,\beta=1,2$, the $\theta^{\alpha a}$ are fermionic\footnote{It is sometimes convenient to think about the three  bosonic R-symmetry variables $y^{(ab)}$ as coordinates of a three dimensional auxiliary space on which the (complexification) of the R-symmetry act as the (complexification) of the conformal symmetry in three dimensions. } and $\epsilon^{\alpha \beta}$ is the antisymmetric tensor. 
The supergroup $OSP(4^*|4)$ acts projectively on $X$ as $X\mapsto (A X+B)(CX+D)^{-1}$, where $A,B,C,D$ are constrained by the condition that $X$ remains graded symmetric under the group action. 
The action of $g=\begin{psmallmatrix} A & B\\C & D\end{psmallmatrix}\in OSP(4^*|4)$ on the half-BPS superprimary $\mathcal{D}_k$ is given by\footnote{
The superdeterminant is defined as 
\begin{equation}
\text{sdet}\begin{pmatrix}
\mathsf{a} &\mathsf{b}\\
\mathsf{c} &\mathsf{d}
\end{pmatrix}
=\frac{\det(\mathsf{a}-\mathsf{b}\, \mathsf{d}^{-1}\mathsf{c}\
)}{\det(\mathsf{d})}\,.
\end{equation}
}
\begin{equation}
\label{groupactiononDk}
\left( g \circ \mathcal{D}_k\right)(g\cdot X) =  \left(\text{sdet}(C X+D)\right)^{k}\,\mathcal{D}_k(X)\,.
\end{equation}
The superspace for short operators is realized as a quotient and it therefore leads to a natural split of $\mathfrak{sp}(4)$ generators as\footnote{\label{rootssp4}Given an $\mathfrak{sp}(4)$ representation $[a,b]$ and denoting with $\alpha_1$ the simple root associated with the fundamental weight $[1,0]$ and with $\alpha_2$ that associated with $[0,1]$, then the correspondence between the entries of $\mathfrak{R}_A^{\,\,\,B}$ and the positive root generators in a Cartan-Weyl basis is as follows:
\begin{align}
\mathfrak{R}^+_{11}\leftrightarrow 2\alpha_1+\alpha_2\,, \qquad
\mathfrak{R}^+_{12}\leftrightarrow \alpha_1+\alpha_2\,, \qquad
\mathfrak{R}^+_{22}\leftrightarrow \alpha_2\,, \qquad
\mathfrak{R}_1^{\,\,\,2}\leftrightarrow \alpha_1\,, 
\end{align}
while the Cartan subalgebra is generated by $\mathfrak{R}_1^{\,\,\,1}$ and $\mathfrak{R}_2^{\,\,\,2}$, whose generators on highest weight states of irreducible representations $[a,b]$ are
\begin{align}
\mathfrak{R}_1^{\,\,\,1}=a+b\,, \qquad
\mathfrak{R}_2^{\,\,\,2}=b\,.
\end{align}}
\begin{align}
\label{Rsplit}
\mathfrak{R}_A^{\,\,\,B}\,\,\rightarrow\,\,
\begin{pmatrix}
\mathfrak{R}_a^{\,\,\,b} \,\,&\,\, \mathfrak{R}^{+}_{ab}\\[8pt]
\mathfrak{R}_{-}^{ab} \,\,&\,\, - \mathfrak{R}_b^{\,\,\,a}
\end{pmatrix}\,,
\end{align}
with $a,b,\ldots\in\{1,2\}$. In a similar way, the supercharges naturally split as
\begin{align}
\label{QSsplit}
\mathfrak{Q}_{\alpha A}\,\,\rightarrow\,\, \left(\mathfrak{Q}_{\alpha a}\,,\quad \tilde{\mathfrak{Q}}_{\alpha}^a\right)\,, \qquad
\mathfrak{S}^{\alpha A}\,\,\rightarrow\,\, \left(\mathfrak{S}^{\alpha a}\,,\quad \tilde{\mathfrak{S}}^{\alpha}_a\right)\,.
\end{align}
We are now ready to introduce differential operators that implement the infinitesimal form of \eqref{groupactiononDk}, which we denote with capital letters from the Latin alphabet and read\footnote{Note that to implement the symmetrization in the indices of $y^{ab}$, we use the following rule for derivatives
\begin{align}
\frac{\partial}{\partial y^{ab}}y^{cd}=\tfrac{1}{2}(\delta_a^c\,\delta_b^d+\delta_a^d\,\delta_b^c)\,.
\end{align}
We summarize our conventions for superspace contractions in Appendix \ref{app:notation}.}
\begin{align}
\label{GeneratorsSHORT}
\begin{split}
P&=\frac{\partial}{\partial t}\,, \qquad
D=-t\,\frac{\partial}{\partial t}-\tfrac{1}{2}\,\theta^{\alpha a}\,\frac{\partial}{\partial \theta^{\alpha a}}- k\,, \qquad 
K=t^2\,\frac{\partial}{\partial t}+t\,\theta^{\alpha a}\,\frac{\partial}{\partial \theta^{\alpha a}}-(\theta^2)^{ab}\,\frac{\partial}{\partial y^{ab}}+2\, k\,t\,,\\[5pt]
L_{\alpha}^{\,\,\,\beta}&=\theta^{\beta a}\,\frac{\partial}{\partial \theta^{\alpha a}}-\tfrac{1}{2}\,\delta_{\alpha}^{\,\,\,\beta}\,\theta^{\gamma c}\,\frac{\partial}{\partial \theta^{\gamma c}}\,, \qquad
R_a^{\,\,\,b}=2\,y^{bc}\,\frac{\partial}{\partial y^{ac}}+\theta^{\gamma b}\,\frac{\partial}{\partial \theta^{\gamma a}}- k\,\delta_a^{\,\,\,b}\,,\\[5pt]
R^+_{ab}&=-2\,\frac{\partial}{\partial y^{ab}}\,, \qquad
R_-^{ab}=2\,y^{ac}\,y^{bd}\,\frac{\partial}{\partial y^{cd}}+(\theta^2)^{ab}\,\frac{\partial}{\partial t}+\left(y^{ac}\,\theta^{\gamma b}+y^{bc}\,\theta^{\gamma a}\right)\,\frac{\partial}{\partial \theta^{\gamma c}}-2\, k\,y^{ab}\,,\\[5pt]
Q_{\alpha a}&=\frac{\partial}{\partial \theta^{\alpha a}}\,, \qquad
\tilde{Q}_{\alpha}^a=y^{ab}\,\frac{\partial}{\partial \theta^{\alpha b}}+\epsilon_{\alpha\beta}\,\theta^{\beta a}\,\frac{\partial}{\partial t}\,,  \qquad
\tilde{S}^{\alpha}_a=-2\,\theta^{\alpha b}\,\frac{\partial}{\partial y^{ab}}+t\,\epsilon^{\alpha\beta}\,\frac{\partial}{\partial \theta^{\beta a}}\,,\\[5pt]
S^{\alpha a}&=-\theta^{\alpha a}\,t\,\frac{\partial}{\partial t}+2\,y^{ab}\,\theta^{\alpha c}\,\frac{\partial}{\partial y^{bc}}-t\,\epsilon^{\alpha\beta}\,y^{ab}\,\frac{\partial}{\partial \theta^{\beta b}}-\theta^{\gamma a}\,\theta^{\alpha c}\,\frac{\partial}{\partial \theta^{\gamma c}}-2\, k\,\theta^{\alpha a}\,.
\end{split}
\end{align}
Recall that we are thinking of these generators as acting on the superspace coordinates of a short superconformal multiplet $\mathcal{D}_k$, and the symbol $k$ appearing in \eqref{GeneratorsSHORT} refers to the $[0,k]$ representation of $\mathfrak{sp}(4)$ in which the superconformal primary of $\mathcal{D}_k$ transforms. The (anti-)commutation relations after the splittings \eqref{Rsplit} and \eqref{QSsplit} can in principle be read off directly from (\ref{ospalgebra_BB}-\ref{ospalgebra_BF}), but we list them here for the reader's convenience. For instance, the $\mathfrak{sp}(4)$ algebra now reads
\begin{equation}
\begin{gathered}
\text{[}R^+_{ab},R^+_{cd}]=0=[R_-^{ab},R_-^{cd}]\,, \quad
[R_a^{\,\,\,b},R_c^{\,\,\,d}]=\delta_a^{\,\,\,d}\,R_c^{\,\,\,b}-\delta_c^{\,\,\,b}\,R_a^{\,\,\,d}\,,\\
[R_a^{\,\,\,b},R^+_{cd}]=-\delta_c^{\,\,\,b}\,R^+_{ad}-\delta_d^{\,\,\,b}\,R^+_{ac}\,, \quad
[R_a^{\,\,\,b},R_-^{cd}]=\delta_a^{\,\,\,c}\,R_-^{bd}+\delta_a^{\,\,\,d}\,R_-^{bc}\,,\\
[R^+_{ab},R_-^{cd}]=
-\delta_a^{\,\,\,c}\,R_b^{\,\,\,d}
-\delta_b^{\,\,\,d}\,R_a^{\,\,\,c}
-\delta_a^{\,\,\,d}\,R_b^{\,\,\,c}
-\delta_b^{\,\,\,c}\,R_a^{\,\,\,d}\,.
\end{gathered}
\end{equation}
It is also interesting to list the commutation relations between $\mathfrak{sp}(4)$ and fermionic generators
\begin{equation}
\begin{gathered}
\text{[}R^+_{ab},Q_{\gamma c}]=0=[R_-^{ab},\tilde{Q}_{\gamma}^c]\,, \quad
[R^+_{ab},\tilde{Q}_{\gamma}^c]=-\delta_a^{\,\,\,c}\,Q_{\gamma b}-\delta_b^{\,\,\,c}\,Q_{\gamma a}\,,\quad
[R_-^{ab},Q_{\gamma c}]=-\delta_c^{\,\,\,a}\,\tilde{Q}_{\gamma}^b-\delta_c^{\,\,\,b}\,\tilde{Q}_{\gamma}^a\,,\\
[R_-^{ab},S^{\gamma c}]=0=[R^+_{ab},\tilde{S}^{\gamma}_c]\,, \quad
[R_-^{ab},\tilde{S}^{\gamma}_c]=+\delta_c^{\,\,\,a}\,S^{\gamma b}+\delta_c^{\,\,\,b}\,S^{\gamma a}\,,\quad
[R^+_{ab},S^{\gamma c}]=+\delta_a^{\,\,\,c}\,\tilde{S}^{\gamma}_b+\delta_b^{\,\,\,c}\,\tilde{S}^{\gamma}_a\,,\\
[R_a^{\,\,\,b},Q_{\gamma c}]=-\delta_c^{\,\,\,b}\,Q_{\gamma a}\,, \quad
[R_a^{\,\,\,b},\tilde{Q}_{\gamma}^c]=\delta_a^{\,\,\,c}\,\tilde{Q}_{\gamma}^b\,, \quad
[R_a^{\,\,\,b},\tilde{S}^{\gamma}_c]=-\delta_c^{\,\,\,b}\,\tilde{S}^{\gamma}_a\,, \quad
[R_a^{\,\,\,b},S_{\gamma c}]=\delta_a^{\,\,\,c}\,S^{\gamma b}\,, 
\end{gathered}
\end{equation}
as well as the anti-commutators
\begin{equation}
\begin{gathered}
\{Q_{\alpha a},Q_{\beta b}\}=0=\{\tilde{Q}_{\alpha}^a,\tilde{Q}_{\beta}^b\}\,, \quad
\{Q_{\alpha a},\tilde{Q}_{\beta}^b\}=-\delta_a^{\,\,\,b}\,\epsilon_{\alpha\beta}\,P\,,\\
\{S^{\alpha a},S^{\beta b}\}=0=\{\tilde{S}^{\alpha}_a,\tilde{S}^{\beta}_b\}\,, \quad
\{S^{\alpha a},\tilde{S}^{\beta}_b\}=+\delta_b^{\,\,\,a}\,\epsilon^{\alpha\beta}\,K\,,\\
\{Q_{\alpha a},S^{\beta b}\}=\delta_\alpha^{\,\,\,\beta}\,\delta_a^{\,\,\,b}\,D+\delta_\alpha^{\,\,\,\beta}\,R_a^{\,\,\,b}-\delta_a^{\,\,\,b}\,L_\alpha^{\,\,\,\beta}\,,\quad
\{ Q_{\alpha a},\tilde{S}^{\beta}_b\}=\delta_\alpha^{\,\,\,\beta}\,R^+_{ab}\,, \\
\{\tilde{Q}_{\alpha}^a,\tilde{S}^{\beta}_b\}=\delta_\alpha^{\,\,\,\beta}\,\delta^a_{\,\,\,b}\,D-\delta_\alpha^{\,\,\,\beta}\,R^a_{\,\,\,b}-\delta^a_{\,\,\,b}\,L_\alpha^{\,\,\,\beta}\,,\quad
\{ \tilde{Q}_{\alpha}^a,S^{\beta b}\}=\delta_\alpha^{\,\,\,\beta}\,R_-^{ab}\,.
\end{gathered}
\end{equation}
Finally, we also give an equivalent expression for the quadratic Casimir operator \eqref{ospCasimir} in terms of the new generators:
\begin{align}
\mathfrak{C}_{\mathfrak{osp}(4^*|4)}=D^2-\frac{1}{2}\{P,K\}+\frac{1}{2}\,L_{\alpha}^{\,\,\,\beta}\,L_{\beta}^{\,\,\,\alpha}-\frac{1}{2}\,R_{a}^{\,\,\,b}\,R_{b}^{\,\,\,a}-\frac{1}{4}\{R^+_{ab},R_-^{ab}\}-\frac{1}{2}\,[Q_{\alpha a},S^{\alpha a}]-\frac{1}{2}\,[\tilde{Q}_{\alpha}^a,\tilde{S}^{\alpha}_a]\,,
\end{align}
which can be checked to have eigenvalue zero on $\mathcal{D}_k$ representations, in agreement with \eqref{ospCasimir_eigenvalue}.

Finite transformations can be obtained by integrating the action of the generators \eqref{GeneratorsSHORT}, or by constructing a matrix realization of $\mathfrak{osp}(4^*|4)$ and studying the projective action of group elements on the coordinates tensor $X$ (see \eqref{XmatrixShort} and below). A distinguished group element, which cannot be obtained by exponentiating infinitesimal generators, is the inversion, given by the supermatrix
\begin{align}
\eta=\begin{pmatrix}
0&1\\
-\Sigma &0
\end{pmatrix}\,,
\end{align}
where we recall that $\Sigma=\text{diag}(-1,-1,1,1)$. In terms of this, we have the conditions
\begin{align}
\mathfrak{g}\in \mathfrak{osp}(4^*|4) \Leftrightarrow \mathfrak{g}^{\text{st}}\eta+\eta\mathfrak{g}=0\,, \qquad
G\in OSP(4^*|4)\Leftrightarrow G^{\text{st}}\eta G=\eta\,,
\end{align}
and notice in particular that $\eta \in OSP(4^*|4)$. $\eta$ has the property that it exchanges positive and negative roots of $\mathfrak{osp}(4^*|4)$: for example $\eta^{\text{st}}\mathfrak{P}\eta=\mathfrak{K}$. The projective action of $\eta$ on $X$ gives the graded-symmetric matrix $(-\Sigma X)^{-1}$, from which one can read off
\begin{align}\label{Xinverse}
X^{-1}=
\begin{pmatrix}
t^{\text{inv}}\epsilon_{\alpha\beta} & \epsilon_{\alpha\beta}(\theta^{\text{inv}})^{\beta b}\epsilon_{ba}\\
 \epsilon_{ab}(\theta^{\text{inv}})^{b\beta}\epsilon_{\beta\alpha} &  \epsilon_{ac}(y^{\text{inv}})^{c d}\epsilon_{db}
\end{pmatrix}\,,
\end{align}
where we have introduced
\begin{align}\label{Xinverseentries}
\begin{split}
t^{\text{inv}}&=\left(t-\frac{\theta^{\alpha a}\theta^{\beta b}y^{cd}\,\epsilon_{\alpha\beta}\,\epsilon_{ac}\,\epsilon_{bd}}{2\,\det(y)}\right)^{-1}\,,\\
(\theta^{\text{inv}})^{\alpha a}&=\frac{\theta^{\alpha b}y^{ac}\,\epsilon_{bc}-\tfrac{1}{3t}\,(\theta^3)^{\alpha a}}{t\,\det(y)}\,,\\
(y^{\text{inv}})^{a b}&=\left[\epsilon_{ac}\left(y^{cd}+\tfrac{1}{t}(\theta^2)^{cd}\right)\epsilon_{db}\right]^{-1}=\frac{(t\,t^{\text{inv}}-2)^2\,y^{ab}-\tfrac{1}{t}\,\theta^{\alpha a}\theta^{\beta b}\epsilon_{\alpha\beta}}{\det(y)}\,.
\end{split}
\end{align}

The half-BPS supermultiplets $\mathcal{D}_k$ are described by superfields in this superspace subject to certain irreducibility constraints. In the case of the displacement supermultiplet $\mathcal{D}_1$, the condition on the associated superfield $\Phi(X)$ is given by
\begin{equation}
\label{Phiconstraint}
\frac{\partial}{\partial y^{(ab}}\frac{\partial}{\partial y^{cd)}}\Phi=0\,,
\quad
\frac{\partial}{\partial \theta^{\alpha(a}}\frac{\partial}{\partial y^{bc)}}\Phi=0\,,
\quad
\left(\frac{1}{2}\epsilon^{\alpha \beta}\frac{\partial}{\partial \theta^{\alpha a}}\frac{\partial}{\partial \theta^{\beta b}}-
\frac{\partial}{\partial t}\frac{\partial}{\partial y^{a b}}
\right)\Phi=0\,,
\end{equation}
where $(a\cdots c)$ denotes total symmetrization.
These conditions can be thought of as a $\mathfrak{sl}(2|2)\subset \mathfrak{osp}(4^*|4)$ covariantization of the first equation which singles out the five-dimensional representation of $\mathfrak{sp}(4)$ out of arbitrary functions of $y^{ab}$. The constraint \eqref{Phiconstraint} can be solved in terms of the conformal primaries in the decomposition of $\mathcal{D}_1$ given in eq. \eqref{D1Structure}, as
\begin{equation}
\label{Phisolveconstraint}
\Phi(X)=D_{\varphi} \varphi(t,y)+\theta^{\alpha a}D_{\Psi}\Psi_{\alpha a}(t,y)+(\theta^2)^{\alpha \beta}\,f_{\alpha \beta}(t)\,,
\end{equation}
where
\begin{align}
\begin{split}
D_{\varphi}&=1-\tfrac{1}{2}\,(\theta^2)^{ab}\tfrac{\partial}{\partial t}\tfrac{\partial}{\partial y^{ab}}+\tfrac{1}{72}\, \theta^4\det\left(\tfrac{\partial}{\partial y^{ab}}\right) \tfrac{\partial^2}{\partial t^2}\,, \\
D_{\Psi}&=1-\tfrac{1}{3}\,(\theta^2)^{ab}\tfrac{\partial}{\partial t}\tfrac{\partial}{\partial y^{ab}}\,,\\
\end{split}
\end{align}
and we have introduced
\begin{equation}
\label{varhpiandPsidef}
%\begin{aligned}
\varphi(t,y)=\phi(t)+y^{ab} \phi_{ab}(t)+\text{det}(y) \bar{\phi}(t)\,,
\qquad
\Psi_{\alpha a}(t,y)=\psi_{\alpha a}(t)+\epsilon_{ac} y^{cd} \tilde{\psi}_{\alpha d}(t)\,.
%\end{aligned}
\end{equation}

\subsection{Realization of long multiplets in superspace}

We also introduce an ordinary superspace whose coordinates are associated with translations $\mathfrak{P}$ and supersymmetries $\mathfrak{Q}$. We denote these by\footnote{The $\mathcal{N}=2$ version was used in \cite{Gimenez-Grau:2019hez}.} $\mathbf{t}=(t,\Theta^{\alpha A})$, with $\alpha=1,2$, $A=1,\dots,4$. The action of $OSP(4|4)$ on these coordinates follows from the usual coset construction -- see, {\it e.g.}, \cite{Liendo:2015cgi} for more details. The transformation properties of a long superprimary are 
 \begin{equation}
\label{groupactiononLongk}
\left( g \circ \mathcal{L}^{\h}_{s,[a,b]}\right)(g\cdot \mathbf{t}) =  \Omega^{\h}\,\rho_s(h)\,\rho_{[a,b]}(H)
\mathcal{L}^{\Delta}_{s,[a,b]}(\mathbf{t}) \,,
\end{equation}
where $(\Omega, h,H) \in \mathbb{R} \times SU(2)\times SP(4)$ depend on $(\mathbf{t}, g)$ as given by the coset construction and $\rho_s$, $\rho_{[a,b]}$ denote the relevant representation of $SU(2)\times SP(4)$. As for the short superspace, we can introduce differential operators implementing the infinitesimal version of \eqref{groupactiononLongk}, which we denote with calligraphic letters and read
\begin{align}\label{GeneratorsLONG}
\begin{split}
\mathcal{P}=&\frac{\partial}{\partial t}\,, \qquad
\mathcal{D}=-t\,\frac{\partial}{\partial t}-\tfrac{1}{2}\,\Theta^{\alpha A}\,\frac{\partial}{\partial \Theta^{\alpha A}}- h\,, \\
\mathcal{K}=&\left(t^2-\tfrac{1}{8}\,\Theta^4\right)\frac{\partial}{\partial t}+\left(t\,\Theta^{\alpha A}+\tfrac{1}{2}\,(\Theta^3)^{\alpha A}\right)\frac{\partial}{\partial \Theta^{\alpha A}}+2\, h\,t+\tfrac{1}{2}\epsilon_{\beta\gamma}\,(\Theta^2)^{\alpha\beta}\,\mathbf{L}_{\alpha}^{\,\,\,\gamma}-\tfrac{1}{2}\Omega_{BC}\,(\Theta^2)^{AB}\,\mathbf{R}_A^{\,\,\,C},\\
\mathcal{L}_{\alpha}^{\,\,\,\beta}=&\Theta^{\beta A}\,\frac{\partial}{\partial \Theta^{\alpha A}}-\tfrac{1}{2}\,\delta_{\alpha}^{\,\,\,\beta}\,\Theta^{\gamma C}\,\frac{\partial}{\partial \Theta^{\gamma C}}+\mathbf{L}_{\alpha}^{\,\,\,\beta}\,,\\ 
\mathcal{R}_A^{\,\,\,B}=&\Theta^{\alpha B}\,\frac{\partial}{\partial \Theta^{\alpha A}}+\Omega_{AC}\,\Omega^{BD}\,\Theta^{\alpha C}\,\frac{\partial}{\partial\Theta^{\alpha D}}+\mathbf{R}_{A}^{\,\,\,B}\,,\\
\mathcal{Q}_{\alpha A}=&\frac{\partial}{\partial\Theta^{\alpha A}}+\tfrac{1}{2}\,\epsilon_{\alpha\beta}\,\Omega_{AB}\,\Theta^{\beta B}\,\frac{\partial}{\partial t}\,,\\
\mathcal{S}^{\alpha A}=&-\tfrac{1}{4}\,\left(2\,t\,\Theta^{\alpha A}+(\Theta^3)^{\alpha A}\right)\,\frac{\partial}{\partial t}-\left(t\,\epsilon^{\alpha\beta}\,\Omega^{AB}+\Theta^{\beta A}\,\Theta^{\alpha B}-\tfrac{1}{2}\,\Omega^{AB}\,(\Theta^2)^{\alpha\beta}\right)\,\frac{\partial}{\partial\Theta^{\beta B}}- h\,\Theta^{\alpha A}\\
&-\Theta^{\beta A}\,\mathbf{L}_{\beta}^{\,\,\,\alpha}+\Theta^{\alpha B}\,\mathbf{R}_B^{\,\,\,A}\,.
\end{split}
\end{align}
Here $\mathbf{L}_{\alpha}^{\,\,\,\beta}$ and $\mathbf{R}_{A}^{\,\,\,B}$ could be represented as matrices in the suitable representation of $\mathfrak{su}(2)$ and $\mathfrak{sp}(4)$ (respectively), but we find it more convenient to saturate all indices with polarizations, each of them associated to the positive roots of the corresponding Lie algebras. In particular, the spin-$s$ representation of $\mathfrak{su}(2)$ can be represented in terms of homogeneous polynomials\footnote{As well-known,  $\mathfrak{sl}(2)$ admits irreducible representations in terms of differential operators acting on polynomials in one variable, but it is sometimes more convenient to consider a representation in terms of polynomials in two variable. The price to pay is that such representation is reducible: it can be written as a direct sum of irreducible representations, corresponding to vector spaces of homogeneous polynomials.} in two variables $\vv_{\alpha}$ of degree $s$, with the highest weight represented by $(\vv_1)^s$ and the lowest weight by $(\vv_2)^s$. In such representation $\mathbf{L}_{\alpha}^{\,\,\,\beta}$ reads
\begin{align}
\mathbf{L}_{\alpha}^{\,\,\,\beta}=-\vv_{\alpha}\,\frac{\partial}{\partial \vv_{\beta}}+\tfrac{1}{2}\delta_{\alpha}^{\,\,\,\beta}\,\vv_{\gamma}\,\frac{\partial}{\partial \vv_{\gamma}}\,.
\end{align}
On the other hand, for $\mathfrak{sp}(4)$ we already introduced the polarizations $y^{ab}$ associated to three of the positive roots as described in footnote \ref{rootssp4} and corresponding to representations with Dynkin labels $[0,b]$. So in principle we need only one more variable, which is necessary to define representations $[a,b]$ with $a\neq 0$ and is associated to the simple root $\alpha_1$ whose fundamental weight is $[1,0]$. However, as in the case of $\mathfrak{su}(2)$ it is convenient to introduce instead two variables $\ww_a$ and irreducible representations given by homogeneous polynomials in $\ww_a$. Using the five variables $(y^{ab},\ww_a)$, we can write the $\mathfrak{sp}(4)$ generators as\footnote{This is the same as the familiar conformal algebra in three dimensions after identifying $y$ with the three space-time coordinates.}
\begin{align}
\label{SP4generatorsGeneric}
\begin{split}
\mathbf{R}^+_{ab}&=-2\frac{\partial}{\partial y^{ab}}\,,\quad
\mathbf{R}_a^{\,\,\,b}=2\,y^{bc}\,\frac{\partial}{\partial y^{ac}}-\tfrac{1}{2}(a+2b)\,\delta_a^{\,\,\,b}+\mathbf{r}_a^{\,\,\,b}\,,\\
\mathbf{R}_-^{ab}&=2\,y^{ac}\,y^{cd}\,\frac{\partial}{\partial y^{cd}}-(a+2b)\,y^{ab}+y^{ac}\mathbf{r}_c^{\,\,\,b}+y^{cb}\mathbf{r}_c^{\,\,\,a}\,, 
\end{split}
\end{align}
where
\begin{equation}
 \mathbf{r}_a^{\,\,\,b}=-\ww_a\,\frac{\partial}{\partial \ww_b}+\tfrac{1}{2}\delta_{a}^{\,\,\,b}\,\ww_c\,\frac{\partial}{\partial \ww_c}\,.
\end{equation}
Note that the conventions are chosen in such a way as to match those used for the generators in the short superspace \eqref{GeneratorsSHORT}. In other words, an $\mathfrak{osp}(4^*|4)$ transformation on a correlation function involving both short and long operators is implemented simply by acting with the sum of the generators corresponding to that transformation defined in the respective superspaces. In particular, when considering expressions that depend on coordinates living in both superspaces, it will be useful to split the eight fermionic coordinates $\Theta^{\alpha A}$ into two sets of four, as
\begin{align}\label{Theta_updown}
\Theta^{\alpha A}\,\,\,\to\,\,\,\left((\Theta_u)^{\alpha a},\,(\Theta_d)^{\alpha}_a\right)\equiv \left(\Theta^{\alpha a},\,\Theta^{\alpha a+2}\right)\,,
\end{align} 
with $a\in\{1,2\}$, where $\Theta_u$ and $\Theta_d$ transform as $(\mathbf{2},\mathbf{2}_1)$ and $(\mathbf{2},\mathbf{2}_{-1})$ of $\mathfrak{su}(2)\oplus\mathfrak{u}(2)_R$, where $\mathfrak{u}(2)_R\subset \mathfrak{sp}(4)$ is generated by $\mathfrak{R}_a^{\,\,\,b}$ in \eqref{Rsplit} and the subscript on the dimension of the $\mathfrak{u}(2)_R$ representation is the charge under the $\mathfrak{u}(1)_R\subset \mathfrak{u}(2)_R$ generated by $\mathfrak{R}_1^{\,\,\,1}+\mathfrak{R}_2^{\,\,\,2}$. This implies, in particular, that $\mathfrak{sp}(4)$ invariants should contain the same number of $\Theta_u$ and $\Theta_d$ coordinates. On the other hand, $\mathfrak{R}^+_{ab}$ maps $\Theta_u$ to $\Theta_d$ and annihilates $\Theta_d$, while the opposite is true for $\mathfrak{R}_-^{ab}$.

For some applications it will be useful to obtain expressions for certain finite transformations, rather than just the infinitesimal version which can easily be obtained from the generators \eqref{GeneratorsLONG}. One way to do this is to use again a coset formulation of the superspace, but since we will only be interested in certain type of transformations we will use a different approach which uses the superconformal inversion. In particular, for most applications we will be interested in singlet long multiplets $\mathcal{L}^{\h}_{0,[0,0]}$ and we focus on the finite version of translations ($\mathcal{P}$), special conformal transformations ($\mathcal{K}$), super-translations ($\mathcal{Q}$) and special super-conformal transformations ($\mathcal{S}$). Directly from the generators one can easily read off 
\begin{align}
\begin{split}
&e^{a\mathcal{P}}\,t=t+a\,,\qquad
e^{a\mathcal{P}}\,\Theta=\Theta\,,\\
&e^{\xi^{\alpha A}\mathcal{Q}_{\alpha A}}\,t=t+\tfrac{1}{2}\epsilon_{\alpha\beta}\Omega_{AB}\xi^{\alpha A}\Theta^{\beta B}\,, \qquad 
e^{\xi^{\alpha A}\mathcal{Q}_{\alpha A}}\,\Theta=\Theta+\xi\,.
\end{split}
\end{align}
To work out the finite action of $\mathcal{K}$ and $\mathcal{S}$ it is useful to introduce the inversion $I$, which acts on the coordinates $\mathbf{t}$ as
\begin{align}\label{inversecoordinates}
I\,t:=\frac{t}{t^2+\tfrac{1}{8}\Theta^4}\,, \qquad
I\,\Theta^{\alpha A}=\frac{(\sigma^3)^{\alpha}_{\,\,\,\beta}(t\,\Theta^{\beta A}-\tfrac{1}{2}(\Theta^3)^{\beta A})}{t^2+\tfrac{1}{8}\Theta^4}\,,
\end{align}
where $\sigma^3=\text{diag}(1,-1)$ is the third Pauli matrix and the inversion is such that $I^2=1$ and
\begin{align}\label{inversiongenerators}
e^{b\mathcal{K}}=I\,e^{a \mathcal{P}}\,I\,, \qquad
e^{\psi_{\alpha A}\mathcal{S}^{\alpha A}}=I\,e^{\xi^{\alpha A}\mathcal{Q}_{\alpha A}}\,I\,,
\end{align}
where the parameters in the two pairs of transformations are related by
\begin{align}
b=-a\,, \qquad \psi_{\alpha A}=-\Omega_{AB}\epsilon_{\alpha\beta}(\sigma^3)^{\beta}_{\,\,\,\gamma}\xi^{\gamma B}\,.
\end{align}
Using the relations \eqref{inversiongenerators} and the explicit expressions \eqref{inversecoordinates} for the inverse coordinates, one can read off the finite transformations
\begin{align}
\begin{split}
&e^{b\mathcal{K}}\,t=\frac{t^{(K)}}{1-b\,t^{(K)}}\,, \qquad
e^{b\mathcal{K}}\,\Theta=\frac{\Theta+\tfrac{b}{1-b\,t}\Theta^3}{1-b\,t^{(K)}}\,,\qquad t^{(K)}:=t-\frac{b}{8(1-b\,t)}\Theta^4\,,
\end{split}
\end{align}
while the finite action of $\mathcal{S}$ is more complicated and we shall not give an explicit expression here. However, it should be clear that choosing $a=-I\,t$ and $\xi=-I\,\Theta$ one can send a generic superspace point $\mathbf{t}$ ``to infinity''.

To conclude this section let us spend a few words on how the $\mathfrak{su}(2)\oplus \mathfrak{sp}(4)$ variables $\vv_\alpha$, $\ww_a$ and $y^{ab}$ can be used as polarizations for long operators. States in fundamental representation $\mathbf{2}$ ($s=1$) of $\mathfrak{su}(2)$ are simply described by the two monomials $\vv_\alpha$ for $\alpha\in\{1,2\}$. The latter transforms as
\begin{align}
\mathbf{L}_{\alpha}^{\,\,\,\beta}\vv_{\gamma}=-\delta_{\gamma}^{\,\,\,\beta}\vv_{\alpha}+\tfrac{1}{2}\,\delta_{\alpha}^{\,\,\,\beta}\vv_{\gamma}\,,
\end{align}
and it will be useful to introduce the invariant arising in the contraction of two fundamental representations:
\begin{align}
\vv^2_{ij}:=\epsilon^{\alpha\beta}\,(v_i)_{\alpha}\,(v_j)_{\beta}\,.
\end{align}
For $\mathfrak{sp}(4)$ we will give the states associated to the representations $\mathbf{4}=[1,0]$ and $\mathbf{5}=[0,1]$, which can be used to build the other irreps\footnote{Note that the weights $[\omega_a,\omega_b]$ of states in $\mathfrak{sp}(4)$ irreps in our basis are read off from the eigenvalues of $[-\mathbf{R}_1^{\,\,\,1}+\mathbf{R}_2^{\,\,\,2},-\mathbf{R}_2^{\,\,\,2}]$.}. For $[1,0]$ we introduce the vector
\begin{align}
\mathcal{W}_A=\{\ww_a\,,\,y^{ab}\,\ww_b\}_A\,,
\end{align}
which transforms as
\begin{align}
\mathbf{R}_{A}^{\,\,\,B}\mathcal{W}_C=-\delta_C^{\,\,\,B}\mathcal{W}_A-\Omega_{AC}\,\Omega^{BD}\,\mathcal{W}_D\,,
\end{align}
and can be used to form the two-point invariant in $[1,0]\otimes [1,0]$
\begin{align}
\ww^2_{ij}:=\Omega^{AB}\,(\mathcal{W}_1)_A\,(\mathcal{W}_2)_B=-(w_i)_a\,(y_i^{ab}-y_j^{ab})\,(w_j)_b\,.
\end{align}
Finally, for $[0,1]$ we introduce the tensor
\begin{align}
\mathcal{Y}_A^{\,\,\,B}=\frac{1}{\sqrt{2}}
\begin{pmatrix}
\epsilon_{ca}\,y^{cb} & \epsilon_{ab}\\
\epsilon^{ab}\,\det(y) & \epsilon_{cb}\,y^{ac}
\end{pmatrix}\,,
\end{align}
where notice that the position of the indices $a,b,\ldots\in\{1,2\}$ reflects the decomposition \eqref{Rsplit}. The polarization $\mathcal{Y}_A^{\,\,\,B}$ transforms as
\begin{align}
\mathbf{R}_A^{\,\,\,B}\mathcal{Y}_C^{\,\,\,D}=\delta_A^{\,\,\,D}\,\mathcal{Y}_C^{\,\,\,B}+\delta_C^{\,\,\,B}\,\Omega^{DE}\,\Omega_{AF}\,\mathcal{Y}_E^{\,\,\,F}+\Omega^{BD}\,\Omega_{AE}\,\mathcal{Y}_C^{\,\,\,E}+\Omega_{AC}\,\Omega^{DE}\,\mathcal{Y}_E^{\,\,\,B}\,,
\end{align}
and can be used to form the two-point invariant arising in the tensor product $[0,1]\otimes [0,1]$
\begin{align}
y^2_{ij}:=\tr(\mathcal{Y}_i \mathcal{Y}_j)=\det(y_i-y_j)\,.
\end{align}

\section{Superconformal kinematics}\label{sec:kinematics}

\subsection{Correlation functions of $\mathcal{D}_k$ operators}

Let us start by examining two-, three- and four-point functions of $\tfrac{1}{2}$-BPS $\mathcal{D}_k$ operators. This will be mostly a review of \cite{Liendo:2018ukf}. Two-point functions are entirely fixed by superconformal symmetry and we normalize them in such a way that
\begin{equation}\label{twoPOINTfunction}
\langle \mathcal{D}_k(1)\mathcal{D}_l(2)\rangle\,=\,\delta_{k,l}(12)^k\,,
\end{equation}
where the propagators $(ij)$ are defined as
\begin{align}
(12)\equiv \frac{\det( y_{12}^{ab}+t_{12}^{-1}\, (\theta_{12}^2)^{ab})}{t_{12}^2}= \frac{y^2_{12}}{t_{12}^2}+\frac{\theta_{12}y_{12}\theta_{12}}{t_{12}^3}+\frac{\theta^4_{12}}{2\,t_{12}^4}\,,
\end{align}
where we have introduced the two quantities
\begin{align}\label{thetaytheta,theta4}
\begin{split}
\theta_{12}y_{12}\theta_{12}&=\epsilon_{ac}\epsilon_{bd}(\theta_{12}^2)^{ab}\,y_{12}^{cd}=\epsilon_{ac}\epsilon_{bd}\epsilon_{\alpha\beta}\theta_{12}^{\alpha a}y_{12}^{cd}\,\theta_{12}^{\beta b}\,,\\
\theta^4_{12}&=\epsilon_{ac}\epsilon_{bd}\epsilon_{\alpha\beta}\epsilon_{\gamma\delta}\theta_{12}^{\alpha a}\theta_{12}^{\beta b}\theta_{12}^{\gamma c}\theta_{12}^{\delta d}\,.
\end{split}
\end{align}
For the two-point function of the $\mathcal{D}_1$ multiplet we can compare \eqref{twoPOINTfunction} with the expansion of the superfield $\Phi(X)$ associated to $\mathcal{D}_1$, introduced in \eqref{Phisolveconstraint}. This allows to find the two-point functions between components of $\mathcal{D}_1$:
\begin{align}
\label{twopointD1components}
\begin{split}
\langle\varphi(t_1,y_1)\varphi(t_2,y_2)\rangle&=\frac{y_{12}^2}{t_{12}^2}\,,\quad
\langle \Psi_{\alpha a}(t_1,y_1)\Psi_{\beta b}(t_2,y_2)\rangle=-2\frac{\epsilon_{\alpha\beta}\,\epsilon_{ac}\,\epsilon_{bd}\,y_{12}^{cd}}{t_{12}^3}\,,\\
\langle f_{\alpha\beta}(t_1)f_{\gamma\delta}(t_2)\rangle&=-3\frac{\epsilon_{\alpha\gamma}\epsilon_{\beta\delta}+\epsilon_{\alpha\delta}\epsilon_{\beta\gamma}}{4\,t_{12}^4}\,.
\end{split}
\end{align}
Similarly, three-point functions of half-BPS operators are fixed by superconformal invariance, up to their normalization or OPE coefficients
\begin{equation}\label{3POINTfunctionDs}
\langle \mathcal{D}_{k_1}(1)\mathcal{D}_{k_2}(2)\mathcal{D}_{k_3}(3)\rangle\,=\,\mathsf{C}_{123}\,
(12)^\frac{k_1+k_2-k_3}{2}\,(13)^\frac{k_1+k_3-k_2}{2}\,(23)^\frac{k_2+k_3-k_1}{2}\,.
\end{equation}
While this expression has the correct transformation properties for any $k_1,k_2,k_3$, the requirement of polynomiality in the R-symmetry variables implies that $k_3\,\in\,\{|k_1-k_2|,|k_1-k_2|+2,\dots,k_1+k_2\}$, or in other words that the representation $[0,k_3]$ of $\mathfrak{sp}(4)$ appears in the tensor product $[0,k_1]\otimes [0,k_2]$ (see Appendix \ref{app:operules}).

Starting from four points one can construct superconformal invariants, so that correlators are only determined up to arbitrary functions of cross ratios. At four points, the cross ratios $(\chi,\zeta_1,\zeta_2)$ are given by the eigenvalues of the supermatrix\footnote{Notice that $\chi$ is a degenerate eigenvalue.}
\begin{equation}
Z=X_{12}^{}X_{13}^{-1}X_{34}^{}X_{24}^{-1}\,,
\qquad 
X_{ij}:=X_{i}-X_{j}\,,
\end{equation}
with $X_i$ as in \eqref{XmatrixShort}. From now on we will set the fermionic coordinates of all operators to zero and, by an abuse of notation, we shall refer to $(ij)$ and $(\chi,\zeta_1,\zeta_2)$ as the result of setting fermions to zero in the corresponding superspace expressions. With this in mind, for a four-point function of $\mathcal{D}_k$ operators we can write
\begin{equation}
\label{eq: correlation function in K and A}
\langle  \mathcal{D}_{k_1}(1)\mathcal{D}_{k_2}(2)\mathcal{D}_{k_3}(3)\mathcal{D}_{k_4}(4)\rangle \,=\,
K_{\{k_1,k_2,k_3,k_4\}}\,
\mathcal{G}_{\{k_1,k_2,k_3,k_4\}}(\chi,\zeta_1,\zeta_2)\,,
\end{equation}
where $m_1+\cdots+m_4$ is even due to R-symmetry and the prefactor reads
\begin{equation}\label{K_1234}
K_{\{k_1,k_2,k_3,k_4\}}\,=\,
(12)^{\frac{1}{2}(k_1+k_2)}
(34)^{\frac{1}{2}(k_3+k_4)}
\left(\frac{(14)}{(24)}\right)^{\frac{1}{2}(k_1-k_2)}
\left(\frac{(13)}{(14)}\right)^{\frac{1}{2}(k_3-k_4)}\,,
\end{equation}
while the bosonic expression for the cross ratios is
\begin{equation}
\label{eq: definition of the chi and xi variables}
\chi=\frac{t_{12}t_{34}}{t_{13}t_{24}}\,,\qquad 
\zeta_1\zeta_2=\frac{y_{12}^2y_{34}^2}{y_{13}^2y_{24}^2}\,,
\qquad
(1-\zeta_1)(1-\zeta_2)=\frac{y_{14}^2y_{23}^2}{y_{13}^2y_{24}^2}\,.
\end{equation}
The expression \eqref{eq: correlation function in K and A} is formally superconformal for any function $\mathcal{G}$, but the superspace constraints further restrict its form.
A first condition comes from the requirement that, when the fermionic coordinates are set to zero, the correlator \eqref{eq: correlation function in K and A} has a polynomial dependence on the R-symmetry variables $y_i^{ab}$. The second condition can be understood as the requirement that there is no singularity in the R-symmetry coordinates when the fermionic coordinates are turned on. This leads to the superconformal Ward identities (SCWI)
\begin{equation}
\label{eq: Ward identities}
\left(\partial_{\zeta_1} \mathcal{G}+\tfrac{1}{2}\,\partial_{\chi}\mathcal{G}\right)_{\big| \zeta_1=\chi}\,=\,
\left(\partial_{\zeta_2} \mathcal{G}+\tfrac{1}{2}\,\partial_{\chi}\mathcal{G}\right)_{\big| \zeta_2=\chi}\,=\,0\,.
\end{equation}
Notice that these conditions imply that $\mathcal{G}(\chi,\chi,\chi)$ is a constant, independent of $\chi$. The algebraic structure associated to this restriction is discussed in \cite{Liendo:2016ymz,Liendo:2018ukf} and will be reviewed with new results in \cite{Ferrero:2023gnu}. The SCWI can be solved in terms of this constant and a certain number of functions of $\chi$ depending on the weights $\{k_1,k_2,k_3,k_4\}$.\footnote{After introducing the extremality degree of the correlator (see, {\it e.g.}, \cite{Zhou:2017zaw,Alday:2021odx})
\begin{equation}
E(k_1,k_2,k_3,k_4)=
\begin{cases}
%%(k_{\text{tot}}-2 k_{text{max})
\tfrac{1}{2} (k_{\text{tot}}-2 k_{\text{max}})  &  \text{if}\,\,\, k_{\text{max}}+k_{\text{min}} \geq \tfrac{1}{2}k_{\text{tot}} \\
k_{\text{min}} & \text{otherwise}
\end{cases}\,,
\end{equation}
where $k_{\text{min}}$ and $k_{\text{max}}$ are the minimal and maximal value among $k_i$ and $k_{\text{tot}}=k_1+k_2+k_3+k_4$ (with $k_{\text{tot}}$ even), the number of such functions is given by $\tfrac{1}{2} E(E+1)$ if $E\geq 0$ and zero otherwise.
} 

For the four-point function of the superdisplacement operator, a convenient parametrization is given in terms of a constant $\mathsf{f}$ and a function $f(\chi)$ by
\begin{align}\label{1111}
\frac{\langle \mathcal{D}_{1} \mathcal{D}_{1} \mathcal{D}_{1} \mathcal{D}_{1}\rangle}{\langle \mathcal{D}_{1} \mathcal{D}_{1}\rangle \langle \mathcal{D}_{1} \mathcal{D}_{1}\rangle}\,=
\mathsf{f}\,\sX+\mathbb{D}\ffcorrshort(\chi)\,,
\qquad 
\mathbb{D}=v_1+v_2- v_1v_2\,\chi^2\partial_{\chi}\,,
%\quad
%v_i=\chi^{-1}-\myzeta_{i}^{-1}\,.
\end{align}
where $v_i=\chi^{-1}-\myzeta_{i}^{-1}$ and $\mathfrak{X} \equiv \frac{\chi^{2}}{\myzeta_{1} \myzeta_{2}}$.

\subsection{Correlation functions involving long operators}\label{sec:longcorrelators}

We now discuss two-, three- and four-point functions involving mixtures of short and long multiplets. Crucially, we consider such structures in superspace, so that for instance multiple independent conformal structures are present already for three-point functions of scalars, as opposed to ordinary bosonic CFT. While we discuss only certain three-point function in details, we collect some comments on more general OPE rules in Appendix \ref{app:operules}.

\paragraph{Two-point functions of long operators.}
Let us begin with the simplest case of a two-point function between two long operators $\mathcal{L}^{\h}_{s,[a,b]}$ in the same representation of $\mathfrak{osp}(4^*|4)$ (the two-point function would be otherwise vanishing). We can write
\begin{align}\label{twopointLong}
\langle \mathcal{L}^{\h}_{s,[a,b]}(1)\,\mathcal{L}^{\h}_{s,[a,b]}(2)\rangle=\frac{(\vv^2_{12})^s\,(\ww_1y_{12}\ww_2)^a\,(y^2_{12})^b}{(\det \mathbf{t}_{12})^{\h-a-b+s/2}}\,\big(t_{12}^{1,[0,0]}\big)^s\,\big(t_{12}^{0,[1,0]}\big)^a\,\big(t_{12}^{0,[0,1]}\big)^{-3a-2b}\,,
\end{align}
where we have introduced the $2\times 2$ matrix 
\begin{align}
\mathbf{t}_{12}^{\alpha \beta}=\hat{t}_{12}\,\epsilon^{\alpha\beta}+\frac{1}{2}(\Theta_{12}^2)^{\alpha\beta}\,,\qquad
\hat{t}_{12}=t_1-t_2+\tfrac{1}{2}(\Theta\Theta)_{12}\,,
\end{align}
using the definitions of Appendix \ref{app:notation} as well as $\Theta_{12}=\Theta_1-\Theta_2$. We have also introduced the combinations
\begin{align}
\begin{split}
t_{12}^{1,[0,0]}&=\hat{t}_{12}+\tfrac{1}{2}\,(\vv^2_{12})^{-1}\,(\Theta_{12}^2)^{\alpha\beta}\,(v_1)_{\alpha}\,(v_2)_{\beta}\,,\\
t_{12}^{0,[1,0]}&=\hat{t}_{12}-(\ww^2_{12})^{-1}\,(\Theta_{12}^2)^{AB}\,(\mathcal{W}_1)_{A}\,(\mathcal{W}_2)_{B}-3\,(y^2_{12})^{-1}\,(\mathcal{Y}_1)_{AB}\,(\Theta^2_{12})^{BC}\,(\mathcal{Y}_2)_C^{\,\,\,A}\,,\\
t_{12}^{0,[0,1]}&=\hat{t}_{12}-\,(y^2_{12})^{-1}\,(\mathcal{Y}_1)_{AB}\,(\Theta^2_{12})^{BC}\,(\mathcal{Y}_2)_C^{\,\,\,A}\,.
\end{split}
\end{align}

\paragraph{The three-point function $\langle \mathcal{D}_1\mathcal{D}_1 \mathcal{L}^{\h}_{0,[0,0]}\rangle$.} From now on we focus on long operators that are singlets of $\mathfrak{su}(2)\oplus \mathfrak{sp}(4)$. Let us start by quoting the result, that is (up to an overall normalization, which is the corresponding OPE coefficient)
\begin{equation}
\label{DDLresult}
\langle \mathcal{D}_1(1)\mathcal{D}_1(2) \mathcal{L}^{\h}_{0,[0,0]}(3)\rangle\,=
\,\langle \mathcal{D}_1(1)\mathcal{D}_1(2)\rangle\,\left(\omega_{12,3}\right)^{-\h}\,,
\end{equation}
where the factor
\begin{equation}
\label{OmegaExpanded}
\omega_{12,3}=\frac{t_{31}t_{23}}{t_{12}}+\text{fermions}\,,
\end{equation}
is constructed below, see (\ref{Y123entries}-\ref{omega123_def}). To prove \eqref{DDLresult} we can start by rewriting 
\begin{align}
\langle \mathcal{D}_1\mathcal{D}_1 \mathcal{L}\rangle= \langle \mathcal{D}_1\mathcal{D}_1\rangle\, F_{\mathcal{D}_1\mathcal{D}_1 \mathcal{L}}(X_1,X_2;\mathbf{t}_3)\,.
\end{align}
The function $F_{\mathcal{D}_1\mathcal{D}_1 \mathcal{L}}$ is invariant under the chain of transformations that sends the long point to zero with a super-translation and one of the short points to infinity\footnote{This is because the factor $ \langle \mathcal{D}_1\mathcal{D}_1\rangle$ carries the weight of the short operators, while the conformal factor $\Omega$ associated with the second transformation in \eqref{Fthreesteps} according to the general expression \eqref{groupactiononLongk} is one.}, namely
\begin{equation}
\label{Fthreesteps}
F_{\mathcal{D}_1\mathcal{D}_1 \mathcal{L}}(X_1,X_2;\mathbf{t}_3)=F_{\mathcal{D}_1\mathcal{D}_1 \mathcal{L}}(\widetilde{X}_1,\widetilde{X}_2;\mathbf{0})=F_{\mathcal{D}_1\mathcal{D}_1 \mathcal{L}}(Y_{12,3},\infty;\mathbf{0})\,,
\end{equation}
where\footnote{
The first transformation can be derived by recalling that the coset representative of a long point takes the form
\begin{equation}
\begin{pmatrix}
A  & B \\
C &  D
\end{pmatrix}=
\begin{pmatrix}
A^{}_{\mathbf{t}} & 0 \\
0 &  \left(A^{\text{st}}_{\mathbf{t}} \right)^{-1}
\end{pmatrix}
\begin{pmatrix}
1  & B_{\mathbf{t}} \\
0 &  1
\end{pmatrix}\,.
\end{equation}
It is easy to check using the form of the generators given in \eqref{GeneratorsSHORT} that the matrix entries of $\widetilde{X}_i$ are invariant under all eight Q-supersymmetries.
}
\begin{align}
\label{Xtildeidef}
\widetilde{X}_i&=A^{}_{\mathbf{t_3}}\left(X_{i}-B_{\mathbf{t_3}}\right)A_{\mathbf{t_3}}^{st}\,,
\qquad
A_{\mathbf{t}}=
\left(\begin{smallmatrix}
1 &-\Theta_d \\
0  & 1
\end{smallmatrix}\right)\,,
\quad
B_{\mathbf{t}}=
\left(\begin{smallmatrix}
\hat{t}\, \epsilon & \Theta_u \\
-\Theta_u^{\text{t}}  &  0
\end{smallmatrix}\right)\,,
\\
(\Theta_u)^{\alpha a}&= \Theta^{\alpha\, a}\,,
\qquad 
(\Theta_d)^{\alpha}_{a}= \Theta^{\alpha\, a+2}\,,
\qquad
\hat{t}=t+\tfrac{1}{2}\epsilon_{\alpha \beta} (\Theta_u)^{\alpha a}(\Theta_d)^{\beta}_{a}\,,
\\
\label{Y123def}
Y_{12,3} :& =\left(\widetilde{X}_{1}^{-1}-\widetilde{X}_{2}^{-1}\right)^{-1}\,=\,
A_{\mathbf{t_3}}^{}\left(B_{\mathbf{t_3}}-X_{1}\right)X_{12}^{-1}\left(X_{2}-B_{\mathbf{t_3}}\right) A_{\mathbf{t_3}}^{st}\,,
\end{align}
Notice that $Y_{12,3} =-Y_{21,3}$.
The transformations that leave the short point $\infty$ and the long point $\mathbf{0}$ fixed still act on the matrix $Y_{12,3}$. These are $U(2)_R \times SU(2)_{\text{spin}}\times \mathbb{R}_{\mathfrak{D}}$, as well as an Abelian factor generated by $R^+_{ab}$ and $\tilde{S}^{\alpha}_a$, see \eqref{QSsplit}, \eqref{Rsplit} and \eqref{GeneratorsSHORT}. This leftover symmetry allows to fix the form of $F_{\mathcal{D}_1\mathcal{D}_1 \mathcal{L}}$. We write the matrix entries of $Y_{12,3}$, defined in \eqref{Y123def}, as
\begin{equation}
\label{Y123entries}
Y_{12,3}=\begin{pmatrix} 
\epsilon\,\omega_{12,3}  & \xi_{12,3}^{} \\
-\xi^{\text{t}}_{12,3} & \widetilde{y}_{12,3}
\end{pmatrix}\,,
\end{equation}
where the explicit expressions of $\xi_{12,3}$ and $\widetilde{y}_{12,3}$ are not relevant here (but can be deduced from \eqref{Y123def}), while $\omega_{12,3}$ will be derived shortly and is given in \eqref{omega123_def}. Note that invariance under $R^+_{ab}$ and $\tilde{S}^{\alpha}_a$ implies that the function $F_{\mathcal{D}_1\mathcal{D}_1 \mathcal{L}}$ in \eqref{Fthreesteps} is independent of the entries $\xi_{12,3}^{}$ and $ \widetilde{y}_{12,3}$ and by its scaling weight must be as given in \eqref{DDLresult}.

To be more explicit, we can write 
\begin{align}
\widetilde{X}_i=\begin{pmatrix}
\widetilde{t}_i\,\epsilon^{\alpha\beta} & \widetilde{\theta}_i^{\alpha b}\\
-\widetilde{\theta}^{\beta a}_i & \widetilde{y}_i^{ab}
\end{pmatrix}\,,
\end{align}
where
\begin{align}
\begin{split}
\widetilde{t}_{i}&=t_i-t_3-\theta_i^{\alpha a}(\Theta_{3,d})^{\beta}_a\epsilon_{\alpha\beta}+\tfrac{1}{2}(\Theta_{3,d})^{\alpha}_a(\Theta_{3,d})^{\beta}_b y_i^{ab}\epsilon_{\alpha\beta}+\tfrac{1}{2}(\Theta_{3,u})^{\alpha a}(\Theta_{3,d})^{\beta}_a\epsilon_{\alpha\beta}\,,\\
\widetilde{\theta}_{i}^{\alpha a}&=\theta_i^{\alpha a}-(\Theta_{3,u})^{\alpha a}-y_i^{ab}(\Theta_{3,d})^{\alpha}_b\,, \qquad \widetilde{y}_i=y_i\,.
\end{split}
\end{align}
Using the expression of $X^{-1}$ given in \eqref{Xinverse} and \eqref{Xinverseentries} one can then compute, in terms of the above,
\begin{align}\label{omega123_def}
\omega_{12,3}=\widetilde{t}_1\,\widetilde{t}_2\,\widetilde{t}^{\text{inv}}_{21}-\tfrac{1}{2}\epsilon_{\alpha\beta}\epsilon_{ab}(\widetilde{t}_1\,\widetilde{\theta}_2^{\alpha a}+\widetilde{t}_2\,\widetilde{\theta}_1^{\alpha a})\,(\widetilde{\theta}_{21}^{\text{inv}})^{\beta b}-\tfrac{1}{2}\epsilon_{\alpha\beta}\epsilon_{ac}\epsilon_{bd}\widetilde{\theta}_1^{\alpha a}\,\widetilde{\theta}_2^{\beta b}\,(\widetilde{y}_{21}^{\text{inv}})^{cd}\,.
\end{align}

Let us now pause to comment on a technical point that will be relevant in the study of $\langle \mathcal{D}_1\mathcal{D}_1\mathcal{D}_2\mathcal{L}\rangle$ four-point functions, and especially for the applications of \cite{Ferrero:2023gnu}. Notice that despite the presence of a unique superconformal structure in \eqref{DDLresult}, one can interpret that result as the contribution of various operators in a long multiplet $\mathcal{L}^{\h}_{0,[0,0]}$ to the OPE between two superconformal primaries of the $\mathcal{D}_1$ multiplet, once the fermionic coordinates associated to the latter are set to zero. Because of our interest in $\langle \mathcal{D}_1\mathcal{D}_1\mathcal{D}_2\mathcal{L}\rangle$ correlators, which we shall study later, let us focus on the superconformal descendants of $\mathcal{L}^{\h}_{0,[0,0]}$ in the representation $\{0,[0,2]\}$: as clear from \eqref{L00Structure} there is only one such operator at level four, thus having dimension $\h+2$. Because of the presence of a unique structure in \eqref{DDLresult}, the OPE coefficient between two operators $\phi^a$ and this particular component of the $\mathcal{L}^{\h}_{0,[0,0]}$ multiplet must be related to the OPE coefficient with the superprimary and here we make this relation explicit\footnote{This is an instance of a well-known phenomenon, see {\it e.g.} \cite{Dolan:2001tt}.}. Let $\mathscr{O}(t,\Theta)$ be a superfield describing a multiplet $\mathcal{L}^{\h}_{0,[0,0]}$. The two components we are interested in are its superconformal primary
\begin{align}
\label{Otcomponent}
\mathcal{O}(t):=\mathscr{O}(t,\Theta=0)\,,\quad \text{with}\quad 
%\{s,[a,b],\h\}=\{0,[0,0],\h_{\mathcal{O}}\}
[a,b]_s^{\h}=[0,0]_0^{\h_{\mathcal{O}}}\,,
\end{align}
and the superconformal descendant
\begin{align}
\label{Ottildecomponent}
\widetilde{\mathcal{O}}(t,y):=\left[\delta^{(0|4)}(\QforOtilde(y))\mathscr{O}(t,\Theta)\right]_{\Theta=0}\,,\quad \text{with}\quad
% \{s,[a,b],\h\}=\{0,[0,2],\h_{\mathcal{O}}+2\}
[a,b]_s^{\h}=[0,2]_0^{\h_{\mathcal{O}}+2}\,,\,,
\end{align}
where 
\begin{equation}
\label{Qofydef}
\QforOtilde^a_{\alpha}(y)=\mathcal{Q}_{\alpha\,a+2}- y^{ab}\mathcal{Q}_{\alpha\,b}\,,
\qquad
\delta^{(0|4)}(\QforOtilde)=
\QforOtilde^1_{1}
\QforOtilde^2_{1}
\QforOtilde^1_{2}
\QforOtilde^2_{2}\,.
\end{equation}
With these definitions, expanding the two-point function \eqref{twopointLong} for singlet operators, we find that 
\begin{align}
\langle \mathcal{O}_1(t_1)\mathcal{O}_2(t_2)\rangle=\,\mathsf{g}_{\mathcal{O}_1\mathcal{O}_2}\,\frac{1}{\left(t_{12}^2\right)^{\h}}\,, \qquad
\langle \widetilde{\mathcal{O}}_1(t_1,y_1) \widetilde{\mathcal{O}}_2(t_2,y_2)\rangle=
\,\widetilde{\mathsf{g}}_{\widetilde{\mathcal{O}}_1\widetilde{\mathcal{O}}_2}\,\frac{\det y_{12}}{\left(t_{12}^2\right)^{\h+2}}\,,
\end{align}
where $\h$ is the (common) conformal dimension of $\mathcal{O}_1$ and $\mathcal{O}_2$ and the normalizations of two-point functions are related by
\begin{align}
\label{gOvsgOtilde}
\widetilde{\mathsf{g}}_{\widetilde{\mathcal{O}}_1\widetilde{\mathcal{O}}_2}&\,= 
\h^2(\h^2-1)\,
 \mathsf{g}_{\mathcal{O}_1\mathcal{O}_2}\,.
\end{align}
Similarly, expanding the three-point function \eqref{DDLresult} we find
\begin{align}
\label{phiphiO3point}
\begin{split}
\frac{\langle \phi(t_1,y_1) \phi(t_2,y_2)\mathcal{O}(t_3)\rangle}{\langle \phi(t_1,y_1) \phi(t_2,y_2)\rangle}&=\,\mu_{11\mathcal{O}}\,
\left(\frac{t_{12}}{t_{31}t_{23}}\right)^{\h}\,,\\
\frac{\langle \phi(t_1,y_1) \phi(t_2,y_2)\widetilde{\mathcal{O}}(t_3,y_3)\rangle}{\langle \phi(t_1,y_1) \phi(t_2,y_2)\rangle}&=\mu_{11\widetilde{\mathcal{O}}}\frac{\det y_{31}\det y_{23}}{\det y_{12}}\,
\left(\frac{t_{12}}{t_{31}t_{23}}\right)^{\h+2}\,,
\end{split}
\end{align}
where $\h$ is the conformal dimension of $\mathcal{O}(t_3)$ and 
\begin{align}
\label{mu11Otildevsmu11O}
\mu_{11\widetilde{\mathcal{O}}}\,=\,
\h(\h+1)\,
\mu_{11\mathcal{O}}\,,
\end{align}
As a consistency check we can reproduce the coefficients $c_9=\tfrac{\h+1}{\h-1}$ in the superconformal blocks (\ref{longsuperblock}-\ref{longsuperblockcoeff}) with $[a,b]=[0,0]$ from $\widetilde{\mathsf{g}}_{\widetilde{\mathcal{O}}_1\widetilde{\mathcal{O}}_2}^{-1}\,\mu_{11\widetilde{\mathcal{O}}}^2=\tfrac{\h+1}{\h-1}\,\mathsf{g}_{\mathcal{O}\mathcal{O}}^{-1}\,\mu_{11\mathcal{O}}^2$.

To end this discussion, let us make a remark about more three-point functions involving two short and one long multiplet, that is
\begin{align}\label{SSLrestriction}
\langle \mathcal{D}_{k_1}(1)\mathcal{D}_{k_2}(2)\mathcal{L}^{\h}_{0,[a,b]}(3)\rangle\,.
\end{align}
In principle the allowed values of the Dynkin labels $[a,b]$ for the $\mathfrak{sp}(4)$ representation of the third operator are all those that can appear in the tensor product $[0,k_1]\otimes [0,k_2]$, see \eqref{sp4[0,b][0,d]}, and in particular the maximum value of the sum $a+b$ allowed by $\mathfrak{sp}(4)$ selection rule is $k_1+k_2$. However, studying the superspace expression of \eqref{SSLrestriction} in the extremal case where $a+b=k_1+k_2$ reveals that the three-point function develops an ``harmonic singularity'' unless the third multiplet is actually of short type (see \cite{Eden:2001ec,Arutyunov:2001qw,Eden:2001wg,Ferrara:2001uj}). This is completely analogous to a well-known phenomenon that happens in higher dimensions as well and it was originally observed for 4d $\mathcal{N}=4$ superconformal symmetry in \cite{Eden:2001ec}, 4d $\mathcal{N}=2$ in \cite{Arutyunov:2001qw}, 6d $\mathcal{N}=(2,0)$ in \cite{Eden:2001wg} and more generally for the maximal superconformal algebras in each dimension between 3d and 6d in \cite{Ferrara:2001uj}. Hence, if we insist on the third multiplet being of long type, there is a further restriction on the allowed values of $[a,b]$ which can be expressed as
\begin{align}
a+b\le k_1+k_2-2\,.
\end{align}

\paragraph{The three-point function $\langle \mathcal{D}_2\, \mathcal{L}^{\h_2}_{0,[0,0]}\, \mathcal{L}^{\h_3}_{0,[0,0]}\rangle$.} The reason why we are interested in this three-point function is again related to $\langle \mathcal{D}_1\mathcal{D}_1\mathcal{D}_2\mathcal{L}\rangle$ correlators, which we shall analyze later from a kinematical point of view and will play a crucial role in \cite{Ferrero:2023gnu}. Note that while the correlator between the superconformal primaries of these multiplets is obviously zero because of R-symmetry selection rules, the full three-point function, including super-descendants is not zero.
This fact can be easily understood in superspace, where there is a unique, up to normalization, superconformal structure. In the following, we will present this structure in two different superconformal frames.
%
% since we consider the three-point function in superspace there are generally speaking various conformal structures built out of fermionic coordinates, encoding the three-point functions between the various conformal primaries in the three supermultiplets, with generally independent OPE coefficients.
% In this case, there is precisely one structure which, according to the choice of superconformal frame, can be thought of as responsible for different OPE coefficients as we now discuss. Moreover, since we are ultimately only interested in the three-point function evaluated in specific frames, we will not determine the superconformal covariant expression for $\langle \mathcal{D}_2\, \mathcal{L}^{\h_1}_{0,[0,0]}\, \mathcal{L}^{\h_2}_{0,[0,0]}\rangle$.
 The first frame that we consider is
\begin{align}
\mathscr{F}_1:\quad \Theta_2=\Theta_3=0\,,
\end{align}
where the superfields associated to the long operators are restricted to their superconformal primary and the only active fermionic coordinates explore the components of the $\mathcal{D}_2$ multiplet. From \eqref{DkStructure} (with $k=2$) we deduce the presence of a unique operator with $\{s,[a,b]\}=\{0,[0,0]\}$ at level four, which is the only contribution to the three-point function. We thus have a unique structure, which in the chosen frame reads
\begin{align}\label{SLLframe1}
\left.\langle \mathcal{D}_2(1)\, \mathcal{L}^{\h_2}_{0,[0,0]}(2)\, \mathcal{L}^{\h_3}_{0,[0,0]}(3)\rangle\right|_{\mathscr{F}_1}=\frac{1}{t_{23}^{\h_2+\h_3}}\left(\frac{t_{31}}{t_{12}}\right)^{\h_2-\h_3}\,\delta^{(0|4)}\left(\frac{t_{23}}{t\_{12}t_{31}}\theta_1^{\alpha a}\right)\,,
\end{align}
and can be thought of as representing the exchange of the R-symmetry singlet conformal primary in the $\mathcal{D}_2$ multiplet between the superconformal primaries of the long multiplets $\mathcal{L}^{\h_2}_{0,[0,0]}$ and $\mathcal{L}^{\h_3}_{0,[0,0]}$. In \eqref{SLLframe1}, $\delta^{(0|4)}$ is defined as in \eqref{Qofydef}.

A different perspective, which will actually be more useful for $\langle \mathcal{D}_1\mathcal{D}_1\mathcal{D}_2\mathcal{L}\rangle$ four-point functions, can be obtained by considering the frame
\begin{align}
\mathscr{F}_2:\quad \theta_1=\Theta_2=0\,,
\end{align}
where the only contributions to the three-point function arise from $\{0,[0,2]\}$ superconformal descendants in the $\mathcal{L}^{\h_3}_{0,[0,0]}$ multiplet. As evident from \eqref{L00Structure} there is only one such operator, appearing again at level four. In this frame we then have the similar-looking expression
\begin{align}\label{SLLframe2}
\left.\langle \mathcal{D}_2(1)\, \mathcal{L}^{\h_2}_{0,[0,0]}(2)\, \mathcal{L}^{\h_3}_{0,[0,0]}(3)\rangle\right|_{\mathscr{F}_2}=\frac{1}{t_{23}^{\h_2+\h_3}}\left(\frac{t_{31}}{t_{12}}\right)^{\h_2-\h_3}\,\delta^{(0|4)}\left(t_{31}^{-1}((\Theta_{d,3})^{\alpha}_{a}-\epsilon_{bc} y_3^{ab}(\Theta_{u,3})^{\alpha c})\right)\,,
\end{align}
where $\Theta_3$ has been split as in \eqref{Theta_updown}. This second perspective reflects the structure of the direct channel OPE in $\langle \mathcal{D}_1\mathcal{D}_1\mathcal{D}_2\mathcal{L}\rangle$: the $\mathcal{D}_1\times \mathcal{D}_1$ OPE clearly contains $\mathcal{L}_{0,[0,0]}$ multiplets, but it could at first seem surprising that such multiplets can also appear in $\mathcal{D}_2\times \mathcal{L}_{0,[0,0]}$. However, there is no contradiction since, as clear from the presence of a $\delta^{(0|4)}$ in \eqref{SLLframe2}, it is a descendant and not the primary of $\mathcal{L}_{0,[0,0]}$ to be exchanged in the latter OPE. 

\paragraph{The four-point function $\langle \mathcal{D}_1\mathcal{D}_1 \mathcal{D}_2\, \mathcal{L}^\h_{0,[0,0]}\rangle$.} As anticipated in \cite{Ferrero:2021bsb}, superconformal symmetry implies that the four-point correlator $\langle \mathcal{D}_1\mathcal{D}_1 \mathcal{D}_2\, \mathcal{L}^\h_{0,[0,0]}\rangle$ is fixed in terms of a single function $F(\chi)$. This is easy to understand using the following argument. With a superconformal transformation we can set the fermionic coordinates of all but the first point to zero\footnote{An indication that this is the case comes from the fact that  the number of fermionic coordinates that we want to set to zero, namely $4+4+8$, coincides with the number of fermionic generators in $\mathfrak{osp}(4^*|4)$. It is indeed possible to find an explicit transformation that achieves this.}. Once this is done, the dependence on the remaining fermionic coordinates $\theta^{\alpha a}_1$ is determined by the constraint satisfied by $\Phi=\mathcal{D}_1$, see \eqref{Phiconstraint}.
More explicitly we have that
\begin{equation}
\label{D1D1D2Lwithonlyfirstfermions}
\begin{aligned}
\langle \Phi(1)\varphi(t_2,y_2) \mathcal{D}_2(3)|_{\theta_3=0} \mathcal{O}_{\h}(t_4)\rangle\,&=\,
D_{\varphi}^{(1)}
\langle \varphi(t_1,y_1)\varphi(t_2,y_2) \mathcal{D}_2(3)|_{\theta_3=0}  \mathcal{O}_{\h}(t_4)\rangle \\
\,&=\,D_{\varphi}^{(1)}
\left(
%(13)(23) \left(\Omega_{12,4}\right)^{-\Delta}\big{|}_{\text{ferm}=0}\,
\frac{y_{13}^2 y_{23}^2}{t_{13}^2t_{23}^2}\left(\frac{t_{12}}{t_{14}t_{24}}\right)^{\h}
F(\chi)\right)\,,
\end{aligned}
\end{equation}
where $ \mathcal{O}_{\h}(t)$ is the superconformal primary of the long supermultiplet $\mathcal{L}^{\h}_{0,[0,0]}$ and $\chi$ is given in \eqref{eq: definition of the chi and xi variables}. In the first equality in \eqref{D1D1D2Lwithonlyfirstfermions} we used the explicit solution of the constraints given in \eqref{Phisolveconstraint} and notice that $\Psi(t_1,y_1)$ and $f(t_1)$ cannot contribute to this  correlator due to $\mathfrak{sp}(4)$ selection rules. The second equality is an immediate consequence of conformal and R-symmetry. We notice that the last expression in \eqref{D1D1D2Lwithonlyfirstfermions} simplifies drastically when the forth point is sent to zero and the third to infinity, since in this frame $D_{\varphi}^{(1)}$ acts as the identity.
 
While this argument shows quickly that the correlator $\langle \mathcal{D}_1\mathcal{D}_1 \mathcal{D}_2\, \mathcal{L}^\h_{0,[0,0]}\rangle$ is fixed in terms of a single function, we still need some work to find the expression for this correlator in a general frame. We will use this result in the next subsection to derive the superconformal blocks of this four-point function as eigenfunctions of the relevant Casimir operator. We write the four-point function in question as\footnote{
Notice that, using \eqref{OmegaExpanded}, we have that 
\begin{equation}
\label{OmegainvminusOmegainvfootnote}
\omega_{23,4}^{-1}-\omega_{13,4}^{-1}=\tfrac{t_{12}}{t_{14} t_{24}} +\text{fermions}\,.
\end{equation}
This factor coincides with $-\omega_{12,4}$ when all the fermions are set to zero.
Replacing the factor \eqref{OmegainvminusOmegainvfootnote} with  $-\omega_{12,4}$ would correspond to a redefinition of  $\mathsf{F}(\text{invariants})$. The one chosen in \eqref{D1D1D2LongfirstEquation} appears to be more convenient. } 
\begin{equation}
\label{D1D1D2LongfirstEquation}
\langle \mathcal{D}_1(1)\mathcal{D}_1(2)\mathcal{D}_2(3) \mathcal{L}^{\h}_{0,[0,0]}(4)\rangle\,=(13)(23) 
\left(\omega_{23,4}^{-1}-\omega_{13,4}^{-1}\right)^{\h}
\mathsf{F}(\text{invariants})\,,
\end{equation}
where the combinations $(ij)$ and $\omega_{ij,k}$ were introduced in \eqref{twoPOINTfunction} and \eqref{Y123entries} respectively.
The superconformal invariants, that appear as argument of the function $\mathsf{F}$, can be understood by going to the frame 
\begin{equation}
\label{framefor112L}
\{X_1,X_2,X_3,\mathbf{t}_4\}\,\mapsto\,
\{W:=\left(\begin{smallmatrix}t_W \epsilon & \theta_W\\-\theta_W & y_W\end{smallmatrix}\right), \left(\begin{smallmatrix}\epsilon & 0\\0 & 0\end{smallmatrix}\right), \infty,\mathbf{0}\}\,,
\end{equation}
with a superconformal transformation. After this is done we still need to impose the $U(2)_R \times SU(2)_{\text{spin}}$ symmetry that preserves the frame. The latter implies that the invariants are $t_{W}$ and the nilpotent combination\footnote{The conventions for the contractions of indices is fixed as $\theta^{}_W y_W^{-1} \theta^{\text{t}}_W=\mathcal{I}\,\epsilon$.} $\mathcal{I}:=(\theta^{}_W)^2 y_W^{-1}$, so that $\mathsf{F}(\text{invariants})=F_0(t_W)+\mathcal{I}\,F_1(t_W)+\mathcal{I}^2\,F_2(t_W)$. While this object is formally superconformal invariant for any $F_0, F_1,F_2$, we will now see that the constraint \eqref{Phiconstraint} implies that the functions $F_1$ and $F_2$ are determined in terms of $F_0$. 
To do so we use a little trick: we match \eqref{D1D1D2LongfirstEquation} with the last formula in  \eqref{D1D1D2Lwithonlyfirstfermions} when all the fermionic coordinated but the one of the first operator are set to zero and use the explicit form of $D_{\phi}^{(1)}$ given in \eqref{Phisolveconstraint}. The matching is easily done in the frame above since $D^{(1)}_{\phi}$ acts on something which is independent of $y_1$ so it reduces to the identity\footnote{In fact when  $X_3 \rightarrow \infty$ the  factor $(13)(23)$ becomes independent of $X_1$.}. Concerning the factor $\omega_{23,4}^{-1}-\omega_{13,4}^{-1}$ in the frame \eqref{framefor112L}, it is easily extracted from the matrices $Y_{23,4}$ and $Y_{13,4}$,  see \eqref{Y123entries} and \eqref{Y123def}, which in this frame are just given by $W$ and $ \left(\begin{smallmatrix}\epsilon & 0\\0 & 0\end{smallmatrix}\right)$ respectively so that 
\begin{equation}
\left(
\omega_{23,4}^{-1}-\omega_{13,4}^{-1}
\right)
\big{|}_{\text{frame  \eqref{framefor112L} }}=
1-t_W^{-1}
\,.
\end{equation}
By imposing that \eqref{D1D1D2LongfirstEquation}  coincides with  \eqref{D1D1D2Lwithonlyfirstfermions} in this frame we get that 
\begin{equation}
\label{122LFinvariants}
%\left(
%F_0(t_W)+\mathcal{I}\,F_1(t_W)+\mathcal{I}^2\,F_2(t_W)
%\right)
\mathsf{F}(\text{invariants})
=
F(1-t_W)\,,
\end{equation}
where we used the fact that $\chi=1-t_W+\text{fermions}$.
To conclude our analysis we need the covariant expression for $t_W$. To find it, it is enough to construct an object which is superconformal invariant and coincides with $t_W$ in the frame  \eqref{D1D1D2LongfirstEquation}. This is given by  
\begin{equation}\label{tW_general}
t_W=\frac{\omega_{13,4}}{\omega_{23,4}}
=\frac{t_{14} t_{23}}{t_{13} t_{24}}
+\text{fermions}
\,.
\end{equation}

\paragraph{The four-point function $\langle \mathcal{D}_1\mathcal{D}_1 \mathcal{L}_{0,[0,0]}\, \mathcal{L}_{0,[0,0]}\rangle$.}

Let us now consider the four-point function $\langle\mathcal{D}_1(1)\mathcal{D}_1(2)\mathscr{O}(3)\mathscr{O}(4)\rangle$ in superspace, where $\mathscr{O}(i)$ are superfields representing long supermultiplets $\mathcal{L}^{\h_i}_{0,[0,0]}$ whose superconformal primaries $\mathcal{O}_{\h_i}(t_i)$ are singlets of $\mathfrak{su}(2)\oplus\mathfrak{sp}(4)$. A possible superconformal covariant expression for the correlator is
\begin{align}\label{11LL_superspace}
\langle\mathcal{D}(1)\mathcal{D}(2)\mathscr{O}(3)\mathscr{O}(4)\rangle=(12)\,\frac{1}{(\det \mathbf{t}_{34})^{\h_3}}\,\omega_{12,4}^{\h_3-\h_4}\,\mathsf{H}(\text{invariants})\,,
\end{align}
where $\mathsf{H}$ is a function of cross-ratios that can be formed with the superspace coordinates, some of which can be nilpotent. To better understand the structure of $\mathsf{H}$ and the number of independent functions of bosonic cross-ratios that it depends on, it is convenient to choose a frame where $\mathbf{t}_3=0$ (using the generators $\mathcal{P}$ and $\mathcal{Q}_{\alpha A}$), $X_2^{-1}=0$ (using $K$, $R^{ab}$ and $S^{\alpha a}$) and finally $t_1=1$ (using $D$), $y_1^{ab}=0$ (using $R_{ab}$ ) and $\theta_1^{\alpha a}=0$ (using $\tilde{S}^{\alpha}_a$). One is then left with (effective) coordinates $\hat{t}_4$ and $\hat{\Theta}_4$ and $\mathsf{H}$ must be a function of these that is invariant under the group $U(2)_R\otimes SU(2)_{\text{spin}}$ generated by $R_a^{\,\,\,b}$ and $L_{\alpha}^{\,\,\,\beta}$ that preserves this frame. As in \eqref{Theta_updown} one can split the fermionic coordinates as $\hat{\Theta}_4^{\alpha A}=((\hat{\Theta}_{4,u})^{\alpha a},\,(\hat{\Theta}_{4,d})^{\alpha}_a)$, where $\hat{\Theta}_{4,u}$ and $\hat{\Theta}_{4,d}$ have opposite $U(1)_R$ charge under $R_a^{\,\,\,a}$ but otherwise in the same representation of $SU(2)_R\otimes SU(2)_{\text{spin}}$ (where we are thinking of $U(2)_R$ as $U(1)_R\otimes SU(2)_R$). Hence, nilpotent cross-ratios with $2n$ fermionic coordinates can be found by contracting every representation formed taking $n$ powers of $\hat{\Theta}_{4,u}$ with the conjugate representation formed with $n$ powers of $\hat{\Theta}_{4,d}$. This leads to
\begin{align}\label{Hinvariants}
\mathsf{H}(\text{invariants})=\sum_{i=0}^5\,\rho_i\,H_i(\mathcal{X})\,,
\end{align}
where a possible choice is
\begin{align}
\begin{split}
\rho_0=1\,,\quad
\rho_1=\epsilon_{\alpha\beta}(\hat{\Theta}_{4,u})^{\alpha a}\,(\hat{\Theta}_{4,d})^{\beta}_a\,, \quad
\rho_2=(\rho_1)^2\,,\quad
\rho_3=(\hat{\Theta}_4)^4\,, \quad
\rho_4=(\rho_1)^3\,,\quad
\rho_5=(\rho_1)^4\,,
\end{split}
\end{align}
while $\mathcal{X}$ is a bosonic cross-ratio that could in principle be any function of $\hat{t}_4$ and we choose $\mathcal{X}=\hat{t}_4$, which is such that
\begin{align}
\mathcal{X}=\frac{t_{12}\,t_{34}}{t_{13}\,t_{24}}+\text{fermions}=\chi+\text{fermions}\,.
\end{align}
So far, the arguments that we used apply to any four-point function of the type $\langle \mathcal{D}_p\mathcal{D}_q\mathcal{L}_{0,[0,0]}\mathcal{L}_{0,[0,0]}\rangle$, thus proving that in general these are determined by six independent structures. However, when specializing to the case $p=q=1$ one has to take into account that $\mathcal{D}_1$ is constrained by the extra-shortening conditions \eqref{Phiconstraint}. Imposing such conditions on \eqref{11LL_superspace} reduces the number of independent functions of $\mathcal{X}$ from six to three. The counting can be understood in two ways, which are in a sense related to the two possible OPE channels, as we shall discuss in the next subsection.

\subsection{Superconformal blocks}

Let us now discuss the superconformal blocks associated to the three types of four-point functions discussed above.

\paragraph{Superconformal blocks for $\langle \mathcal{D}_{k_1}\mathcal{D}_{k_2}\mathcal{D}_{k_3}\mathcal{D}_{k_4}\rangle$.} The selection rules for the OPE in this type of correlators can be easily obtained from the discussion of three-point functions given in the previous subsections, see also Appendix \ref{app:operules}. For completeness we recall the result here, which reads\footnote{We are disregarding the presence of semi-short multiplet, which as we shall argue in Section \ref{sec:operators} do not play a role at strong coupling.}
\begin{align}\label{OPEpqKINEM}
\mathcal{D}_{k_1}\times \mathcal{D}_{k_2}=\mathcal{I}\,\delta_{k_1,k_2}+\sum_{k_3=|k_1-k_2|}^{k_1+k_2}\mathcal{D}_{k_3}+\sum_{\h} \sum_{n=0}^{\text{min}(k_1,k_2)}\sum_{m=0}^{n-1}\mathcal{L}^{\h}_{0,[2\,\text{min}(k_1,k_2)-2n,|k_1-k_2|+2m]}\,.
\end{align}
where $\mathcal{I}$ is the identity operator. This OPE already appeared in \cite{Liendo:2018ukf}, where the derivation of the associated superconformal blocks was also highlighted. The idea is, as usual in these cases, to make an ansatz for the superconformal blocks in terms of bosonic conformal blocks associated to the exchanged super-descendants of each superconformal primary and demand that the SCWI \eqref{eq: Ward identities} are satisfied. We review this in Appendix \ref{app:blocks}, where complete results for the superconformal blocks are given, extending those obtained in \cite{Liendo:2018ukf}.

\paragraph{Superconformal blocks for $\langle \mathcal{D}_1\,\mathcal{D}_1\,\mathcal{D}_2\,\mathcal{L}^{\DeltaExt}_{0,[0,0]}\rangle$.} In the previous section we have shown how this type of four-point functions is fixed in terms of a single function of one variable, which we have expressed in terms of the effective coordinate $t_W$ whose expression can be obtained from \eqref{tW_general}. We will obtain the superconformal blocks for the two OPE channels that we label ``direct'' (when $t_1\to t_2$) and ``crossed'' (when $t_1\to t_4$). From the OPE rules (see Appendix \ref{app:operules} for more details)
\begin{equation}\label{D1LOPEs}
\mathcal{D}_1\times \mathcal{L}^{\DeltaExt}_{0,\left[0,0\right]}=
\underline{\mathcal{D}_1} \oplus \sum_{\h} 
\left(
\underline{\mathcal{L}^{\h}_{0,\left[0,1\right]}}
\oplus
\mathcal{L}^{\h}_{1,\left[1,0\right]}
\oplus
\mathcal{L}^{\h}_{2,\left[0,0\right]}
\right)\,,
\end{equation}
\begin{equation}\label{D2LOPEs}
\mathcal{D}_2\times \mathcal{L}^{\DeltaExt}_{0,\left[0,0\right]}=
\underline{\mathcal{D}_2} \oplus \sum_{\h} 
\left(
\underline{\mathcal{L}^{\h}_{0,\left[0,0\right]}}
\oplus
\mathcal{L}^{\h}_{1,\left[1,0\right]}
\oplus
\mathcal{L}^{\h}_{0,\left[2,0\right]}
\oplus
\mathcal{L}^{\h}_{2,\left[0,1\right]}
\oplus
\mathcal{L}^{\h}_{1,\left[1,1\right]}
\oplus
\mathcal{L}^{\h}_{0,\left[0,2\right]}
\right)\,,
\end{equation}
and \eqref{OPEpqKINEM} we can read off the exchanged supermultiplets in the two channels, which are those underlined in \eqref{D2LOPEs} for the direct channel and those underlined in \eqref{D1LOPEs} for the crossed one.

To obtain the blocks, one needs to act with the superconformal Casimir operator on two sites, which we will take to be $X_1,X_2$ for the direct channel and $X_2,X_3$ for the crossed one. We are then interested in the expression of \eqref{D1D1D2LongfirstEquation} in two frames where only the relevant coordinates are generic, and to this end it is useful to record the following:
\begin{equation}
\begin{aligned}
t_{W}\big{|}_{\{3,\mathbf{4}\}\mapsto\{\infty,\mathbf{0}\} }\,&=\frac{t_1}{t_2}\,,
\qquad\,\,
 \left( \omega_{23,4}^{-1}-\omega_{13,4}^{-1}\right)\big{|}_{\{3,\mathbf{4}\}\mapsto\{\infty,\mathbf{0}\} }\,=\,
 t_2^{-1}-t_1^{-1}\,,
 \\
t_{W}\big{|}_{\{1,\mathbf{4}\}\mapsto\{\infty,\mathbf{0}\} }\,&=
\frac{t_3}{\widehat{t}_{23}}\,,
\qquad
 \left( \omega_{23,4}^{-1}-\omega_{13,4}^{-1}\right)\big{|}_{\{1,\mathbf{4}\}\mapsto\{\infty,\mathbf{0}\} }\,=\,
  t_3^{-1}-\widehat{t}_{23}^{-1}\,,
\end{aligned}
\end{equation}
where $\widehat{t}_{23}$ is the ``$t$'' component of the matrix $X_2^{} X^{-1}_{23} X_3^{}$, which reads
\begin{align}
\widehat{t}_{23}=t_2 t_3 t_{23}^{\mathrm{inv}}-\frac{1}{2}\left(t_2 \theta_3^{\alpha a}+t_3 \theta_2^{\alpha a}\right)\left(\theta_{23}^{\text {inv }}\right)^{\beta b} \epsilon_{\alpha \beta} \epsilon_{a b}-\frac{1}{2} \theta_2^{\alpha a} \theta_3^{\beta b}\left(y_{23}^{\text {inv }}\right)^{c d} \epsilon_{\alpha \beta} \epsilon_{a c} \epsilon_{b d}\,,
\end{align}
with the components of $X^{\text{inv}}$ given in \eqref{Xinverseentries}. With these expressions at hand, it is a tedious but straightforward computation\footnote{We do this using an implementation of anticommuting coordinates in $\mathtt{Mathematica}$, for which we are thankful to Aleix Gimenez-Grau.} to obtain an expression for the Casimir equations and derive the blocks. We denote the solutions to such equations with $\mathfrak{F}_{d,\mathcal{O}}$ for the exchange of a supermultiplet $\mathcal{O}$ in the direct channel and $\mathfrak{F}_{c,\mathcal{O}}$ for the exchange of a supermultiplet $\mathcal{O}$ in the crossed channel. The results read, in the direct channel ($\chi \to 0$)
\begin{align}\label{directblocks112L}
\begin{split}
\mathfrak{F}_{d,\mathcal{D}_2}&=(1-\chi)^{\DeltaExt}\,\chi^{-\DeltaExt}\,{}_2F_1(2,\DeltaExt,4;\chi)\,,\\
\mathfrak{F}_{d,\h}&=(1-\chi)^{\DeltaExt}\,\chi^{\h-\DeltaExt}\,{}_2F_1(2+\h,\DeltaExt+\h,2(2+\h);\chi)\,,
\end{split}
\end{align}
where we have used the shorthand $\mathfrak{F}_{d,\h}$ for the exchange of $\mathcal{L}^{\h}_{0,\left[0,0\right]}$, which is the only type of long multiplet exchanged in the direct channel. On the other hand, in crossed channel ($\chi\to 1$) we have
\begin{align}\label{crossedblocks112L}
\begin{split}
\mathfrak{F}_{c,\mathcal{D}_1}&=\chi^{-\DeltaExt}\,,\\
\mathfrak{F}_{c,\h}&=(1-\chi)^{\h-1}\,\chi^{4}\,{}_2F_1(5+\h,3+\DeltaExt+\h,2(2+\h);1-\chi)\,,
\end{split}
\end{align}
where similarly we have used the shorthand $\mathfrak{F}_{c,\h}$ for the exchange of $\mathcal{L}^{\h}_{0,\left[0,1\right]}$, which is the only type of long multiplet exchanged in the crossed channel. 

\paragraph{Superconformal blocks for $\langle \mathcal{D}_1\,\mathcal{D}_1\,\mathcal{L}^{\h_3}_{0,[0,0]}\,\mathcal{L}^{\h_4}_{0,[0,0]}\rangle$.}

As in the previous case, we compute the superconformal blocks by considering the expression of the four-point function in two different frames, where the fermionic coordinates of two operators at the time are set to zero. One can then act with the Casimir operator associated with the remaining two points, thus obtaining differential equations whose solutions (under suitable boundary conditions) represent the superconformal blocks in the direct ($t_1\to t_2$) and crossed ($t_1\to t_4$) channel. Let us consider the two cases separately.

The representations exchanged in the direct channel coincide with those appearing in the $\mathcal{D}_1\times \mathcal{D}_1$ OPE, with one difference: as we shall see, the three independent functions of $\mathcal{\chi}$ appearing in \eqref{Hinvariants} (once the constraint on $\mathcal{D}_1$ is imposed) correspond to the fact that when a long multiplet is exchanged in $\mathcal{L}_{0,[0,0]}\times\mathcal{L}_{0,[0,0]}$, all three singlets appearing in \eqref{L00Structure} contribute with independent coefficients. Let us see this in more detail. We set $\Theta_3=\Theta_4=0$ and in principle act directly on \eqref{11LL_superspace} with $\widehat{\mathfrak{C}}_{12}$ after the shortening constraints are imposed. However, there is a quicker way to do this, which also highlights why there are only three independent functions of the bosonic cross-ratio. One just needs to recall that \eqref{Phiconstraint} admits a simple solution which we give in \eqref{Phisolveconstraint}. This allows us to write
\begin{align}\label{PhiPhiOOdirectexpansion}
\begin{split}
\langle\mathcal{D}_1(1)\mathcal{D}_1(2)\mathcal{O}_{\h_3}(t_3)\mathcal{O}_{\h_4}(t_4)\rangle&=
D_{\varphi}^{(1)}D_{\varphi}^{(2)}\langle\varphi(t_1,y_1)\varphi(t_2,y_2)\mathcal{O}_{\h_3}(t_3)\mathcal{O}_{\h_4}(t_4)\rangle\\
&+
\theta_1^{\alpha a}\theta_2^{\beta b}D_{\Psi}^{(1)}D_{\Psi}^{(2)}\langle\Psi_{\alpha a}(t_1,y_1)\Psi_{\beta b}(t_2,y_2)\mathcal{O}_{\h_3}(t_3)\mathcal{O}_{\h_4}(t_4)\rangle\\
&+(\theta_1^2)^{\alpha\beta}(\theta_2^2)^{\delta\gamma}\langle f_{\alpha\beta}(t_1)f_{\gamma\delta}(t_2)\mathcal{O}_{\h_3}(t_3)\mathcal{O}_{\h_4}(t_4)\rangle\,,
\end{split}
\end{align}
where we have expanded $\langle\mathcal{D}_1(1)\mathcal{D}_1(2)\mathcal{O}_{\h_3}(t_3)\mathcal{O}_{\h_4}(t_4)\rangle$ in terms of correlators between the components of the superfield $\Phi$ and notice that crossed-terms between $\phi$, $\Psi$ and $f$ vanish because of R symmetry. This not only shows that there are only three independent functions once \eqref{Phiconstraint} is taken into account, but also gives an explicit characterization of such functions in terms of the components of the $\mathcal{D}_1$ supermultiplet. In particular, we have
\begin{align}\label{phipsif_4points}
\begin{split}
\langle\varphi(t_1,y_1)\varphi(t_2,y_2)\mathcal{O}_{\h_3}(t_3)\mathcal{O}_{\h_4}(t_4)\rangle&=\frac{y_{12}^2}{t_{12}^2\,t_{34}^{2\h_3}}\left(\frac{t_{41}t_{24}}{t_{12}}\right)^{\h_3-\h_4}H^{(\varphi)}(\chi)\,,\\
\langle\Psi_{\alpha a}(t_1,y_1)\Psi_{\beta b}(t_2,y_2)\mathcal{O}_{\h_3}(t_3)\mathcal{O}_{\h_4}(t_4)\rangle&=\frac{\epsilon_{\alpha\beta}\,\epsilon_{ac}\,\epsilon_{bd}\,y_{12}^{cd}}{t_{12}^3\,t_{34}^{2\h_3}}\left(\frac{t_{41}t_{24}}{t_{12}}\right)^{\h_3-\h_4}H^{(\Psi)}(\chi)\,,\\
\langle f_{\alpha\beta}(t_1)f_{\gamma\delta}(t_2)\mathcal{O}_{\h_3}(t_3)\mathcal{O}_{\h_4}(t_4)\rangle&=\frac{\epsilon_{\alpha\gamma}\,\epsilon_{\beta\delta}+\epsilon_{\alpha\delta}\,\epsilon_{\beta\gamma}}{8\,t_{12}^4\,t_{34}^{2\Delta_3}}\left(\frac{t_{41}t_{24}}{t_{12}}\right)^{\h_3-\h_4}H^{(f)}(\chi)\,,
\end{split}
\end{align}
where $\chi=\frac{t_{12}t_{34}}{t_{13}t_{24}}$ is the usual cross-ratio. The superconformal blocks can be obtained acting with the superconformal Casimir on $X_1$ and $X_2$ and solving the differential equations that arise from the independent fermionic structures. Given that there are three independent functions of $\chi$ in \eqref{PhiPhiOOdirectexpansion} one has three independent superblocks $\mathfrak{H}^{(\varphi)}$, $\mathfrak{H}^{(\Psi)}$ and $\mathfrak{H}^{(f)}$. The result has a simple expression in terms of bosonic conformal blocks adapted to the prefactor in the first line of \eqref{phipsif_4points}, which is given by
\begin{align}
g^{(\h_3,\h_4)}_{d,\h}(\chi)=\chi^{\h+\h_3-\h_4}\,{}_2F_1(\h,\h+\h_3-\h_4,2\h;\chi)\,.
\end{align}
For long operators, we have
\begin{align}
\begin{split}
\mathfrak{H}^{(\varphi)}_{d,\h}(\chi)&=a_{d,\h}\,g^{(\h_3,\h_4)}_{d,\h}(\chi)+b_{d,\h}\,g^{(\h_3,\h_4)}_{d,\h+2}(\chi)+c_{d,\h}\,g^{(\h_3,\h_4)}_{d,\h+4}(\chi)\,,\\
\mathfrak{H}^{(\Psi)}_{d,\h}(\chi)&=(\h-2)\,a_{d,\h}\,g^{(\h_3,\h_4)}_{d,\h}(\chi)-\frac{5}{2}\,b_{\h}\,g^{(\h_3,\h_4)}_{d,\h+2}(\chi)-(\h+5)\,c_{d,\h}\,g^{(\h_3,\h_4)}_{d,\h+4}(\chi)\,,\\
\mathfrak{H}^{(f)}_{d,\h}(\chi)&=-(\h-2)(\h-3)\,a_{d,\h}\,g^{(\h_3,\h_4)}_{d,\h}(\chi)+\frac{5}{3}(\h+4)(\h-1)\,b_{\h}\,g^{(\h_3,\h_4)}_{d,\h+2}(\chi)\\
&\,\,\,\,\,\,-(\h+5)(\h+6)\,c_{d,\h}\,g^{(\h_3,\h_4)}_{d,\h+4}(\chi)\,,
\end{split}
\end{align}
while the superblocks for short multiplets $\mathcal{D}_2$ can be obtained from this by setting $\h=0$:
\begin{align}
\mathfrak{H}^{(\alpha)}_{d,\mathcal{D}_2}(\chi)=\left.\mathfrak{H}^{(\alpha)}_{d,\h}(\chi)\right|_{\h\to 0}\,,\quad \alpha=\varphi,\Psi,f\,.
\end{align}
In the above, $a_{\h}$, $b_{\h}$ and $c_{\h}$ represent products of OPE coefficients between external operators (in the direct channel) and components of the exchanged multiplet that are singlets under $\mathfrak{su}(2)\oplus \mathfrak{sp}(4)$. As one can argue from \eqref{L00Structure}, there are three such operators: the superprimary, of dimension $\h$ (related to $a_{\h}$), a level-four descendant of dimension $\h+2$ (related to $b_{\h}$) and a level-eight descendant of dimension $\h+4$ (related to $c_{\h}$).

The OPE rules for the crossed channel are read off from\begin{align}\label{D1xL00_OPE}
\mathcal{D}_1\times \mathcal{L}^{\h_1}_{0,[0,0]}=\mathcal{D}_1\oplus \sum_{\h_2}\left(\mathcal{L}^{\h_2}_{0,[0,1]}\oplus\mathcal{L}^{\h_2}_{1,[1,0]}\oplus\mathcal{L}^{\h_2}_{2,[0,0]}\right)\,,
\end{align}
see Appendix \ref{app:operules} for more details. One then expects three distinct superconformal blocks, one for each of the three exchanged representations, once again in agreement with the fact that there are three independent functions for this correlator. To study the superconformal Casimir equation for this OPE channel we set $X_2^{-1}=0$ and $\mathbf{t}_3=0$ and then act with the superconformal Casimir $\widehat{\mathfrak{C}}_{14}$ on the first and last operator. In practice, one can follow the procedure outlined in the previous subsection and only keep the dependence of the effective coordinates $\hat{t}_4$ and $\hat{\Theta}_4$ on $X_1$, while choosing the frame above for the other two points. To express the result, it is useful to introduce the following quantities
\begin{align}
\begin{split}
(\theta_1\Theta_{4,d})&=\epsilon_{\alpha\beta}\,\theta_1^{\alpha a}\,(\Theta_{4,d})^{\beta}_a\,, \quad
(\Theta_{4,u}\Theta_{4d})=\epsilon_{\alpha\beta}\,(\Theta_{4,u})^{\alpha a}\,(\Theta_{4,d})^{\beta}_a\,,\\
(\theta_1^2\Theta_{4,d})^{\alpha a}&=\epsilon_{\beta\gamma}\,\theta_1^{\beta b}\theta_1^{\gamma a}(\Theta_{4,d})^{\alpha}_b\,,\quad
(\theta_1\Theta_{4,d}^2)^{\alpha}_a=\epsilon_{\beta\gamma}\,\theta_1^{\beta b}(\Theta_{4,d})^{\alpha}_b\,(\Theta_{4,d})^{\beta}_a\,,\\
(\theta_1\Theta_{4,u}\Theta_{4,d})_A^{\alpha a}&=\epsilon_{\beta\gamma}\,\theta_1^{\beta b}\,\Theta_{4,u}^{\gamma a}\,(\Theta_{4,d})^{\alpha}_b\,, \quad
(\theta_1\Theta_{4,u}\Theta_{4,d})_B^{\alpha a}=\epsilon_{\beta \gamma}\,\theta_1^{\beta a}\,\Theta_{4,u}^{\gamma a}\,(\Theta_{4,d})^{\alpha}_b\,, \\
(\theta_1^2\Theta_{4,d}^2)&=\epsilon_{\alpha\beta}\,(\theta_1^2\Theta_{4,d})^{\alpha a}\,(\Theta_{4,d})^{\beta}_a\,, \quad
(\theta_1\Theta_{4,u}\Theta_{4,d}^2)=\epsilon_{\alpha\beta}\,(\theta_1\Theta_{4,u}\Theta_{4,d})^{\alpha a},(\Theta_{4,d})^{\beta}_a\,.
\end{split}
\end{align}
To find $\hat{t}_4$ and $\hat{\Theta}_4$, one first sets $\theta_1=0$ (using $\tilde{S}^{\alpha}_a$) and defines effective coordinates for points 1 and 4 given by
\begin{align}
\begin{split}
\tilde{t}_1&=t_1\,, \quad
\tilde{y}_1^{ab }=y_1^{ab}+3\,t_1^{-1}\,(\theta_1^2)^{ab}\,,\\
\tilde{t}_4&=t_4\,\left[1+\left(\tfrac{(\theta_2\Theta_{4,d})}{2t_2}-\tfrac{(\theta_1\Theta_{4,u}\Theta_{4,d}^2)}{2t_2t_4}-\tfrac{(\theta_2\Theta_{4,d})(\Theta_{4,u}\Theta_{4,d}}{4t_2t_4}\right)\left(1+\tfrac{(\theta_2\Theta_{4,d})}{t_2}-\tfrac{(\theta_2\Theta_{4,d})^2}{2t_2^2}\right)+\tfrac{(\theta_1^2\Theta_{4,d}^2)}{2t_2^2}\right]\,,\\
(\tilde{\Theta}_{4,d})^{\alpha}_a&=(\Theta_{4,d})^{\alpha}_a\,\left(1-\tfrac{(\theta_2\Theta_{4,d})^2}{2t_2^2}\right)+\tfrac{(\theta_2\Theta_{4,d}^2)^{\alpha}_a}{t_2}\,,\\
\tilde{\Theta}_{4,u}^{\alpha a}&=\Theta_{4,u}^{\alpha a}\,\left(1-\tfrac{\theta_2\Theta_{4,d})^2}{2t_2^2}+\tfrac{(\theta_1^2\Theta_{4,d}^2)}{2t_2^2}-\tfrac{\theta_2\Theta_{4,d})^3}{2t_2^3}-\tfrac{\theta_2\Theta_{4,d})^4}{16t_2^4}\right)-\tfrac{(\theta_2\Theta_{4,u}\Theta_{4,d})_A^{\alpha a}}{t_2}\,\left(1+\tfrac{2\,(\theta_2\Theta_{4,d})}{t_2}\right)\\
&-\tfrac{t_4}{t_2}\theta_2^{\alpha a}\,\left(1+\tfrac{(\Theta_{4,u}\Theta_{4,d})}{2t_4}+\tfrac{(\Theta_{4,u}\Theta_{4,d})(\theta_2\Theta_{4,d})}{t_2t_4}+\tfrac{(\theta_2^2\Theta_{4,d}^2}{2t_2^2}-\tfrac{(\theta_2\Theta_{4,d})^3}{2t_2^3}+\tfrac{3(\Theta_{4,u}\Theta_{4,d})(\theta_2\Theta_{4,d})^2}{4t_2^2t_3}\right)\\
&+\tfrac{t_4}{t_2^2}(\theta_2^2\Theta_{4,d})_B^{\alpha a}\,\left(1+\tfrac{(\theta_2\Theta_{4,d})}{t_2}-\tfrac{(\Theta_{4,u}\Theta_{4,d})}{2t_4}\right)-\tfrac{(\theta_2\Theta_{4,u}\Theta_{4,d})_B^{\alpha a}}{t_2}\,.
\end{split}
\end{align}
Then, we set $\tilde{y}_1^{ab}=0$ using $R^+_{ab}$ and find
\begin{align}
\bar{t}_1=\tilde{t}_1\,, \quad
\bar{t}_4=\tilde{t}_4\,, \quad
 (\bar{\Theta}_{4,d})^{\alpha}_a=(\tilde{\Theta}_{4,d})^{\alpha}_a\,,\quad
\bar{\Theta}_{4,u}^{\alpha a}=\tilde{\Theta}_{4,u}^{\alpha a}-\tilde{y}_1^{ab}\,(\tilde{\Theta}_{4,d})^{\alpha}_{b}\,,
\end{align}
and with a dilatation we set $\bar{t}_1=1$, which leads to the desired effective coordinates for point 4:
\begin{align}\label{that,Thetahat}
\hat{t}_4=\bar{t}_4/\bar{t}_1\,, \quad
\hat{\Theta}_4=\bar{\Theta}_4/\bar{t}_1^{1/2}\,.
\end{align}
Now that we have an expression for \eqref{11LL_superspace} where all coordinates relative to points 1 and 4 are explicit, we can in principle act with the superconformal Casimir on these two points to derive the superconformal blocks. However, before that one should impose the constraint \eqref{Phiconstraint} on the short points. A way to do this, while at the same time connecting the parametrization \eqref{11LL_superspace} with that in \eqref{phipsif_4points}, is to go one step back from \eqref{that,Thetahat}, restore the dependence on $X_1$, and then compare \eqref{11LL_superspace} to \eqref{phipsif_4points} after setting $\Theta_4=0$. The computation is tedious but conceptually straightforward and leads to expressions for all functions $H_i(\chi)$ in \eqref{Hinvariants} in terms of $H^{(\phi)}(\chi)$, $H^{(\Psi)}(\chi)$ and $H^{(f)}(\chi)$ (and their derivatives). For example, we have
\begin{align}
\begin{split}
H_1(\chi)&=H^{(\varphi)}(\chi)\,, \quad
H_2(\chi)=\frac{-(1-\chi)(2H^{(\Psi)}(\chi)+4H^{(\varphi)}(\chi)-\chi(2-\chi)\,\partial_{\chi}H^{(\varphi)}(\chi)}{\chi}\,, \\
H_4(\chi)&=\frac{(1-\chi)^2}{16\chi^2}(H^{(f)}(\chi)+6H^{(\varphi)}(\chi))+\left[\frac{(1-\chi)}{16\chi}(-4+5\chi+\chi^2)\partial_{\chi}-\frac{(1-\chi)^3}{16}\partial_{\chi}^2\right]\,H^{(\varphi)}(\chi)\,,
\end{split}
\end{align}
while the other relations will not be displayed as they are more complicated and not necessary for what follows. Indeed, using the relations above and acting with the Casimir $\widehat{\mathfrak{C}}_{14}$ on the four-point function \eqref{11LL_superspace} one can obtain superconformal blocks. In this channel, the OPE for each of the three functions $H^{(\alpha)}(\chi)$ ($\alpha=\varphi,\Psi,f$) contains contributions from three different superconformal blocks, corresponding to the three representations exchanged in \eqref{D1xL00_OPE}. The blocks are conveniently expressed in terms of a bosonic conformal block adapted to the prefactor in \eqref{11LL_superspace},
\begin{align}
g_{c,\h}^{(\h_0,\h_3,\h_4)}(\chi)=(\chi-1)^{\h-\h_0-\h_3}\,\chi^{2\h_0+\h_3-\h_4}\,{}_2F_1(\h+\h_0-\h_3,\h+\h_0-\h_4,2\h;1-\chi)\,,
\end{align}
where $\h_0$ represents the (protected) dimension of the first two operators and its value will be taken to be $\h_0=1,3/2,2$ for $H^{(\varphi)}(\chi)$, $H^{(\Psi)}(\chi)$ and $H^{(f)}(\chi)$, respectively. Each of the three correlators receives contributions from the three exchanged representations, with superconformal blocks of the schematic form
\begin{align}
\begin{split}
\mathfrak{H}_{c,\{\h,0,[0,1]\}}^{(\alpha)}&=\sum_{k=0}^{\mathtt{x}(\alpha)} x^{(\alpha)}_k\,g_{c,\h+k}^{(\h_{\alpha},\h_3,\h_4)}(\chi)\,,\\
\mathfrak{H}_{c,\{\h,1,[1,0]\}}^{(\alpha)}&=\sum_{k=0}^{\mathtt{y}(\alpha)} y^{(\alpha)}_k\,g_{c,\h+k}^{(\h_{\alpha},\h_3,\h_4)}(\chi)\,,\\
\mathfrak{H}_{c,\{\h,2,[0,0]\}}^{(\alpha)}&=\sum_{k=0}^{\mathtt{z}(\alpha)} z^{(\alpha)}_k\,g_{c,\h+k}^{(\h_{\alpha},\h_3,\h_4)}(\chi)\,,
\end{split}
\end{align}
where $\alpha=\varphi,\Psi,f$ (with $\h_{\varphi}=1$, $\h_{\Psi}=3/2$ and $\h_{f}=2$) and the coefficients $x^{(\alpha)}_k$, $y^{(\alpha)}_k$ and $z^{(\alpha)}_k$ are rational functions of $\h$, $\h_3$ and $\h_4$, whose expression is given in the $\mathtt{Mathematica}$ notebook attached to the arXiv submission of this paper. In the expressions above $\mathtt{x}(\alpha)$, $\mathtt{y}(\alpha)$ and $\mathtt{z}(\alpha)$ are a cutoffs for the sum which depends on the value of $\alpha$ as follows
\begin{align}
\begin{split}
\mathtt{x}(\alpha)=\{4,3,2\}_{\alpha}\,, \qquad
\mathtt{y}(\alpha)=\{3,4,3\}_{\alpha}\,, \qquad
\mathtt{z}(\alpha)=\{2,3,4\}_{\alpha}\,.
\end{split}
\end{align}

\section{Spectrum of local operators on the line}\label{sec:spectrum}

While the discussion of the previous sections applies to any 1d theory invariant under $\mathfrak{osp}(4^*|4)$, we now switch gears and turn our attention to theories that can be defined as Wilson lines in 4d $\mathcal{N}=4$ SYM. We shall focus on the problem of determining the spectrum of operators  supported on the line defect in these theories, when a free theory description is available: this could be at weak coupling, where one can use the gauge theory Lagrangian, or at strong coupling, via the AdS/CFT correspondence. 

\subsection{Wilson line defects in $\mathcal{N}=4$ super Yang-Mills}

Let us start by considering the $\tfrac{1}{2}$-BPS Wilson line in $\mathcal{N}=4$ SYM with gauge group $G=SU(N)$ in a generic representation $R$ of $G$, connecting two points 1 and 2\footnote{The discussion of this subsection actually holds more generally for Wilson lines that are not necessarily supersymmetric, in generic gauge theories.}
\begin{align}
W_R^{(1\to 2)}=\text{P}\,e^{\int_1^2 dt(i A_{\mu}\dot{x}^{\mu}+\Phi^6|\dot{x}|)}\,,
\end{align}
where $A_{\mu}=\sum_a A^a_{\mu} \,(T^a_R)^i_{\,\,\,j}$, $a=1,\dots N^2-1$ are adjoint indices, $i,j=1,\dots ,\text{dim}R$ are indices in $R$ (or its conjugate $\bar{R}$) and $T^a_R$ are the generators of $SU(N)$ in the representation $R$. Then $W_R^{(1\to 2)}$ naturally transforms under two copies of the gauge group in the representation $R\otimes \bar{R}$, as follows
\begin{align}
W_R^{(1\to 2)}\to U_R(1)\,W_R^{(1\to 2)}\,U_R(2)^{\dagger}\,.
\end{align}
The line defect CFTs we are interested in are defined by insertions of local operators $\mathcal{O}_i(t_i)$ along a Wilson loop defined on a straight line or a circle, with their correlators defined by\footnote{The ``trace'' in this equation should be interpreted schematically, as the operation of extracting a singlet out of the tensor product of various representations. We explain how this is done in detail in the rest of this subsection.}
\begin{align}\label{defectcorrelators}
\begin{split}
\langle \langle \mathcal{O}_1(t_1)\dots  \mathcal{O}_n(t_n)\rangle \rangle&=\frac{1}{\langle \mathcal{W}_R\rangle}\left(\frac{1}{\text{dim}R}\left\langle \tr_R\text{P}[ \mathcal{O}_1(t_1)\,W_R^{(1\to 2)}\,\mathcal{O}_2(t_2)\,W_R^{(2\to 3)}\,\dots\,\mathcal{O}_n(t_n)\,W_R^{(n\to 1)} \right\rangle\right)\\
&=\frac{1}{\langle \mathcal{W}_R\rangle}\left(\frac{1}{\text{dim}R}\left\langle \tr_R\text{P}[ \mathcal{O}_1(t_1)\,\dots\,\mathcal{O}_n(t_n)\,W_R^{(1\to 1)} \right\rangle\right)\,,
\end{split}
\end{align}
where we have defined the circular Wilson loop operator
\begin{align}
\mathcal{W}_R=\frac{1}{\text{dim} R}\tr_R\text{P}\,W_R^{(1\to 1)}\,.
\end{align}
The question we are interested in, for the sake of counting the states in a given defect theory, is what type of insertions define good, gauge invariant local operators in the 1d CFT. A first, obvious aspect that we would nonetheless like to stress is that since one is taking a trace {\it after} taking the product of $\mathcal{O}_i$, local operators are not required to be gauge invariant (as opposed to ordinary gauge theory), but we will simply require them to transform covariantly under gauge transformations, in a certain representation $R_i$:
\begin{align}
\mathcal{O}_i(t_i)\to U_{R_i}(i)\,\mathcal{O}_i(t_i)\,.
\end{align}
With this requirement, it is clear that the quantity inside the squared brackets in the first line of \eqref{defectcorrelators} transforms covariantly under $n$ copies of the gauge group, one for each operator insertion. In particular, at each site we find the tensor product of three $SU(N)$ representations, $R_i\otimes R\otimes \bar{R}$, where the first factor comes from the operator $\mathcal{O}_i$ while the other two from the start and end point of a Wilson line and $\bar{R}$ denotes the conjugate representation to $R$. The trace over the representation $R$ then yields a gauge-invariant observable if and only if
\begin{align}\label{representationrule}
R_i\otimes R\otimes \bar{R}\supset \mathbf{1}\qquad \forall i\,,
\end{align}
where $\mathbf{1}$ is the singlet representation. In the gauge theory description, the local operators $\mathcal{O}_i$ are words built in terms of ``fundamental letters'', corresponding to the fields in the Lagrangian\footnote{Note that the letter associated to the gauge field is not the connection itself, which does {\it not} transform covariantly under the action of $G$, but rather its field strength.}, and different choices of $R$ lead to different possible ways of constructing acceptable local operators on the line: see for instance section 2 of \cite{Giombi:2020amn} for a nice discussion of this point.

\subsection{Fundamental Wilson loop in the planar limit}

From now on we specialize to the case where $R$ is the fundamental representation $F$ of $SU(N)$, of dimension $N$, and we take the limit $N\to\infty$ with fixed 't Hooft coupling $\lambda=g^2_{YM}\,N$. We will study the spectrum of the theory in two opposing regimes: $\lambda=0$, where operators are built in terms of elementary fields of a gauge theory, and $\lambda=\infty$, where the states of the CFT are described holographically by free fields in AdS. While there is a vast literature on the subject, we would like to emphasize that this analysis is particularly useful in clarifying aspects of mixing problems that arise when considering an analytic bootstrap approach to this defect CFTs \cite{Liendo:2016ymz,Liendo:2018ukf,Ferrero:2021bsb}, or when studying the spectrum of the theory with integrability-based methods \cite{Grabner:2020nis,Cavaglia:2021bnz,Cavaglia:2022qpg}. 

\subsubsection{Weak coupling}

We begin with the free theory limit $\lambda=0$, where we use gauge theory language to discuss the construction of operators. Note that for $R=F$ we have
\begin{align}\label{FtimesFbar}
F\otimes \bar{F}=\mathbf{1}+{\mathbf{adj}}\,,
\end{align}
so that the local operators $\mathcal{O}_i$ of the 1d CFT can either be singlets or in the adjoint representation of $G$. We are interested in the planar limit, where as well-known it is useful to organize operators according to the number of traces appearing in their definition. The singlet in \eqref{FtimesFbar} can be realized by taking traces of products of elementary fields (and products thereof), so operators of this type always contain at least one trace. On the other hand, the adjoint in \eqref{FtimesFbar} can be obtained either by taking products of adjoint operators with open indices (with no traces involved in the definition), or multiplying the latter by operators in the singlet sector (thus adding a certain number of traces). We then obtain a similar hierarchy of traces to the bulk $\mathcal{N}=4$ SYM theory, but with a notable difference: in the adjoint sector of \eqref{FtimesFbar} one can have operators with open indices, that is no traces at all. We shall refer to these, which will play a crucial role in the following, as {\it open-trace} (OT) operators. Their importance stems from the fact that
\begin{center}
\it open-trace operators form a closed subsector under the OPE in the planar limit.
\end{center}
By this we mean that OPE coefficient of the form $\langle OT\, OT \,OT\rangle$ are of ${O}(1)$ at large $N$, while $\langle OT\, OT \,\mathcal{O}\rangle$ for any other $\mathcal{O}$ is suppressed by powers of $N$, if $\mathcal{O}$ contains traces in its definition. This situation should be contrasted with bulk operators in the planar limit of ordinary gauge theories, where the OPE between single-trace operators contains both single- and higher-trace operators. This allows one to study the theory in the limit of infinite $N$, consistently neglecting all $1/N$ corrections.

The discussion above shows that the defect CFT is well-defined for $N=\infty$, where the only states are those corresponding to OT operators at $\lambda=0$. We would now like to develop a prescription to count such states. The letters that we can use to build such states are the fields of $\mathcal{N}=4$ SYM, transforming in the adjoint representation of $SU(N)$, and their derivatives, restricted to the line. One should, however, be particularly careful with derivatives: while derivatives along the line generate descendants from the 1d CFT perspective, derivatives orthogonal to the line generated new primaries with non-trivial transverse spin. For each fundamental field of $\mathcal{N}=4$ SYM one has therefore an infinite tower of 1d primaries of increasing transverse spin. A convenient way to count these states is to use characters, as we describe in detail in Appendix \ref{app:counting}. To summarize, we introduce four fugacities for $\mathfrak{osp}(4^*|4)$ and characters associated to the bosonic subalgebra as follows:
\begin{itemize}
\item $\mathfrak{sl}(2)$: we use a fugacity $q$ to count the conformal dimension, where a state of dimension $\h$ carries a factor of $q^{2\h}$.
\item $\mathfrak{su}(2)$: we use a fugacity $z$ and characters $\chi^{\mathfrak{su}(2)}_s(z)$ for finite-dimensional representations of dimension $s+1$.
\item $\mathfrak{sp}(4)$: we use two fugacities $x,y$ and characters $\chi^{\mathfrak{sp}(4)}_{[a,b]}(x,y)$ for finite-dimensional representations of dimension $\tfrac{1}{6}(1+a)(1+b)(2+a+b)(3+a+2b)$.
\end{itemize}
All states in the theory can be obtained by taking suitable products of states in the free vector multiplet of $\mathcal{N}=4$ SYM, after suitably restricting the fields to a line. The way to do this is described in Appendix \ref{app:counting-weak} and here we just quote the result that from the defect CFT perspective the character of the free vector multiplet reads\footnote{Here $\Phi$, $\Psi$ and $F$ represent the components of a free $\mathcal{N}=4$ vector multiplet: the scalars, fermions and field strength respectively.}
\begin{align}
\begin{split}
{\chi}(\mathbb{F})=(1+\chi^{\mathfrak{sp}(4)}_{[0,1]}(x,y))\,\chi(\Phi)+2\,\chi^{\mathfrak{sp}(4)}_{[1,0]}(x,y)\,\chi(\Psi)+2\,\chi(F)\,,
\end{split}
\end{align}
where 
\begin{align}
\begin{split}
\chi(\Phi)&=q^2\,\left(1+\sum_{n=1}^{\infty}q^{2n}\,\chi^{\mathfrak{su}(2)}_n(z)^2\right)\,,\\
\chi(\Psi)&=q^3\,\left(\chi^{\mathfrak{su}(2)}_1(z)+\sum_{n=1}^{\infty}q^{2n}\,\chi^{\mathfrak{su}(2)}_n(z)\,\chi^{\mathfrak{su}(2)}_{n+1}(z)\right)\,,\\
\chi(F)&=q^4\,\left(\chi^{\mathfrak{su}(2)}_2(z)+\sum_{n=1}^{\infty}q^{2n}\,\chi^{\mathfrak{su}(2)}_n(z)\,\chi^{\mathfrak{su}(2)}_{n+2}(z)\right)\,.
\end{split}
\end{align}
The open-trace operators that are relevant for the planar theory at weak coupling are built simply by taking products of the fundamental letters contained in ${\chi}(\mathbb{F})$ where, since we are multiplying $SU(N)$ matrices, the order of the insertions is important and therefore we do not have to implement any (anti-)symmetrization over the letters that form a given word. By the state-operator correspondence we can define a vector space $\mathbb{V}_{\mathbb{F}}$ containing the states associated with the operators in ${\chi}(\mathbb{F})$ and open-trace operators of a given length $L$ are words built using $L$ letters in $\mathbb{V}_{\mathbb{F}}$,. Hence, the full Hilbert space of states associated with open-trace operators at weak coupling can be built summing over such sectors at fixed length:
\begin{align}
\mathcal{H}^{\text{weak}}=\bigoplus_L\mathcal{H}^{\text{weak}}_L\,,\qquad
\mathcal{H}^{\text{weak}}_L=\underbrace{\mathbb{V}_{\mathbb{F}}\otimes\dots\otimes \mathbb{V}_{\mathbb{F}}}_{L\,\,\text{times}}\,,
\end{align}
and correspondingly the partition function reads
\begin{align}\label{Zweak_expression}
\mathcal{Z}_{\text{weak}}=\sum_{k=0}^{\infty}\chi(\mathbb{F})^k\,.
\end{align}
Each $\mathcal{H}^{\text{weak}}_L$ (and therefore the full $\mathcal{H}^{\text{weak}}$) can be decomposed into irreducible representations $\mathcal{R}$ of $\mathfrak{osp}(4^*|4)$ as
\begin{align}
\mathcal{H}^{\text{weak}}_L=\bigoplus_{\mathcal{R}}\mathtt{d}_L^{\text{weak}}(\mathcal{R})\otimes \mathcal{R}\,,
\end{align}
where $\mathtt{d}_L^{\text{weak}}(\mathcal{R})$ are multiplicity spaces: generally speaking, there is more than one way to construct a word of length $L$ in the representation $\mathcal{R}$, using the letters in $\mathbb{F}$, and the number of ways to do so is the dimension of the degeneracy space $\mathtt{d}_L^{\text{weak}}(\mathcal{R})$. Such dimension can be obtained by expanding the partition function \eqref{Zweak_expression} in characters of $\mathfrak{osp}(4^*|4)$, which are discussed in appendix  \ref{app:counting-supermultiplets}. We collect the results for the degeneracy of long and semi-short multiplets of given dimension and representation (up to $\h=8$) in Table \ref{tab:longdegeneraciesweak}.
\begin{table}[!ht]
\centering
 \begin{tabular}{| c || c|c|c|c|c|c|c|c|c|c|c|c|c|c|c|} 
 \hline
$\Delta$&
 $1$ & 
$2$  &
$3$ &
$4$  & 
$5$ &
$6$ &
$7$ &
$8$   \\
   \hline
    \hline
      $[0,0]$ & $1 $   & $ 2$    & $6 $  & $25 $   & $128 $  &  $758 $   & $4986$ &  $35550$ \\
      \hline
        $[0,1]$ & $-$   & $2$    & $6$  & $28$   & $167$  &  $1134$   & $8386$ &  $65995$ \\
      \hline
        $[0,2]$ & $-$   & $-$    & $3$  & $12$   & $76$  &  $588$   & $4972$  & $44260$ \\
      \hline
        $[0,3]$ & $-$   & $-$    & $-$  & $4$   & $20$  &  $160$   & $1520$ & $15388$ \\
      \hline
        $[0,4]$ & $-$   & $-$    & $-$  & $-$   & $5$  &  $30$   & $290$ & $3265$ \\
      \hline
        $[0,5]$ & $-$   & $-$    & $-$  & $-$   & $-$  &  $6$   & $42$ & $476$ \\
      \hline
        $[0,6]$ & $-$   & $-$    & $-$  & $-$   & $-$  &  $-$   & $7$ & $56$\\
      \hline
        $[0,7]$ & $-$   & $-$    & $-$  & $-$   & $-$  &  $-$   & $-$ & $8$\\
        \hline
      \hline
        $[2,0]$ & $-$   & $-$    & $4$  & $16$   & $128$  &  $1038$   & $8638$ &  $74000$ \\
      \hline
        $[2,1]$ & $-$   & $-$    & $-$  & $11$   & $55$  &  $570$   & $5750$ &  $57595$ \\
      \hline
        $[2,2]$ & $-$   & $-$    & $-$  & $-$   & $21$  &  $126$   & $1602$ &  $19341$ \\
      \hline
      $[2,3]$ & $-$   & $-$    & $-$  & $-$   & $-$  &  $34$   & $238$ &  $3584$ \\
      \hline
      $[2,4]$ & $-$   & $-$    & $-$  & $-$   & $-$  &  $-$   & $50$ &  $400$ \\
      \hline
       $[2,5]$ & $-$   & $-$    & $-$  & $-$   & $-$  &  $-$   & $-$ &  $69$ \\
      \hline
      \hline
       $[4,0]$ & $-$   & $-$    & $-$  & $-$   & $15$  &  $90$   & $1270$ &  $15947$ \\
      \hline
       $[4,1]$ & $-$   & $-$    & $-$  & $-$   & $-$  &  $46$   & $322$ &  $5516$ \\
      \hline
       $[4,2]$ & $-$   & $-$    & $-$  & $-$   & $-$  &  $-$   & $98$ &  $784$ \\
      \hline
\end{tabular}
\caption{Spectrum of spin-less, long and semi-short multiplets at weak coupling. For the $\mathfrak{sp}(4)$ representations that are not listed, no supermultiplets with those quantum numbers are present at weak coupling.}
\label{tab:longdegeneraciesweak}
\end{table}
We can also consider the counting problem distinguishing states according to their length: we focus on the case of singlet supermultiplets, for which the results are summarized in Table \ref{tab:longdegeneraciesfixedlength_weak}.
\begin{table}[!ht]
\centering
 \begin{tabular}{| c || c | c|c| c|c|c|c|c|} 
\hline
$\Delta$
 & $1$& $2$  & $3$  &$4$ &$5$ &$6$ & $7$ &  $8$   \\
   \hline
   \hline
    $L=1$ & 1  & $-$ &  $-$ & $-$ &  $-$ & $-$ &    $-$
  &  $-$  \\
   \hline
   $L=2$ & $-$  & $2$ &  $2$ & $3$ &  $3$ & $4$ &    $4$
  &  $5$  \\
   \hline
   $L=3$ & $-$  & $-$ &  $4$ & $12$ &  $33$ & $64$ &    $120$
  &  $195$  \\
   \hline
   $L=4$ & $-$  & $-$ &  $-$ & $10$ &  $65$ & $288$ &    $896$
  &  $2371$  \\
   \hline
   $L=5$ & $-$  & $-$ &  $-$ & $-$ &  $27$ & $320$ &    $2144$
  &  $9745$  \\
   \hline
   $L=6$ & $-$  & $-$ &  $-$ & $-$ &  $-$ & $82$ &    $1554$
  &  $14790$  \\
  \hline
   $L=7$ & $-$  & $-$ &  $-$ & $-$ &  $-$ & $-$ &    $268$
  &  $7504$  \\
   \hline
   $L=8$ & $-$  & $-$ &  $-$ & $-$ &  $-$ & $-$ &    $-$
  &  $940$  \\
   \hline
\end{tabular}
\caption{Long and semi-short multiplets in the singlet representation of $\mathfrak{sp}(4)\oplus \mathfrak{su}(2)$ classified according to their length, at weak coupling.}
\label{tab:longdegeneraciesfixedlength_weak}
\end{table}
Let us make some general remarks on the spectrum. First we observe that, at least in the planar limit, there is exactly one open-trace $\tfrac{1}{2}$-BPS multiplet of type $\mathcal{D}_k$ for each $k$. As we shall discuss soon, the same happens at strong coupling. It is interesting to observe that the other type of absolutely protected multiplet, $B_1[1,b]_0^{1+b}$ in the notation of \cite{Agmon:2020pde}, do not appear in the planar limit at weak coupling (and therefore they will not appear at strong coupling either, as we shall prove). Besides these, we observe long multiplets as well as all possible types of semi-short multiplets, although with some restrictions on the allowed values of the Dynkin labels: in particular, it seems that a given multiplet whose superconformal primary has transverse spin $s$ and $\mathfrak{sp}(4)$ Dynkin labels $[a,b]$ can only appear if $a+2b+s\in 2\mathbb{N}$, which then in particular rules out the $B_1[1,b]_0^{1+b}$ multiplets.

Finally, an interesting question that can be addressed when semi-short multiplets are present in the spectrum in a certain region of the conformal manifold ($\lambda=0$ in this case) is whether they are all allowed to recombine into long multiplets or not, following the recombination rules \eqref{recombination}. As we show in Appendix \ref{app:counting-weak}, the degeneracies of the semi-short supermultiplets are such that all recombinations are allowed to happen, and it is commonly believed that they will in fact occur if this is the case. This is also confirmed by the study of the spectrum at strong coupling, which will be carried out in the next subsection, where we will find that for $\lambda=\infty$ only $\tfrac{1}{2}$-BPS multiplets of type $\mathcal{D}_k$ and long multiplets are present, but no semi-short ones.

\subsubsection{Strong coupling}
\label{sec:specStrong}

The other regime that we are interested in is that of strong coupling ($\lambda=\infty$), where the $\mathcal{N}=4$ SYM Lagrangian description of the theory is of course no longer valid but one has an alternative Lagrangian description via AdS/CFT, in terms of an open string in AdS$_5\times S^5$ fluctuating around an AdS$_2\subset$AdS$_5$ minimal surface \cite{Giombi:2017cqn}. The elementary fields in the Lagrangian sit in the $\mathcal{D}_1$ multiplet \eqref{D1Structure}, where $\phi$ parametrizes fluctuations along the $S^5$ and $f$ describes fluctuations along the AdS$_5$ directions that are orthogonal to AdS$_2$. At $\lambda=\infty$ one can neglect the interactions and is left with a free Lagrangian in AdS$_2$ for the fields in $\mathcal{D}_1$. Once again we can use the state-operator correspondence and associate to $\mathcal{D}_1$ a vector space $\mathbb{V}_{\Phi}$ in a canonical way, which contains the fundamental letters we use to build operators/words. Note that, as opposed to the case of weak coupling, such fundamental letters are fields in AdS$_2$ and therefore in particular are singlets under the gauge group. States of length $L$ are then built as graded-symmetrized tensor products of states in $\mathbb{V}_{\Phi}$ and the full Hilbert space is obtained as a sum of such fixed-length words:
\begin{align}
\mathcal{H}^{\text{strong}}=\bigoplus_L\mathcal{H}^{\text{strong}}_L\,,\qquad
\mathcal{H}^{\text{strong}}_L=(\underbrace{\mathbb{V}_{\Phi}\otimes\dots\otimes \mathbb{V}_{\Phi}}_{L\,\,\text{times}})^{S_L}\,,
\end{align}
where $S_L$ denotes graded-symmetrization. Correspondingly, one can build the partition function at strong coupling in terms of the character of the $\mathcal{D}_1$ multiplet, which we recall from Appendix \ref{app:counting}
\begin{align}
\chi(\mathcal{D}_1)(q;x,y;z)=\chi^{\mathfrak{sl}(2)}_1(q)\,\chi^{\mathfrak{sp}(4)}_{[0,1]}(x,y)+\chi^{\mathfrak{sl}(2)}_{3/2}(q)\,\chi^{\mathfrak{sp}(4)}_{[1,0]}(x,y)\,\chi^{\mathfrak{su}(2)}_{1}(z)+\chi^{\mathfrak{sl}(2)}_2(q)\,\chi^{\mathfrak{su}(2)}_{2}(z)\,.
\end{align}
However, as opposed to what happens at weak coupling, we are no longer dealing with matrix-valued fields and we have to implement the graded-symmetrization $S_L$. This is conveniently done by means of the plethystic exponential
\begin{align}
\text{PE}[f](x)=\text{exp}\left(\sum_{k=1}^{\infty}\frac{f(x^k)}{k}\right)\,,
\end{align}
which length by length implements the symmetrization. We can then write the partition function at strong coupling as
\begin{align}\label{Zstrong_expression}
\mathcal{Z}_{\text{strong}}=\widetilde{\text{PE}}[\chi(\mathcal{D}_1)]\,
\end{align}
where $\widetilde{\text{PE}}$ is a slightly modified version of the plethystic exponential which implements graded-symmetrization, {\it i.e.} symmetrizes products of bosons and anti-symmetrizes products of fermions. More details on the explicit construction of $\mathcal{Z}_{\text{strong}}$ in terms of plethystic exponential can be found around eq. \eqref{Zstrong_PE}. Each $\mathcal{H}^{\text{strong}}_L$ (and therefore the full $\mathcal{H}^{\text{strong}}$) can be decomposed into irreducible representations $\mathcal{R}$ of $\mathfrak{osp}(4^*|4)$ as
\begin{align}\label{strongRdecomposition}
\mathcal{H}^{\text{strong}}_L=\bigoplus_{\mathcal{R}}\mathtt{d}_L^{\text{strong}}(\mathcal{R})\otimes \mathcal{R}\,,
\end{align}
where $\mathtt{d}_L^{\text{strong}}(\mathcal{R})$ are multiplicity spaces, with the same meaning as those introduced at weak coupling. Their dimension can be obtained by expanding the partition function \eqref{Zstrong_expression} in characters of $\mathfrak{osp}(4^*|4)$, which are discussed in appendix  \ref{app:counting-supermultiplets}. We give more details of the construction in Appendix \ref{app:counting-strong}, while here we collect the results. In Table \ref{tab:longdegeneraciesstrong} we give the degeneracy of states of fixed dimension and representation, regardless of their length.
\begin{table}[!ht]
\centering
 \begin{tabular}{| c || c | c|c| c|c|c|c|c|c|c|c|c|c|} 
 \hline
$\Delta$
 & $2$  & 
$3$  &
$4$ &
$5$ &
$6$ &
 $7$ &
  $8$ 
  &  $9$ &  $10$ 
  &   $11$ &  $12$
   &   $13$ &  $14$  \\
   \hline
    \hline
   $[0,0]$
 & $1$%_{(\ell^2)}$
 & 
$0$  &
$2$%_{(\ell^2+\ell^4)}$ 
&
$0$ &
$4$ &
 $1$ &
  $9$ 
   &  $5$  &  $21$ 
     & $20$  & $55$  
      & $65$   & $149$ \\
   \hline
      $[0,1]$
 & $-$  & 
$1$  &
$1$ &
$2$ &
$2$ &
 $6$ &
  $6$ &  $15$ & $24$
    & $43$ &  $75$
    &  $140$ &  $240$\\
   \hline
         $[0,2]$
 & $-$  & 
$-$  &
$1$ &
$1$ &
$3$ &
 $2$ &
  $8$
  &  $8$ & $22$ 
    & $31$ & $68$
    & $108$   & $223$ \\
   \hline
            $[0,3]$
 & $-$  & 
$-$  &
$-$ &
$1$ &
$1$ &
 $3$ &
  $3$& $8$ & $10$
    & $24$ & $38$
    & $77$  & $134$   \\
   \hline
            $[0,4]$
 & $-$  & 
$-$  &
$-$ &
$-$ &
$1$ &
 $1$ &
  $3$
  &  $3$ & $9$ 
    & $10$ & $26$ 
    &   $40$&  $84$\\
   \hline
               $[0,5]$
 & $-$  & 
$-$  &
$-$ &
$-$ &
$-$ &
 $1$ &
  $1$ 
  &  $3$ & $3$
    & $9$ & $11$ 
    &   $26$& $42$ \\
   \hline
                 $[0,6]$
 & $-$  & 
$-$  &
$-$ &
$-$ &
$-$ &
 $-$ &
  $1$ 
  &  $1$ & $3$
    & $3$ & $9$
    & $11$  & $27$  \\
   \hline
                 $[0,7]$
 & $-$  & 
$-$  &
$-$ &
$-$ &
$-$ &
 $-$ &
  $-$ 
  &  $1$ & $1$
    &  $3$ &  $3$ 
    &  $9$ & $11$ \\
   \hline
                 $[0,8]$
 & $-$  & 
$-$  &
$-$ &
$-$ &
$-$ &
 $-$ &
  $-$ 
  &  $-$ & $1$
    &  $1$ &  $3$
    &  $3$ & $9$  \\
   \hline
    \hline
                  $[2,0]$
 & $-$  & 
$-$  &
$-$ &
$1$ &
$0$ &
 $3$ &
  $3$
  &  $10$ & $14$
    & $37$  & $57$  
    & $128$  & $215$  \\
   \hline
                     $[2,1]$
 & $-$  & 
$-$  &
$-$ &
$-$ &
$1$ &
 $1$ &
  $3$
  &  $6$  & $14$
    & $27$& $57$  
    & $111$  & $221$ \\
   \hline
                        $[2,2]$
 & $-$  & 
$-$  &
$-$ &
$-$ &
$-$ &
 $1$ &
  $1$
  &  $4$  & $6$
    & $17$& $31$
    &  $71$ &   $134$ \\
   \hline
                           $[2,3]$
 & $-$  & 
$-$  &
$-$ &
$-$ &
$-$ &
 $-$ &
  $1$
  &  $1$  & $4$
    & $7$& $17$ 
    &  $34$ & $75$  \\
   \hline
                           $[2,4]$
 & $-$  & 
$-$  &
$-$ &
$-$ &
$-$ &
 $-$ &
  $-$
  &  $1$  & $1$ 
    & $4$& $7$ 
    &  $18$ & $34$ \\
   \hline
                           $[2,5]$
 & $-$  & 
$-$  &
$-$ &
$-$ &
$-$ &
 $-$ &
  $-$
  &  $-$  & $1$ 
    &  $1$ &  $4$
    & $7$   &  $18$ \\
   \hline
      \hline
                           $[4,0]$
 & $-$  & 
$-$  &
$-$ &
$-$ &
$-$ &
 $-$ &
  $1$
  &  $0$  & $3$
    & $4$& $14$
    &  $21$ &  $60$ \\
     \hline
                           $[4,1]$
 & $-$  & 
$-$  &
$-$ &
$-$ &
$-$ &
 $-$ &
  $-$
  &  $1$  & $1$
    & $3$& $7$
    &  $18$  &  $36$\\
     \hline
                           $[4,2]$
 & $-$  & 
$-$  &
$-$ &
$-$ &
$-$ &
 $-$ &
  $-$
  &  $-$  & $1$
    & $1$& $4$
    &  $7$ & $21$   \\
   \hline
\end{tabular}
\caption{Spectrum of spin-less, long multiplets at strong coupling. For the $\mathfrak{sp}(4)$ representations that are not listed, no supermultiplets with those quantum numbers are present at strong coupling.}
\label{tab:longdegeneraciesstrong}
\end{table}
On the other hand, we can also study the degeneracy of states for each fixed length. This is possible for any representation, but we are mostly interested in the case of singlets, for which we collect the results in Table \ref{tab:longdegeneraciesfixedlength_strong}.
\begin{table}[!ht]
\centering
 \begin{tabular}{| c || c | c|c| c|c|c|c|c|c|c|c|c|c|} 
\hline
$\Delta$
 & $2$  & $3$  &$4$ &$5$ &$6$ & $7$ &  $8$ 
  &  $9$ &  $10$  &   $11$ &  $12$  &   $13$ &  $14$  \\
   \hline
    \hline
    $L=2$ & 1  & $-$ & 1 & $-$ & 1 & $-$ &   1
  &  $-$ &  1  &  $-$ &  1  & $-$  &  1  \\
   \hline
   $L=3$ &  $-$ & $-$ & $-$ & $-$ & $-$ & $-$ &   $-$
  & $-$  &  $-$  & $-$  &  $-$  & $-$  &  $-$  \\
   \hline
   $L=4$ & $-$  & $-$ & 1 &$-$ & 2 & 1 & 4 &   2
  &  6 & 4   & 9  & 6   & 12      \\
   \hline
   $L=5$ & $-$  &$-$  &  $-$& $-$ &$-$  &$-$  &   $-$
  &  1 &  1  &  3 & 3   & 6  & 7   \\
   \hline
   $L=6$ &  $-$ & $-$ &$-$  &$-$  & 1 & $-$  &   3
  &  2 &   9 & 8  & 22   & 22  &  46  \\
   \hline
   $L=7$ & $-$  & $-$ &$-$  &$-$  &$-$  &$-$  & $-$  
  & $-$  &  $-$  &2   &  3  &  11 &  17  \\
   \hline
   $L=8$ &  $-$ & $-$ &$-$  & $-$ &$-$  &$-$  &   1
  &  $-$ &  3  & 3  &  13  & 15  &  44  \\
   \hline
   $L=9$ &  $-$ & $-$ &$-$  & $-$ &$-$  &$-$  &   $-$
  &  $-$ &  $-$  & $-$  &  $-$  & 2  &  4  \\
   \hline
   $L=10$ &  $-$ & $-$ &$-$  & $-$ &$-$  &$-$  &   $-$
  &  $-$ &  1  & $-$  &  3  & 3  &  14  \\
   \hline
   $L=11$ &  $-$ & $-$ &$-$  & $-$ &$-$  &$-$  &   $-$
  &  $-$ &  $-$  & $-$  &  $-$  & $-$  &  $-$  \\
   \hline
   $L=12$ &  $-$ & $-$ &$-$  & $-$ &$-$  &$-$  &   $-$
  &  $-$ &  $-$  & $-$  &  1  & $-$  &  3  \\
   \hline
   $L=13$ &  $-$ & $-$ &$-$  & $-$ &$-$  &$-$  &   $-$
  &  $-$ &  $-$  & $-$  &  $-$  & $-$  &  $-$  \\
   \hline
   $L=14$ &  $-$ & $-$ &$-$  & $-$ &$-$  &$-$  &   $-$
  &  $-$ &  $-$  & $-$  &  $-$  & $-$  &  1  \\
   \hline
\end{tabular}
\caption{Long multiplets in the singlet representation of $\mathfrak{sp}(4)\oplus \mathfrak{su}(2)$ classified according to their length, at strong coupling.}
\label{tab:longdegeneraciesfixedlength_strong}
\end{table}

\begin{figure}[hbt!]
\begin{center}
        \includegraphics[width=0.7\textwidth]{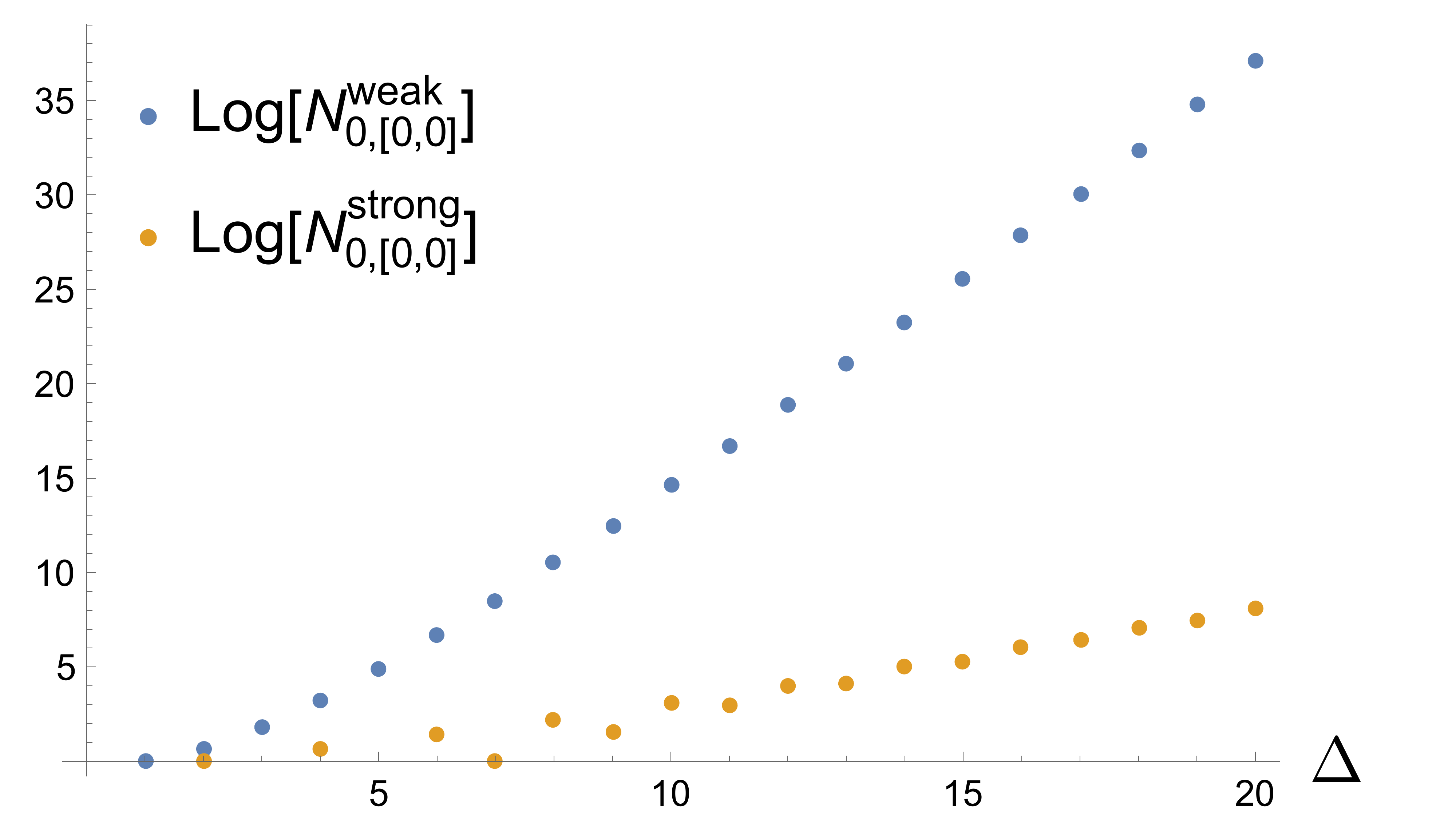}
\caption{Logarithm of the number of long multiplets in the singlet representation of $\mathfrak{sp}(4)\oplus \mathfrak{su}(2)$ for dimension $1\le\D\le20$, at weak and strong coupling.}
\label{fig:N00_Weak_vs_Strong}
\end{center}
\end{figure}

As opposed to the regime of weak coupling, we find that in the spectrum at strong coupling there are no semi-short multiplets of any kind. The only representations $\mathcal{R}$ appearing in the expansion \eqref{strongRdecomposition} are $\tfrac{1}{2}$-BPS $\mathcal{D}_k$ multiplets and long multiplets, as anticipated in Section \ref{sec:superalgebra}, which is why in sections \ref{sec:superalgebra} and \ref{sec:kinematics} we chose to focus on these two types of multiplets only. Moreover, as in the weak coupling regime, there is exactly one $\mathcal{D}_k$ multiplet for each $k$. It is particularly interesting to compare the growth with $\D$ in the number of states for a given representation $\mathcal{R}$ with weights $\omega(\mathcal{R})=\{\D,s,[a,b]\}$ at weak and strong coupling, that is $N^{\text{weak}}_{\D,s,[a,b]}=\text{dim}[\mathtt{d}^{\text{weak}}(\mathcal{R}))]$ and $N^{\text{strong}}_{\D,s,[a,b]}=\text{dim}[\mathtt{d}^{\text{strong}}(\mathcal{R}))]$, where we are now considering operators of all possible lengths $L$. From the various tables collected in this section, it is immediately clear that such growth is much faster at weak than at strong coupling. As an example, we collect in Table \ref{tab:N00_Weak_vs_Strong} we collect the number of singlet long (or semi-short) multiplets of any length for $1\le \D\le 11$ at weak and strong coupling. Both seem to have exponential behavior, as shown by the plots in Figure \ref{fig:N00_Weak_vs_Strong}, but the growth at weak coupling is significantly faster.
\begin{table}[!ht]
\centering
 \begin{tabular}{| c || c |c |c |c |c |c |c |c |c |c |c|} 
 \hline
$\D$ & 1& 2& 3& 4& 5& 6& 7& 8& 9& 10& 11\\
\hline\hline
$N_{0,[0,0]}^{\text{weak}}$ & 1 & 2 & 6 & 25 & 128 & 758 & 4986 & 35550 & 270289 & 2166106 &18140558\\
\hline
$N_{0,[0,0]}^{\text{strong}}$ & 0 & 1 & 0 & 2 & 0 & 4 & 1 & 9 & 5 & 21 & 20 \\
\hline
\end{tabular}
\caption{Number of long multiplets in the singlet representation of $\mathfrak{sp}(4)\oplus \mathfrak{su}(2)$ for dimension $1\le\D\le 11$, at weak and strong coupling.}
\label{tab:N00_Weak_vs_Strong}
\end{table}
The dramatic difference in the number of states for each given representation between weak and strong coupling can be ascribed to the higher number of fundamental letters in the gauge theory description of the Wilson line defect theory. It is important to remark that since all states described in Figure \ref{fig:N00_Weak_vs_Strong} are associated with open-trace operators, if we follow their conformal dimension from weak to strong coupling the latter increases for each state but always remains finite. Simply, given any state at weak coupling, one can follow its dimension in a continuous way while the coupling is increased up to infinity: at strong coupling, the state has a dimension which is higher by a certain {\it finite} amount. As a consequence of this fact, the spectrum at strong coupling is more sparse, or in other words for each given dimension the degeneracy is lower than at weak coupling. For a visual representation of this, we refer the reader to Figure 2 of \cite{Cavaglia:2021bnz} (see also \cite{Cavaglia:2022qpg}), where the spectrum of the fundamental Wilson line theory in the planar limit is studied using the quantum spectral curve. The situation should be contrasted with, {\it e.g.}, the Konishi operator in the full $\mathcal{N}=4$ SYM theory, whose dimension diverges at strong coupling in the 't Hooft limit (since it is dual to a string state).

As we discuss in Appendix \ref{app:counting-strong}, for singlet supermultiplets of length $L$ we find the following asymptotic formula for large $\D$,
\begin{align}\label{Nstrong00L}
N^{\text{strong}}_{0,[0,0],L}\sim \frac{c_L}{(L-2)!}\D^{L-2}\,,\qquad (\D\to \infty)\,,
\end{align}
where $c_L$ are numerical coefficients (independent of $\D$) that are given in \eqref{c_L_HWGS} for some values of $L$. The asymptotic expression of $N^{\text{strong}}_{0,[0,0]}$ will be given by a sum over \eqref{Nstrong00L} for all values of $L$, which we expect to give rise to an exponential behavior in $\D$ (possibly times a polynomial). We could not guess the expression for $c_L$ as a function of $L$ and it was therefore not possible to perform the sum explicitly. 

To conclude, let us give explicit results for singlet operators of length $L=2,3,4$ at strong coupling. This will be particularly useful for the analysis of the mixing problem of \cite{Ferrero:2023gnu}. For $L=2$ we find exactly one state per dimension, while there are no singlets of $L=3$:
\begin{align}
N^{\text{strong}}_{0,[0,0],L=2}=1\,, \quad
N^{\text{strong}}_{0,[0,0],L=3}=0\,.
\end{align}
The most interesting case is that of states with $L=4$, which we build explicitly in the next section. Here the number of states of fixed dimension $\D$ grows quadratically with $\D$, in agreement with \eqref{Nstrong00L}, although with a difference between even and odd $\D$:
\begin{align}
\label{NstrongL4000}
N^{\text{strong}}_{0,[0,0],L=4}=
\begin{cases}
\lfloor \left(\tfrac{\D}{4}\right)^2\rfloor\,, \quad &\D \ge 4\,\,\,\text{even}\,,\\
\lfloor \left(\tfrac{\D-3}{4}\right)^2\rfloor\,, \quad &\D \ge 3\,\,\,\text{odd}\,.
\end{cases}
\end{align}

\section{Construction of components of  supermultiplets as  composite operators}\label{sec:operators}

In this section we address the question of computing correlation functions in the free theory at strong coupling. This problem seems to be trivial since in this limit we have a free  CFT in which all operators are composites made of the $(5+3|8)$ operators $(\varphi,\Psi,f)$, together with their derivatives, and any correlator can be computed by Wick contractions. In practice, it becomes quickly complicated due to the fact that one needs to first determine the explicit form of superprimary operators, and possibly some of their superdescendants. The complexity of this problem grows with the length $L$ of the composite operator we consider  and this is related to the large degeneracies discussed in the previous section. The case of $L=2$ is well known and is particularly simple due to the absence of degeneracies. We are mostly interested in the case of long operators $\mathcal{L}^{\Delta}_{0,[0,0]}$ of length $L=4$ with $\Delta$ even. For the applications in  \cite{Ferrero:2023gnu}, we will need to efficiently\footnote{In  \cite{Ferrero:2021bsb} we reached the case $\Delta= 26$ with intensive computer time. Since then the method has been systematized to the form we present here, allowing to reach higher values of $\Delta$.} compute the following correlators in the free theory:
\begin{subequations}
\label{correlatorsWeneedFREE}
\begin{align}
\label{correlatorsWeneedFREE1}
\langle \mathcal{D}_1 \mathcal{D}_1 \mathcal{D}_2\mathcal{O}^{(2)}\rangle\,,
\qquad &\qquad
\langle \mathcal{D}_1 \mathcal{D}_1 \mathcal{D}_2\mathcal{O}^{(4)}\rangle\,,
\\
\label{correlatorsWeneedFREE2}
\langle  \mathcal{D}_2\mathcal{O}^{(2)}_A\widetilde{\mathcal{O}}^{(2)}_B\rangle\,,
\qquad
\langle  \mathcal{D}_2\mathcal{O}&^{(2)}_A\widetilde{\mathcal{O}}^{(4)}_B\rangle\,,
\qquad
\langle  \mathcal{D}_2\mathcal{O}^{(4)}_A\widetilde{\mathcal{O}}^{(4)}_B\rangle\,,
\\
\label{correlatorsWeneedFREE3}
\langle \widetilde{\mathcal{O}}^{(2)}_A\widetilde{\mathcal{O}}^{(2)}_B\rangle\,,
\qquad&\qquad
\langle  \widetilde{\mathcal{O}}^{(4)}_B\widetilde{\mathcal{O}}^{(4)}_B\rangle\,,
\end{align}
\end{subequations}
where the label in parenthesis denotes the length $(L)$ of the operators and we use the letter $\mathcal{O}$ to denote the superprimary of a $\mathcal{L}^{\Delta}_{0,[0,0]}$ multiplet, while $\widetilde{\mathcal{O}}$ refers to the unique conformal primary in the $\{0,[0,2]\}$ representation of $\mathfrak{su}(2)\oplus \mathfrak{sp}(4)$ present in a $\mathcal{L}^{\Delta}_{0,[0,0]}$ multiplet at level four. This notation was introduced in \eqref{Ottildecomponent}, while in \cite{Ferrero:2021bsb} we referred to this component as $Q^4_{0,[0,2]}\mathcal{O}$ -- for its uniqueness, see \eqref{L00Structure}. Of course, to compute \eqref{correlatorsWeneedFREE} we need to first determine the explicit expression of the various operators involved in terms of the elementary fields $(\varphi,\Psi,f)$ and their derivatives. It should be remarked that while we presented explicit components of the supercorrelators in \eqref{correlatorsWeneedFREE}, each of them possess a unique completion in superspace, see \eqref{twopointLong}, \eqref{SLLframe1}, \eqref{SLLframe2}, \eqref{D1D1D2LongfirstEquation} and \eqref{122LFinvariants} respectively. So far as two-point functions are concerned, normalizing them to unity involves fixing the normalization of the operators that we are going to construct. This turns out to give rise to simpler expressions\footnote{This is particularly evident in the case of $L=2$ operators.} for two-point functions of the $\widetilde{\mathcal{O}}$ components rather than for the norms of the associated superconformal primaries $\mathcal{O}$, hence why we are listing those in \eqref{correlatorsWeneedFREE3}, but the two can be related using \eqref{gOvsgOtilde}.

One should keep in mind that in each of the expressions in \eqref{correlatorsWeneedFREE} we have the freedom to choose the long operators within its degeneracy space.  More abstractly, one should think about correlators involving $n$ long operators as multilinear maps acting on the $n$-fold tensor product of the degeneracy spaces associated with each long operator. As an example, consider the two- and three-point functions in \eqref{correlatorsWeneedFREE}: in both cases, the kinematic part is completely fixed and the correlators contain two long operators belonging to $\mathcal{L}^{\Delta}_{0,[0,0]}$ multiplets. Hence, all correlators listed in \eqref{correlatorsWeneedFREE2} and \eqref{correlatorsWeneedFREE3} can be viewed as maps from the product of degeneracies spaces $\mathsf{d}_{L_1}(\Delta_1)\times \mathsf{d}_{L_2}(\Delta_2)$ to the complex numbers. In the case of two-point functions this map is diagonal in $\Delta$ and $L$ and moreover is non-degenerate. This  suggests a clear geometric interpretation. For example, the three-point functions $v_{\widetilde{\mathcal{O}}_2}(\mathcal{O}_1)\propto \langle  \mathcal{D}_2\mathcal{O}_1\widetilde{\mathcal{O}}_2\rangle$ with $\Delta_{\widetilde{\mathcal{O}}_2}$ fixed can be thought of as a collection of vectors (labelled by the choice of $\mathcal{O}_1$) in the degeneracy space to which $\widetilde{\mathcal{O}}_2$ belongs. For the applications in \cite{Ferrero:2023gnu} it will be important to  understand which set of choices of $\mathcal{O}_1$ provides a basis of this degeneracy space.

Before diving in the technical discussion presented in this section, let us emphasize that the explicit construction of operators and systematic of computation of the correlators \eqref{correlatorsWeneedFREE} turned out to be the main bottleneck of the calculation that leads to the results presented in  \cite{Ferrero:2023gnu}.

%--------------------------------------------------------------------------------------
\paragraph{Highlights of the strategy.} This section is organized as follows:

\begin{itemize}
\item In Section \ref{sec:toyexamples}  we discuss how to build $\mathfrak{sl}(2)$ primaries as composite operators and how to determine their correlation functions explicitly.
Along the way we will take a closer look at $L=3$ operators and discuss possible choices of basis in the degeneracy spaces. The $L=3$ case is considered in detail not only for pedagogical reasons but also because it immediately provides a construction of $L=3$ operators in long supermultiplets of type $\mathcal{L}_{0,[0,1]}^{\Delta}$.
\item In Section \ref{sec:generalStrategySUSY} we introduce a general strategy to determine the explicit form of certain components of supermultiplets realized as composite operators.  These components are linear combinations of conformal primaries with fixed $\Delta$ and R-symmetry representation. The specific linear combination is fixed (up to general linear transformations in degeneracy space) by requiring that the correlator $\langle \Phi(1)\dots \Phi(L) \mathscr{O}^{(L)}\rangle$ computed in the free theory, is invariant under the action of the superconformal group.
\item In Section \ref{sec:L2longOperators} we apply this strategy to the well-known case of $L=2$ operators, which are non degenerate. This is useful to obtain some of the results used in \cite{Ferrero:2023gnu}, and serves as a warm up for the more interesting and complicated case of $L=4$.
\item In Section \ref{sec:L4operatorsConstraint} we consider $L=4$ composite operators transforming in the $\{[0,0],0\}$ representation of R-symmetry with $\Delta$ even. 
This case is considerably more complicated so that the discussion is further organized as follows:
\begin{itemize}
\item[(A)]  In Section \ref{subsec:constraint5ptsuperpr} we show that superconformal symmetry fixes all the components of the five-point correlator $\langle
\Phi(X_1) 
\Phi(X_2) 
\Phi(X_3) 
\Phi(X_4) 
\mathcal{O}(t_5)
\rangle$, where $\mathcal{O}$ is the superprimary of a long multiplet $\mathcal{L}^{\Delta}_{0,[0,0]}$, in terms of the correlator involving the five superprimaries. The latter is fixed in terms of three functions of two bosonic cross ratios. We provide explicit expressions for all component correlators.
\item[(B)]  In Section \ref{subsec:constraint5ptsuperdesc} we give explicit formulas for the components of the five-point function $\langle
\Phi(X_1) 
\Phi(X_2) 
\Phi(X_3) 
\Phi(X_4) 
\widetilde{\mathcal{O}}(t_5,y_5)\rangle$ in terms of the three functions that determine the correlators in the point (A) above.
\item[(C)] In Sections \ref{sec:operatorsforthesuperprimary} and \ref{sec:operatorsforthesuperdescendant} we discuss the explicit construction of operators of the type $\mathcal{O}^{(4)}$ and $\widetilde{\mathcal{O}}^{(4)}$ -- see \eqref{Opieces} and \eqref{Otildepieces} -- in terms of the elementary fields $(\varphi,\Psi,f)$ and their derivatives. We will summarize the various building blocks that are necessary to build such operators and then proceed to provide the necessary tools to compute correlators involving such building blocks in Sections  \ref{sec:operatorsforthesuperprimary} and \ref{sec:operatorsforthesuperdescendant}. The explicit construction of \eqref{Opieces} and \eqref{Otildepieces} is based on an ansatz for them as a sum of $\mathfrak{sl}(2)$ conformal primaries built out of $(\varphi,\Psi,f)$ and derivatives, and the construction of correlators involving this building blocks allows to determine which precise linear combination of conformal primaries is such that the constraints derived in (A) and (B) hold. Once the form of $\mathcal{O}^{(4)}$ and  $\widetilde{\mathcal{O}}^{(4)}$ is fixed by this procedure, we can finally compute the correlators of interest, that we summarize in equation \eqref{correlatorsWeneedFREE}.
\end{itemize}
\item Some of the technical steps used in the construction described above are discussed in the appendices. In particular, it is useful to construct projectors to extract irreducible representations of the R-symmetry $\mathfrak{sp}(4)$ from composite operators and we discuss how to do this in Appendix \eqref{app:Rsymmprojectors}. Moreover, we find it useful to proceed by first computing correlation functions involving composite operators without derivatives, since correlators of composite operators with derivatives can be obtained from the former by the action of suitable differential operators. We describe how to systematically build such differential operators in such a way that they give rise to conformal primaries in Appendix \eqref{app:actingwithderivatives}.
\end{itemize}

\subsection{Construction of conformal primaries}
\label{sec:toyexamples}

Let us first consider an illustrative simplified example where the only free field is a complex scalar $\phi(t)$ of dimension one,  with two-point function\footnote{These fields can be considered as two components of the five scalar fields of the $\mathcal{D}_1$ supermultiplet, compare to \eqref{varhpiandPsidef}.}
\begin{equation}
\langle \phi(t_1)\bar{\phi}(t_2)\rangle=\frac{1}{(t_1-t_2)^2}\,.
\end{equation}
As a further simplification, we restrict our attention to composite operators that maximize the $U(1)$ charge for given length. Let us proceed with increasing length $L$.

\paragraph{The case $L=2$.} This is well known, but let us recall how it works as a warm up. One starts from an ansatz  for the composite operator of the form
\begin{equation}
\label{L2OusingD}
\mathcal{O}_M^{(L=2)}(t)=  D_{M/2}(\partial_1,\partial_2)\, :\phi(t_1) \phi(t_2): \Big{|}_{t_1=t_2=t}\,,
\qquad
D_{M/2}(\partial_1,\partial_2):=\sum_{\ell=0}^M c_{M,m} \partial_1^\ell \,\partial_2^{M-\ell}
\end{equation}
where $:\dots:$ denotes normal ordering in the free theory while the label $M$ represents the total number of derivatives, related to the conformal dimension by $\Delta=2+M$. Because of Bose symmetry, the coefficients should be chosen to satisfy $c_{M,\ell}=c_{M,M-\ell}$. For each even $M=2n$ there is a unique superconformal primary, which up to normalization\footnote{In this normalization we have $D_{2n}(\partial_1,0)=\partial_1^{2n}$ and all the coefficients in \eqref{D2nintermsofDERIVATIVES} are integers.} is given by \eqref{L2OusingD} where the differential operators are fixed to be
\begin{align}
\label{D2nintermsofDERIVATIVES}
\begin{split}
D_{n}(\partial_1,\partial_2)&=
(2n+1)!
 \sum_{\ell=0}^{2n}(-1)^{\ell}
\binom{2n}{\ell}
\frac{\partial_1^\ell}{(\ell+1)!} \,\frac{\partial_2^{2n-\ell}}{(2n-\ell+1)!}\,,\\
&=
(\partial_1+\partial_2)^{2n}\,\frac{C^{(\frac{3}{2})}_{2n}(\tfrac{\partial_1-\partial_2}{\partial_1+\partial_2})}{C^{(\frac{3}{2})}_{2n}(1)}\,,
\end{split}
\end{align}
where $C_b^{(a)}(z)$ are Gegenbauer polynomials. In this normalization we can compute the simplest three-point functions and the matrix of norms to be 
\begin{subequations}
\label{twoandthreeL2}
\begin{align}
\langle  \bar{\phi}(t_1)\bar{\phi}(t_2) \mathcal{O}_{M=2n}^{(L=2)}(t_3)\rangle^{(0)}\,&=\,
2 (2n+1)!\times 
\frac{t_{12}^{2n}}{t_{13}^{2(n+1)}t_{23}^{2(n+1)}}\,,
\\
\langle  \bar{ \mathcal{O}}_{M_1=2n_1}^{(L=2)}(t_1) \mathcal{O}_{M_2=2n_2}^{(L=2)}(t_2)\rangle^{(0)}\,&=\,
%\frac{1}{((2n_1+1)!)^{2}}\times
  2\frac{(4n_1+1)!}{(n_1+1)} \times 
\frac{\delta_{n_1,n_2}}{t_{12}^{2(2n_1+1)}}\,.
\end{align}
\end{subequations}
In particular $\frac{(4n_1+1)!}{(n_1+1)}$ are integers as they should in our normalization.

%%%%%%%%%%%%%%%%%%%%%%%%%%%%%%%%%%%%%%%%
\paragraph{Composites made of $\Phi_1$ and  $\Phi_2$.}
The discussion above can be generalized to construct composite operators made of two scalar operators $\Phi_1$ and $\Phi_2$ of dimension $\D_1$ and $\D_2$, in a generalized free theory where all correlators can be computed by Wick contractions starting from the two-point function
\begin{equation}
\langle
\Phi_1(t_1) \Phi_1(t_2)
\rangle=\frac{1}{(t_1-t_2)^{2\Delta_1}}\,,
\qquad
\langle
\Phi_2(t_1) \Phi_2(t_2)
\rangle=\frac{1}{(t_1-t_2)^{2\Delta_2}}\,.
\end{equation}
In this case, the conformal primaries take the form\footnote{For the case of double-trace operators in higher dimensions see {\it e.g. } Appendix C of  \cite{Fitzpatrick:2011dm}.}  
\begin{equation}
\label{L2OusingK}
\mathcal{O}_K(t)= \mathsf{D}_K^{(\Delta_1,\Delta_2)}(\partial_1,\partial_2)\, :\Phi_1(t_1) \Phi_2(t_2): \Big{|}_{t_1=t_2\equiv t}\,.
\end{equation}
where
\begin{equation}
\label{DoperatorsTomakeSinglets}
\mathsf{D}_K^{(\Delta_1,\Delta_2)}(\partial_1,\partial_2)=
 \sum_{\ell=0}^{K}(-1)^{\ell}
\binom{K}{\ell}
\frac{\partial_1^\ell}{\Gamma(2 \Delta_1+\ell)} \,\frac{\partial_2^{K-\ell}}{\Gamma(2 \Delta_2+K-\ell)}\,.
\end{equation}
Note that we used a slightly different normalization compared to \eqref{D2nintermsofDERIVATIVES} so that $D_{n}=(2n+1)!\mathsf{D}^{(1,1)}_{2n}$.
\vspace{0.2cm}

\noindent{\emph{Remark}:} It is also interesting to consider the well-known expression \eqref{D2nintermsofDERIVATIVES} from a different perspective -- see {\it e.g.} \cite{Braun:2003rp}. One can consider a dual description of a given irreducible highest weight representation, in which the 1d special conformal generator $\mathfrak{K}$ is realized as $\partial_{z}$ acting on the space of polynomials in $z$. In this picture it is easy to find the primaries in the $L$-fold tensor product of such representations: they are given by homogeneous polynomials in $L$ variables $z_1,\dots,z_L$ which are invariant under the shift $z_i\mapsto z_i+\lambda$. To go back to the original description in which the translation operator $\mathfrak{P}$ is realized as $\partial_t$ on each irreducible representation one has to apply the map
\begin{equation}
\label{zellmap}
z^\ell \rightarrow
 \frac{\left(\partial_{t}\right)^{\ell}}{\Gamma(\ell+2\Delta)}\,.
\end{equation}
In the example of $L=2$, the primaries in the  dual description  are simply given by $(z_2-z_1)^K$ and the map  \eqref{zellmap} gives
\begin{equation}
(z_2-z_1)^K=\sum_{\ell=0}^K(-1)^{\ell}z_1^{\ell}\,z_2^{K-\ell} \binom{K}{\ell}
\rightarrow 
\mathsf{D}_K^{(\Delta_1,\Delta_2)}(\partial_1,\partial_2)\,,
\end{equation}
where the differential operator on the right is defined in \eqref{DoperatorsTomakeSinglets} with $\partial_i=\partial_{t_i}$. This description can be used also for $L>2$,  but at the moment we did not obtain any advantage from employing it so it will not be used further. \qed

\vspace{0.2cm}
As an example of the use of the operator \eqref{DoperatorsTomakeSinglets}, we can compute the simplest three-point functions  to get
\begin{subequations}
\label{2and3GFT}
\begin{align}
\langle  \bar{\Phi}_1(t_1)\bar{\Phi}_2(t_2) \mathcal{O}_{K}(t_3)\rangle^{(0)}\,&=\,
\frac{1}{\Gamma(2\Delta_1)}\frac{1}{\Gamma(2\Delta_2)}
\times 
\frac{1}{t_{13}^{2\Delta_1}t_{23}^{2\Delta_2}}
\left(\frac{t_{12}}{t_{13}t_{23}}
\right)^{K}\,,
\\
\langle  \bar{ \mathcal{O}}_{K_1}(t_1) \mathcal{O}_{K_2}(t_2)\rangle^{(0)}\,&=\,
%\frac{1}{((2n_1+1)!)^{2}}\times
\frac{1}{\Gamma(2\Delta_1)^2}
\frac{1}{\Gamma(2\Delta_2)^2}
\frac{
\frac{\Gamma(K+1)\Gamma(2(\Delta_1+\Delta_2)+2K-1)}{\Gamma(2(\Delta_1+\Delta_2)+K-1)}
}{
\frac{\Gamma(2\Delta_1+K)}{\Gamma(2\Delta_1)}
\frac{\Gamma(2\Delta_2+K)}{\Gamma(2\Delta_2)}
}
 \times 
\frac{(-1)^K\delta_{K_1,K_2}}{t_{12}^{2(\Delta_1+\Delta_2+K)}}\,,
\end{align}
\end{subequations}
where $K=K_1=K_2$. Apart from the previously discussed normalization, the two- and three-point functions \eqref{2and3GFT} differ from \eqref{twoandthreeL2} by additional factors of 2 due to the fact that \eqref{twoandthreeL2} uses identical operators.

The differential operator \eqref{DoperatorsTomakeSinglets} can be used to construct composite operators of higher length $L$ as well as composite operators made of  the other components of $\mathcal{D}_1$, namely $\psi$ and $f$. 
We will now illustrate the procedure in the case of operators of $L=3$ made of a single complex scalar $\phi(t)$ of conformal dimension $\Delta=1$.
This is is not only a toy model but corresponds to the $[0,3]^{\Delta+2}_0$ components of $\mathcal{L}^{\Delta}_{0,[0,1]}$ supermultiplet.

\paragraph{$L=3$ operators made of $\phi$.}
This case allow us to discuss the significant complications due to operator degeneracies before dealing with supersymmetry.
The degeneracies in this case grow linearly with the dimension $\Delta$ of the operator and  are give by
\begin{align}
\text{dim}[\mathsf{d}_3(\Delta)]
=
\begin{cases}
\lfloor \tfrac{\Delta}{6}\rfloor\,, \quad &\Delta \ge 4\,\,\,\text{even}\,,\\
\lfloor \tfrac{\Delta+4}{6}\rfloor\,, \quad &\Delta \ge 3\,\,\,\text{odd}\,,
\end{cases}
\qquad
\sum_{\Delta=0}^{\infty}\, \text{dim}[\mathsf{d}_3(\Delta)] q^{\Delta}=\frac{q^3}{(1-q^2)(1-q^3)}\,.
\end{align}
A possible choice of (overcomplete) basis of conformal primary $L=3$ operators made of a single complex scalar $\phi(t)$ is
\begin{equation}
\label{OL3phi3operators}
\!\!\mathcal{O}_{\phi^3}^{\{K,n\}}(t)=
\mathsf{D}_K^{(1,2+2n)}(\partial_1,\partial_2+\partial_3)
\mathsf{D}_{2n}^{(1,1)}(\partial_2,\partial_3)
\,
:\phi(t_1)\phi(t_2)\phi(t_3) :
\Big{|}_{t_i\equiv t}
\,.
%(2n+1)!
%\frac{1}{2}\frac{1}{(2n+1)!}
\end{equation}
Varying the integers $K$ and $n$, these operators span the degeneracy space of dimension $\mathsf{d}_3(\Delta)$ for $\Delta=3+K+2n$.
This choice of basis\footnote{See also Appendix A of \cite{Braun:2003rp}.} does not seem to be particularly convenient since it is overcomplete due the the Bose symmetry of the field $\phi(t)$. This does not stop us from using \eqref{OL3phi3operators} to compute the four-point functions, which can be done using the basic identity \eqref{acton3general} and gives
\begin{align}
\label{fourpointphiphiphiOL3}
\begin{split}
\langle \bar{\phi}(t_1)
\bar{\phi}(t_2)\bar{\phi}(t_3)&\mathcal{O}_{\phi^3}^{\{K,n\}}(t_4)\rangle=\\
\frac{2}{t_{14}^2t_{24}^2t_{34}^2}\left(\frac{t_{32}}{t_{42}t_{43}}\right)^{2n+K}
&
\!\left(
\mathcal{A}(\tfrac{1}{1-\chi})+
\left(\tfrac{1}{1-\chi}\right)^{2n+K}
\mathcal{A}(1-\chi)+
\left(\tfrac{\chi}{\chi-1}\right)^{2n+K}
\mathcal{A}(\chi^{-1})
\right)\,,
\end{split}
\end{align}
where $\mathcal{A}:=\mathcal{A}^{(2n,K)}_{1,1,1}$ is given in  \eqref{AforL3}.
Starting from \eqref{fourpointphiphiphiOL3} we can further compute the three-point function of type $\langle \bar{\phi}\bar{\mathcal{O}}^{(L=2)}\mathcal{O}^{(L=3)}\rangle$ and, following the discussion below \eqref{2and3GFT}, the matrix of norms for $L=3$ primaries.

As we stressed, the basis \eqref{OL3phi3operators} is overcomplete, and it is an interesting to consider the construction of a complete basis of dimension $\text{dim}[\mathsf{d}_3(\Delta)]$ for $L=3$ operators which has some nice properties. We denote the operators in such basis as 
\begin{equation}
\widehat{\mathcal{O}}_{\Delta,\alpha}\,,
\qquad
\alpha=1,\dots, \text{dim}[\mathsf{d}_3(\Delta)]\,,
\end{equation}
To construct them, we adopt a recursive procedure: fix a value for the dimension $\Delta$ and say that we want to define a complete basis for $\mathsf{d}_3(\Delta)$. This can be done by analyzing the OPE OPE between the elementary field $\phi$ and previously constructed basis elements having dimension $\Delta'<\Delta$. The construction works as follows 
\begin{align}
\begin{split}
\widehat{\mathcal{O}}_{\Delta,1 } \,\,& \text{is the operator in the $\phi \times \mathcal{O}^{(L=2)}_{\Delta'=2}$ OPE} \,,
\\
\widehat{\mathcal{O}}_{\Delta,2 } \,\,& \text{is the operator in the $\phi \times \mathcal{O}^{(L=2)}_{\Delta'=4}$ OPE orthogonal to $\widehat{\mathcal{O}}_{\Delta,1 }$}\,,
\\
\widehat{\mathcal{O}}_{\Delta,3 } \,\,& \text{is the operator in the $\phi \times \mathcal{O}^{(L=2)}_{\Delta'=6}$ OPE orthogonal to $\widehat{\mathcal{O}}_{\Delta,1 }$ and $\widehat{\mathcal{O}}_{\Delta,2 }$}   \,,
\\
\vdots \,\,\,\,\,\,\,\,& \qquad  \qquad  \qquad   \qquad   \qquad      \qquad \vdots   \qquad  \qquad  \qquad  \qquad     \qquad  \qquad  \vdots \qquad   \qquad  \qquad   \qquad  
\end{split}
\end{align}
where $ \mathcal{O}^{(L=2)}_{\Delta}$ are defined as $\mathsf{D}^{(1,1)}_{\Delta} :\phi(t_1) \phi(t_2): \big{|}_{t_i=t}$ and the leftover normalization ambiguity of $\widehat{\mathcal{O}}_{\Delta,\alpha}$ is fixed by the requirement that the three-point functions
\begin{equation}
\label{C0threenormis}
\mathsf{C}^{(0)}_{\phi \, \mathcal{O}^{(L=2)}_{\Delta'=2n'}\widehat{\mathcal{O}}_{\Delta,\alpha=1+n'}}=1\,.
\end{equation} 
In this basis, the matrix of norms is diagonal by construction. By means of an explicit computation we find that, in the subspace of operators of dimension $\Delta$, the two-point functions produce the inner product 
\begin{equation}
\mathsf{g}_{\alpha\beta}^{(L=3)}(\Delta)=v_{\alpha}(\Delta)\,\delta_{\alpha\beta}\,,
\qquad \alpha,\beta=1,\dots, \text{dim}[\mathsf{d}_3(\Delta)]\,,
\end{equation}
where,  for even $\Delta$,
\begin{equation}
v_{\alpha}(\Delta)=
\frac{\Gamma(2\Delta-1)(2\alpha-1)(4\alpha-1)\Gamma(2\alpha)\Gamma(2\alpha+1)}{
2^{4\alpha-1}
\Gamma(\Delta-2\alpha)\Gamma(\Delta+2\alpha+2)
(\alpha-\tfrac{\Delta}{2})_{2\alpha}
(\alpha+\tfrac{\Delta+3}{2})_{2\alpha-2}
}
\left(\frac{2\alpha+\Delta}{2\alpha-\Delta}\right)\,,
\end{equation}
while for odd $\Delta$ we have the more complicated expression
\begin{equation}
v_{\alpha}(\Delta)=
\frac{\Gamma(2\Delta-1)\left(\Gamma(2\alpha+1)\right)^2
\frac{f_{\alpha}(\Delta)}{f_{\alpha+1}(\Delta)}}{
2^{4\alpha+1}
(4\alpha-3)
\Gamma(\Delta-2\alpha)\Gamma(\Delta+2\alpha-1)
(\alpha+\tfrac{\Delta}{2})_{2\alpha-2}
(\alpha-\tfrac{\Delta-1}{2})_{2\alpha-2}
(2\alpha+\Delta-1)(2\alpha-\Delta)^2
}\,,
\end{equation}
with
\begin{equation}
f_{\alpha}(\Delta)=
{}_4F_3(1-\alpha,1-\alpha,\alpha-\tfrac{\Delta}{2},\alpha+\tfrac{\Delta}{2}-\tfrac{1}{2};\tfrac{5}{2}-2\alpha,\alpha,\alpha;1)\,.
\end{equation}
It is interesting to observe that if one computes $v_{\alpha}(\Delta)$ for fixed $\Delta$ and $\alpha$ not in the set $\{1,2,\dots,	\text{dim}[\mathsf{d}_3(\Delta)]\}$ one finds either $0$ or $\infty$, signaling the non existence of the corresponding operator. For the three-point functions in this basis we employ the notation
\begin{equation}
\lambda_{\Delta,\alpha}(h):=
\mathsf{C}^{(0)}_{\phi \, \mathcal{O}^{(L=2)}_h\widehat{\mathcal{O}}_{\Delta,\alpha}}\,.
\end{equation} 
For $\Delta$ even, the condition \eqref{C0threenormis} fixes $\lambda_{\Delta,\alpha}(2\alpha)=1$, while $\lambda_{\Delta,\alpha}(h)=0$ for $h\leq \Delta$ or $2\alpha \leq h$. In the remaining cases, we find 
\begin{align}
\begin{split}
\lambda_{\Delta,\alpha}(h)&=
\frac{2(h-\Delta)\Gamma(h+\Delta)\Gamma(2+\alpha+\tfrac{\Delta}{2})}{\Gamma(2-\alpha+\tfrac{\Delta}{2})\Gamma(3+2\alpha+\Delta)\Gamma(h)}
\frac{(2\alpha-1)(4\alpha-1)\Gamma(2\alpha+1)}{(\alpha-\tfrac{\Delta}{2})_{2\alpha}}\times \,
\\
&
\qquad \qquad\qquad
{}_4F_3(2-2\alpha,1+2\alpha,1+\tfrac{\Delta}{2}-\tfrac{h}{2},1+\tfrac{\Delta}{2}+\tfrac{h}{2};2,2-\alpha+\tfrac{\Delta}{2},\tfrac{3+\Delta}{2}+\alpha,;1)\,.
\end{split}
\end{align}
For odd $\Delta$ we could not find a closed form expression for $\lambda_{\Delta,\alpha}(h)$.

As a check, or as an alternative derivation of the results above, we can expand the four-point function $\langle\phi\, \mathcal{O}^{(L=2)}_{\h_1}\bar{\phi}\, \bar{\mathcal{O}}^{(L=2)}_{\h_2} \rangle$ in conformal blocks to extract the OPE coefficients.  Up to a prefactor chosen as in \eqref{sl24pt} and in the normalization of \eqref{DoperatorsTomakeSinglets}, we find that this four-point function is given by
\begin{equation}
\delta_{\h_1,\h_2}\,\chi^{\h_1+1}\mathcal{N}_{\h_1}
+
\tfrac{4 \chi}{(\chi-1)^2}\left(\tfrac{\chi}{\chi-1}\right)^{\h_1}(1-\chi)^{\h_2}
\tfrac{\Gamma(\h_1+\h_2-2)}{\Gamma(\h_1)\Gamma(\h_2)}
\,
{}_2F_1(1-\h_2,2-\h_2,3-\h_1-\h_2,\tfrac{1}{1-\chi})\,,
\end{equation}
where $\mathcal{N}_{\Delta}=\frac{4 \Delta\Gamma(2\Delta-2)}{\Gamma(\Delta+1)^2}$. This expression can be expanded in conformal blocks with coefficients
\begin{equation}
a^{(h_1,h_2)}_{\Delta}=\sum_{\alpha=1}^{\mathsf{d}_3(\Delta)} \frac{1}{v_\alpha(\Delta)}\lambda_{\Delta,\alpha}(h_1)\lambda_{\Delta,\alpha}(h_2)\,,
\end{equation}
as expected. We leave a more systematic study of an orthogonal basis of operators and associated OPE coefficients for operators of $L\geq 4$ to future work.

%---------------------------------------------------------------------------------------------------------

\subsection{Outline of the general strategy to build components of supermultiplets}
\label{sec:generalStrategySUSY}
%----

In this section we present the general strategy to construct certain components of supermultiplets as composite operators made of $(\varphi, \Psi, f)$ in the free theory at strong coupling. For fixed length $L$, the operators in a given superconformal multiplet are completely characterized by the requirement that the $(L+1)$-point correlation function\footnote{The notation $\langle \dots \rangle^{(0)}$ indicates that the correlator is computed by Wick contractions.}
\begin{equation}
\langle  \Phi(X_1)\dots \Phi(X_L)\,\mathscr{O}^{(L)}\rangle^{(0)}\,,
\end{equation}
is annihilated by the generators of $\mathfrak{osp}(4^*|4)$ in the appropriate representation.
In practice, we will only be interested in constructing certain specific components of a given superconformal multiplet $\mathcal{L}_{s,[a,b]}^{\h}$ among those listed in \eqref{L[a,b]_s} (see \eqref{L00Structure} for the case $a=b=s=0$). We adopt the previously introduced notation to denote with $\mathscr{O}^{(L)}$ the operator associated with the whole multiplet in superspace, and we shall be interested in the construction of the superconformal primary of the multiplet, transforming in the $[a,b]^{\h}_s$ representation, as well as its (unique) superconformal descendant in the  $[a,b+2]^{\h+2}_s$ representation. Following the notation introduced in Section \ref{sec:kinematics}, we will refer to such two components as $\mathcal{O}^{(L)}$ and $\widetilde{\mathcal{O}}^{(L)}$ respectively, defined by
\begin{equation}
\label{OandOtildereported}
\mathcal{O}(t):=\mathscr{O}(t,\Theta=0)\,,\qquad
\widetilde{\mathcal{O}}(t,y):=\left[\delta^{(0|4)}(\QforOtilde(y))\mathscr{O}(t,\Theta)\right]_{\Theta=0}\,,
%\quad \text{with}\quad \{s,[a,b],\h\}=\{0,[0,2],\h_{\mathcal{O}}+2\}\,,
\end{equation}
which trivially generalizes \eqref{Otcomponent} and \eqref{Ottildecomponent} to generic long multiplets. The correlators involving these components are still invariant under the bosonic subalgebra of the full superconformal algebra $\mathfrak{osp}(4^*|4)$, but are invariant only 
with respect to half of the fermionic generators. More precisely\footnote{Note that we dropped ${}^{(0)}$ in \eqref{invariancefor5pt} and \eqref{invariancefor5pt2}, since they represent kinematical statements which hold in the interacting theory as well.}
\begin{align}
\label{invariancefor5pt}
\langle  \Phi(X_1)\dots \Phi(X_L)\,\mathcal{O}(t)\rangle &\qquad \text{is invariant under $\mathcal{S}^{\alpha}_{A}+t \epsilon^{\alpha \beta} \mathcal{Q}_{A\beta}$}\,,
\\
\label{invariancefor5pt2}
\langle \Phi(X_1)\dots \Phi(X_L)\,\widetilde{\mathcal{O}}(t,y)\rangle &\qquad \text{is invariant under $\QforOtilde^{a}_{\alpha}(y)$,  $\SforOtilde^{\alpha a}(t,y)$}\,,
\end{align}
where $\QforOtilde^{a}_{\alpha}(y)$  is defined in \eqref{Qofydef} using \eqref{QSsplit}, while we have introduced the combination
\begin{align}\label{SforOtilde}
\SforOtilde^{\alpha a}(t,y)=\mathcal{S}^{\alpha a}+t\, y^{ab} \epsilon^{\alpha \beta} \mathcal{Q}_{\beta b}\,.
\end{align} 
The invariance of  the correlators above with respect to the combinations of fermionic generators given in \eqref{invariancefor5pt} and \eqref{invariancefor5pt2} follows from the fact that these combinations act trivially on $\mathcal{O}(t)$ and $\widetilde{\mathcal{O}}(t,y)$ respectively, as we now explain more in detail.

First, consider the correlator in \eqref{invariancefor5pt}. Since $\mathcal{O}$ is a superconformal primary, by definition $\mathcal{S}^{\alpha}_{A}\,\mathcal{O}(0)=0$. Applying a finite translation then translates this expression into the condition \eqref{invariancefor5pt}. Let us now move to the statement \eqref{invariancefor5pt2}. Given the presence of the fermionic delta function in the definition of $\widetilde{\mathcal{O}}(t,y)$, the invariance under $\QforOtilde^{a}_{\alpha}(y)$ follows trivially. Moreover, the combination $\SforOtilde^{\alpha a}(t,y)$ in \eqref{SforOtilde} was defined in such a way as to commute with $\QforOtilde^{a}_{\alpha}$, from which it immediately follows that \eqref{invariancefor5pt2} should hold. For practical purposes, it is enough to impose invariance under half of the fermionic generators in \eqref{invariancefor5pt} and \eqref{invariancefor5pt2}, since the invariance under the remaining half follows automatically as a consequence of the commutation relations with the bosonic generators, if the correlators are constructed in such a way as to be invariant under those (which they will, since these are the simplest conditions to impose). Recalling the action of the supercharges on short operators, which we have presented in \eqref{GeneratorsSHORT}, the conditions \eqref{invariancefor5pt} and \eqref{invariancefor5pt2} can be rewritten as
\begin{align}
\label{SinvarianceLplus1correlator}
\sum_{i=1}^L
\left(\tilde{S}_a^{\alpha}(i)+t \epsilon^{\alpha \beta}Q_{\beta a}(i) \right)\,\,\,
\langle \Phi(1)\dots \Phi(L)\,\mathcal{O}(t)\rangle\,\,\,
&
=\,
\left(\mathbf{s}^-+\mathbf{s}^+\right)\langle \Phi(1)\dots \Phi(L)\,\mathcal{O}(t)\rangle\,=0\,,\\
\label{QyinvarianceLplus1correlator}
\sum_{i=1}^L\left(\tilde{Q}^a_{\alpha}(i)-y^{ab}\,Q_{\alpha b}(i) \right)\langle \Phi(1)\dots \Phi(L)\,\widetilde{\mathcal{O}}(t,y)\rangle\,
&=\,
\left(\mathbf{q}^-+\mathbf{q}^+\right)\langle \Phi(1)\dots \Phi(L)\,\widetilde{\mathcal{O}}(t,y)\rangle\,=0\,,
\end{align}
where
\begin{equation}
\mathbf{s}^-=
\sum_{i=1}^L\,\epsilon^{\alpha \beta}(t_i-t) \frac{\partial}{\partial \theta_i^{\beta a}}\,,
\qquad
\mathbf{s}^+=-2\sum_{i=1}^L\,\theta_i^{\alpha b}\frac{\partial}{\partial y_i^{ab}}\,.
\end{equation}
%%%
\begin{equation}
\mathbf{q}^-=
\sum_{i=1}^L\,(y_i-y)^{ab}\frac{\partial}{\partial \theta_i^{\alpha b}}\,,
\qquad
\mathbf{q}^+=\sum_{i=1}^L\,\epsilon_{\alpha \beta }\theta_i^{\beta a}\frac{\partial}{\partial t_i}\,.
\end{equation}

In the free theory at strong coupling, the composite operators $\mathcal{O}$ and $\widetilde{\mathcal{O}}$ will take the form of a linear combination of all possible conformal primaries with the correct scaling dimension and R-symmetry quantum numbers built out of $(\varphi,\Psi,f)$ and derivatives. The specific linear combination can be constrained by computing the correlators $\langle \Phi(1)\dots \Phi(L)\,\mathcal{O}(t)\rangle$ and $\langle \Phi(1)\dots \Phi(L)\,\widetilde{\mathcal{O}}(t,y)\rangle$ using Wick contractions and imposing 
\eqref{SinvarianceLplus1correlator} and \eqref{QyinvarianceLplus1correlator}. The technical complexity of this (conceptually straightforward) procedure increases with the length $L$ of the long multiplet that one wishes to study. The case of $L=2$ is particularly simple since one only has two compute a three-point function -- see the result \eqref{DDLresult} -- and will be discussed in subsection \ref{sec:L2longOperators}. On the other hand, the case of $L=4$ (which is the one we are mostly interested in) is significantly more complicated and will be discussed in subsection \ref{sec:L4operatorsConstraint}, which will take most of the remainder of this Section. Our strategy for $L=4$ will be to first express the two correlators appearing in \eqref{SinvarianceLplus1correlator} and \eqref{QyinvarianceLplus1correlator} for generic $L=4$ in terms of three functions of two cross ratios, and translate \eqref{SinvarianceLplus1correlator} and \eqref{QyinvarianceLplus1correlator} into differential conditions on such functions. Later, we shall impose that the explicit expression for these correlators computed from an ansatz for $\mathcal{O}$ and $\widetilde{\mathcal{O}}$ in terms of conformal primaries is such that it satisfies these differential conditions. We will not consider $L>4$.

\subsection{$L=2$ long operators}
\label{sec:L2longOperators}

Let us outline the strategy in the simplest case of $L=2$ operators\footnote{This is the 1d analogue of the well-known case of 4d $\mathcal{N}=4$ superconformal symmetry, see {\it e.g.} \cite{Henn:2005mw}.}. As explained in Section \ref{sec:specStrong}, in this case the only long operators are of type $\mathcal{L}^{\h}_{0,[0,0]}$ with $\h \geq 2$ and are non-degenerate. First, let us recall the constraints on the relevant component of $L+1=3$-point functions, starting from the one with $\mathcal{O}(t)$:
\begin{subequations}
\label{PhiPhiOL2pr}
\begin{align}
\langle \varphi(1)\varphi(2)\mathcal{O}(t_3)\rangle\,= &
\, C_{11\mathscr{O}}\,
\frac{y_{12}^2}{t_{12}^2}\,
\left(\frac{t_{12}}{t_{31}t_{23}}\right)^{\h}\,,
\\
\langle \theta\Psi(1)\,\theta \Psi(2)\mathcal{O}(t_3)\rangle\,= &\,
\,-(\h-2)\, C_{11\mathscr{O}}\,
\frac{\theta_{1}y_{12}\theta_{2}}{t_{12}^3}\,
\left(\frac{t_{12}}{t_{31}t_{23}}\right)^{\h}\,,
\\
\langle \theta^2\!f(1)\,\theta^2\!f(2)\mathcal{O}(t_3)\rangle\,= &
\,-\frac{1}{4} (\h-2)(\h-3)\, C_{11\mathscr{O}}\,
\frac{\theta_{1}^2\theta_{2}^2}{t_{12}^4}\,
\left(\frac{t_{12}}{t_{31}t_{23}}\right)^{\h}\,,
\end{align}
\end{subequations}
where 
\begin{equation}
\theta_1y_{12}\theta_2=\epsilon_{\alpha\beta}\epsilon_{ac}\epsilon_{bd}\theta_1^{\alpha a}y_{12}^{cd}\theta_2^{\beta b}\,,\qquad\,\,\,\,
\theta_1^2\theta_2^2=\epsilon_{ab}\epsilon_{cd}\epsilon_{\alpha\gamma}\epsilon_{\beta\delta}\theta_1^{\alpha a}\theta_1^{\beta b}\theta_2^{\gamma c}\theta_2^{\delta d}\,,
\end{equation}
and we defined $\theta\Psi=\theta^{\alpha a}\Psi_{\alpha a}$ and $\theta^2 f=\epsilon_{ab}\theta^{\alpha a}\theta^{\beta b}f_{\alpha\beta}$.
Note that we reported only the non-vanishing components of $\langle \Phi \Phi \mathcal{O}\rangle$.
These expressions can be extracted from the three-point function \eqref{DDLresult} and follow from supersymmetry.
Alternatively, they can be uniquely fixed by recalling that the proportionality between three-point functions of descendants and primaries are polynomials in $\h$ with the appropriate zeroes. The overall $\h$ independent part is in turn fixed by comparing to the case $\h=0$ given in \eqref{twopointD1components}.

The next step is to write the most general (up to overall normalization) $L=2$ operator which is a conformal primary and R-symmetry singlet, namely
\begin{equation}
\label{OL2superprimaryansatz}
\mathcal{O}^{(L=2)}_{\Delta=2+2n}(t)=
\mathsf{D}_{2n}^{(1,1)}
S_{\varphi^2}(t_1,t_2)
+\,a_n\,
\mathsf{D}_{2n-1}^{(3/2,3/2)}
S_{\Psi^2}(t_1,t_2)
+\,b_n\,
\mathsf{D}_{2n-2}^{(2,2)}
S_{f^2}(t_1,t_2)
\Big{|}_{t_i=t}
\,,
\end{equation}
with
\begin{subequations}
\label{SingletsBIlinears}
\begin{align}
S_{\varphi^2}(t_1,t_2)&=\,\,:(\varphi^2)^{[0,0]}(t_1,t_2):\,,
\\
S_{\Psi^2}(t_1,t_2)&=\,\,:(\Psi^2)^{[0,0]}(t_1,t_2):\,,
\\
S_{f^2}(t_1,t_2)&=\,\,:(f^2)^{[0,0]}(t_1,t_2):\,,
\end{align}
\end{subequations}
where $(\varphi^2)^{[0,0]}$, $(\Psi^2)^{[0,0]}$ and $(f^2)^{[0,0]}$ are given in \eqref{phi2singlet}, \eqref{ALLfermionicbilinearcompact}, \eqref{f2bilinear} respectively and the operators  $\mathsf{D}_{K}^{(\Delta_1,\Delta_2)}=\mathsf{D}_{K}^{(\Delta_1,\Delta_2)}(\partial_1,\partial_2)$ are defined in  \eqref{DoperatorsTomakeSinglets}. To fix the coefficients $a_n$ and $b_n$ we compute the three-point functions \eqref{PhiPhiOL2pr} for  \eqref{OL2superprimaryansatz} using the generating correlators
\begin{subequations}
\label{PhiPhiOL2prGENERATING}
\begin{align}
\label{PhiPhiOL2prGENERATINGphiphi}
\langle \varphi(1)\varphi(2)S_{\varphi^2}(t_3,t_4)\rangle^{(0)}\,= &
\,\, y^2_{12}
\left(\tfrac{1}{t^2_{13} t^2_{24}}
+\tfrac{1}{t^2_{14} t^2_{23}}
\right)\,,
\\
\label{PhiPhiOL2prGENERATINGpsipsi}
\langle \theta\Psi(1)\,\theta \Psi(2)S_{\Psi^2}(t_3,t_4)\rangle^{(0)}\,= &\,
\, -6\,
\theta_{1}y_{12}\theta_{2}\,
\left(\tfrac{1}{t^3_{13} t^3_{24}}
-\tfrac{1}{t^3_{14} t^3_{23}}
\right)\,,
\\
\label{PhiPhiOL2prGENERATINGff}
\langle \theta^2\!f(1)\,\theta^2\!f(2)S_{f^2}(t_3,t_4)\rangle^{(0)}\,= &\,
\,-\tfrac{9}{4}\,
\theta_{1}^2\theta_{2}^2\,
\left(\tfrac{1}{t^4_{13} t^4_{24}}
+\tfrac{1}{t^4_{14} t^4_{23}}
\right)\,,
\end{align}
\end{subequations}
and match the coefficients against \eqref{PhiPhiOL2pr}.
This gives the conditions
\begin{equation}
a_n= \tfrac{2}{3}\,(2n)\,,
\qquad
b_n=4\,(2n)(2n-1)\,,
\end{equation}
thus fixing completely the free parameters in the ansatz \eqref{OL2superprimaryansatz} for the $L=2$ superconformal primaries. We also find $C_{11\mathscr{O}}=2$. Since the normalization in \eqref{OL2superprimaryansatz} is still arbitrary, though, we should also compute the associated two-point functions. We do so by using again three generating correlators
\begin{subequations}
\label{PhiPhiOL2prGENERATINGforNORMS}
\begin{align}
\label{PhiPhiOL2prGENERATINGforNORMSm0}
\langle  S_{\varphi^2}(t_1,t_2)\, S_{\varphi^2}(t_3,t_4)\rangle^{(0)}\,= &
\,\,\,\,\,\,\,5\,\,\,
\left(\tfrac{1}{t^2_{13} t^2_{24}}
+\tfrac{1}{t^2_{14} t^2_{23}}
\right)\,,
\\
\langle S_{\Psi^2}(t_1,t_2)\,S_{\Psi^2}(t_3,t_4)\rangle^{(0)}\,= &-72
\left(\tfrac{1}{t^3_{13} t^3_{24}}
-\tfrac{1}{t^3_{14} t^3_{23}}
\right)\,,
\\
\langle S_{f^2}(t_1,t_2)\, S_{f^2}(t_3,t_4)\rangle^{(0)}\,= &\,
\,\,\,\tfrac{27}{4}\,\,\,
\left(\tfrac{1}{t^4_{13} t^4_{24}}
+\tfrac{1}{t^4_{14} t^4_{23}}
\right)\,.
\end{align}
\end{subequations}
Acting with differential operators as dictated by \eqref{PhiPhiOL2prGENERATINGforNORMS} leads to
\begin{equation}
\label{gOLis2}
 \mathsf{g}_{\mathcal{O}\mathcal{O}}\,=\,
 \frac{4\,\Gamma(2 \Delta+2)}{( \Delta-1)\Gamma( \Delta+1)\Gamma( \Delta+3)}\,,\qquad
 \Delta=2+2n\,.
\end{equation}

We now move to the $\widetilde{\mathcal{O}}$ component. In this case, the ansatz contains only one term
\begin{equation}
\label{OtildeL2form}
\widetilde{\mathcal{O}}^{(L=2)}_{\Delta=4+2n}(t,y)=\,\,
c_n\,
\mathsf{D}_{2n+2}^{(1,1)}(\partial_1,\partial_2)
:\varphi(t_1,y) \varphi(t_2,y):\Big{|}_{t_i=t}\,.
\end{equation}
We have introduced an overall normalization coefficient $c_n$ since the norm of $\widetilde{\mathcal{O}}^{(L=2)}$ is no longer arbitrary after that of ${\mathcal{O}}^{(L=2)}$ is fixed, (by \eqref{OL2superprimaryansatz}) as a consequence of \eqref{gOvsgOtilde}. The coefficient $c_n$ can be obtained by comparing the three-point function $\langle \varphi\varphi \widetilde{\mathcal{O}} \rangle$ obtained using \eqref{OtildeL2form} to the expression that follows from supersymmetry found in \eqref{mu11Otildevsmu11O}. This results in $c_n=(2n+2)(2n+3)$. As a consistency check one can verify that the norms satisfy \eqref{gOvsgOtilde}.

There are other three- and four-point functions that we will need in \cite{Ferrero:2023gnu} and, once the operators are constructed as we just discussed, they can be computed in a straightforward way with similar methods. We summarize the interesting results here:
\begin{equation}
\langle \mathcal{D}_1(1) \mathcal{D}_1(2)\mathcal{D}_2(3)
\mathcal{O}_{\Delta}^{(L=2)}(t_4) \rangle^{(0)}\,=\,
\frac{y_{13}^2}{t_{13}^2}
\frac{y_{23}^2}{t_{23}^2}
\left(\frac{t_{12}}{t_{14}t_{24}}\right)^{\Delta}\,
4\left(
\left(\frac{\chi-1}{\chi}\right)^{\Delta}
+
\left(\frac{1}{\chi}\right)^{\Delta}
\right)
\end{equation}
and
\begin{equation}
\langle \mathcal{D}_2(t_1,y_1) 
\mathcal{O}_{\Delta_1}^{(L=2)}(t_2)
 \widetilde{\mathcal{O}}_{\Delta_2+2}^{(L=2)}(t_3,y_3) \rangle^{(0)}\,=\,
 \frac{4 \Gamma(\Delta_1+\Delta_2)}{\Gamma(\Delta_1)\Gamma(\Delta_2)}
 \left(\frac{y_{13}^2}{t_{13}^2}\right)^{2}\,
 \frac{1}{t_{23}^{\Delta_1+\Delta_2}}
  \left(\frac{t_{31}}{t_{12}}\right)^{\Delta_1-\Delta_2}\,,
\end{equation}
where $\mathcal{D}_2=\varphi^2$.
%------------------------------------------------------------------------------------------------------------------------
\subsection{$L=4$  long operators of type $\mathcal{L}^{\h}_{0,[0,0]}$}
\label{sec:L4operatorsConstraint}

In the following we will generalize the construction above to operators of length four. The steps and the logic are exactly the same, but there are two important technical difficulties to address:
\begin{itemize}
\item[(i)] To construct the operators $\mathcal{O}$ and $\widetilde{\mathcal{O}}$ of $L=2$ in the free theory we used the fact that all the component three-point functions of $\langle \Phi \Phi \mathscr{O}\rangle$ are fixed in terms of a unique constant $C_{11\mathcal{O}}$. The ones we used are given in  \eqref{PhiPhiOL2pr} and \eqref{mu11Otildevsmu11O}. To construct the operators $\mathcal{O}$ and $\widetilde{\mathcal{O}}$ of $L=4$, we will use similar constraints for the components of the five-point function $\langle \Phi \Phi\Phi \Phi \mathscr{O}\rangle$. We will find that superconformal symmetry implies that all the components are fixed in terms of the correlator of the five superprimaries\footnote{
This is not surprising and is related to the fact the multiplet $\mathcal{D}_1$ is ultra-short. While we prove this fact below, at the moment we do not have a quick explanation for it along the line of the one presented for the correlator  \eqref{D1D1D2LongfirstEquation}.
},which can be expressed in terms of three functions of two cross ratios. The complete derivation of this fact will occupy Sections 
\ref{subsec:constraint5ptsuperpr}, \ref{subsec:constraint5ptsuperdesc}, as well as an ancillary $\mathtt{Mathematica}$ file included with the arXiv submission.
\item[(ii)]  Superconformal primaries of $L=2$ are non-degenerate. So the corresponding operators are uniquely fixed up to their normalization that we fixed arbitrarily. On the other hand, as discussed in Section  \ref{sec:specStrong}, $L=4$ operators are degenerate. We will be interested in operators of type $\mathcal{L}^{\h}_{0,[0,0]}$, whose degeneracy is given in \eqref{NstrongL4000}, and we will consider only the case in which $\h\equiv \Delta$ is even which is the case of interest to address the mixing problem in \cite{Ferrero:2023gnu}.
\end{itemize}

%------------------------------------------------------------------------------------------------------------------------
\subsubsection{Superconformal constraints on the five-point function}

The construction of $L=4$ operators in the free theory at strong coupling is considerably more involved compared to the $L=2$ case.
Following the strategy outlined above, the first step is to solve the constraints of superconformal symmetry on the five-point function 
\begin{equation}
\label{G5withaLONGfull}
\langle
\Phi(X_1) 
\Phi(X_2) 
\Phi(X_3) 
\Phi(X_4) 
\mathscr{O}(\mathbf{t}_5)
\rangle\,,
\end{equation}
where $\mathscr{O}$ is an operator transforming in the  $\mathcal{L}^{\h}_{0,[0,0]}$ representation and $\mathbf{t}=(t,\Theta)$.
We will show below that all the components, in the sense of an expansion in the Grassmann variables, of \eqref{G5withaLONGfull} are completely determined in terms of the three functions entering the correlator of the superprimaries
\begin{equation}
\label{phi4terminSuperPR5pointfun}
\langle
\varphi(1)
\varphi(2)
\varphi(3)
\varphi(4) 
\mathcal{O}(5)
\rangle
\,=\, \frac{y^2_{12}}{t_{12}^2}
 \frac{y^2_{34}}{t_{34}^2}\,
\left(
\frac{t_{34}^2}{t_{35}^2t_{4 5}^2}
\right)^{\h/2}\!
F_{\varphi\varphi\varphi\varphi}^{(12)}+
\text{two more}\,,
\end{equation}
where ``two more'' refers to terms with $(2\leftrightarrow 3)$ and $(2\leftrightarrow 4)$, as in \eqref{psipsipsipsiOtildeF}, and
$\varphi(i)\equiv \varphi(t_i,y_i)$, $\mathcal{O}(5)\equiv \mathcal{O}(t_5)=\mathscr{O}(t_5,\Theta=0)$ and
$F_{\varphi\varphi\varphi\varphi}^{(12)}$, $F_{\varphi\varphi\varphi\varphi}^{(13)}$, $F_{\varphi\varphi\varphi\varphi}^{(14)}$ are functions of two cross ratios\footnote{The functions $F_{\varphi\varphi\varphi\varphi}^{(12)}$, $F_{\varphi\varphi\varphi\varphi}^{(13)}$, $F_{\varphi\varphi\varphi\varphi}^{(14)}$ of course depend on the choice of operator $\mathcal{O}$, but we suppressed this dependence from the notation.}. The analysis of this part is purely kinematical and holds beyond the free theory at strong coupling, for this reason we dropped the superscript $(0)$ from correlation functions.

%------------------------------------------------------------------------------------------------------------------------------------------
\subsubsection{Constraints on the five-point function with a superprimary}
\label{subsec:constraint5ptsuperpr}

Let us consider the five-point function
\begin{equation}
\label{G5othercomp}
G^{(5)}_{\mathcal{O}}:=
\langle
\Phi(X_1) 
\Phi(X_2) 
\Phi(X_3) 
\Phi(X_4) 
\mathcal{O}(t_5)
\rangle\,,
\end{equation}
where $\mathcal{O}(t_5)$ is the superprimary of a long  $\mathcal{L}_{s=0,[0,0]}^{\h}$ supermultiplet, see \eqref{OandOtildereported}.
We would like to solve all the constraints that superconformal symmetry implies on this correlator.
Recall that since $\mathcal{D}_1$ is an ultra-short multiplet, its associated superfield $\Phi(X)$ is subject to the constraints listed in \eqref{Phiconstraint}. To satisfy those conditions, we proceed by inserting in \eqref{G5othercomp} the solution of the constraints in terms of $\varphi, \Psi, f$ given in \eqref{Phisolveconstraint}. The operators $\varphi, \Psi, f$ are conformal primaries with definite transformation properties under $\mathfrak{sp}(4)\oplus \mathfrak{su}(2)$ and their correlation functions are constrained by this symmetry. The component with all $\theta_i=0$ is given by \eqref{phi4terminSuperPR5pointfun} and we now proceed by analyzing how the correlators between components appearing at higher orders in the expansion in the Grassman variables are related to the three functions in \eqref{phi4terminSuperPR5pointfun}. 

The first non-vanishing component, apart from the one generated by the action of $D^{(\varphi)}$ entering \eqref{Phisolveconstraint}, is 
\begin{equation}
\label{phi2psi2terminSuperPR5pointfun}
\langle
\theta\Psi(1)
\theta \Psi(2) 
\varphi(3)
\varphi(4)
\mathcal{O}(5)
\rangle
\,=\,
\frac{1}{t_{12}^3 t_{34}^2}
\left(
\frac{t_{34}^2}{t_{35}^2t_{4 5}^2}
\right)^{\h/2}
\left(
\mathcal{I}_{12,34}\,
F^{(12),34}_{\Psi \Psi \varphi\varphi}+(3 \leftrightarrow 4)
%\mathcal{I}^{12,4 5} y^2_{35}\,
%F^{(\Psi^2\varphi^2)}(\{t\})
\right)\,,
\end{equation}
and five more contributions corresponding to the six choices of distributing two $\Psi$ operators on the four points.
Above we used the notation $\Psi(i)\equiv \Psi(t_i,y_i)$ and we introduced the R-symmetry invariants 
\begin{equation}
\label{IRsymmdef}
\mathcal{I}_{12,34}=\epsilon_{\alpha \beta} \theta_1^{a\alpha}\theta_2^{b \beta}(\epsilon y_{13}\epsilon y_{34}\epsilon y_{42}\epsilon)_{ab}\,,
\end{equation}
which satisfy the identities
\begin{equation}
\mathcal{I}_{12,34}+\mathcal{I}_{12,43}=
y_{34}^2\,\theta_1y_{12}\theta_2\,,
\quad
\mathcal{I}_{12,34}+\mathcal{I}_{21,43}=0\,,
\end{equation}
and finally $F^{(12),34}_{\Psi \Psi \varphi\varphi}$ and $F^{(12),43}_{\Psi \Psi \varphi\varphi}$ are functions of the two bosonic cross ratios.

We now need to impose supersymmetry via equation \eqref{SinvarianceLplus1correlator}.
The linear term in $\theta$'s of \eqref{SinvarianceLplus1correlator} contains three terms: $\mathbf{s}^+$ acting on \eqref{phi4terminSuperPR5pointfun}, 
 $\mathbf{s}^-$ acting on \eqref{phi2psi2terminSuperPR5pointfun} (together with suitable permutations) and finally a contribution due to the action of $\mathbf{s}^-$ on the $\theta^2$ term of $D^{(\varphi)}$'s acting on  \eqref{phi4terminSuperPR5pointfun}. It turns out that the linear equation obtained in this way fixes the twelve functions in \eqref{phi2psi2terminSuperPR5pointfun} and permutations in terms of the three functions determining the correlator of superprimaries \eqref{phi4terminSuperPR5pointfun}. 
 Here we present the contribution from $F_{\varphi\varphi\varphi\varphi}^{(12)}$
\begin{subequations}
\label{L4primaryFpsipsiphiphi}
\begin{align}
F^{(12),34}_{\Psi \Psi \varphi\varphi}&=
F^{(12),43}_{\Psi \Psi \varphi\varphi}=
\left(\mathfrak{d}^{(2,2)}+2\right)
F_{\varphi\varphi\varphi\varphi}^{(12)}\,,
\\
F^{(34),12}_{\Psi \Psi \varphi\varphi}&=
F^{(34),21}_{\Psi \Psi \varphi\varphi}=
\left(\tfrac{(\chi_1-\chi_2)^2}{\chi_1^2\chi_2^2}\right)^{\h/2}
\left(\mathcal{U}\,\tilde{\mathfrak{d}}
+(2-\h)\right)
F_{\varphi\varphi\varphi\varphi}^{(12)}\,,
\\
F^{(13),24}_{\Psi \Psi \varphi\varphi}&=
\left(\tfrac{(\chi_1-\chi_2)^2}{\chi_1^2(1-\chi_2)^2}\right)^{(\h-2)/2}\,\,\,\,\,\,\chi_1^{-3}\,\,\,\,\,
\,
\mathfrak{d}^{(0,0)}
F_{\varphi\varphi\varphi\varphi}^{(12)}\,,
\\
F^{(14),23}_{\Psi \Psi \varphi\varphi}&=
\left(\tfrac{(\chi_1-\chi_2)^2}{\chi_2^2(1-\chi_1)^2}\right)^{(\h-2)/2}\,\,\,\,\,\,\chi_2^{-3}\,\,\,\,\,
\,
\mathfrak{d}^{(0,0)}
F_{\varphi\varphi\varphi\varphi}^{(12)}\,,
\\
F^{(23),14}_{\Psi \Psi \varphi\varphi}&=
\left(\tfrac{(\chi_1-\chi_2)^2}{\chi_1^2}\right)^{(\h-2)/2}\,\left(\tfrac{1-\chi_1}{\chi_1}\right)^3
\,
\mathfrak{d}^{(0,0)}
F_{\varphi\varphi\varphi\varphi}^{(12)}\,,
\\
F^{(24),13}_{\Psi \Psi \varphi\varphi}&=
\left(\tfrac{(\chi_1-\chi_2)^2}{\chi_2^2}\right)^{(\h-2)/2}\,\left(\tfrac{1-\chi_2}{\chi_2}\right)^3
\,
\mathfrak{d}^{(0,0)}
F_{\varphi\varphi\varphi\varphi}^{(12)}\,,
\\
F^{(13),42}_{\Psi \Psi \varphi\varphi}&=
F^{(14),32}_{\Psi \Psi \varphi\varphi}=
F^{(23),41}_{\Psi \Psi \varphi\varphi}=
F^{(24),31}_{\Psi \Psi \varphi\varphi}=
0\,,
\end{align}
\end{subequations}
the contribution of $F_{\varphi\varphi\varphi\varphi}^{(13)}$ and $F_{\varphi\varphi\varphi\varphi}^{(14)}$ is obtained by applying suitable permutations.
Here we have chosen as independent cross ratios
\begin{equation}
\chi_1=\frac{t_{12}t_{35}}{t_{13}t_{25}}\,,
\qquad
\chi_2=\frac{t_{12}t_{45}}{t_{14}t_{25}}\,,
\end{equation}
and we introduced the differential operators
\begin{equation}
\label{somenotationds}
\mathfrak{d}^{(\alpha,\beta)}:=\tfrac{\chi_1(\chi_1-\alpha)}{2}\partial_{\chi_1}+\tfrac{\chi_2(\chi_2-\beta)}{2}\partial_{\chi_2}\,,
\qquad
\tilde{\mathfrak{d}}:=\tfrac{\chi_1^2\partial_{\chi_1} -\chi_2^2\partial_{\chi_2} }{2}\,,
\qquad
\mathcal{U}=
\tfrac{\chi_1-\chi_2}{\chi_1 \chi_2}\,.
\end{equation}
notice that $\mathfrak{d}^{(0,0)}\,\mathcal{U}=\mathcal{U}\,\mathfrak{d}^{(0,0)}$ and $\tilde{\mathfrak{d}}\,\mathcal{U}=\mathcal{U}\,\tilde{\mathfrak{d}}+1$.

At order $\theta^4$ we have, apart from total time derivatives of previous terms, the new contributions
\begin{subequations}
\begin{align}
\label{ffphiphi}
\langle
\theta^2\! f(1)\,
\theta^2\! f(2) 
\varphi(3)
\varphi(4)
\mathcal{O}(5)
\rangle
&\,=\,
\frac{1}{t_{12}^4 t_{34}^2}
\left(
\frac{t_{34}^2}{t_{35}^2t_{4 5}^2}
\right)^{\h/2}
\mathcal{K}_{12,34}\,
F^{(12)}_{ff \varphi \varphi}
%\mathcal{I}^{12,4 5} y^2_{35}\,
%F^{(\Psi^2\varphi^2)}(\{t\})
\,,
\\
\label{fPsiPsiphi}
\langle
\theta^2\! f(1)\,
\theta \Psi(2) 
\theta \Psi(3)
\varphi(4)
\mathcal{O}(5)
\rangle
&\,=\,
\frac{t_{24} t_{34}}{t_{23}^4 t_{14}^4}
\left(
\frac{t_{34}^2}{t_{35}^2t_{4 5}^2}
\right)^{\h/2}
\mathcal{L}_{1,23,4}\,
F^{1(23)4}_{f\Psi \Psi \varphi}
%\mathcal{I}^{12,4 5} y^2_{35}\,
%F^{(\Psi^2\varphi^2)}(\{t\})
\,,\\
\label{Psi4}
\langle
\theta\Psi(1)
\theta \Psi(2) 
\theta \Psi(3)
\theta \Psi(4)
\mathcal{O}(5)
\rangle
&\,=\,
\frac{1}{t_{12}^3 t_{34}^3}
\left(
\frac{t_{34}^2}{t_{35}^2t_{4 5}^2}
\right)^{\h/2}
\left(
\mathcal{H}^{[1]}_{12,34}\,
F^{[1](12)}_{\Psi \Psi \Psi \Psi }+
\mathcal{H}^{[2]}_{12,34}\,
F^{[2](12)}_{\Psi \Psi \Psi \Psi }
%\mathcal{I}^{12,4 5} y^2_{35}\,
%F^{(\Psi^2\varphi^2)}(\{t\})
\right)+\text{two more}\,,
\end{align}
\end{subequations}
where
\begin{subequations}
\begin{align}
\mathcal{K}_{12,34}&=
\tfrac{1}{4}
\left(
\epsilon_{ab} \theta_1^{a\sigma_1}\theta_1^{b\gamma_1}
\right)
\left(
\epsilon_{cd} \theta_2^{c\sigma_2}\theta_2^{d\gamma_2}
\right)
\epsilon_{\gamma_1\gamma_2}\epsilon_{\sigma_1\sigma_2}\,
y^2_{34}\,,
\\
\label{L1234def}
\mathcal{L}_{1,23,4}&=
\tfrac{1}{2}\epsilon_{\alpha \gamma} \epsilon_{\beta \sigma}\,
\left(\epsilon_{ab} \theta_1^{a\alpha}\theta_1^{b\beta}\right)
\theta_2^{c\gamma}\theta_3^{d\sigma}
\left(\epsilon y_{24} \epsilon y_{34} \epsilon \right)_{cd}\,,
\\
\mathcal{H}^{[1]}_{12,34}&=
\left( y_{12}^{a_1 b_1}\epsilon_{a_1 c_1}\epsilon_{b_1 d_1} \theta_1^{c_1\gamma_1}\theta_2^{d_1\gamma_2}\right)
\left( y_{34}^{a_2 b_2}\epsilon_{a_2 c_2}\epsilon_{b_2 d_2} \theta_3^{c_2\sigma_1}\theta_4^{d_2\sigma_2}\right)
\epsilon_{\gamma_1\gamma_2}\epsilon_{\sigma_1\sigma_2}\,,
\\
\mathcal{H}^{[2]}_{12,34}&=
\left( y_{12}^{a_1 b_1}\epsilon_{a_1 c_1}\epsilon_{b_1 d_1} \theta_1^{c_1\gamma_1}\theta_2^{d_1\gamma_2}\right)
\left( y_{34}^{a_2 b_2}\epsilon_{a_2 c_2}\epsilon_{b_2 d_2} \theta_3^{c_2\sigma_1}\theta_4^{d_2\sigma_2}\right)
\epsilon_{\gamma_1\sigma_2}\epsilon_{\sigma_1\gamma_2}\,.
\end{align}
\end{subequations}
There are six functions from permutations of \eqref{ffphiphi}, twelve functions from permutations of \eqref{fPsiPsiphi} and six functions in \eqref{Psi4}.
Equation \eqref{SinvarianceLplus1correlator} at order $\theta^3$ fixes them all in terms of the three functions in  \eqref{phi4terminSuperPR5pointfun}.
The contribution from $F_{\varphi\varphi\varphi\varphi}^{(12)}$ reads
\begin{subequations}
\label{L4primarytheta4part1}
\begin{align}
F^{(12)}_{ff \varphi \varphi}=& -
\left(\tfrac{1}{3}
\mathfrak{d}^{(0,0)}
\mathfrak{d}^{(0,0)}+
(\mathfrak{d}^{(2,2)}+2)(\mathfrak{d}^{(2,2)}+3)
\right)F_{\varphi\varphi\varphi\varphi}^{(12)}\,,
\\
F^{(34)}_{ff \varphi \varphi}=& 
-\mathcal{U}^h
\left(
\tfrac{1}{3}
\mathfrak{d}^{(0,0)}
\mathcal{U}^2\,
\mathfrak{d}^{(0,0)}
+(\mathcal{U}\,\tilde{\mathfrak{d}}+\h-2 )
(\mathcal{U}\,\tilde{\mathfrak{d}}+\h-3 )
\right)F_{\varphi\varphi\varphi\varphi}^{(12)}\,,
\\
F^{(13)}_{ff \varphi\varphi}=&
-\tfrac{4}{3}
\left(\tfrac{(\chi_1-\chi_2)^2}{\chi_1^2(1-\chi_2)^2}\right)^{(\h-2)/2}\,\,\,\,\,\chi_1^{-4}\,\,\,\,\,
\,\mathfrak{d}^{(0,0)}
\mathfrak{d}^{(0,0)}
F_{\varphi\varphi\varphi\varphi}^{(12)}\,,
\\
F^{(14)}_{ff \varphi\varphi}=&
-\tfrac{4}{3}
\left(\tfrac{(\chi_1-\chi_2)^2}{\chi_2^2(1-\chi_1)^2}\right)^{(\h-2)/2}\,\,\,\,\,\chi_2^{-4}\,\,\,\,\,
\,\mathfrak{d}^{(0,0)}
\mathfrak{d}^{(0,0)}
F_{\varphi\varphi\varphi\varphi}^{(12)}\,,
\\
F^{(23)}_{ff \varphi\varphi}=&
-\tfrac{4}{3}
\left(\tfrac{(\chi_1-\chi_2)^2}{\chi_1^2}\right)^{(\h-2)/2}\,\left(\tfrac{1-\chi_1}{\chi_1}\right)^4
\,\mathfrak{d}^{(0,0)}
\mathfrak{d}^{(0,0)}
F_{\varphi\varphi\varphi\varphi}^{(12)}\,,
\\
F^{(24)}_{ff \varphi\varphi}=&
-\tfrac{4}{3}
\left(\tfrac{(\chi_1-\chi_2)^2}{\chi_2^2}\right)^{(\h-2)/2}\,\left(\tfrac{1-\chi_2}{\chi_2}\right)^4
\,\mathfrak{d}^{(0,0)}
\mathfrak{d}^{(0,0)}
F_{\varphi\varphi\varphi\varphi}^{(12)}\,,
\end{align}
\end{subequations}
%-------------------------------------------------------------------------------------------------------------------------------
\begin{subequations}
\label{L4primarytheta4part2}
\begin{align}
F^{1(23)4}_{f\Psi \Psi \varphi }=& 
\,\tfrac{2}{3}
\left(
\tfrac{(1-\chi_1)^4}{\chi_1 (\chi_1-\chi_2)^3 (1-\chi_2)}\right)
\left(\mathfrak{d}^{(3,3)}+\tfrac{9}{2}\right)
\mathfrak{d}^{(0,0)}F_{\varphi\varphi\varphi\varphi}^{(12)}\,,
\\
F^{1(24)2}_{f\Psi \Psi \varphi }=& 
\,\tfrac{2}{3}
\left(
\tfrac{(1-\chi_2)^4}{\chi_2 (\chi_2-\chi_1)^3 (1-\chi_1)}\right)
\left(\mathfrak{d}^{(3,3)}+\tfrac{9}{2}\right)
\mathfrak{d}^{(0,0)}F_{\varphi\varphi\varphi\varphi}^{(12)}\,,
%%%
\\
F^{1(34)2}_{f\Psi \Psi \varphi }=& 
-\tfrac{4}{3}
\left(
\tfrac{(\chi_1-\chi_2)^2}{\chi_1^2  (1-\chi_2)^2}\right)^{1+h/2}
\left(
\tfrac{\chi_1(1-\chi_2)}{\chi_2 (1-\chi_1)}
\right)
\mathfrak{d}^{(0,0)}
\mathfrak{d}^{(0,0)}F_{\varphi\varphi\varphi\varphi}^{(12)}\,,
\\
F^{2(34)1}_{f\Psi \Psi \varphi }=& 
-\tfrac{4}{3}\,
\left(
\tfrac{(\chi_1-\chi_2)^2}{\chi_1^2}\right)^{1+h/2}\,\,\,\,\,\,\,\,\,
\left(
\tfrac{\chi_1}{\chi_2}
\right)\,\,\,\,\,\,\,\,
\mathfrak{d}^{(0,0)}
\mathfrak{d}^{(0,0)}F_{\varphi\varphi\varphi\varphi}^{(12)}\,,
\\
F^{3(12)4}_{f\Psi \Psi \varphi }=& 
-\tfrac{4}{3}\,
\left(
\tfrac{(\chi_1-\chi_2)^2}{\chi_1^2(1-\chi_2)^2}\right)^{1+h/2}\,\,\,
\left(
1-\chi_2
\right)\,\,\,\,
\mathfrak{d}^{(0,0)}
\mathfrak{d}^{(0,0)}F_{\varphi\varphi\varphi\varphi}^{(12)}\,,
\\
F^{4(12)3}_{f\Psi \Psi \varphi }=& 
-\tfrac{4}{3}\,
\left(
\tfrac{(\chi_1-\chi_2)^2}{\chi_2^2(1-\chi_1)^2}\right)^{1+h/2}\,\,\,\,
\left(
1-\chi_1
\right)\,\,\,\,
\mathfrak{d}^{(0,0)}
\mathfrak{d}^{(0,0)}F_{\varphi\varphi\varphi\varphi}^{(12)}\,,
%%%
\\
F^{2(13)4}_{f\Psi \Psi \varphi }=& 
-\tfrac{4}{3}\,
\left(
\tfrac{(1-\chi_2)^4}{\chi_1 (\chi_1-\chi_2)^3}\right)\,\,\,
\left(\mathfrak{d}^{(3/2,3/2)}+\tfrac{9}{4}\right)
\mathfrak{d}^{(0,0)}F_{\varphi\varphi\varphi\varphi}^{(12)}\,,
\\
F^{2(14)3}_{f\Psi \Psi \varphi }=& 
-\tfrac{4}{3}\,
\left(
\tfrac{(1-\chi_1)^4}{\chi_2 (\chi_2-\chi_1)^3}\right)\,\,\,
\left(\mathfrak{d}^{(3/2,3/2)}+\tfrac{9}{4}\right)
\mathfrak{d}^{(0,0)}F_{\varphi\varphi\varphi\varphi}^{(12)}\,,
\\
F^{3(14)2}_{f\Psi \Psi \varphi }=& 
-\tfrac{2}{3}\left(\tfrac{(\chi_1-\chi_2)}{\chi_1(1-\chi_2)}\right)^{h-3}
\left(
\tfrac{(1-\chi_1)^4}{\chi_1^4(1-\chi_2)^4}\right)
\left(\mathcal{U}\,\tilde{\mathfrak{d}}_2+\tfrac{3}{2}(h-2)\right)
\mathfrak{d}^{(0,0)}F_{\varphi\varphi\varphi\varphi}^{(12)}\,,
\\
F^{3(24)1}_{f\Psi \Psi \varphi }=& 
-\tfrac{2}{3}\left(\tfrac{(\chi_2-\chi_1)}{\chi_1}\right)^{h-3}\,\,\,\,\,
\left(
\tfrac{(1-\chi_2)^4}{\chi_1^4}\right)\,\,\,
\left(\mathcal{U}\,\tilde{\mathfrak{d}}_2+\tfrac{3}{2}(h-2)\right)
\mathfrak{d}^{(0,0)}F_{\varphi\varphi\varphi\varphi}^{(12)}\,,
\\
F^{4(13)2}_{f\Psi \Psi \varphi }=& 
+\tfrac{2}{3}\left(\tfrac{(\chi_1-\chi_2)}{\chi_2(1-\chi_1)}\right)^{h-3}\,
\left(
\tfrac{(1-\chi_2)^4}{\chi_2^4(1-\chi_1^4)}\right)\,\,
\left(\mathcal{U}\,\tilde{\mathfrak{d}}_3+\tfrac{3}{2}(h-2)\right)
\mathfrak{d}^{(0,0)}F_{\varphi\varphi\varphi\varphi}^{(12)}\,,
\\
F^{4(23)1}_{f\Psi \Psi \varphi }=& 
+\tfrac{2}{3}\left(\tfrac{(\chi_2-\chi_1)}{\chi_2}\right)^{h-3}\,\,\,\,
\left(
\tfrac{(1-\chi_1)^4}{\chi_2^4}\right)\,\,\,\,
\left(\mathcal{U}\,\tilde{\mathfrak{d}}_3+\tfrac{3}{2}(h-2)\right)
\mathfrak{d}^{(0,0)}F_{\varphi\varphi\varphi\varphi}^{(12)}\,,
\end{align}
\end{subequations}
%-------------------------------------------------------------------------------------------------------------------------------
\begin{subequations}
\label{L4primarytheta4part3}
\begin{align}
F^{[1](12)}_{\Psi \Psi \Psi  \Psi  }=& 
\left(-\tfrac{1}{3} \mathcal{U}\, \mathfrak{d}^{(0,0)} \mathfrak{d}^{(0,0)}-
(\mathcal{U}\,\tilde{\mathfrak{d}}+\h-2)(\mathfrak{d}^{(2,2)}+2)\right)
F_{\varphi\varphi\varphi\varphi}^{(12)}\,,
\\
F^{[2](12)}_{\Psi \Psi \Psi  \Psi  }=& 
\tfrac{2}{3} \mathcal{U}\, \mathfrak{d}^{(0,0)} \mathfrak{d}^{(0,0)}
F_{\varphi\varphi\varphi\varphi}^{(12)}\,,
\\
F^{[1](13)}_{\Psi \Psi \Psi  \Psi  }=& 
F^{[2](13)}_{\Psi \Psi \Psi  \Psi  }=
+\tfrac{2}{3}\left(\tfrac{(\chi_1-\chi_2)^2}{\chi_1^2(1-\chi_2)^2}\right)^{h/2-1}\,
\left(\tfrac{(1-\chi_2)}{\chi_1^3\,\chi_2}\right)
\mathfrak{d}^{(0,0)} \mathfrak{d}^{(0,0)}
F_{\varphi\varphi\varphi\varphi}^{(12)}\,,
\\
F^{[1](14)}_{\Psi \Psi \Psi  \Psi  }=& 
F^{[2](14)}_{\Psi \Psi \Psi  \Psi  }=
+\tfrac{2}{3}\left(\tfrac{(\chi_1-\chi_2)^2}{\chi_2^2(1-\chi_1)^2}\right)^{h/2-1}\,
\left(\tfrac{(1-\chi_1)}{\chi_2^3\,\chi_1}\right)
\mathfrak{d}^{(0,0)} \mathfrak{d}^{(0,0)}
F_{\varphi\varphi\varphi\varphi}^{(12)}\,,
\end{align}
\end{subequations}
%-------------------------------------------------------------------------------------------------------------------------------
where, as in \eqref{L4primaryFpsipsiphiphi}, we displayed only the contribution of $F_{\varphi\varphi\varphi\varphi}^{(12)}$ and the differential operators are defined in \eqref{somenotationds} together with 
\begin{equation}
\tilde{\mathfrak{d}}_2=\tfrac{1}{2}\chi_1^2\partial_{\chi_1} -\chi_2^2\partial_{\chi_2}\,,
\qquad
\tilde{\mathfrak{d}}_3=\chi_1^2\partial_{\chi_1} -\tfrac{1}{2}\chi_2^2\partial_{\chi_2}\,.
\end{equation}
To complete the discussion one would need to determine, by looking at higher orders in the $\theta$ expansion of \eqref{SinvarianceLplus1correlator}, the components
\begin{equation}
\label{notcomputedcomponentswithsuperprimary}
\langle
\theta^2\! f(1)\,
\theta^2\! f(2) 
\theta\Psi(3)
\theta\Psi(4)
\mathcal{O}(5)
\rangle\,,
\qquad
\langle
\theta^2\! f(1)\,
\theta^2\! f(2) 
\theta^2\! f(3)
\theta^2\! f(4)
\mathcal{O}(5)
\rangle\,,
\end{equation}
which are the only non-vanishing ones due to R-symmetry.
Since we will not need them here we will not compute them.
What is clear is that they can be expressed in terms of the three functions in \eqref{phi4terminSuperPR5pointfun}.
This is because we can set, for example, $\theta_3=\theta_4=0$ and $\Theta=0$ with a superconformal transformation.

%---------------------------------------------------------------------------------------------------------
\subsubsection{Constraints on the five-point function with a superdescendant}
\label{subsec:constraint5ptsuperdesc}

Let us consider the five-point function
\begin{equation}
\label{G5desc}
G^{(5)}_{\widetilde{\mathcal{O}}}:=
\langle
\Phi(X_1) 
\Phi(X_2) 
\Phi(X_3) 
\Phi(X_4) 
\widetilde{\mathcal{O}}(t_5,y_5)
\rangle\,,
\end{equation}
where $\widetilde{\mathcal{O}}(t_5,y_5)$ is the descendant of a long  $\mathcal{L}_{s=0,[0,0]}^{\h}$ supermultiplet defined in \eqref{OandOtildereported}.
The goal of this section is to show that all the components of \eqref{G5desc} can be obtained  in terms of the three functions entering the correlator of superprimaries given in \eqref{phi4terminSuperPR5pointfun}. The explicit expressions in the parametrization given in \eqref{5pointsuperdescphi4comp}, \eqref{PsiPsiphiphitildeOcomponent} and \eqref{moretildeOcomponentstheta4}  can be found in \eqref{FtildefromF} and the $\mathtt{Mathematica}$ file included with the arXiv submission. 

\paragraph{The $\langle
\varphi
\varphi
\varphi
\varphi 
\widetilde{\mathcal{O}}
\rangle$ component.} Let us start from the component obtained by setting all the $\theta_i$'s to zero in \eqref{G5desc}, namely
\begin{equation}
\label{5pointsuperdescphi4comp}
\langle
\varphi(1)
\varphi(2)
\varphi(3)
\varphi(4) 
\widetilde{\mathcal{O}}(5)
\rangle
\,=\,\sum_{1\le i<j\le 4} \frac{y^2_{ij}}{t_{ij}^2}
 \frac{y^2_{k5}}{t_{k5}^2}
 \frac{y^2_{\ell5}}{t_{\ell5}^2}
\left(
\frac{t_{k\ell}^2}{t_{k5}^2t_{\ell 5}^2}
\right)^{\h/2}
\,\tilde{F}_{\varphi\varphi\varphi\varphi}^{(ij)}
%(\tfrac{t_{ij}t_{k 5}}{t_{ik}t_{j5}},\tfrac{t_{ij}t_{\ell 5}}{t_{i\ell}t_{j5}})
\,,
\end{equation}
where $\varphi(i)\equiv\varphi(t_i,y_i)$ and  $\widetilde{\mathcal{O}}(5)\equiv\widetilde{\mathcal{O}}(t_5,y_5)$. The sum in \eqref{5pointsuperdescphi4comp} contains six terms associated to the six singlets in the tensor product  $[0,1]^{\otimes 4}\otimes [0,2]$, see Table \ref{tab:lsingletsforcorrelatorwithThildeO}, in contrast with the correlator involving only superprimaries \eqref{phi4terminSuperPR5pointfun} which contains three terms. These six functions of two cross ratios can be extracted in the following way. First, one notices that super-translation invariance immediately implies
\begin{equation}
\label{G5withaLONGfullBIS}
G^{(5)}_{\mathscr{O}}
:=\langle
\Phi(X_1) 
\Phi(X_2) 
\Phi(X_3) 
\Phi(X_4) 
\mathscr{O}(\mathbf{t}_5)
\rangle\,=
\langle
\Phi(\widetilde{X}_1) 
\Phi(\widetilde{X}_2) 
\Phi(\widetilde{X}_3) 
\Phi(\widetilde{X}_4) 
\mathscr{O}(\mathbf{0})
\rangle\,,
\end{equation}
where $\widetilde{X}_i$ are defined as in \eqref{Xtildeidef} with $\mathbf{t}_3$  replaced by $\mathbf{t}_5$ and $\mathscr{O}(\mathbf{0})=\mathcal{O}(0)$.
From the correlator \eqref{G5withaLONGfullBIS}  we can extract \eqref{5pointsuperdescphi4comp} by setting the fermionic variables $\theta$ of the short operators to zero and using the definition \eqref{Ottildecomponent} with the action of the supersymmetry generators on long operators given in \eqref{GeneratorsLONG}. 
Using the splitting \eqref{Theta_updown} and conventions from Appendix \ref{app:notation}, this results in\footnote{Here we have set $y_5=0$ and $\Theta_{5,u}=0$ for convenience. The general expression of $\QforOtilde(y)$ acting on a long supermultiplet is 
\begin{equation}
\QforOtilde^a_{\alpha}(y)=
\Big{(}
\frac{\partial}{\partial (\Theta_d)^{\alpha}_a}-y^{ab} \frac{\partial}{\partial (\Theta_u)^{\alpha b}}
\Big{)}
+\tfrac{1}{2}\epsilon_{\alpha \beta}
\Big{(}
(\Theta_u)^{\beta a}+
y^{ab}(\Theta_d)^{\beta}_b
\Big{)}
\frac{\partial}{\partial t}\,.
\end{equation}
}
\begin{equation}
\label{extractingphi4OtildeFromG5insuper}
\langle
\varphi(1)
\varphi(2)
\varphi(3)
\varphi(4) 
\widetilde{\mathcal{O}}(t_5,0)
\rangle
\,=\,
\delta^{(0|4)}(\partial_{\Theta_d})
G^{(5)}_{\mathscr{O}}\Big{|}_{\theta_i=0,\Theta_{5,u}=0,y_5=0}\,.
\end{equation}
Thanks to the analysis of Section \ref{subsec:constraint5ptsuperpr} and the equality in \eqref{G5withaLONGfullBIS}, we know $G^{(5)}_{\mathscr{O}}$.
More precisely, the right hand side of \eqref{G5withaLONGfullBIS} is expressed, upon using the solution of the constraints on $\Phi$ given in $\eqref{Phisolveconstraint}$, in terms of the components \eqref{phi4terminSuperPR5pointfun}, \eqref{phi2psi2terminSuperPR5pointfun}, \eqref{ffphiphi}, \eqref{fPsiPsiphi}, \eqref{Psi4} and \eqref{notcomputedcomponentswithsuperprimary}, which in turns are expressed in terms of the three functions in  \eqref{phi4terminSuperPR5pointfun}. See \eqref{L4primaryFpsipsiphiphi}, \eqref{L4primarytheta4part1}, \eqref{L4primarytheta4part2}, \eqref{L4primarytheta4part3} 
for the contribution of $F_{\varphi\varphi\varphi\varphi}^{(12)}$.
The final ingredient to compute \eqref{extractingphi4OtildeFromG5insuper} is the expression of $\widetilde{X}_i$ when all fermionic variables but $\Theta_{5,d}$ are set to zero.
Using the definition \eqref{Xtildeidef} with $\mathbf{t}_3$  replaced by $\mathbf{t}_5$ we get
\begin{equation}
\widetilde{X}_i\big{|}_{\theta_i=0,\Theta_{5,u}=0}=
\begin{pmatrix} 
\epsilon\,t_{i5}-\Theta_{5,d}^{} y_i\Theta_{5,d}^t  &\,\,\,\, -\Theta_{5,d}y_i \\
y_i\Theta_{5,d}^t&\,\, y_i
\end{pmatrix}\,.
\end{equation}
This means that in order to obtain the correlator \eqref{extractingphi4OtildeFromG5insuper} we need to start from the correlator \eqref{G5othercomp},
make the substitution $t_i\mapsto t_i+\Theta_{5,d}y_i\Theta_{5,d}^t$, $i=1,\dots,4$, $\theta_i\mapsto -\Theta_{5,d}y_i$ and extract the $(\Theta_{5,d})^4$ component.
This procedure gives the relations
%---------------------------------------------------------
\begin{subequations}
\label{FtildefromF}
\begin{align}
\tilde{F}_{\varphi\varphi\varphi\varphi}^{(12)}&=
D_1D_2\,F_{\varphi\varphi\varphi\varphi}^{(12)}\,,
\\
\tilde{F}_{\varphi\varphi\varphi\varphi}^{(13)}&=
\left(\tfrac{(\chi_1-\chi_2)^2}{\chi_1^2(1-\chi_2)^2}\right)^{h/2}
\left(
2(\chi_1-1)+D_1
\right)\partial_{\chi_1}F_{\varphi\varphi\varphi\varphi}^{(12)}\,,
\\
\tilde{F}_{\varphi\varphi\varphi\varphi}^{(14)}&=
\left(\tfrac{(\chi_1-\chi_2)^2}{\chi_2^2(1-\chi_1)^2}\right)^{h/2}
\left(
2(\chi_2-1)+D_1
\right)\partial_{\chi_2}F_{\varphi\varphi\varphi\varphi}^{(12)}\,,
\\
\tilde{F}_{\varphi\varphi\varphi\varphi}^{(23)}&=
(1-\chi_1)^2
\left(\tfrac{(\chi_1-\chi_2)^2}{\chi_1^2}\right)^{h/2}
\left(
1-D_2
\right)\partial_{\chi_1}F_{\varphi\varphi\varphi\varphi}^{(12)}\,,
\\
\tilde{F}_{\varphi\varphi\varphi\varphi}^{(24)}&=
(1-\chi_2)^2
\left(\tfrac{(\chi_1-\chi_2)^2}{\chi_2^2}\right)^{h/2}
\left(
1-D_2
\right)\partial_{\chi_2}F_{\varphi\varphi\varphi\varphi}^{(12)}\,,
\\
\tilde{F}_{\varphi\varphi\varphi\varphi}^{(34)}&=
(\chi_1-\chi_2)^2
\left(\tfrac{(\chi_1-\chi_2)^2}{\chi_1^2\chi_2^2}\right)^{h/2}
\partial_{\chi_1}
\partial_{\chi_2}F_{\varphi\varphi\varphi\varphi}^{(12)}\,,
\end{align}
\end{subequations}
where
\begin{equation}
D_1:=
\h+1+2\,\mathfrak{d}^{(1,1)} \,,
\qquad
D_2:=
h-2\,\mathfrak{d}^{(0,0)}\,,
\end{equation}
and the differential operators $\mathfrak{d}^{(\alpha,\beta)}$ are defined in \eqref{somenotationds}.

\paragraph{The remaining components from \eqref{QyinvarianceLplus1correlator}.} To obtain the remaining components of the correlator \eqref{G5desc}, we impose the Q-invariance condition\footnote{It should be remarked that Q-invariance imposes conditions on the six functions entering \eqref{5pointsuperdescphi4comp} as well. They are automatically solved by \eqref{FtildefromF}.}  \eqref{QyinvarianceLplus1correlator}.
The analysis here follows similar steps to the one presented in Section \ref{subsec:constraint5ptsuperpr}: we express $\Phi(X)$ in terms of its components  $\varphi, \Psi, f$  using the parametrization  $\eqref{Phisolveconstraint}$, we write the most general correlators for $\{\varphi, \Psi, f\}$ with $\widetilde{\mathcal{O}}$ with the correct conformal and R-symmetry covariance (see Table \ref{tab:lsingletsforcorrelatorwithThildeO}) and, finally, we impose $ \eqref{QyinvarianceLplus1correlator}$ in a $\theta$ expansion. In this way we obtain the correlators of the various components in terms of the three functions in \eqref{phi4terminSuperPR5pointfun}.
At order $\theta^2$ we have the new components
%
%------------------
\begin{align}
\label{PsiPsiphiphitildeOcomponent}
\begin{split}
\langle
\theta\Psi(1)
\theta\Psi(2) 
\varphi(3)
\varphi(4)
\widetilde{\mathcal{O}}(5)
\rangle
\,=\,\qquad\qquad
\qquad\qquad\qquad\qquad\qquad
\\
=
\frac{1}{t_{12}^3 t_{34}^2}
\left(
\frac{t_{34}^2}{t_{35}^2t_{4 5}^2}
\right)^{(\h+2)/2}
\left(
\left(
\mathcal{I}^{}_{12,35}\, y^2_{45}\,
\tilde{F}^{(12)34}_{\Psi \Psi \varphi \varphi}+(3 \leftrightarrow 4)
\right)
+
(\theta_1  y_{12} \theta_2)\,
  y^2_{35} y^2_{45}\,
\tilde{F}^{(12)}_{\Psi \Psi \varphi \varphi}
\right)\,,
\end{split}
\end{align}
where $\mathcal{I}_{12,34}$ is defined in \eqref{IRsymmdef}.
At order $\theta^4$ we have the new components
%------------------
\begin{subequations}
\label{moretildeOcomponentstheta4}
\begin{align}
\langle
\theta^2\!f(1)
\theta^2\!f(2) 
\varphi(3)
\varphi(4)
\widetilde{\mathcal{O}}(5)
\rangle
\,=\,&
\frac{1}{t_{12}^4 t_{34}^2}
\left(
\frac{t_{34}^2}{t_{35}^2t_{4 5}^2}
\right)^{(\h+2)/2}
\,
\theta_1^2\theta_2^2
\,y_{35}^2 y_{45}^2\,
\tilde{F}^{(12)}_{ff \varphi \varphi}\,,\\
%--------
\langle
\theta^2\! f(1)\,
\theta \Psi(2) 
\theta \Psi(3)
\varphi(4)
\widetilde{\mathcal{O}}(5)
\rangle
\,=\,&
\frac{t_{24} t_{34}}{t_{23}^4 t_{14}^4}
\left(
\frac{t_{34}^2}{t_{35}^2t_{4 5}^2}
\right)^{(\h+2)/2}
\mathcal{L}_{1,23,5}\,y^2_{45}\,
\tilde{F}^{1(23)4}_{f\Psi \Psi \varphi}\,,\\
%--------------
%Are the three structures here just (?)
%\begin{equation}
%\mathcal{J}_{12,5}\mathcal{J}_{34,5}
%\qquad
%\mathcal{J}_{13,5}\mathcal{J}_{24,5}
%\qquad
%\mathcal{J}_{14,5}\mathcal{J}_{23,5}
%\end{equation}
\langle
\theta\Psi(1)
\theta \Psi(2) 
\theta \Psi(3)
\theta \Psi(4)
\widetilde{\mathcal{O}}(5)
\rangle
\,=\,& 
\frac{1}{t_{12}^3 t_{34}^3}
\left(
\frac{t_{34}^2}{t_{35}^2t_{4 5}^2}
\right)^{(\h+2)/2}\!
\mathcal{J}_{12,5}\mathcal{J}_{34,5}\,
\tilde{F}^{(12)}_{\Psi \Psi \Psi \Psi }
%\mathcal{I}^{12,4 5} y^2_{35}\,
%F^{(\Psi^2\varphi^2)}(\{t\})
+\text{two more}\,\,
\nonumber
\\
& \qquad\qquad \qquad+\,\,\,
\frac{1}{t_{12}^3 t_{34}^3}
\left(
\frac{t_{34}^2}{t_{35}^2t_{4 5}^2}
\right)^{(\h+2)/2}\,\widehat{\mathcal{J}}\,
\tilde{F}_{\Psi \Psi \Psi \Psi }\,,
\label{psipsipsipsiOtildeF}
\end{align}
\end{subequations}
where again ``two more'' refers to two copies of the first term which can be obtained from the first term on the l.h.s. of \eqref{psipsipsipsiOtildeF}  by replacing $(2\leftrightarrow 3)$ and $(2\leftrightarrow 4)$, respectively. The combination $\mathcal{L}_{1,23,4}$ was defined  in \eqref{L1234def} and
\begin{equation}
\label{J123def}
\mathcal{J}_{12,3}=
\epsilon_{\alpha \beta}\theta_1^{a\alpha}\theta_2^{b\beta}
(\epsilon y_{13}\epsilon y_{23 }\epsilon)_{ab}\,,
\end{equation}
with $\mathcal{J}_{12,3}+\mathcal{J}_{21,3}=0$.
Finally, $\widehat{\mathcal{J}}$, which is totally symmetric under permutations of the first four points is defined in \eqref{Jhatdef}.
It is easy to see that all other possible component contributions to  \eqref{G5desc} vanish due to R-symmetry.
The expressions for the functions entering \eqref{PsiPsiphiphitildeOcomponent} and \eqref{moretildeOcomponentstheta4} in terms of the functions $F^{(12)}_{\varphi\varphi\varphi\varphi}$, $F^{(13)}_{\varphi\varphi\varphi\varphi}$, $F^{(14)}_{\varphi\varphi\varphi\varphi}$ are collected in the $\mathtt{Mathematica}$ file included with the arXiv submission.

\begin{table}[!ht]
\centering
 \begin{tabular}{| c || c | c|c| c|c|c|c|c|} 
\hline
Correlator
 & $\mathfrak{sp}(4)$ tensor product & $\mathfrak{sp}(4)$  singlets   &  $\mathfrak{su}(2)$  singlets    & number of choices      \\
   \hline
   \hline
    $\langle \varphi \varphi \varphi \varphi \widetilde{\mathcal{O}} \rangle $
     &  $[0,1]^{\otimes 4}\otimes [0,2]$   & $6$ &  $1$ & $1$ 
      \\
   \hline
   $\langle\Psi \Psi\varphi \varphi \widetilde{\mathcal{O}} \rangle $ 
   & $[1,0]^{\otimes 2}\otimes[0,1]^{\otimes 2}\otimes [0,2]$  & $3$ &  $1$ & $6$
 \\
   \hline
      $\langle ff\varphi \varphi \widetilde{\mathcal{O}} \rangle $ 
   & $[0,1]^{\otimes 2}\otimes [0,2]$  & $1$ &  $1$ & $6$
 \\
   \hline
         $\langle f\Psi\Psi \varphi \widetilde{\mathcal{O}} \rangle $ 
   & $[1,0]^{\otimes 2}\otimes [0,1]\otimes [0,2]$  & $1$ &  $1$ & $12$
 \\
   \hline
            $\langle \Psi\Psi\Psi \Psi \widetilde{\mathcal{O}} \rangle $ 
   & $[1,0]^{\otimes 4}\otimes [0,2]$  & $2$ &  $2$ & $1$
 \\
   \hline
\end{tabular}
\caption{The total number of functions is $6+3\times 6+6 + 12+2\times 2=46$. }
\label{tab:lsingletsforcorrelatorwithThildeO}
\end{table}

\subsubsection{Building blocks for $L=4$ superprimary $\mathcal{O}$}
\label{sec:operatorsforthesuperprimary}

We write a given $L=4$ superprimary $\mathcal{O}(t)$ as a linear combination of conformal primaries 
\begin{align}
\label{Opieces}
\mathcal{O}(t)\,\,&=
\mathcal{O}_{\varphi^4}
+
\mathcal{O}_{\Psi^2\varphi^2}
+
  {\color{red} \mathcal{O}_{\Psi^4}}
+
\mathcal{O}_{f^2\varphi^2}
+
\mathcal{O}_{f\Psi^2\varphi}
+
 \color{blue}{\mathcal{O}_{f^2\Psi^2}}
+
\color{blue}{\mathcal{O}_{f^4}}\,.
\end{align}
To compute the correlators that are needed in \cite{Ferrero:2023gnu}, we will not need the explicit form of the terms in {\color{red} red}  and {\color{blue} blue} in equation \eqref{Opieces}, so we will not determine these terms. It follows from the analysis of the five-point function that once the piece $\mathcal{O}_{\varphi^4}$ is chosen (and this choice corresponds to picking a vector in the degeneracy space, whose dimension is given in \eqref{NstrongL4000}) all the other pieces are fixed in terms of this choice.
We will now discuss a choice for $\mathcal{O}_{\varphi^4}$ and  how the remaining parts are fixed.
Notice that below we dropped the $(0)$ from $\langle \dots \rangle$, but all correlators in this section are computed in the free theory by Wick contractions.

\paragraph{The $\mathcal{O}_{\varphi^4}$ component.}
Let us define the following composite operators
\begin{equation}
\label{OL4phi4operators}
\!\!\mathcal{O}_{\varphi^4}^{\{K,n_1,n_2\}}(t)=
\mathsf{D}_K^{(2+2n_1,2+2n_2)}(\partial_1+\partial_2,\partial_3+\partial_4)
\mathsf{D}_{2n_1}^{(1,1)}(\partial_1,\partial_2)\mathsf{D}_{2n_2}^{(1,1)}(\partial_3,\partial_4)
S_{\varphi^4}(t_1,t_2,t_3,t_4)\Big{|}_{t_i=t}
\,,
%(2n+1)!
%\frac{1}{2}\frac{1}{(2n+1)!}
\end{equation}
where\footnote{The letter $S$ is chosen to remind that these are R-symmetry singlets.} 
\begin{equation}
S_{\varphi^4}(t_1,t_2,t_3,t_4)\,= \, :(\varphi^2)^{[0,0]}(t_1,t_2)(\varphi^2)^{[0,0]}(t_3,t_4):
\end{equation}
and $(\varphi^2)^{[0,0]}$ is given in\footnote{Notice that the projector $P=\mathsf{P}^{[0,1][0,1]}_{[0,0]}$ entering \eqref{phi2singlet} is normalized in such a way that $P_{12} y^2_{13}y^2_{24}=y^2_{34}$. With this choice of normalization $P_{12} y^2_{12}=5$.
} \eqref{phi2singlet}.
The operators \eqref{OL4phi4operators} provide a basis for the primary operators upon requiring (from Bose symmetry) that $n_1\geq n_2$ and $K$ even when $n_1=n_2$. They have conformal dimension $\h=4+2n_1+2n_2+K$ and, with this range of labels, they are linearly independent.
To compute the five-point function \eqref{phi4terminSuperPR5pointfun} for the operator \eqref{OL4phi4operators} we will act with the appropriate differential operators on the following generating correlator
%\begin{subequations}
\begin{align}
\label{phiphiphiphiS}
\begin{split}
%\begin{align}
\langle \varphi(1)\varphi(2)\varphi(3)\varphi(4)
S_{\varphi^4}(t_5,t_6,t_7,t_8)\rangle =
\qquad
\qquad
\qquad
\qquad
\qquad
&
\\
=
y^2_{12}y^2_{34}\,
\left(
\left(\tfrac{1}{t_{15}^2 t_{26}^2}
+
\tfrac{1}{t_{16}^2 t_{25}^2}
\right)
\left(\tfrac{1}{t_{37}^2 t_{48}^2}
+
\tfrac{1}{t_{38}^2 t_{47}^2}
\right)
+
\left(\tfrac{1}{t_{17}^2 t_{28}^2}
+
\tfrac{1}{t_{18}^2 t_{27}^2}
\right)
\left(\tfrac{1}{t_{35}^2 t_{46}^2}
+
\tfrac{1}{t_{36}^2 t_{45}^2}
\right)
\right)&+\text{two more}\,,
\end{split}
\end{align}
where the ``two more'' terms are obtained exchanging point 2 with 3 and 2 with 4 respectively.
%\end{align}
%\end{subequations}
%
The correlator $\langle S S \rangle $ is obtained from \eqref{phiphiphiphiS} by further applying the projector $(\mathsf{P}_{12})^{[0,1][0,1]}_{[0,0]}(\mathsf{P}_{34})^{[0,1][0,1]}_{[0,0]}$, which boils down to the replacement
\begin{equation}
y^2_{12}y^2_{34}\mapsto 25\,,
\quad
y^2_{13}y^2_{24}\mapsto 5\,,
\quad
y^2_{14}y^2_{23}\mapsto 5\,.
\end{equation}
In Appendix \ref{app:actingwithderivatives} we explain  how to compute the correlators
\begin{equation}
\label{5pointCOMPUTEvarphiOKn1n2MainTEXT}
\langle \varphi(1) \varphi(2)  \varphi(3)  \varphi(4)
\mathcal{O}_{\varphi^4}^{\{K,n_1,n_2\}}(t_5)\rangle\,,
\end{equation}
 from \eqref{phiphiphiphiS} and the definition  \eqref{OL4phi4operators}.
The explicit result takes the form \eqref{phi4terminSuperPR5pointfun} with $F^{(12)}_{\varphi\varphi\varphi \varphi}$ given in \eqref{F12fromAA}, while the remaining $F^{(13)}_{\varphi\varphi\varphi \varphi}$ and $F^{(14)}_{\varphi\varphi\varphi \varphi}$ can be obtained by Bose symmetry.
We will not present a closed formula for the two-point function
\begin{equation}
\langle 
\mathcal{O}_{\varphi^4}^{\{K,n_1,n_2\}}(t_1)\mathcal{O}_{\varphi^4}^{\{K',n_1',n_2'\}}(t_2)\rangle\,,
\end{equation}
but for fixed labels it can be obtained by applying the appropriate R-symmetry projectors and differential operators in \eqref{OL4phi4operators} to the 5-point function \eqref{5pointCOMPUTEvarphiOKn1n2MainTEXT}.

%-------------------------------------------------------------------------------
\paragraph{The $\mathcal{O}_{\Psi^2 \varphi^2}$ component.}
In this case, extracting the R-symmetry singlets requires a bit more work. 
First we decompose the bilinears $:\Psi  \Psi :$ and $:\varphi \varphi :$ according to the rules\footnote{Recall that our conventions for Dynkin labels are such that $[1,0]=\mathbf{4}$, $[0,1]=\mathbf{5}$.} 
\begin{align}
([1,0]\otimes [1,0])_S
&=[2,0]
\qquad
\qquad
\,\,\,\,\,\,\,\,\,\,\,
([1,0]\otimes [1,0])_A=[0,1]\oplus [0,0]\,,
\\
([0,1]\otimes [0,1])_S
&=[0,2]\oplus[0,0]\,,
\,\,\,\,\,\,\,\,\,\,\,\,
([0,1]\otimes [0,1])_A=[2,0]\,.
\end{align}
This implies that there are two R-symmetry singlets of the schematic form $\Psi^2\varphi^2$: one, denoted by $S^{(1)}$, is anti-symmetric in the exchange of $t_1$ with $t_2$ and symmetric in the exchange of $t_3$ with $t_4$ while the other, which we call $S^{(2)}$, has opposite symmetry properties. Their explicit expressions are given by 
\begin{align}
S^{(1)}_{\Psi^2 \varphi^2}(t_1,t_2,t_3,t_4):&=(\Psi^2)^{[0,0]}(t_1,t_2) (\varphi^2)^{[0,0]}(t_3,t_4)\,,
\\
S^{(2)}_{\Psi^2 \varphi^2}(t_1,t_2,t_3,t_4):&=
%P^{[2,0]}
\left(
\mathsf{P}_{12}\right)^{[2,0],[2,0]}_{[0,0]}
(\Psi^2)^{[2,0],0}(t_1,t_2;y_1,\ww_1)(\varphi^2)^{[2,0]}(t_2,t_4;y_2,\ww_2)\,,
\end{align}
where the field bilinears as well as the projectors $\mathsf{P}$ are collected in Appendix \ref{app:Rsymmprojectors}.
From these definitions we compute the relevant generating correlators as
\begin{subequations}
\label{generatingcorrpsi2phi2O}
\begin{align}
\langle \theta\Psi(1) \theta \Psi(2)  \varphi(3)\varphi(4)\,
S^{(1)}_{\Psi^2 \varphi^2}(t_5,t_6,t_7,t_8)\rangle 
&=
\left(\tfrac{1}{t_{15}^3 t_{26}^3}
-
\tfrac{1}{t_{16}^3 t_{25}^3}
\right)
\left(\tfrac{1}{t_{37}^2 t_{48}^2}
+
\tfrac{1}{t_{38}^2 t_{47}^2}
\right)\,A
\,,
\\
\langle \theta\Psi(1) \theta \Psi(2)  \varphi(3)\varphi(4)\,
S^{(2)}_{\Psi^2 \varphi^2}(t_5,t_6,t_7,t_8)\rangle 
&=
\left(\tfrac{1}{t_{15}^3 t_{26}^3}
+
\tfrac{1}{t_{16}^3 t_{25}^3}
\right)
\left(\tfrac{1}{t_{37}^2 t_{48}^2}
-
\tfrac{1}{t_{38}^2 t_{47}^2}
\right)B\,,
\end{align}
\end{subequations}
with 
\begin{equation}
A=-6\,
\theta_{1}y_{12}\theta_{2}\,y_{34}^2\,,\qquad
B= 2\,(\mathcal{I}_{12,34}-\mathcal{I}_{12,43})\,,
\end{equation}
where $\mathcal{I}$ is defined in \eqref{IRsymmdef}.
Notice that $A$ and $B$ are linear combinations of the two  R-symmetry structures appearing in \eqref{phi2psi2terminSuperPR5pointfun}, as they should.
Projecting \eqref{generatingcorrpsi2phi2O} onto the appropriate R-symmetry singlets we obtain the generating correlators for two-point functions:
\begin{align}
\langle
S^{(1)}_{\Psi^2 \varphi^2}(t_1,t_2,t_3,t_4)\,
S^{(1)}_{\Psi^2 \varphi^2}(t_5,t_6,t_7,t_8)\rangle 
&=-360\,
\left(\tfrac{1}{t_{15}^3 t_{26}^3}
-
\tfrac{1}{t_{16}^3 t_{25}^3}
\right)
\left(\tfrac{1}{t_{37}^2 t_{48}^2}
+
\tfrac{1}{t_{38}^2 t_{47}^2}
\right)\,,
\\
\langle
S^{(1)}_{\Psi^2 \varphi^2}(t_1,t_2,t_3,t_4)\,
S^{(2)}_{\Psi^2 \varphi^2}(t_5,t_6,t_7,t_8)\rangle 
&=0\,,
\\
\langle
S^{(2)}_{\Psi^2 \varphi^2}(t_1,t_2,t_3,t_4)\,
S^{(2)}_{\Psi^2 \varphi^2}(t_5,t_6,t_7,t_8)\rangle 
&=\,\,2560
\left(\tfrac{1}{t_{15}^3 t_{26}^3}
+
\tfrac{1}{t_{16}^3 t_{25}^3}
\right)
\left(\tfrac{1}{t_{37}^2 t_{48}^2}
-
\tfrac{1}{t_{38}^2 t_{47}^2}
\right)\,.
\end{align}
Furthermore, starting from $S^{(1)}_{\Psi^2 \varphi^2}$ and $S^{(2)}_{\Psi^2 \varphi^2}$ we can construct the conformal primaries of scaling dimension $6+K+2n_1+2n_2$, whose expressions are
\begin{subequations}
\label{psi2phi2confpri}
\begin{align}
\mathcal{O}_{\Psi^2 \varphi^2}^{\{K,n_1,n_2\},(1)}:=
&
\mathsf{D}_K^{(4+2n_1,2+2n_2)}(\partial_1+\partial_2,\partial_3+\partial_4)
\mathsf{D}_{2n_1+1}^{(3/2,3/2)}(\partial_1,\partial_2)\mathsf{D}_{2n_2}^{(1,1)}(\partial_3,\partial_4)
S^{(1)}_{\Psi^2 \varphi^2}\Big{|}_{t_i=t}\,,
\\
\mathcal{O}_{\Psi^2 \varphi^2}^{\{K,n_1,n_2\},(2)}:=
&
\mathsf{D}_K^{(3+2n_1,3+2n_2)}(\partial_1+\partial_2,\partial_3+\partial_4)
\mathsf{D}_{2n_1}^{(3/2,3/2)}(\partial_1,\partial_2)\mathsf{D}_{2n_2+1}^{(1,1)}(\partial_3,\partial_4)
S^{(2)}_{\Psi^2 \varphi^2}\Big{|}_{t_i=t}
\end{align}
\end{subequations}
where we have taken into account the restrictions coming from Bose/Fermi symmetry to have non-vanishing operators.Now that we have built all the conformal primaries of the form $\partial^{\#} \Psi^2 \varphi^2$ which are R-symmetry singlets, we have to determine which ones can appear in \eqref{Opieces}, contributing to the completion of $\partial^{\#}  \varphi^4$ in  \eqref{OL4phi4operators} to a superconformal primary. This is done by computing \eqref{phi2psi2terminSuperPR5pointfun}
and requiring that it satisfies \eqref{L4primaryFpsipsiphiphi} with $F_{\varphi\varphi\varphi\varphi}^{(12)}$ fixed\footnote{Recall that  \eqref{L4primaryFpsipsiphiphi}  contains only the contribution from $F_{\varphi\varphi\varphi\varphi}^{(12)}$ and we should add the contributions of $F_{\varphi\varphi\varphi\varphi}^{(13)}$ and $F_{\varphi\varphi\varphi\varphi}^{(14)}$ as well.} by the choice \eqref{OL4phi4operators}. In Appendix \ref{app:actingwithderivatives} we present a closed formula for
\begin{equation}
\langle \theta\Psi(1) \theta \Psi(2)  \varphi(3)\varphi(4)\,
\mathcal{O}_{\Psi^2 \varphi^2}^{\{K,n_1,n_2\},(i)}(t_5)\rangle\,,
\qquad i=1,2\,,
\end{equation}
which facilitates such computation.

%-------------------------------------------------------------------------------
\paragraph{The $\mathcal{O}_{\Psi^4}$ component.}
In this case there are six terms. As discussed after equation, \eqref{Opieces} for our purposes we will not use the explicit form of this operator so we will not derive it here.
%-------------------------------------------------------------------------------
\paragraph{The $\mathcal{O}_{f^2 \varphi^2}$ component.}
This case is quite simple with only one singlet
\begin{equation}
S_{f^2 \varphi^2}(t_1,t_2,t_3,t_4)=(f^2)^{[0,0],0}(t_1,t_2)\,
(\varphi^2)^{[0,0]}(t_3,t_4)\,,
\end{equation}
which is separately symmetric in the exchange of $t_1$ with $t_2$ and of $t_3$ with $t_4$. From this definition we compute the generating correlator
\begin{equation}
\langle \theta^2\!f(1)\, \theta^2\!f(2)  \varphi(3)\varphi(4)\,
S_{f^2 \varphi^2}(t_5,t_6,t_7,t_8)\rangle 
=
-\tfrac{9}{4}
\left(\tfrac{1}{t_{15}^4 t_{26}^4}
+
\tfrac{1}{t_{16}^4 t_{25}^4}
\right)
\left(\tfrac{1}{t_{37}^2 t_{48}^2}
+
\tfrac{1}{t_{38}^2 t_{47}^2}
\right)\,\theta^2_1\theta^2_2\,y^2_{34}\,,
\end{equation}
which can be compared to \eqref{PhiPhiOL2prGENERATINGphiphi} and  \eqref{PhiPhiOL2prGENERATINGff}, and
\begin{equation}
\langle
S_{f^2 \varphi^2}(t_1,t_2,t_3,t_4)\,
S_{f^2 \varphi^2}(t_5,t_6,t_7,t_8)\rangle 
=
\tfrac{135}{4}
\left(\tfrac{1}{t_{15}^4 t_{26}^4}
+
\tfrac{1}{t_{16}^4 t_{25}^4}
\right)
\left(\tfrac{1}{t_{37}^2 t_{48}^2}
+
\tfrac{1}{t_{38}^2 t_{47}^2}
\right)\,.
\end{equation}
Next, to determine the contribution of the form $\partial^{\#} f^2 \varphi^2$ to  \eqref{Opieces} for the operator $\partial^{\#} \varphi^4$ given in  \eqref{OL4phi4operators}, we compute \eqref{ffphiphi} for an appropriate linear combination of operators of the form
\begin{equation}
\mathsf{D}_K^{(4+2n_1,2+2n_2)}(\partial_1+\partial_2,\partial_3+\partial_4)
\mathsf{D}_{2n_1}^{(2,2)}(\partial_1,\partial_2)\mathsf{D}_{2n_2}^{(1,1)}(\partial_3,\partial_4)
S_{f^2 \varphi^2}\Big{|}_{t_i=t}\,,
\end{equation}
with fixed conformal dimensions, see Appendix \ref{app:actingwithderivatives},
and impose that it satisfies \eqref{L4primarytheta4part1}.

%-------------------------------------------------------------------------------
\paragraph{The $\mathcal{O}_{f \Psi^2 \varphi}$ component.}
Also in this case there is a single term 
\begin{equation}
S_{f \Psi^2\varphi}(t_1,t_2,t_2,t_4)=
\epsilon^{\alpha \gamma}
\epsilon^{\beta \delta}
\left(\mathsf{P}_{12}\right)^{[0,1],[0,1]}_{[0,0]}
:f_{\alpha \beta}(t_1)
(\Psi^2)_{\gamma \delta}^{[0,1],2}(t_2,t_3,y_1)
\varphi(t_4,y_2):\,.
\end{equation}
where bilinears and projectors are found in Appendix \ref{app:Rsymmprojectors}.
The generating correlators are
\begin{equation}
\langle
\theta^2\! f(1)\,
\theta \Psi(2) 
\theta \Psi(3)
\varphi(4)
S_{f \Psi^2\varphi}(t_5,t_6,t_7,t_8)
\rangle
\,=\,
\tfrac{12}{t_{15}^4}
\left(\tfrac{1}{t_{26}^3 t_{37}^3}
+
\tfrac{1}{t_{36}^3 t_{27}^3}
\right)
\tfrac{1}{t_{48}^2}\,
\mathcal{L}_{1,23,4}\,
\end{equation}
where $\mathcal{L}_{1,23,4}$ is defined in \eqref{L1234def}  and
\begin{equation}
\langle
S_{f \Psi^2\varphi}(t_1,t_2,t_3,t_4)
S_{f \Psi^2\varphi}(t_5,t_6,t_7,t_8)
\rangle=
\tfrac{180}{t_{15}^4}
\left(\tfrac{1}{t_{26}^3 t_{37}^3}
+
\tfrac{1}{t_{36}^3 t_{27}^3}
\right)
\tfrac{1}{t_{48}^2}\,.
\end{equation}
As for the previous operators, to determine the contribution of the form $\partial^{\#} f \Psi^2\varphi$ to  \eqref{Opieces} for the operator $\partial^{\#} \varphi^4$ given in  \eqref{OL4phi4operators}, we compute \eqref{fPsiPsiphi} for an appropriate linear combination of operators of the form
\begin{equation}
\mathsf{D}_K^{(3+n_1,3+2n_2)}(\partial_1+\partial_4,\partial_2+\partial_3)
\mathsf{D}_{n_1}^{(2,1)}(\partial_1,\partial_4)\mathsf{D}_{2n_2}^{(3/2,3/2)}(\partial_2,\partial_3)
S_{f \Psi^2\varphi}\Big{|}_{t_i=t}\,,
\end{equation}
with fixed conformal dimensions, see Appendix \ref{app:actingwithderivatives},
and impose that it satisfies \eqref{L4primarytheta4part2}.
%-----

\subsubsection{Building blocks for the $L=4$ superdescendant $\widetilde{\mathcal{O}}$}
\label{sec:operatorsforthesuperdescendant}

Similarly to the previous subsection, we make an ansatz for a $L=4$ superdescendant $\widetilde{\mathcal{O}}(t,y)$ as a linear combination of conformal primaries as
\begin{align}
\label{Otildepieces}
\widetilde{\mathcal{O}}(t,y)&=
\widetilde{\mathcal{O}}_{\varphi^4}
+
\widetilde{\mathcal{O}}_{\Psi^2\varphi^2}
+
\widetilde{\mathcal{O}}_{\Psi^4}
+
\widetilde{\mathcal{O}}_{f^2\varphi^2}
+
\widetilde{\mathcal{O}}_{f\Psi^2\varphi}\,,
\end{align}
and in the following we analyze the various components one at the time.

%-------------------------------------------------------------------------------
\paragraph{The $\widetilde{\mathcal{O}}_{\varphi^4}$ component.}
In this case there is a single composite operator made of $\varphi$ which transforms in the $[0,2]$ representation\footnote{Recall that $[0,2]$  is the symmetric traceless representation of $SO(5)$.}
\begin{equation}
T_{\varphi^4}(t_1,t_2,t_3,t_4,y)=
\,\,
:(\varphi^2)^{[0,2]}(t_1,t_2,y)(\varphi^2)^{[0,0]}(t_3,t_4):
\end{equation}
where
\begin{equation}
(\varphi^2)^{[0,2]}(t_1,t_2,y)=
\,\,
:\varphi(t_1,y)\varphi(t_2,y):\,.
\end{equation}
Using the definition of the composites $(\varphi^2)^{[0,2]}$ and $(\varphi^2)^{[0,0]}$ we compute the generating correlator
\begin{align}
\begin{split}
\langle  \varphi(1)   \varphi(2) \varphi(3)\varphi(4)\,&
T_{\varphi^4}(t_5,t_6,t_7,t_8,y_5)\rangle =\sum_{\sigma \in S_4}
\frac{y_{\sigma_1 5}^2}{t_{\sigma_1 5}^2}
\frac{y_{\sigma_2 5}^2}{t_{\sigma_2 6}^2}
\frac{y_{\sigma_3 \sigma_4}^2}{t_{\sigma_3 7}^2t_{\sigma_4 8}^2}\,=
\\
=&
\sum_{\sigma \in S_4/\mathbb{Z}_2\times \mathbb{Z}_2 }
y_{\sigma_1 5}^2 y_{\sigma_2 5}^2
y_{\sigma_3 \sigma_4}^2
\left(
\tfrac{1}{t_{\sigma_1 5}^2 t_{\sigma_2 6}^2}+\tfrac{1}{t_{\sigma_1 6}^2 t_{\sigma_2 5}^2}
\right)
\left(
\tfrac{1}{t_{\sigma_3 7}^2 t_{\sigma_4 8}^2}+\tfrac{1}{t_{\sigma_3 8}^2 t_{\sigma_4 7}^2}
\right)\,,
\end{split}
\end{align}
and
\begin{equation}
\langle 
T_{\varphi^4}(t_1,t_2,t_3,t_4,y_1)
T_{\varphi^4}(t_5,t_6,t_7,t_8,y_2)\rangle =(y_{12}^2)^2
\left(5
\mathsf{T}_1+
\mathsf{T}_2
\right)\,,
\end{equation}
where
\begin{equation}
\mathsf{T}_1=
\left(\tfrac{1}{t_{1 5}^2 t_{2 6}^2}+\tfrac{1}{t_{1 6}^2 t_{2 5}^2}\right)
\left(\tfrac{1}{t_{3 7}^2 t_{4 8}^2}+\tfrac{1}{t_{3 8}^2 t_{4 7}^2}\right)\,,
\end{equation}
\begin{equation}
\mathsf{T}_2=
\left(\tfrac{1}{t_{1 5}^2 t_{3 6}^2}+\tfrac{1}{t_{1 6}^2 t_{3 5}^2}\right)
\left(\tfrac{1}{t_{2 7}^2 t_{4 8}^2}+\tfrac{1}{t_{2 8}^2 t_{4 7}^2}\right)+\text{three more}\,.
\end{equation}
The three extra terms in the last equation are obtained exchanging points $1$ with $2$, $3$ with $4$, as well as both exchanges at once.
If we take $\mathcal{O}_{\varphi^4}$ as in \eqref{OL4phi4operators}, the corresponding $\widetilde{\mathcal{O}}_{\varphi^4}$
is obtained by taking a linear combination of operators of the form
\begin{equation}
\mathsf{D}_K^{(2+2n_1,2+2n_2)}(\partial_1+\partial_2,\partial_3+\partial_4)
\mathsf{D}_{2n_1}^{(1,1)}(\partial_1,\partial_2)\mathsf{D}_{2n_2}^{(1,1)}(\partial_3,\partial_4)
T_{\varphi^4}(t_1,t_2,t_3,t_4)\Big{|}_{t_i=t}
\end{equation}
with appropriate scaling dimension, computing \eqref{5pointsuperdescphi4comp} for this choice of $\widetilde{\mathcal{O}}_{\varphi^4}$ and imposing that it satisfies \eqref{FtildefromF}, where $F_{\varphi\varphi\varphi\varphi}$ is computed using  $\mathcal{O}_{\varphi^4}$
given in  \eqref{OL4phi4operators} .

%-------------------------------------------------------------------------------
\paragraph{The $\widetilde{\mathcal{O}}_{\Psi^2\varphi^2}$ component.}
In this case there are three combinations transforming in the $[0,2]$ representation
\begin{align}
T^{(1)}_{\Psi^2\varphi^2}(t_1,t_2,t_3,t_4,y)
&=\,\,
:(\Psi^2)^{[0,0],0}(t_1,t_2)(\varphi^2)^{[0,2]}(t_3,t_4,y):
\\
T^{(2)}_{\Psi^2\varphi^2}(t_1,t_2,t_3,t_4,y)
&=\,\,
\left(
\mathsf{P}_{12}\right)^{[2,0],[0,2]}_{[0,2]}
:(\Psi^2)^{[2,0],0}(t_1,t_2,y_1,\ww_1)(\varphi^2)^{[0,2]}(t_3,t_4,y_2):
\\
T^{(3)}_{\Psi^2\varphi^2}(t_1,t_2,t_3,t_4,y)
&=\,\,
\left(
\mathsf{P}_{12}\right)^{[2,0],[2,0]}_{[0,2]}
:(\Psi^2)^{[2,0],0}(t_1,t_2,y_1,\ww_1)(\varphi^2)^{[2,0]}(t_3,t_4,y_2):
\end{align}
Using these definitions together with the explicit expressions for the projectors and  field bilinears  collected in Appendix \ref{app:Rsymmprojectors} one obtains the generating correlators
\begin{subequations}
\label{PsiPsiphiphiT}
\begin{align}
\langle \theta\Psi(1) \theta \Psi(2)  \varphi(3)\varphi(4)\,
T^{(1)}_{\Psi^2\varphi^2}(t_5,t_6,t_7,t_8,y_5)\rangle 
&=
-6
\,I_1\,
\left(\tfrac{1}{t_{1 5}^3 t_{2 6}^3}-\tfrac{1}{t_{1 6}^3 t_{2 5}^3}\right)
\left(\tfrac{1}{t_{3 7}^2 t_{4 8}^2}+\tfrac{1}{t_{3 8}^2 t_{4 7}^2}\right)
\,,
\\
\langle \theta\Psi(1) \theta \Psi(2)  \varphi(3)\varphi(4)\,
T^{(2)}_{\Psi^2\varphi^2}(t_5,t_6,t_7,t_8,y_5)\rangle
&=
\,\,\,3
\,I_2\,
\left(\tfrac{1}{t_{1 5}^3 t_{2 6}^3}+\tfrac{1}{t_{1 6}^3 t_{2 5}^3}\right)
\left(\tfrac{1}{t_{3 7}^2 t_{4 8}^2}+\tfrac{1}{t_{3 8}^2 t_{4 7}^2}\right)
\,,
\\
\langle \theta\Psi(1) \theta \Psi(2)  \varphi(3)\varphi(4)\,
T^{(3)}_{\Psi^2\varphi^2}(t_5,t_6,t_7,t_8,y_5)\rangle
&=
\,\,\,4
\,I_3\,
\left(\tfrac{1}{t_{1 5}^3 t_{2 6}^3}+\tfrac{1}{t_{1 6}^3 t_{2 5}^3}\right)
\left(\tfrac{1}{t_{3 7}^2 t_{4 8}^2}-\tfrac{1}{t_{3 8}^2 t_{4 7}^2}\right)\,,
\end{align}
\end{subequations}
where  the R-symmetry structures are defined as
 \begin{equation}
 I_1=\theta_1y_{12}\theta_2\, y_{35}^2 y_{45}^2\,,
 \quad
  I_2=y^2_{45}\,\mathcal{I}_{12,35}+y^2_{35}\, \mathcal{I}_{12,45}+4\, I_1\,,
 \quad
  I_3=y^2_{45}\,\mathcal{I}_{12,35}-y^2_{35}\, \mathcal{I}_{12,45}\,,
 \end{equation}
and $\mathcal{I}$ is defined in \eqref{IRsymmdef}.
Notice that they are linear combinations of the three structures in \eqref{PsiPsiphiphitildeOcomponent}.
Under the $\mathbb{Z}_2\times \mathbb{Z}_2$ transformations that exchange point 1 with 2 and 3 with 4 respectively, the three structures $I_1,I_2,I_3$ transform as
$(-,+)$, $(+,-)$ and $(+,+)$ as they should, to compensate the transformation properties of the $t$ dependent part in \eqref{PsiPsiphiphiT}.
We also collect the generating correlators used to compute norms
\begin{subequations}
\label{PsiPsiphiphiTforNORMS}
\begin{align}
\langle T^{(1)}_{\Psi^2\varphi^2}(t_1,t_2,t_3,t_4,y_1)
T^{(1)}_{\Psi^2\varphi^2}(t_5,t_6,t_7,t_8,y_2)\rangle 
\!&=\!
-72%
\left(\tfrac{1}{t_{1 5}^3 t_{2 6}^3}-\tfrac{1}{t_{1 6}^3 t_{2 5}^3}\right)
\left(\tfrac{1}{t_{3 7}^2 t_{4 8}^2}+\tfrac{1}{t_{3 8}^2 t_{4 7}^2}\right)
(y_{12}^2)^2\,,
\\
\langle T^{(2)}_{\Psi^2\varphi^2}(t_1,t_2,t_3,t_4,y_1)
T^{(2)}_{\Psi^2\varphi^2}(t_5,t_6,t_7,t_8,y_2)\rangle
\!&=\!
-720\!
\left(\tfrac{1}{t_{1 5}^3 t_{2 6}^3}+\tfrac{1}{t_{1 6}^3 t_{2 5}^3}\right)
\left(\tfrac{1}{t_{3 7}^2 t_{4 8}^2}+\tfrac{1}{t_{3 8}^2 t_{4 7}^2}\right)
(y_{12}^2)^2\,,
\\
\langle 
T^{(3)}_{\Psi^2\varphi^2}(t_1,t_2,t_3,t_4,y_1)
T^{(3)}_{\Psi^2\varphi^2}(t_5,t_6,t_7,t_8,y_2)\rangle
\!&=\!
-768\!
\left(\tfrac{1}{t_{1 5}^3 t_{2 6}^3}+\tfrac{1}{t_{1 6}^3 t_{2 5}^3}\right)
\left(\tfrac{1}{t_{3 7}^2 t_{4 8}^2}-\tfrac{1}{t_{3 8}^2 t_{4 7}^2}\right)\,
(y_{12}^2)^2\,,
\end{align}
\end{subequations}
while $\langle T^{(i)}T^{(j)}\rangle$ vanishes when $i\neq j$.
%-------------------------------------------------------------------------------
\paragraph{The $\widetilde{\mathcal{O}}_{\Psi^4}$ component.}
In this case there are four combinations transforming in the $[0,2]$ representation, see Table \ref{tab:lsingletsforcorrelatorwithThildeO}
\begin{align}
T^{(1)}_{\Psi^4}(t_1,t_2,t_3,t_4,y)
&=\,\,
:(\Psi^2)^{[0,1],0}(t_1,t_2,y)(\Psi^2)^{[0,1],0}(t_3,t_4,y):
\\
T^{(2)}_{\Psi^4}(t_1,t_2,t_3,t_4,y)
&=
\epsilon^{\alpha \gamma}
\epsilon^{\beta \delta}
:(\Psi^2)_{\alpha \beta}^{[0,1],2}(t_1,t_2,y)(\Psi^2)_{\gamma \delta}^{[0,1],2}(t_3,t_4,y):
\\
T^{(3)}_{\Psi^4}(t_1,t_2,t_3,t_4,y)
&=
\left(
\mathsf{P}_{12}\right)^{[2,0],[2,0]}_{[0,2]}
:(\Psi^2)^{[2,0],0}(t_1,t_2,y_1,\ww_1)(\Psi^2)^{[2,0],0}(t_3,t_4,y_2,\ww_2):
\\
T^{(4)}_{\Psi^4}(t_1,t_2,t_3,t_4,y)
&=
\epsilon^{\alpha \gamma}
\epsilon^{\beta \delta}
\left(
\mathsf{P}_{12}\right)^{[2,0],[2,0]}_{[0,2]}\!
:\!(\Psi^2)_{\alpha \beta}^{[2,0],2}(t_1,t_2,y_1,\ww_1)(\Psi^2)_{\gamma \delta}^{[2,0],2}(t_3,t_4,y_2,\ww_2):
\end{align}
Using these definitions together with the explicit expressions for the projectors and  field bilinears  collected in Appendix \ref{app:Rsymmprojectors}  one obtains the generating correlators
%---------------------------------------------------------------------------
\begin{subequations}
\label{PsiPsiPsiPsiT}
\begin{align}
\langle \theta\Psi(1) \theta \Psi(2)\theta \Psi(3)\theta \Psi(4)\,
T^{(1)}_{\Psi^4}(t_5,t_6,t_7,t_8,y_5)\rangle 
&=
\mathsf{T}_{12,34}^{(-)}\,
A_{12,34}^{(--|+)}+\text{two more}\,,
\\
\langle \theta\Psi(1) \theta \Psi(2)\theta \Psi(3)\theta \Psi(4)\,
T^{(2)}_{\Psi^4}(t_5,t_6,t_7,t_8,y_5)\rangle 
&=
\mathsf{T}_{12,34}^{(+)}\,
A_{12,34}^{(++|+)}
+\text{two more}\,,
\\
\langle \theta\Psi(1) \theta \Psi(2)\theta \Psi(3)\theta \Psi(4)\,
T^{(3)}_{\Psi^4}(t_5,t_6,t_7,t_8,y_5)\rangle 
&=
\tfrac{4}{3}
\mathsf{T}_{12,34}^{(+)}\,
B_{12,34}^{(++|+)}
+\text{two more}\,,
\\
\langle \theta\Psi(1) \theta \Psi(2)\theta \Psi(3)\theta \Psi(4)\,
T^{(4)}_{\Psi^4}(t_5,t_6,t_7,t_8,y_5)\rangle 
&=
\mathsf{T}_{12,34}^{(-)}\,
B_{12,34}^{(--|+)}
+\text{two more}\,.
\end{align}
\end{subequations}
%---------------------------------------------------------------------------
The notation deserves some explanation.
Firstly we defined the factors depending on the coordinate on the line as
\begin{equation}
\mathsf{T}_{12,34}^{(\pm)}:=
\left(\tfrac{1}{t_{1 5}^3 t_{2 6}^3}\pm \tfrac{1}{t_{1 6}^3 t_{2 5}^3}\right)
\left(\tfrac{1}{t_{3 7}^3 t_{4 8}^3}\pm \tfrac{1}{t_{3 8}^3 t_{4 7}^3}\right)+
\left(\tfrac{1}{t_{3 5}^3 t_{4 6}^3}\pm \tfrac{1}{t_{3 6}^3 t_{4 5}^3}\right)
\left(\tfrac{1}{t_{1 7}^3 t_{2 8}^3}\pm \tfrac{1}{t_{1 8}^3 t_{2 7}^3}\right)\,.
\end{equation}
Next we define the R-symmetry structures
\begin{subequations}
\begin{align}
A_{12,34}^{(--|+)}&
=\mathcal{J}_{12,5}\mathcal{J}_{34,5}\,,
\\
A_{12,34}^{(++|+)}&
=\epsilon^{\alpha \gamma}\epsilon^{\beta \delta} (\mathcal{J}_{\alpha\beta})_{12,5}(\mathcal{J}_{\gamma \delta})_{34,5}\,,
\end{align}
\end{subequations}
where $\mathcal{J}$ and $\mathcal{J}_{\alpha\beta}$ are defined in \eqref{J123def} and \eqref{J123defwithspin} respectively. 
The $B$-structures are related to the $A$-structures as follows
\begin{subequations}
\begin{align}
B_{12,34}^{(++|+)}&
=+A_{12,34}^{(++|+)}
-2 A_{13,24}^{(++|+)}
-2 A_{14,23}^{(++|+)}\,,
\\
B_{12,34}^{(--|+)}&
=-A_{12,34}^{(--|+)}
-2 A_{13,24}^{(--|+)}
+2 A_{14,23}^{(--|+)}\,.
\end{align}
\end{subequations}
The signs in parenthesis indicate the transformation properties under the $(\mathbb{Z}_2)^3$ that exchange points $1$ and $2$, $3$ and $4$ and $(1,2)$ with $(3,4)$ respectively.
Finally ``two more'' in \eqref{PsiPsiPsiPsiT} refers to the terms that need to be added to make the correlator invariant under the permutation of the first four points\footnote{Notice that the whole supercoordinates $(t,y,\theta)$ need to be permuted.}. 

The generating correlators  \eqref{PsiPsiPsiPsiT}  are written in terms of six structures $A_{12,34}^{(\pm\pm|+)},
 A_{13,24}^{(\pm\pm|+)}, 
 A_{14,23}^{(\pm\pm|+)}$, but only four are linearly independent.
 This fact was anticipated in \eqref{psipsipsipsiOtildeF} where only four structures, corresponding to the multiplicity of the $[0,2]_0$ representations in  $[1,0]_1^{\otimes 4}$, appear.
 Since the action of the permutation group $\mathfrak{S}_4$ commutes with the action of $\mathfrak{sp}(4)\oplus \mathfrak{su}(2)$ we can decompose the multiplicity space in representation of $\mathfrak{S}_4$. A simple analysis using plethysms\footnote{This can be done efficiently using \cite{LiE1992}, to which we also refer for further details.} reveals that this four-dimensional space decomposes as the trivial representation, the sign representation and the two-dimensional irreducible representation. The associated structures are given by
 \begin{subequations}
\begin{align}
\widehat{\mathcal{J}}\,:=&\,A_{12,34}^{(++|+)}
+ A_{13,24}^{(++|+)}+
A_{14,23}^{(++|+)}\,,
\label{Jhatdef}
\\
&A_{12,34}^{(--|+)}
- A_{13,24}^{(--|+)}
+ A_{14,23}^{(--|+)}\,,
%A_{12,34}^{(--|+)}-A_{14,23}^{(--|+)}=
%-\tfrac{1}{2}
%A_{12,34}^{(++|+)}
%-
%A_{13,24}^{(++|+)}
%-\tfrac{1}{2}
%A_{14,23}^{(++|+)}\,,
%%
%A_{13,24}^{(--|+)}-A_{14,23}^{(--|+)}=
%-\tfrac{1}{2}
%A_{13,24}^{(++|+)}
%-
%A_{12,34}^{(++|+)}
%+\tfrac{1}{2}
%A_{14,23}^{(++|+)}\,,
\\
\label{TwocomponentsForPsi4}
\Bigg{\{}\,
\begin{split}
&A_{12,34}^{(++|+)}-A_{14,23}^{(++|+)}= -\tfrac{1}{2}A_{12,34}^{(--|+)}-A_{13,24}^{(--|+)}-\tfrac{1}{2}A_{14,23}^{(--|+)}\,,\\
& A_{13,24}^{(++|+)}-A_{14,23}^{(++|+)}= -\tfrac{1}{2}A_{13,24}^{(--|+)}-A_{12,34}^{(--|+)}+\tfrac{1}{2}A_{14,23}^{(--|+)}\,.
\end{split}
\end{align}
\end{subequations}
Notice that the second component in \eqref{TwocomponentsForPsi4} is obtained from the first upon exchanging $2$ and $3$.
A basis of invariant is clearly given by $A_{12,34}^{(--|+)}$, $ A_{13,24}^{(--|+)}$, $A_{14,23}^{(--|+)}$ and  $\widehat{\mathcal{J}}$, which is is the choice made in \eqref{psipsipsipsiOtildeF}.

To compute norms we need the generating correlators for two-point functions
\begin{equation}
\langle
T_{\Psi^4}^{(i)}(\underline{t},y_1)T_{\Psi^4}^{(j)}(\underline{t}',y_2)
\rangle
=
(y_{12}^2)^2
\begin{pmatrix}
64\mathsf{X}^{(-)} &-96\mathsf{Y}^{(+)}_{-}& 384\mathsf{Y}^{(+)}_{-}&-576 4\mathsf{Y}^{(-)}_{-}\\
- & 48 \mathsf{X}^{(+)} &-576 \mathsf{Y}^{(+)}_{-} &  288  \mathsf{Y}^{(-)}_{+}\\
- & - &768 \mathsf{X}^{(+)} & 1152   \mathsf{Y}^{(-)}_{+} \\
- & - &- &576 \mathsf{X}^{(-)} 
\end{pmatrix}_{ij}
\end{equation}
where
%\begin{subequations}
%\begin{align}
\begin{equation}
\mathsf{X}^{(\pm)}:=4\mathsf{T}^{(\pm)}_{12,34}\pm\mathsf{T}^{(\pm)}_{13,24}+\mathsf{T}^{(\pm)}_{14,23}\,,
\qquad
\mathsf{Y}^{(\pm)}_{\sigma}:=\mathsf{T}^{(\pm)}_{13,24}+\sigma \mathsf{T}^{(\pm)}_{14,23}\,,
\end{equation}
and $\underline{t}=\{t_1,t_2,t_3,t_4\}$,  $\underline{t}'=\{t_5,t_6,t_7,t_8\}$.
Notice that  we do not display the lower diagonal entries since they are fixed by symmetry. 

 %-------------------------------------------------------------------------------
\paragraph{The $\widetilde{\mathcal{O}}_{f^2\varphi^2}$ component.}
In this case there is a single composite operator that transforms in the $[0,2]$ representation and is given by
\begin{equation}
T_{f^2 \varphi^2}(t_1,t_2,t_3,t_4,y)=
:(f^2)^{[0,0]}(t_1,t_2)(\varphi^2)^{[0,2]}(t_3,t_4,y):\,.
\end{equation}
The relevant generating correlators are
\begin{equation}
\langle \theta^2f(1) \theta^2f(2) \varphi(3)\varphi(4)T_{f^2 \varphi^2}(\underline{t}',y_5)\rangle=
-\tfrac{9}{4}
\theta_{1}^2\theta_{2}^2\,
y_{35}^2y_{45}^2
\left(\tfrac{1}{t^4_{15} t^4_{26}}
+\tfrac{1}{t^4_{16} t^4_{25}}
\right)
\left(\tfrac{1}{t^4_{37} t^4_{48}}
+\tfrac{1}{t^4_{38} t^4_{47}}
\right)\,,
\end{equation}
and
\begin{equation}
\langle T_{f^2 \varphi^2}(\underline{t},y_1)T_{f^2 \varphi^2}(\underline{t}',y_2)\rangle=
\,\tfrac{27}{4}
(y_{12}^2)^2
\left(\tfrac{1}{t^4_{15} t^4_{26}}
+\tfrac{1}{t^4_{16} t^4_{25}}
\right)
\left(\tfrac{1}{t^4_{37} t^4_{48}}
+\tfrac{1}{t^4_{38} t^4_{47}}
\right)\,.
\end{equation}
%-------------------------------------------------------------------------------
\paragraph{The $\widetilde{\mathcal{O}}_{f \Psi^2\varphi}$ component.}
Also in this case there is a single composite operator
\begin{equation}
T_{f \Psi^2\varphi}(t_1,t_2,t_3,t_4,y)=
\epsilon^{\alpha \gamma}
\epsilon^{\beta \delta}
:f_{\alpha \beta}(t_1)
(\Psi^2)_{\gamma \delta}^{[0,1],2}(t_2,t_3,y)
\varphi(t_4,y):\,,
\end{equation}
and the relevant generating correlators are give by 
\begin{equation}
\langle 
\theta^2f(1)  \theta \Psi(2)\theta \Psi(3) \varphi(4)
 T_{f \Psi^2\varphi}(t_5,t_6,t_7,t_8,y_5)
 \rangle=
 \tfrac{12}{t_{15}^4}
 \left(\tfrac{1}{t^3_{26} t^3_{37}}
+\tfrac{1}{t^3_{36} t^3_{27}}
\right)
\tfrac{1}{t_{48}^2}
y_{45}^2
 \mathcal{L}_{1,23,5}\,,
\end{equation}
where $ \mathcal{L}_{1,23,5}$ is defined in \eqref{L1234def}
and 
\begin{equation}
\langle 
 T_{f \Psi^2\varphi}(\underline{t},y_1)
 T_{f \Psi^2\varphi}(\underline{t}',y_2)
 \rangle=
 \tfrac{3}{t_{15}^4}
 \left(\tfrac{1}{t^3_{26} t^3_{37}}
+\tfrac{1}{t^3_{36} t^3_{27}}
\right)
\tfrac{1}{t_{48}^2}
(y_{12}^2)^2
\,.
\end{equation}

\subsection{Discussion}

In this section we have presented in detail the construction of composite superconformal primary operators belonging to unprotected multiplets, as well as selected superdescendants. While the problem is technical in nature, designing an efficient way to solve it has proved essential for the derivation of the three-loop four-point function of the displacement supermultiplet on the half-BPS Wilson line defect in $\mathcal{N}=4$ SYM, which we have presented for the first time in \cite{Ferrero:2021bsb}. While more details on how the results derived here are employed for that purpose are given in \cite{Ferrero:2023gnu}, an important point that we want to emphasize is that the techniques developed in this paper allowed to successfully solve a mixing problem which is significantly harder than those faced so far in the literature on perturbative CFTs. The main issue is related to the degeneracy of long supermultiplets in the perturbative expansion around strong coupling, which for the observable considered in \cite{Ferrero:2021bsb} and \cite{Ferrero:2023gnu} requires us to ``un-mix'' a number of operators which grows quadratically with their (common) conformal dimension at strong coupling -- see \eqref{NstrongL4000}. This should be contrasted, for instance, with analytic bootstrap studies of the $\epsilon$-expansion in the Wilson-Fisher theory, where a strongly constrained ansatz allowed to by-pass the need to address the mixing problem beyond the first few degenerate sectors \cite{Alday:2017zzv} (see \cite{Henriksson:2022rnm} for a detailed review and an historical account of this problem, as well as the references therein). Another notorious example is the mixing between double-trace long multiplets in holographic CFTs at tree level (in the $1/N$ expansion), where the growth of the degeneracy with the dimension is linear, as opposed to quadratic: an enormous technical simplification compared to the situation investigated in \cite{Ferrero:2021bsb,Ferrero:2023gnu} with the techniques developed here. Indeed, the explicit construction of such operators was not necessary in the various works that have addressed this problem \cite{Aprile:2017bgs,Alday:2017xua,Alday:2020tgi,Alday:2021ajh,Behan:2022uqr,Alday:2022rly} (a more thorough analysis is available for AdS$_5\times S^5$ holography, see \cite{Aprile:2017xsp,Aprile:2018efk,Aprile:2019rep}). Despite the success of these approaches, a more detailed understanding of operators mixing is necessary in order to obtain analogous results at higher perturbative orders. Moreover, such an investigation would be interesting to gain more insight on the perturbative structure of the dilatation operator of $\mathcal{N}=4$ SYM at strong coupling in the planar limit, given the success in the weak coupling version of this problem thanks to integrability \cite{Beisert:2003tq,Beisert:2003yb,Beisert:2004ry}. 

\section{Conclusion}\label{sec:discussion}

In this paper we presented several new results involving non-dynamical aspects of 1d CFTs with $\mathfrak{osp}(4^*|4)$ invariance. We discussed two representations of the superconformal generators, one adapted to absolutely protected $\tfrac{1}{2}$-BPS multiplets and one to generic, long supermultiplets, and discussed kinematical aspects of certain two-, three- and four-point functions between such representations, including explicit results for superconformal blocks. We also provided a description of the content of all unitary, irreducible representations of $\mathfrak{osp}(4^*|4)$ in terms of representations of its bosonic subalgebra.

Despite the crucial role played by SCFTs in our understanding of QFT, and the great importance of superconformal blocks in the conformal bootstrap program, not much is known beyond four-point functions of $\tfrac{1}{2}$-BPS operators for generic SCFTs. This clearly calls for further investigations of the implications of superconformal invariance for more general correlation functions. On the one hand four-point functions are the main observables for most bootstrap studies, at least of the numerical side, and the construction of superconformal blocks for correlators involving non-protected operators would open the way to the study of novel observables that have eluded bootstrap investigations so far. Some work in this direction was carried out in \cite{Cornagliotto:2017dup,Buric:2020buk,Buric:2020qzp} and with this paper we hope to show that 1d CFTs provide an ideal setup where to further develop these ideas, to then export similar strategies to SCFTs in general dimension. Similarly, it would be interesting to develop an understanding of the superconformal kinematics for higher-point functions, which at this stage is not entirely developed even just for $\tfrac{1}{2}$-BPS operators (see \cite{Barrat:2021tpn,Barrat:2022eim,Bliard:2023zpe} for some progress in this direction). In the bosonic case, results have been found for certain topologies at arbitrary number of points, with the case of 1d theories providing the starting point of such investigations \cite{Rosenhaus:2018zqn,Parikh:2019dvm,Fortin:2019zkm,Fortin:2020yjz,Fortin:2020bfq,Fortin:2020zxw,Hoback:2020pgj,Hoback:2020syd,Poland:2021xjs} (see also  \cite{Buric:2021ywo,Buric:2021ttm,Buric:2021kgy} for the relation between higher-point conformal blocks and Gaudin models). Once again, line defects could prove an ideal setup for this type of investigations and we hope to address this problem both from a kinematical and from a dynamical perspective in future work. This could serve as a starting point for the study of more general higher-point functions in $d>1$, which have recently become subject to a greater attention due to the impressive progress in the computations of holographic correlators between more than four operators  \cite{Goncalves:2019znr,Alday:2022lkk,Goncalves:2023oyx,Alday:2023kfm}.\footnote{See also \cite{Barrat:2021tpn,Barrat:2022eim,Bliard:2023zpe} for multi-point results on the Wilson line at weak coupling.}

The other main theme of this paper was the study of the spectrum of local operators in Wilson line defect theories, with particular focus on the fundamental Wilson loop in the planar limit. This is interesting in connection with both the analytic bootstrap program started in \cite{Liendo:2018ukf,Ferrero:2021bsb} and the ``bootstrability'' investigations of \cite{Cavaglia:2021bnz,Cavaglia:2022qpg}: on the one hand it is useful to address the problem of operators mixing, which is a common issue to be faced when studying CFTs perturbatively around a free point, on the other hand an independent counting of states is important from the point of view of integrability to make sure that all operators are accounted for by the quantum spectral curve. In this paper we have focused on the fundamental Wilson line but an interesting problem for the future is certainly that of repeating the counting for other representations, where again one has two holographically dual descriptions that are perturbative in opposite regimes of the 't Hooft coupling. Similarly, it would be interesting to address the problem of $1/N$ corrections. Similarly, it is interesting to consider the counting of states for the half-BPS Wilson line in the ABJM theory \cite{Aharony:2008ug,Drukker:2009hy,Cardinali:2012ru,Bianchi:2017ozk,Drukker:2019bev}, where the line operator has fermionic nature, making the counting at weak coupling not just a trivial generalization of the techniques considered here. This should also be interesting for the bootstrap program initiated in \cite{Bianchi:2020hsz}.

Besides the counting, we have presented an explicit construction of superconformal primaries in terms of a free fields description of the theory in the holographic limit at large 't Hooft coupling, although the methods apply more in general when a free theory description is available. Our main purpose with this construction is to address the problem of mixing between operators with the same quantum numbers at $\lambda=\infty$ and in \cite{Ferrero:2023gnu} we will be more precise about how this is done (see \cite{Ferrero:2021bsb} for a summary of such method). From this perspective, we believe that analogous constructions, in conjunction with the study of external operators that are not absolutely protected, could be relevant to address other types of mixing problems. Some examples where this could be relevant are holographic correlators in theories with maximal \cite{Alday:2020dtb} and half-maximal supersymmetry \cite{Alday:2021odx}, where beyond one loop one should study the mixing between triple- and higher-trace operators, a problem which could possibly be addressed with similar techniques to the ones proposed here and in \cite{Ferrero:2023gnu} -- see \cite{Bissi:2021hjk} for some steps in this direction, where four-point functions with one external $\tfrac{1}{4}$-BPS operator in 4d $\mathcal{N}=4$ SYM are considered. Another example is the $\epsilon$-expansion \cite{Alday:2017zzv,Henriksson:2022rnm}.

\section*{Acknowledgments}

We thank Fernando Alday and Pedro Liendo for collaboration at early stages of this work and for many fruitful discussions. We thank Fernando Alday for sharing with us unpublished notes and are grateful to Alex Gimenez-Grau and Johan Henriksson for sharing with us some Mathematica code and some unpublished results respectively. The work of C.M. has received funding from the European Union’s Horizon 2020 research and innovation program under the Marie Sklodowska Curie grant agreement No 754496. The work of P.F. has received funding from the European Research Council (ERC) under the European Union’s Horizon 2020 research and innovation program (grant agreement No 787185).

\appendix
\addtocontents{toc}{\protect\setcounter{tocdepth}{1}}

\section{Superspace conventions}\label{app:notation}

We collect here some of our conventions and some notation use throughout the paper.

In this paper and in \cite{Ferrero:2023gnu} we use two different symbols to denote conformal dimensions: $h$ and $\Delta$. The intended distinction between the two is that we use $h$ in any equation that is valid non-perturbatively, {\it i.e.} it is not specific to the fact that one is working in a free theory, or anyway in an expansion around a free theory. On the other hand, $\Delta$ is use to denote the dimensions of operators in free theories: in the case of the Wilson line defect CFT, this could be at zero or at infinite coupling. The distinction is of course unimportant for protected operators.

Our convention for the invariant tensors of $\mathfrak{su}(2)$ and $\mathfrak{sp}(4)$ are
\begin{align}
\begin{split}
\epsilon_{\alpha\beta}
&=\begin{pmatrix}
0 & +1 \\
-1&0
\end{pmatrix}_{\alpha\beta}\,,\qquad
\epsilon^{\alpha\beta}=(\epsilon_{\alpha\beta})^{-1}=-\epsilon_{\alpha\beta}\,,
\\
\Omega_{AB}&=
\begin{pmatrix}
0 &0&-1&0\\
0&0&0&-1\\
+1&0&0&0\\
0&+1&0&0
\end{pmatrix}_{AB}\,,
\qquad
\Omega^{AB}=(\Omega_{AB})^{-1}=-\Omega_{AB}\,.
\end{split}
\end{align}
In the main text, we introduced used the following combinations of superspace coordinates
\begin{align}
\begin{split}
(\theta^2)^{ab}&=\epsilon_{\alpha\beta}\,\theta^{\alpha a}\,\theta^{\beta b}\,, \quad
(\theta^2)^{\alpha\beta}=\epsilon_{ab}\,\theta^{\alpha a}\,\theta^{\beta b}\,,  \quad
\theta^4=\epsilon_{\alpha\gamma}\,\epsilon_{\beta\delta}\,(\theta^2)^{\alpha\beta}(\theta^2)^{\gamma\delta}\,,\\
\theta y \theta &=\epsilon_{ac}\epsilon_{bc}(\theta^2)^{ab}y^{cd}\,,\quad 
y^2=\det(y^{ab})\,,\quad \det\left(\frac{\partial}{\partial y^{ab}}\right)=2\,\epsilon^{ac}\,\epsilon^{bd}\,\frac{\partial}{\partial y^{ab}}\,\frac{\partial}{\partial y^{cd}}\,,\\
(\Theta^2)^{\alpha\beta}&=\Omega_{AB}\,\Theta^{\alpha A}\,\Theta^{\beta B}\,, \quad
(\Theta^2)^{AB}=\epsilon_{\alpha\beta}\,\Theta^{\alpha A}\,\Theta^{\beta B}\,,\quad
(\Theta^3)^{\alpha A}=\epsilon_{\beta\gamma}\,\Omega_{BC}\,\Theta^{\alpha B}\,\Theta^{\beta A}\,\Theta^{\gamma C}\,,\\
\Theta^4&=\epsilon_{\alpha\gamma}\,\epsilon_{\beta\delta}\,(\Theta^2)^{\alpha\beta}\,(\Theta^2)^{\gamma\delta}=\epsilon_{\alpha\beta}\,\Omega_{AB}\,(\Theta^3)^{\alpha A}\,\Theta^{\beta B}\,,\quad
(\Theta\Theta)_{ij}=\epsilon_{\alpha\beta}\Omega_{AB}\,\Theta_i^{\alpha A}\,\Theta_j^{\beta B}\,.
\end{split}
\end{align}

\section{Superconformal multiplets of $\mathfrak{osp}(4^*|4)$}\label{app:supermultiplets}

\subsection{Classification of multiplets}

The supermultiplets that are relevant for line defects with $\mathfrak{osp}(4^*|4)$ symmetry are classified in various references \cite{Gunaydin:1990ag,Dorey:2018klg,phdthesis,Agmon:2020pde}. Here we follow Table 8 of \cite{Agmon:2020pde}, which we reproduce here in Table \ref{tab:supermultipletsclassification} for the reader's convenience, adapting their conventions to those used here.
\begin{table}[!ht]
\begin{align*}
\begin{array}{|c|l|l|l|l|l|}
\hline \text { \bf Name } & \text { \bf Primary } & \text {  \bf Unitarity Bound } & \text { \bf BPS } & \text { \bf Null State } & \left.\mathfrak{Q}_{*}^{n} \mid \text { h.w. }\right\rangle=0 \\
\hline
\hline
\,\,L &\,\, [a,b]^{\h}_{s} &\,\, \h>a+b+s/2+1 &\,\, - & \,\,- & \,\,-\\
\hline
\hline
\,\,   A_1 &\,\,  [a,b]^{\h}_{s>0} & \,\,   \h=a+b+s/2+1 & \,\,   \tfrac{1}{8}& \,\,   [a+1,b]^{\h+1/2}_{s-1}  & \,\,  \mathfrak{Q}_{21} \\
\hline
\,\,   & \,\,    [0,b]^{\h}_{s} & \,\,   \h=b+s/2+1 & \,\,   \tfrac{1}{4}& \,\,   [1,b]^{\h+1/2}_{s-1}  & \,\,  \mathfrak{Q}_{21},\,\mathfrak{Q}_{22}    \\
\hline
\,\,   & \,\,    [0,0]^{\h}_{s} & \,\,   \h=s/2+1 & \,\,   \tfrac{1}{2}& \,\,   [1,0]^{\h+1/2}_{s-1}  & \,\,  \mathfrak{Q}_{2A},\,  \\
\hline
\,\,  A_2 & \,\,    [a,b]^{\h}_{0} & \,\,   \h=a+b+1 & \,\,   \tfrac{1}{8}& \,\,   [a+2,b]^{\h+1}_{0}  & \,\,  \mathfrak{Q}_{21}\,\mathfrak{Q}_{11}  \\
\hline
\,\,   & \,\,    [0,b]^{\h}_{0} & \,\,   \h=b+1 & \,\,   \tfrac{1}{4}& \,\,   [2,b]^{\h+1}_{0}  & \,\,  \mathfrak{Q}_{21}\,\mathfrak{Q}_{11} ,\,\mathfrak{Q}_{22}\,\mathfrak{Q}_{11} +\mathfrak{Q}_{21}\,\mathfrak{Q}_{12}   \\
\hline
\,\,   & \,\,   [0,0]^{\h}_{0} & \,\,   \h=1 & \,\,   \tfrac{1}{2}& \,\,   [2,0]^{\h+1}_{0}  & \,\,\mathfrak{Q}_{2A}\,\mathfrak{Q}_{11} +\mathfrak{Q}_{21}\,\mathfrak{Q}_{1A}   \\
\hline
\hline
\,\,  B_1 & \,\,   [a,b]^{\h}_0 & \,\,  \h=a+b  & \,\,    \tfrac{1}{4} & \,\,   [a+1,b]^{\h+1/2}_0 & \,\,  \mathfrak{Q}_{\alpha 1}     \\
\hline
\,\,   & \,\,    [0,b]^{\h}_0 & \,\,  \h=b  & \,\,    \tfrac{1}{2} & \,\,   [1,b]^{\h+1/2}_0 & \,\,  \mathfrak{Q}_{\alpha 1}  ,\,\mathfrak{Q}_{\alpha 2}    \\
\hline
\end{array}
\end{align*}
\caption{Superconformal multiplets of $\mathfrak{osp}(4^*|4)$, from Table 8 of \cite{Agmon:2020pde}.}
\label{tab:supermultipletsclassification}
\end{table}
From \cite{Agmon:2020pde} we can also read off the recombination rules for the decomposition of long multiplets at the unitarity bound $\h_{\star}=a+b+s/2+1$
\begin{align}\label{recombination}
\begin{split}
L[a,b]^{\h\to\h_{\star}}_s\,&=\,A_1[a,b]^{\h_{\star}}_s\,\oplus\,A_1[a+1,b]^{\h_{\star}+1/2}_{s-1}\,, \\
L[a,b]^{\h\to\h_{\star}}_0\,&=\,A_2[a,b]^{\h_{\star}}_0\,\oplus\,B_1[a+2,b]^{\h_{\star}+1}_{0}\,,
\end{split}
\end{align}
from which it follows that the only absolutely protected multiplets are $B_1[a,b]^{a+b}_0$ for $a=0,1$ and $b\in \mathbb{N}$.

In the main test we discussed two types of multiplets, $\mathcal{D}_k$ and $\mathcal{L}^{\h}_{s,[a,b]}$, which are related to those of Table \ref{tab:supermultipletsclassification} by
\begin{align}
\mathcal{D}_k\,\longleftrightarrow\,B_1[0,k]^k_0\,,\qquad
\mathcal{L}^{\h}_{s,[a,b]}\,\longleftrightarrow\,L[a,b]^{\h}_s\,,
\end{align} 
while we never discussed other types of multiplets. The reason is twofold:
\begin{itemize}
\item Semi-short multiplets turn out to be absent from the spectrum at strong coupling, as shown by the counting of states at $\lambda=\infty$ performed in Appendix \ref{app:counting}. However they are present at weak coupling even in the planar limit and in particular the fundamental scalar of $\mathcal{N}=4$ SYM that is coupled to the Maldacena Wilson line (``$\Phi_{\parallel}$'') is the superconformal primary of a semi-short $A_2[0,0]^1_0$ multiplet.
\item The absolutely protected multiplets $B_1[1,b]^{1+b}_0$ are also absent from the spectrum at strong coupling for the same reason, but they also turn out to be absent from the spectrum at weak coupling, at least in the planar limit. 
\end{itemize}

\subsection{Structure of multiplets}

Let us now list the conformal primaries contained in each of the superconformal multiplets listed in table \ref{tab:supermultipletsclassification}. They can be obtained using the techniques of \cite{Cordova:2016emh} but here we derive the tables below using the characters of these multiplets, as we explain in Appendix \ref{app:counting}. Details of the derivation are given there, while here we simply list the states and the associated degeneracies. Note that for each conformal primary (except for the superprimary) we limit to give the $\mathfrak{su}(2)\oplus\mathfrak{sp}(4)$ Dynkin labels, while we omit the $\mathfrak{sl}(2)$ weight $\h$ which is simply raised by $1/2$ for each action of $Q$.

For $\tfrac{1}{4}$-BPS $B_1[a,b]^{a+b}_0$ multiplets ($a>0$) we find the states
\begin{align}\label{B1[a,b]}
\begin{split}
[a,b]^{a+b}_{0} \,\,\mapright Q\,\, &[a-1,b]_1,\,
    [a-1,b+1]_1,\,
    [a+1,b-1]_1\\
  \,\,\mapright {Q^2}\,\, &     [a-2,b]_0,\,[a-2,b+1]_0,\,[a-2,b+2]_0,\,[a,b-1]_0,\,[a,b]_0,\,[a+2,b-2]_0,\\
& [a-2,b+1]_2,\,[a,b-1]_2,\,[a,b]_2\\
 \,\,\mapright {Q^3}\,\, & [a-3,b+1]_1,\,[a-3,b+2]_1,\,[a-1,b-1]_1,\,\mathbf{2}\,[a-1,b]_1,\,[a-1,b+1]_1,\\
 &[a+1,b-2]_1,\,[a+1,b-1]_1,\,[a-1,b]_3\\
  \,\,\mapright {Q^4}\,\, & [a-4,b+2]_0,\,[a-2,b]_0,\,[a-2,b+1]_0,\,[a,b-2]_0,\,[a,b-1]_0,\,[a,b]_0,\\
  & [a-2,b]_2,\,[a-2,b+1]_2,\,[a,b-1]_2\\
  \,\,\mapright {Q^5}\,\, & [a-3,b+1]_1,\,[a-1,b-1]_1,\,[a-1,b]_1\\
  \,\,\mapright {Q^6}\,\, & [a-2,b]_0\,.
\end{split}
\end{align}
In the case $a=0$ the multiplet becomes $\tfrac{1}{2}$-BPS and we find the absolutely protected $B_1[0,b]^b_0\equiv \mathcal{D}_b$ with states
\begin{align}\label{B1[0,b]}
[0,b]^{b}_0\,\,\mapright Q\,\, [1,b-1]_1\,\, \mapright {Q^2} \,\,[0,b-1]_2,\,[2,b-2]_0\,\, \mapright {Q^3}\,\, [1,b-2]_1 \,\,\mapright {Q^4} \,\,[0,b-2]_0\,.
\end{align}

The $\tfrac{1}{8}$-BPS multiplets $A_2[a,b]^{a+b+1}_0$ contains the conformal primaries
\begin{align}\label{A2[a,b]}
\begin{split}
[a,b]^{a+b+1}_{0} \,\,\mapright Q\,\, & [a-1,b]_1,\,[a-1,b+1]_1,\,[a+1,b-1]_1,\,[a+1,b]_1\\
\,\,\mapright {Q^2}\,\, & [a-2,b]_0,\,[a-2,b+1]_0,\,[a-2,b+2]_0,\,[a,b-1]_0,\,\mathbf{2}\,[a,b]_0,\,[a,b+1]_0\\
& [a+2,b-2]_0,\,[a+2,b-1]_0,\,[a-2,b+1]_2,\,[a,b-1]_2,\,\mathbf{2}\,[a,b]_2,\\
&[a,b+1]_2,\,[a+2,b-1]_2\\
\,\,\mapright {Q^3}\,\, & [a-3,b+1]_1,\,[a-3,b+2]_1,\,[a-1,b-1]_1,\,[a-1,b]_1,\,[a-1,b+1]_1,\\
&[a-1,b+2]_1,\,[a+1,b-2]_1,\,\mathbf{3} [a+1,b-1]_1,\,\mathbf{2}\,[a+1,b]_1,\,[a+3,b-2]_1,\\
&[a-1,b]_3,\,[a-1,b+1]_3,\,[a+1,b-1]_3,\,[a+1,b]_3\\
\,\,\mapright {Q^4}\,\, & [a-4,b+2]_0,\,[a-2,b]_0,\,\mathbf{2}\,[a-2,b+1]_0,\,[a-2,b+2]_0,\,[a,b-2]_0,\,\mathbf{2}\,[a,b-1]_0,\\
& \mathbf{3}\,[a,b]_0,\,[a,b+1]_0,\,[a+2,b-2]_0,\,[a+2,b-1]_0,\,[a-2,b]_2,\,\mathbf{2}\,[a-2,b+1]_2,\\
& [a-2,b+2]_2,\,\mathbf{2}\,[a,b-1],\,\mathbf{3}\,[a,b]_2,\,[a,b+1]_2,\,[a+2,b-2]_2,\,[a+2,b-1]_2,\,[a,b]_4\\
\,\,\mapright {Q^5}\,\, & [a-3,b+1]_1,\,[a-3,b+2]_1,\,[a-1,b-1]_1,\,\mathbf{3}\,[a-1,b]_1,\,\mathbf{2}\,[a-1,b+1]_1,\\
&[a+1,b-2]_1,\,\mathbf{2}\,[a+1,b-1]_1,\,[a+1,b]_1,\,[a-1,b]_3,\,[a-1,b+1]_3,\,[a+1,b-1]_3\\
\,\,\mapright {Q^6}\,\, & [a-2,b]_0,\,[a-2,b+1]_0,\,[a,b-1]_0,\,[a,b]_0,\,[a-2,b+1]_2,\,[a,b-1]_2,\,[a,b]_2\\
\,\,\mapright {Q^7}\,\, & [a-1,b]_1\,.
\end{split}
\end{align}
When $a=0$ this becomes $\tfrac{1}{4}$-BPS and contains the states
\begin{align}\label{A2[0,b]}
\begin{split}
[0,b]^{b+1}_0\,\,\mapright {Q}\,\, &[1,b-1]_1,\,[1,b]_1\\
\,\,\mapright {Q^2}\,\, &[0,b]_0,\,[2,b-2]_0,\,[2,b-1]_0,\,[0,b-1]_2,\,[0,b]_2,\,[0,b+1]_2,\,[2,b-1]_2\\
\,\,\mapright {Q^3}\,\, & [1,b-2]_1,\,\mathbf{2}\,[1,b-1]_1,\,[1,b]_1,\,[3,b-2]_1,\,[1,b-1]_3,\,[1,b]_3\\
\,\,\mapright {Q^4}\,\, & [0,b-2]_0,\,[0,b-1]_0,\,[0,b]_0,\,[2,b-2]_0,\,[0,b-1]_2,\,[0,b]_2,\,[2,b-2]_2,\,[2,b-1]_2,\,[0,b]_4\\
\,\,\mapright {Q^5}\,\, & [1,b-2]_1,\,[1,b-1]_1,\,[1,b-1]_3\\
\,\,\mapright {Q^6}\,\, & [0,b-1]_2,
\end{split}
\end{align}
while for $a=b=0$ we have the $\tfrac{1}{2}$-BPS multiplet
\begin{align}\label{A2[0,0]}
\begin{split}
[0,0]^{1}_0\,\,\mapright Q\,\, [1,0]_1\,\, \mapright {Q^2} \,\,[0,1]_2,\,[0,0]_2\,\, \mapright {Q^3}\,\, [1,0]_3 \,\,\mapright {Q^4} \,\,[0,0]_4\,.
\end{split}
\end{align}

The other possible shortening condition gives rise to $\tfrac{1}{8}$-BPS multiplets $A_1[a,b]^{a+b+s/2+1}_s$ ($s>0$) with states
\begin{align}\label{A1[a,b]_s}
\begin{split}
[a,b]^{a+b+s/2+1}_s\,\,\mapright {Q}\,\, &[a-1,b]_{s\pm 1},\,[a-1,b+1]_{s\pm 1},\,[a+1,b-1]_{s\pm 1},\,[a+1,b]_{s+1},\\
\,\,\mapright {Q^2}\,\, &[a-2,b+1]_{s\pm 2},\,[a,b-1]_{s\pm 2},\,[a,b]_{s\pm 2},\,[a-2,b]_s,\,\mathbf{2}\,[a-2,b+1]_s,\\
&[a-2,b+2]_s,\,\mathbf{2}\,[a,b-1],\,\mathbf{3}\,[a,b]_s,\,[a,b+1]_s,\,[a+2,b-2]_s,\,[a+2,b-1]_s\\
& [a,b]_{s+2},\,[a,b+1]_{s+2},\,[a+2,b-1]_{s+2},\\
\,\,\mapright {Q^3}\,\, &[a-1,b]_{s\pm 3},\,[a-3,b+1]_{s\pm 1},\,[a-3,b+2]_{s\pm 1},\,[a-1,b-1]_{s\pm 1},\\
&\mathbf{3}\,[a-1,b]_{s\pm1},\,\mathbf{2}\,[a-1,b+1]_{s\pm 1},\,[a-1,b+2]_{s\pm1},\,\mathbf{2}\,[a+1,b-1]_{s\pm 1}\\
&[a+1,b]_{s\pm 1},\, [a-1,b]_{s+1},\,[a-1,b+1]_{s+1},\,[a-1,b+2]_{s+1},\,[a+1,b-1]_{s+1},\\
&[a+1,b]_{s+1},\,[a+3,b-2]_{s+1},\,[a-1,b+1]_{s+3},\,[a+1,b-1]_{s+3},\,[a+1,b]_{s+3},\\
\,\,\mapright {Q^4}\,\, &[a-2,b]_{s\pm 2},\,[a-2,b+1]_{s\pm 2},\,[a,b-1]_{s\pm 2},\,[a,b]_{s \pm 2},\,[a-4,b-2]_s,\\
&\mathbf{2}\,[a-2,b]_s,\,\mathbf{3}\,[a-2,b+1],\,[a-2,b+2]_s,\,[a,b-2]_s,\,\mathbf{3}\,[a,b-1],\,\mathbf{4}\,[a,b]_s,\\
& [a,b+1]_s,\,[a+2,b-2]_s,\,[a+2,b-1]_s,\,[a-2,b+1]_{s+2},\,[a-2,b+2]_{s+2},\\
&[a,b-1]_{s+2},\,\mathbf{2}\,[a,b]_{s+2},\,[a,b+1]_{s+2},\,[a+2,b-2]_{s+2},\,[a+2,b-1]_{s+2},\,[a,b]_{s+4},\\
\,\,\mapright {Q^5}\,\, & [a-3,b+1]_{s\pm 1},\,[a-1,b-1]_{s\pm 1},\,\mathbf{2}\,[a-1,b]_{s\pm 1},\,[a-1,b+1]_{s\pm 1},\\
&[a+1,b-1]_{s \pm 1},\,[a-3,b+2]_{s+1},\,[a-1,b]_{s+1},\,[a-1,b+1]_{s+1},\,[a+1,b-2]_{s+1},\\
&[a+1,b-1]_{s+1},\,[a+1,b]_{s+1},\,[a-1,b]_{s+3},\,[a-1,b+1]_{s+3},\,[a+1,b-1]_{s+3},\\
\,\,\mapright {Q^6}\,\, & [a-2,b]_s,\,[a-2,b+1]_s,\,[a,b-1]_s,\,[a,b]_s,\,[a-2,b+1]_{s+2},\,[a,b-1]_{s+2},\,[a,b]_{s+2},\\
\,\,\mapright {Q^7}\,\, & [a-1,b]_{s+1},
\end{split}
\end{align}
which for $a=0$ reduces to the $\tfrac{1}{4}$-BPS multiplet
\begin{align}\label{A1[0,b]_s}
\begin{split}
[0,b]^{b+s/2+1}_s\,\,\mapright {Q}\,\, & [1,b-1]_{s\pm1},\,[1,b]_{s+1}\\
\,\,\mapright {Q^2}\,\, & [0,b-1]_{s\pm 2},\,[0,b-1]_s,\,[0,b]_s,\,[2,b-2]_s,\,[2,b-1]_s,\,[0,b]_{s+2},\,[0,b+1]_{s+2},\,[2,b-1]_{s+2}\\
\,\,\mapright {Q^3}\,\, & [1,b-2]_{s\pm 1},\,[1,b-1]_{s\pm 1},\,[1,b-1]_{s+1},\,[1,b]_{s+1},\,[3,b-2]_{s+1},\,[1,b-1]_{s+3},\,[1,b]_{s+3}\\
\,\,\mapright {Q^4}\,\, &[0,b-2]_s,\,[0,b-1]_s,\,[0,b]_s,\,[2,b-2]_s,\,[0,b-1]_{s+2},\,[0,b]_{s+2},\,[2,b-2]_{s+2},\\
& [2,b-1]_{s+2},\,[0,b]_{s+4}\\
\,\,\mapright {Q^5}\,\, & [1,b-2]_{s+1},\,[1,b-1]_{s+1},\,[1,b-1]_{s+3}\\
\,\,\mapright {Q^6}\,\, & [0,b-1]_{s+2},
\end{split}
\end{align}
while for $a=b=0$ leads to the $\tfrac{1}{2}$-BPS multiplet
\begin{align}\label{A1[0,0]_s}
\begin{split}
[0,0]^{s/2+1}_s\,\,\mapright Q\,\, [1,0]_{s+1}\,\, \mapright {Q^2} \,\,[0,1]_{s+2},\,[0,0]_{s+2}\,\, \mapright {Q^3}\,\, [1,0]_{s+3} \,\,\mapright {Q^4} \,\,[0,0]_{s+4}\,.
\end{split}
\end{align}

Finally, we have long multiplets $L[a,b]^{\h}_s$ for $\h>\h_{\star}$, which are not subject to any shortening conditions and contain the states
\begin{align}\label{L[a,b]_s}
\begin{split}
[a,b]^{\h}_s\,\,\mapright {Q}\,\, & [a-1,b]_{s\pm 1},\,[a-1,b+1]_{s\pm 1},\,[a+1,b-1]_{s\pm 1},\,[a+1,b]_{s \pm 1}\\
\,\,\mapright {Q^2}\,\, & [a-2,b]_s,\,[a-2,b+1]_s,\,[a-2,b+2]_s,\,[a,b-1]_s,\,\mathbf{2}\,[a,b]_s,\,[a,b+1]_s,\\
&[a+2,b-2]_s,\,[a+2,b-1]_s,\,[a+2,b]_s,\,[a-2,b+1]_{s,s\pm 2},\,[a,b-1]_{s,s\pm 2},\\
&\mathbf{2}\,[a,b]_{s,s\pm 2},\,[a,b+1]_{s,s\pm 2},\,[a+2,b-1]_{s,s\pm 2}\\
\,\,\mapright {Q^3}\,\, & [a-3,b+1]_{s\pm 1},\,[a-3,b+2]_{s\pm 1},\,[a-1,b-1]_{s\pm 1},\,\mathbf{3}\,[a-1,b]_{s\pm 1},\,\mathbf{3}\,[a-1,b+1]_{s\pm 1},\\
&[a-1,b+2]_{s\pm 1},\,[a+1,b-2]_{s\pm 1},\,\mathbf{3}\,[a+1,b-1]_{s\pm 1},\,\mathbf{3}\,[a+1,b]_{s\pm 1},\\
&[a+1,b+1]_{s\pm 1},\,[a+3,b-2]_{s\pm 1},\,[a+3,b-1]_{s\pm 1},\,[a-1,b]_{s\pm 1,s\pm 3},\,[a-1,b+1]_{s\pm 1,s\pm 3},\\
&[a+1,b-1]_{s\pm 1,s\pm 3},\,[a+1,b]_{s\pm 1,s\pm 3}\\
\,\,\mapright {Q^4}\,\, & [a-4,b+2]_s,\,[a-2,b]_s,\,\mathbf{2}\,[a-2,b+1]_s,\,[a-2,b+2]_s,\,[a,b-2]_s,\,\mathbf{2}\,[a,b-1]_s,\\
&\mathbf{4}\,[a,b]_s,\,\mathbf{2}\,[a,b+1]_s,\,[a,b+2]_s,\,[a+2,b-2]_s,\,\mathbf{2}\,[a+2,b-1]_s,\,[a+2,b]_s,\,[a+4,b-2]_s,\\
&[a-2,b]_{s,s\pm 2},\,\mathbf{2}\,[a-2,b+1]_{s,s\pm 2},\,[a-2,b+2]_{s,s\pm 2},\,\mathbf{2}\,[a,b-1]_{s,s\pm 2},\,\mathbf{3}\,[a,b]_{s,s\pm 2},\\
&\mathbf{2}\,[a,b+1]_{s,s\pm 2},\,[a+2,b-2]_{s,s\pm 2},\,\mathbf{2}\,[a+2,b-1]_{s,s\pm 2},\,[a+2,b]_{s,s\pm 2},\,[a,b]_{s,s\pm 2,s\pm 4}\\
\,\,\mapright {Q^5}\,\, & [a-3,b+1]_{s\pm 1},\,[a-3,b+2]_{s\pm 1},\,[a-1,b-1]_{s\pm 1},\,\mathbf{3}\,[a-1,b]_{s\pm 1},\,\mathbf{3}\,[a-1,b+1]_{s\pm 1},\\
&[a-1,b+2]_{s\pm 1},\,[a+1,b-2]_{s\pm 1},\,\mathbf{3}\,[a+1,b-1]_{s\pm 1},\,\mathbf{3}\,[a+1,b]_{s\pm 1},\\
&[a+1,b+1]_{s\pm 1},\,[a+3,b-2]_{s\pm 1},\,[a+3,b-1]_{s\pm 1},\,[a-1,b]_{s\pm 1,s\pm 3},\,[a-1,b+1]_{s\pm 1,s\pm 3},\\
&[a+1,b-1]_{s\pm 1,s\pm 3},\,[a+1,b]_{s\pm 1,s\pm 3}\\
\,\,\mapright {Q^6}\,\, & [a-2,b]_s,\,[a-2,b+1]_s,\,[a-2,b+2]_s,\,[a,b-1]_s,\,\mathbf{2}\,[a,b]_s,\,[a,b+1]_s,\\
&[a+2,b-2]_s,\,[a+2,b-1]_s,\,[a+2,b]_s,\,[a-2,b+1]_{s,s\pm 2},\,[a,b-1]_{s,s\pm 2},\\
&\mathbf{2}\,[a,b]_{s,s\pm 2},\,[a,b+1]_{s,s\pm 2},\,[a+2,b-1]_{s,s\pm 2}\\
\,\,\mapright {Q^7}\,\, & [a-1,b]_{s\pm 1},\,[a-1,b+1]_{s\pm 1},\,[a+1,b-1]_{s\pm 1},\,[a+1,b]_{s \pm 1}\\
\,\,\mapright {Q^8}\,\, &[a,b]_s.
\end{split}
\end{align}

\section{Counting of states}\label{app:counting}

\subsection{Verma modules and characters}

Here we construct the characters of $\mathfrak{su}(2)$ and $\mathfrak{sp}(4)$ from the associated Verma modules. This is standard material so we will only display the formulas we are interested in without derivation -- for details see, {\it e.g.}, \cite{Fuchs:1997jv}.

For $\mathfrak{su}(2)$ we use a fugacity $z$ associated with the unique fundamental weight and the character for the Verma module of a highest weight representation with Dynkin label $s$ is given by
\begin{align}
\mathcal{V}^{\mathfrak{su}(2)}_s(z)=\frac{z^{s+2}}{z^2-1}\,.
\end{align}
The Weyl group is ${W}^{\mathfrak{su}(2)}=\mathbb{Z}_2$ maps $z \to z^{-1}$, or $s\to -s-2$. When $s\in\mathbb{N}$ the Verma module is reducible, corresponding to finite-dimensional irreducible highest weight representations, whose character is computed by a weighted sum over the images of $\mathcal{V}_s$ under the action of the Weyl group
\begin{align}\label{chisu2}
\chi^{\mathfrak{su}(2)}_s(z)
=\mathcal{V}^{\mathfrak{su}(2)}_s(z)-\mathcal{V}^{\mathfrak{su}(2)}_{-s-2}(z)
=\mathcal{V}^{\mathfrak{su}(2)}_s(z)+\mathcal{V}^{\mathfrak{su}(2)}_{s}(z^{-1})=\frac{z^{2+j}-z^{-j}}{z^2-1}\,.
\end{align}
The dimension of the representation is easily computed by taking $z=1$, which gives $\text{dim}_{\mathfrak{su}(2)}([s])=s+1$, making it clear that in our conventions $s=1$ is the fundamental representation. The expression \eqref{chisu2} also shows explicitly the reflection property
\begin{align}
\chi^{\mathfrak{su}(2)}_{-s-2}(z)=-\chi^{\mathfrak{su}(2)}_s(z)\,,
\end{align}
which must be used when the Dynkin label is negative, and in particular implies $\chi^{\mathfrak{su}(2)}_{-1}(z)=0$. We also record that the characters of $\mathfrak{su}(2)$ obey the orthogonality condition
\begin{align}\label{orthogonalitysu2}
\frac{1}{2\pi i}\oint_{z=0} \frac{\mathrm{d}z}{z}\mu^{\mathfrak{su}(2)}(z)\,\chi^{\mathfrak{su}(2)}_{s_1}(z)\,\chi^{\mathfrak{su}(2)}_{s_2}(z)=\delta_{s_1,s_2}\,,
\end{align}
with respect to the measure
\begin{align}
\mu^{\mathfrak{su}(2)}(z)=-\frac{(1-z^2)^2}{2\,z^2}\,.
\end{align}

For $\mathfrak{sp}(4)$ we fix two simple roots $\alpha_1$ and $\alpha_2$, with the other positive roots given by $\alpha_1+\alpha_2$, $2\,\alpha_1+\alpha_2$ and the Cartan matrix
\begin{align}
K^{\mathfrak{sp}(4)}=
\begin{pmatrix}
2 & -1\\
-2 & 2
\end{pmatrix}\,.
\end{align}
The fundamental weights corresponding to $\alpha_{1,2}$ are $w_1=\alpha_1+\tfrac{1}{2}\,\alpha_2$ and $w_2=\alpha_1+\alpha_2$, to which we associate fugacities $x$ and $y$, respectively. For each positive root $\alpha$ we write $\alpha=\sum_{i=1}^2 n(\alpha)_{i}\,\alpha_i$, from which the character of a Verma module with Dynkin labels $[a,b]$ is computed as
\begin{align}
\begin{split}
\mathcal{V}^{\mathfrak{sp}(4)}_{[a,b]}(x,y)&=x^a\,y^b\,\sum_{N_{\alpha}}\,\prod_{\alpha\in\Phi_+}x^{-N_{\alpha}\sum_{j=1}^2 n(\alpha)_j\,K_{j1}}\,y^{-N_{\alpha}\sum_{j=1}^2 n(\alpha)_j\,K_{j2}}\\
&=\frac{x^{4+a}\,y^{3+b}}{(1-x^2)(x^2-y^2)(y-x^2)(1-y)}\,,
\end{split}
\end{align}
where $\Phi_+$ is the set of (four) positive roots and for each $\alpha\in \Phi_+$ we sum over $N_{\alpha}\in \mathbb{N}$. To obtain the characters for finite-dimensional representations with $a,b\in\mathbb{N}$, we sum again over the weighted action of the Weyl group ${W}^{\mathfrak{sp}(4)}=S_2\ltimes (\mathbb{Z}_2)^2$, which has dimension eight and is generated by combinations of reflections of the two simple roots
\begin{align}
\sigma_1(\alpha_1)=-\alpha_1\,, \quad
\sigma_1(\alpha_2)=2\,\alpha_1+\alpha_2\,, \qquad
\sigma_2(\alpha_1)=\alpha_1+\alpha_2\,, \quad
\sigma_2(\alpha_2)=-\alpha_2\,, 
\end{align} 
from which the action on the fundamental weights $w_{1,2}$ can be easily worked out. It follows that
\begin{align}\label{chisp4}
\begin{split}
\chi^{\mathfrak{sp}(4)}_{[a,b]}(x,y)=\,&
\mathcal{V}^{\mathfrak{sp}(4)}_{[a,b]}(x,y)
-\mathcal{V}^{\mathfrak{sp}(4)}_{[-2-a,1+a+b]}(x,y)
-\mathcal{V}^{\mathfrak{sp}(4)}_{[2+a+2b,-2-b]}(x,y)
+\mathcal{V}^{\mathfrak{sp}(4)}_{[-4-a-2b,1+a+b]}(x,y)\\
&+\mathcal{V}^{\mathfrak{sp}(4)}_{[2+a+2b,-3-a-b]}(x,y)
-\mathcal{V}^{\mathfrak{sp}(4)}_{[a,-3-a-b]}(x,y)
-\mathcal{V}^{\mathfrak{sp}(4)}_{[-4-a-b,b]}(x,y)
+\mathcal{V}^{\mathfrak{sp}(4)}_{[-2-a,-2-b]}(x,y)\\
=\,&\mathcal{V}^{\mathfrak{sp}(4)}_{[a,b]}(x,y)
+\mathcal{V}^{\mathfrak{sp}(4)}_{[a,b]}(x^{-1}\,y,y)
+\mathcal{V}^{\mathfrak{sp}(4)}_{[a,b]}(x,x^2\,y^{-1})
+\mathcal{V}^{\mathfrak{sp}(4)}_{[a,b]}(x^{-1}\,y,x^{-2}\,y)\\
&+\mathcal{V}^{\mathfrak{sp}(4)}_{[a,b]}(x\,y^{-1},x^2\,y^{-1})
+\mathcal{V}^{\mathfrak{sp}(4)}_{[a,b]}(x\,y^{-1},y^{-1})
+\mathcal{V}^{\mathfrak{sp}(4)}_{[a,b]}(x^{-1},x^{-2}\,y)
+\mathcal{V}^{\mathfrak{sp}(4)}_{[a,b]}(x^{-1},y^{-1})\,,
\end{split}
\end{align}
where in the two expressions the action of ${W}^{\mathfrak{sp}(4)}$ on the Dynkin labels $[a,b]$ is paired with that on the fugacities $(x,y)$ in such a way that one can compare the two just by taking the same term in the two sums. The reflection properties of the characters are easily read off from \eqref{chisp4} and in particular give
\begin{align}
\chi^{\mathfrak{sp}(4)}_{[-1,b]}(x,y)=\chi^{\mathfrak{sp}(4)}_{[a,-1]}(x,y)=\chi^{\mathfrak{sp}(4)}_{[-3-2b,b]}(x,y)=\chi^{\mathfrak{sp}(4)}_{[a,-2-a]}(x,y)=0\,,
\end{align}
while the dimension of representations is obtained by taking $x=y=1$ and reads 
\begin{align}
\text{dim}_{\mathfrak{sp}(4)}([a,b])=\frac{1}{6}(1+a)(1+b)(2+a+b)(3+a+2b)\,.
\end{align}
Two representations will be of particular interest: the representation $\mathbf{4}=[1,0]$, which is the defining representation of $\mathfrak{sp}(4)$ or the spinorial of $\mathfrak{so}(5)$, with states and character
\begin{align}
\begin{split}
\begin{tikzpicture}
\node (p1) at (0, 0) {$[1,0]$};
\node (p2) at (2,0) {$[-1,1]$};
\node (p3) at (4,0) {$[1,-1]$};
\node (p4) at (6,0) {$[-1,0]$};
\node (p5) at (0,-0.5) {$x$};
\node (p6) at (2,-0.5) {$x^{-1}\,y$};
\node (p7) at (4,-0.5) {$x\,y^{-1}$};
\node (p8) at (6,-0.5) {$x^{-1}$};
 \draw [-{Stealth}] (p1) -- (p2) node[midway, above] {$-\alpha_1$};
 \draw [-{Stealth}] (p2) -- (p3) node[midway, above] {$-\alpha_2$};
 \draw [-{Stealth}] (p3) -- (p4) node[midway, above] {$-\alpha_1$};
\node (p9) at (9.7,-0.25) {$\Rightarrow \quad
\chi^{\mathfrak{sp}(4)}_{[1,0]}(x,y)=x+\frac{y}{x}+\frac{x}{y}+\frac{1}{x}\,,$};
\end{tikzpicture}
\end{split}
\end{align}
and the representation $\mathbf{5}=[0,1]$ which is the fundamental of $\mathfrak{so}(5)$, with states and character
\begin{align}
\begin{split}
\begin{tikzpicture}
\node (p1) at (0, 0) {$[0,1]$};
\node (p2) at (2,0) {$[2,-1]$};
\node (p3) at (4,0) {$[0,0]$};
\node (p4) at (6,0) {$[-2,1]$};
\node (p5) at (8,0) {$[0,-1]$};
\node (p6) at (0,-0.5) {$y$};
\node (p7) at (2,-0.5) {$x^2\,y^{-1}$};
\node (p8) at (4,-0.5) {$1$};
\node (p9) at (6,-0.5) {$x^{-2}\,y$};
\node (p10) at (8,-0.5) {$y^{-1}$};
 \draw [-{Stealth}] (p1) -- (p2) node[midway, above] {$-\alpha_2$};
 \draw [-{Stealth}] (p2) -- (p3) node[midway, above] {$-\alpha_1$};
 \draw [-{Stealth}] (p3) -- (p4) node[midway, above] {$-\alpha_1$};
  \draw [-{Stealth}] (p4) -- (p5) node[midway, above] {$-\alpha_2$};
\node (p11) at (12,-0.25) {$\Rightarrow \quad
\chi^{\mathfrak{sp}(4)}_{[0,1]}(x,y)=y+\frac{x^2}{y}+1+\frac{y}{x^2}+\frac{1}{y}\,\,.$};
\end{tikzpicture}
\end{split}
\end{align}
To conclude, let us also record the orthogonality property of $\mathfrak{sp}(4)$ characters
\begin{align}\label{orthogonalitysp4}
\frac{1}{(2\pi i)^2}\oint_{x=0}\oint_{y=0} \frac{\mathrm{d}x}{x}\frac{\mathrm{d}y}{y}\mu^{\mathfrak{sp}(4)}(x,y)\,\chi^{\mathfrak{sp}(4)}_{[a,b]}(z)\,\chi^{\mathfrak{sp}(4)}_{[c,d]}(z)=\delta_{a,c}\,\delta_{b,d}\,,
\end{align}
with respect to the measure
\begin{align}
\mu^{\mathfrak{sp}(4)}(x,y)=\frac{(1-x^2)^2\,(x^2-y^2)^2\,(y-x^2)^2\,(1-y)^2}{8\,x^6\,y^4}\,.
\end{align}

The other ingredients that we need are the characters of $\mathfrak{sl}(2)$ and those of the Poincar\'e supercharges. For a representation of $\mathfrak{sl}(2)$ with highest-weight state of dimension $\h$ we have
\begin{align}
\chi^{\mathfrak{sl}(2)}_{\h}(q)=\frac{q^{2\h}}{1-q^2}\,,
\end{align}
where the fugacity $q$ is associated with states of dimension $\h=1/2$. For the Poincar\'e supercharges $\mathfrak{Q}_{\alpha A}\in[1,0]^{\h=1/2}_{s=1}$ ($\alpha=1,2$, $A=1,\dots,4$), following section 3.2.2 of \cite{Agmon:2020pde} we assign weights (and therefore characters $\chi^{\mathfrak{Q}}_{\alpha A}$)
\begin{equation}\label{Qchar}
  \begin{alignedat}{2}
&\mathfrak{Q}_{11}\,[1,0]^{1/2}_{+1}\,:\chi^{\mathfrak{Q}}_{11}=1+q\,x\,z\,, \qquad
&&\mathfrak{Q}_{12}\,[-1,1]^{1/2}_{+1}\,:\chi^{\mathfrak{Q}}_{12}=1+q\,x^{-1}\,y\,z\,, \\
&\mathfrak{Q}_{13}\,[1,-1]^{1/2}_{+1}\,:\chi^{\mathfrak{Q}}_{13}=1+q\,x\,y^{-1}\,z\,, \qquad
&&\mathfrak{Q}_{14}\,[-1,0]^{1/2}_{+1}\,:\chi^{\mathfrak{Q}}_{14}=1+q\,x^{-1}\,z\,, \\
&\mathfrak{Q}_{21}\,[1,0]^{1/2}_{-1}\,:\chi^{\mathfrak{Q}}_{21}=1+q\,x\,z^{-1}\,, \qquad
&&\mathfrak{Q}_{22}\,[-1,1]^{1/2}_{-1}\,:\chi^{\mathfrak{Q}}_{22}=1+q\,x^{-1}\,y\,z^{-1}\,, \\
&\mathfrak{Q}_{23}\,[1,-1]^{1/2}_{-1}\,:\chi^{\mathfrak{Q}}_{23}=1+q\,x\,y^{-1}\,z^{-1}\,, \qquad
&&\mathfrak{Q}_{24}\,[-1,0]^{1/2}_{-1}\,:\chi^{\mathfrak{Q}}_{24}=1+q\,x^{-1}\,z^{-1}\,.
  \end{alignedat}
\end{equation}

\subsection{Characters of $\mathfrak{osp}(4^*|4)$ supermultiplets}\label{app:counting-supermultiplets}

In this section we explain how to construct the characters for all superconformal multiplets of $\mathfrak{osp}(4^*|4)$ allowing, for example, to derive and count the conformal primaries in each multiplet, which we listed in Appendix \ref{app:supermultiplets}. Following \cite{Bianchi:2006ti}, we start with the Verma module character of the highest-weight state in the superconformal multiplet and we generate the Verma module character for the associated $\mathfrak{osp}(4^*|4)$ representation multiplying with characters associated to translations and to the supercharges that do not annihilate the superconformal primary. Finally, to build the characters of the superconformal multiplets, we symmetrize with respect to the action of the Weyl groups of $\mathfrak{su}(2)$ and $\mathfrak{sp}(4)$.

Let us start by defining the Verma module character associated to a given superconformal multiplet, which for generic values of the Dynkin labels reads
\begin{align}
\mathcal{V}^{\h}_{s,[a,b]}(q;x,y;z)=q^{2\h}\,\mathcal{V}^{\mathfrak{sp}(4)}_{[a,b]}(x,y)\,\mathcal{V}^{\mathfrak{su}(2)}_{s}(z)\,\chi^{\mathfrak{P}}(q)\,\frac{\chi^{\mathfrak{Q}}_{\text{All}}(q;x,y;z)}{\chi^{\mathfrak{Q}}_{\text{Null}}(q;x,y;z)}\,,
\end{align}
where 
\begin{align}
\chi^{\mathfrak{P}}(q)=\frac{1}{1-q^2}=\sum_{k=0}^{\infty}q^{2k}\,,\qquad
\chi^{\mathfrak{Q}}_{\text{All}}(q;x,y;z)=\prod_{\alpha=1}^2\prod_{A=1}^4\chi^{\mathfrak{Q}}_{\alpha A}(q;x,y;z)\,,
\end{align}
correspond to the action of all translations and all Poincar\'e supercharges, respectively. On the other hand, we have
\begin{align}
\chi^{\mathfrak{Q}}_{\text{Null}}(q;x,y;z)=\prod_{\alpha,A\,\,\text{s.t.}\,\,\mathfrak{Q}_{\alpha A}\, \ket{\text{h.w. }}=0}\chi^{\mathfrak{Q}}_{\alpha A}(q;x,y;z)\,,
\end{align}
{\it i.e.} dividing by $\chi^{\mathfrak{Q}}_{\text{Null}}(q;x,y;z)$ takes into account that certain supercharges act trivially on the superconformal primary for short representations, as detailed in Table \ref{tab:supermultipletsclassification}. Finally, the character of the supermultiplet is obtained from the Verma module character by symmetrizing with respect to the action of the Weyl group of $\mathfrak{su}(2)$ and $\mathfrak{sp}(4)$, which is implemented by the two operators 
\begin{align}
\begin{split}
\mathfrak{W}^{\mathfrak{su}(2)}\,f(z)\,=\,&f(z)+f(z^{-1})\,,\\
\mathfrak{W}^{\mathfrak{sp}(4)}\,f(x,y)\,=\,&f(x,y)+f(x^{-1}\,y,y)+f(x,x^2\,y^{-1})+f(x^{-1}\,y,x^{-2}\,y)+f(x\,y^{-1},x^2\,y^{-1})\\
&+f(x\,y^{-1},y^{-1})+f(x^{-1},x^{-2}\,y)+f(x^{-1},y^{-1})\,.
\end{split}
\end{align}
We are then ready to define the character of a superconformal multiplet to be
\begin{align}
\chi^{\h}_{s,[a,b]}(q;x,y;z)=\mathfrak{W}^{\mathfrak{su}(2)}\,\mathfrak{W}^{\mathfrak{sp}(4)}\,\mathcal{V}^{\h}_{s,[a,b]}(q;x,y;z)\,.
\end{align}
Of course the final expression depends on the type of multiplet through the shortening conditions and we shall now briefly discuss how this works for each type of shortening.

Let us start from the simplest case of generic long multiplets $L[a,b]^{\h}_s$. In this case there are no shortening conditions, so $\chi^{\mathfrak{Q}}_{\text{Null}}(q;x,y;z)=1$. Since $\chi^{\mathfrak{Q}}_{\text{All}}(q;x,y;z)$ is invariant under the action of $W^{\mathfrak{su}(2)}$ and $W^{\mathfrak{sp}(4)}$, we obtain the simple result that for long multiplets
\begin{align}\label{charlong}
\chi(L)^{\h}_{s,[a,b]}=\chi^{\mathfrak{sl}(2)}_{\h}(q)\,\chi^{\mathfrak{sp}(4)}_{[a,b]}(x,y)\,\chi^{\mathfrak{su}(2)}_{s}(z)\,\chi^{\mathfrak{Q}}_{\text{All}}(q;x,y;z)\,.
\end{align}
We can then expand as a sum of characters of $\mathfrak{su}(2)\oplus\mathfrak{sp}(4)$ and write
\begin{align}
\chi(L)^{\h}_{s,[a,b]}=\chi^{\mathfrak{sl}(2)}_{\h}(q)\,\sum_{k=0}^8 q^k\sum_{\mathcal{O}_k}  \chi^{\mathfrak{sp}(4)}_{[a(\mathcal{O}_k),b(\mathcal{O}_k)]}(x,y)\,\chi^{\mathfrak{su}(2)}_{s(\mathcal{O}_k)}(z)\,,
\end{align}
where $k$ denotes the level in the supermultiplet while the second sum runs over all conformal primary states $\mathcal{O}_k$ present at level $k$, which are listed in eq. \eqref{L[a,b]_s}.

For multiplets $A_1[a,b]^{\h=a+b+s/2+1}_s$ we repeat the exercise with $\chi^{\mathfrak{Q}}_{\text{Null}}(q;x,y;z)=\chi^{\mathfrak{Q}}_{21}$, for generic $a>0$, $b>0$, $s>0$. From now on the Weyl-symmetrizers act non-trivially also on the characters of Poincar\'e supercharges, since not all of them are present due to the shortening condition, so a clean expression like \eqref{charlong} is not immediate to write. However, we can still decompose the result as
\begin{align}\label{charA1}
\chi(A_1)^{a+b+s/2+1}_{s,[a,b]}=\chi^{\mathfrak{sl}(2)}_{a+b+s/2+1}(q)\,\sum_{k=0}^7 q^k\sum_{\mathcal{O}_k}  \chi^{\mathfrak{sp}(4)}_{[a(\mathcal{O}_k),b(\mathcal{O}_k)]}(x,y)\,\chi^{\mathfrak{su}(2)}_{s(\mathcal{O}_k)}(z)\,,
\end{align}
where this time the multiplet only contains states up to level seven since it is generically $\tfrac{1}{8}$-BPS and the states that contribute to each level are those listed in eq. \eqref{A1[a,b]_s}. To obtain the characters in the special cases $a=0$ and $a=b=0$ it is enough to set these to zero in \eqref{charA1}. Using the properties of the characters under Weyl reflections when the Dynkin labels are negative, one can easily derive the lists of states \eqref{A1[0,b]_s} and \eqref{A1[0,0]_s} and moreover it is easy to see that the sum over $k$ truncates to $k=6$ and $k=4$, respectively, corresponding to the fact that the multiplet is $\tfrac{1}{4}$-BPS for $a=0$ and $\tfrac{1}{2}$-BPS for $a=b=0$.

Next, we focus on multiplets $B_1[a,b]^{\h=a+b}_0$, for which $\chi^{\mathfrak{Q}}_{\text{Null}}(q;x,y;z)=\chi^{\mathfrak{Q}}_{11}\,\chi^{\mathfrak{Q}}_{21}$. Again one should act with the Weyl symmetrizer and expand in characters, obtaining 
\begin{align}\label{charB1}
\chi(B_1)^{a+b}_{0,[a,b]}=\chi^{\mathfrak{sl}(2)}_{a+b}(q)\,\sum_{k=0}^6 q^k\sum_{\mathcal{O}_k}  \chi^{\mathfrak{sp}(4)}_{[a(\mathcal{O}_k),b(\mathcal{O}_k)]}(x,y)\,\chi^{\mathfrak{su}(2)}_{s(\mathcal{O}_k)}(z)\,,
\end{align}
where the sum runs over the states listed in eq. \eqref{B1[a,b]}. To obtain the character of the multiplet $B_1[0,b]^{\h=b}_0$ one should repeat the exercise with $\chi^{\mathfrak{Q}}_{\text{Null}}(q;x,y;z)=\chi^{\mathfrak{Q}}_{11}\,\chi^{\mathfrak{Q}}_{21}\,\chi^{\mathfrak{Q}}_{12}\,\chi^{\mathfrak{Q}}_{22}$, with the resulting states listed in \eqref{B1[a,b]}. 

Finally, to obtain the character of $A_2[a,b]^{\h}_0$ multiplets we proceed indirectly, since in this case the shortening condition is enforced by a product of two supercharges, rather than individual ones. Looking at the second of the recombination rules \eqref{recombination} it is clear that one should have
\begin{align}
\chi(A_2)^{a+b+1}_{0,[a,b]}=\chi(L)^{a+b+1}_{0,[a,b]}-\chi(B_1)^{a+b+2}_{0,[a+2,b]}\,,
\end{align}
which gives the desired expression. A consistency check is that $\chi(A_2)^{a+b+1}_{0,[a,b]}$ contains only states up to level 7, as it should be for a $\tfrac{1}{8}$-BPS multiplet, and the resulting states are listed in \eqref{A2[a,b]}. Setting $a=0$ and using the reflection rules of $\mathfrak{sp}(4)$ one obtains a $\tfrac{1}{4}$-BPS multiplet, as expected, with the states listed in \eqref{A2[0,b]}. Repeating the exercise with $a=b=0$ leads to a $\tfrac{1}{2}$-BPS multiplet containing the states given in \eqref{A2[0,0]}. The fact that we obtain the correct shortening conditions when the Dynkin labels are set to zero is a further check of our results.

\subsection{Counting of states at strong coupling}\label{app:counting-strong}

Let us now turn to the problem of counting states at strong coupling and arranging them in supermultiplets of $\mathfrak{osp}(4^*|4)$. This was already addressed in \cite{Liendo:2018ukf} and here we give more details of that construction, while complementing it with some additional results. The counting of states is easily performed from the point of view of the AdS description of the theory, which at $\lambda=\infty$ is simply that of a free theory where all states are generated by taking graded-symmetrized tensor products of states in the $\mathcal{D}_1$ multiplet, which from the bulk perspective contains the coordinates that describe the fluctuations of the superstring in AdS$_5\times S^5$ around a minimal AdS$_2$ worldsheet. The partition function for such a (free) theory is easily computed as
\begin{align}
\mathcal{Z}_{\text{strong}}=\mathcal{Z}_{\varphi}\,\mathcal{Z}_{\Psi}\,\mathcal{Z}_f\,,
\end{align}
where $\mathcal{Z}_{X}$ for $X\in\{\varphi,\Psi,f\}$ is the partition function for the individual components of $\mathcal{D}_1$. To give their expression it is useful to introduce an auxiliary function 
\begin{align}
\chi_{\text{bos}}(a;q^2)=\frac{1}{(a,q^2)_{\infty}}=\prod_{k=0}^{\infty}(1-a\,q^{2k})^{-1}\,,
\end{align}
where $(a,q)_{\infty}$ is a $q$-Pochhammer symbol and $\chi_{\text{bos}}(q^{2\h};q^2)$ is a generating function for all states obtained from a single boson of dimension $\h$, while
\begin{align}
\chi_{\text{ferm}}(a;q^2)=\chi^{-1}_{\text{bos}}(-a;q^2)\,,
\end{align}
is such that $\chi_{\text{ferm}}(q^{2\h};q^2)$ is a generating function for all states obtained from a single fermion of dimension $\h$. From these building blocks one obtains the partition functions for the individual components of $\mathcal{D}_1$ just by taking into account the weights under $\mathfrak{su}(2)\oplus\mathfrak{sp}(4)$. Moreover, we also add a fugacity $\len$ that allows to take into account the {\it length} of states, {\it i.e.} how many times the components of $\mathcal{D}_1$ are used to build a certain state. The results are
\begin{align}
  \begin{split}
\mathcal{Z}_{\varphi}=&\,
\chi_{\text{bos}}(\len\,y\,q^2;q^2)\,
\chi_{\text{bos}}(\len\,x^2\,y^{-1}\,q^2;q^2)\,
\chi_{\text{bos}}(\len\,q^2;q^2)\,
\chi_{\text{bos}}(\len\,x^{-2}\,y\,q^2;q^2)\,
\chi_{\text{bos}}(\len\,y^{-1}\,q^2;q^2)\,,\\
\mathcal{Z}_{\Psi}=&\,
\chi_{\text{ferm}}(\len\,x\,z\,q^3;q^2)\,
\chi_{\text{ferm}}(\len\,x^{-1}\,y\,z\,q^3;q^2)\,
\chi_{\text{ferm}}(\len\,x\,y^{-1}\,z\,q^3;q^2)\,
\chi_{\text{ferm}}(\len\,x^{-1}\,z\,q^3;q^2)\\
&\chi_{\text{ferm}}(\len\,x\,z^{-1}\,q^3;q^2)\,
\chi_{\text{ferm}}(\len\,x^{-1}\,y\,z^{-1}\,q^3;q^2)\,
\chi_{\text{ferm}}(\len\,x\,y^{-1}\,z^{-1}\,q^3;q^2)\,
\chi_{\text{ferm}}(\len\,x^{-1}\,z^{-1}\,q^3;q^2)\,,\\
\mathcal{Z}_{f}=&\,
\chi_{\text{bos}}(\len\,z^2\,q^4;q^2)\,
\chi_{\text{bos}}(\len\,y\,q^4;q^2)\,
\chi_{\text{bos}}(\len\,z^{-2}\,q^4;q^2)\,.\\
  \end{split}
\end{align}
To count the supermultiplets that appear in the theory at strong coupling, one should expand the partition function $\mathcal{Z}$ as a sum over the characters of $\mathfrak{osp}(4^*|4)$ introduced in the previous subsection. It turns out that only the $\mathcal{D}_k$ and $\mathcal{L}^{\h}_{s,[a,b]}$ contribute and we can write
\begin{align}\label{Z=short+long}
\mathcal{Z}_{\text{strong}}=1+\mathcal{Z}^{\text{Short}}_{\text{strong}}+\mathcal{Z}^{\text{Long}}_{\text{strong}}\,,
\end{align}
where
\begin{align}
\mathcal{Z}^{\text{Short}}_{\text{strong}}=\sum_{k=1}^{\infty}\len^k\,\chi_{\mathcal{D}_k}\,,
\end{align}
and note in particular that there is exactly one short multiplet $\mathcal{D}_k$ for each $k$. The non-trivial part is the expansion of $\mathcal{Z}^{\text{Long}}_{\text{strong}}$: in Table \ref{tab:longdegeneraciesstrong} we collect, for instance, some results for the degeneracy of spin-less superprimaries.

Equivalently one can obtain the partition function using the plethystic exponential of the character of the $\mathcal{D}_1$ multiplet. We define the latter by adding an additional fugacity $\sigma$ that distinguishes states with fermionic statistics:
\begin{align}
\chi(\mathcal{D}_1)_{\sigma}(q;x,y;z)=\chi^{\mathfrak{sl}(2)}_1(q)\,\chi^{\mathfrak{sp}(4)}_{[0,1]}(x,y)-\sigma\,\chi^{\mathfrak{sl}(2)}_{3/2}(q)\,\chi^{\mathfrak{sp}(4)}_{[1,0]}(x,y)\,\chi^{\mathfrak{su}(2)}_{1}(z)+\chi^{\mathfrak{sl}(2)}_2(q)\,\chi^{\mathfrak{su}(2)}_{2}(z)\,,
\end{align}
which is the character of the $\mathcal{D}_1$ multiplet for $\sigma=-1$. The fugacity $\sigma$ is useful to write the partition function at strong coupling as a unique plethystic exponential
\begin{align}\label{Zstrong_PE}
\begin{split}
\mathcal{Z}_{\text{strong}}&=\left.\text{PE}^{(\len)}\left[\chi(\mathcal{D}_1)_{\sigma}(q;x,y;z)\right]\right|_{\sigma=-1}=\text{exp}\left[\sum_{k=1}^{+\infty}  \frac{\len^k}{k}\,\chi(\mathcal{D}_1)_{(-1)^k}(q^k;x^k,y^k;z^k)\right]\,,
\end{split}
\end{align}
where notice that we have set $\sigma=-1$ only after taking the plethystic exponential, in such a way that bosonic and fermionic states are taken into account with the correct statistics. Also note that we are still using the fugacity $\len$ to distinguish states of different length.

The plethystic exponential is also a useful tool to compute generating series for states of fixed length and given representation of $\mathfrak{su}(2)\oplus \mathfrak{sp}(4)$. This can be obtained by formally expanding \eqref{Zstrong_PE} in powers of $\len$ and then, for each power, subtracting the descendants dividing by the Verma module character associated to the generators $\mathfrak{P}$ and $\mathfrak{Q}_{\alpha A}$, obtaining so-called {\it highest weight generating series} (HWGS) -- see, {\it e.g.}, \cite{Hanany:2014dia}. In equations, we expand
\begin{align}
\mathcal{Z}_{\text{strong}}=\sum_{L=0}^{\infty}\len^L\,\mathcal{F}_L(q;x,y;z)\,,
\end{align}
where $\mathcal{F}_L$ contains all states at length $L$. Given that no semi-short representations are present, and that only one short multiplet is present for each length, can then obtain the HWGS for long supermultiplets of length $L$, $\text{HWGS}_L$ by subtracting the contribution of short ones as in \eqref{Z=short+long} and suitably removing descendants. We then have
\begin{align}
\text{HWGS}_L=\frac{(1-q^2)}{\chi^{\mathfrak{Q}}_{\text{All}}}\,\left(\mathcal{F}_L-\chi(\mathcal{D}_L)\right)\,.
\end{align}
Each $\text{HWGS}_L$ contains contributions from different representations of $\mathfrak{su}(2)\oplus \mathfrak{sp}(4)$, which can be projected out using \eqref{orthogonalitysu2} and \eqref{orthogonalitysp4}. Just to give some examples, a problem that will be relevant in \cite{Ferrero:2023gnu} is that of counting the number of long supermultiplets of given length in the singlet representation $\{0,[0,0]\}$ of $\mathfrak{su}(2)\oplus \mathfrak{sp}(4)$. The HWGS for such supermultiplets, for the first few values of the length, are given by
\begin{align}
\begin{split}
\text{HWGS}_1^{0,[0,0]}&=0\,,\\
\text{HWGS}_2^{0,[0,0]}&=\frac{q^4}{1-q^4}\,,\\
\text{HWGS}_3^{0,[0,0]}&=0\,,\\
\text{HWGS}_4^{0,[0,0]}&=\frac{q^8(1-q^2+q^4)}{(1-q^2)(1-q^4)(1-q^8)}\,,\\
\text{HWGS}_5^{0,[0,0]}&=\frac{q^{18}(1+q^2+2q^4+q^6+q^8+q^12)}{(1-q^2)(1-q^4)(1-q^6)(1-q^8)(1+q^2+q^4+q^6+q^8)}\,,\\
\text{HWGS}_6^{0,[0,0]}&=\frac{q^12(1-q^2+q^4+q^6+2q^8-q^{10}+3q^{12}+q^{14}+q^{16}-q^{18}+2q^{20})}{(1-q^2)(1-q^4)^2(1-q^{10})(1-q^{12})}\,.
\end{split}
\end{align}
This allows to extract closed-form expressions for the number of singlet supermultiplets of fixed length for each dimension $\D$, $\text{dim}\,\mathtt{d}_L(\mathcal{L}^{\Delta}_{0,[0,0]})$, {\it i.e.} the dimension of degeneracy spaces at fixed quantum numbers and length at strong coupling. For length $L=2,4$, which are the most relevant cases for \cite{Ferrero:2023gnu}, we find
\begin{align}
\begin{split}
\text{dim}\,\mathtt{d}_{L=2}(\mathcal{L}^{\D}_{0,[0,0]})=1\,,\quad
\text{dim}\,\mathtt{d}_{L=4}(\mathcal{L}^{\D}_{0,[0,0]})=
\begin{cases}
\lfloor \big(\tfrac{\D}{4}\big)^2\rfloor \quad \D\ge 4\,\,\,\text{even}\,,\\
\lfloor \big(\tfrac{\D-3}{4}\big)^2\rfloor \quad \D\ge 5\,\,\,\text{odd}\,.
\end{cases}
\end{split}
\end{align}
More in general, it is easier to study the asymptotic of the HWGS at $q\to 1$, from which one can read off the growth in the number of states for large $\D$. In particular, we find
\begin{align}
N^{\text{strong}}_{0,[0,0],L}\sim \frac{c_L}{(L-2)!}\D^{L-2}\,,
\end{align}
where 
\begin{align}\label{c_L_HWGS}
c_L=\left\{\tfrac{1}{2},0,\tfrac{1}{8},\tfrac{7}{120},\tfrac{3}{40},\tfrac{13}{210},\tfrac{473}{8064},\tfrac{6149}{120960},\tfrac{2249}{51840},\tfrac{706303}{19958400},\tfrac{11747}{422400},\tfrac{4355893}{207567360},\tfrac{30162709}{1981324800},\tfrac{192872021}{18162144000},\ldots\right\}\,,
\end{align}
for $L=2,3,\ldots$. For the reader's convenience, we also collect the dimension of degeneracy spaces $\mathtt{d}_L(\mathcal{L}^{\D}_{0,[0,0]})$ at strong coupling for some values of $L$ and $\D$ in Table \ref{tab:longdegeneraciesfixedlength_strong}. Note in particular that the first supermultiplets of {\it even} length $L$ appear with $\D=L$ (for $L\ge 2$), with no degeneracy (their superprimary in schematically of the form $(\varphi^a\varphi_a)^{L/2}$), while for {\it odd} length $L$ the lowest-dimensional supermultiplets appear at dimension $\D=L+4$ (for $L\ge 5$), with no degeneracy for $L=5$ and degeneracy two for higher $L$.

\subsection{Counting of states at weak coupling}\label{app:counting-weak}

For the Maldacena Wilson line in planar $\mathcal{N}=4$ super Yang-Mills at weak coupling, the alphabet is obtained by restricting the fields of the 4d theory to the line. The states are then obtained as products of these fundamental letters, where due to the fact that now the free field description is a gauge theory one and our letters carry adjoint $SU(N)$ indices, the order matters and we no longer have to symmetrize under exchange of the letters.

The first step is then that of taking the character ${\chi}(\mathbb{F})$ of the $\mathcal{N}=4$ free vector multiplet and restrict the operators to the line. Such character is given, for example, in eq. (4.1) of \cite{Bianchi:2006ti} where the restriction to the line is implemented by setting their $SU(2)_L\otimes SU(2)_R$ fugacities $x=\bar{x}=z$ and decomposing $SU(4)$ R-symmetry characters in terms of $SP(4)$ ones. Note that their fugacity $s$ that counts the conformal weight is our $q$. As an example, the four-dimensional character of the conformal multiplet for a single free scalar $\varphi$ is
\begin{align}
\chi^{(4d)}(\varphi)=\frac{q^{2}(1-q^4)}{(1-q^2\,x\,\bar{x})(1-q^2\,x^{-1}\,\bar{x})(1-q^2\,x\,\bar{x}^{-1})(1-q^2\,x^{-1}\,\bar{x}^{-1})}\,,
\end{align}
where the denominator accounts for the action of four-dimensional derivatives, $q^{2}$ accounts for the dimension of the scalar field (one, for a free field) and the factor of $(1-q^4)$ takes care of the null condition $\Box\varphi=0$ for a free field. The restriction to a defect CFT character is achieved by setting $x=\bar{x}=z$, where $z$ is the usual fugacity for transverse spin and yields
\begin{align}
\chi^{(1d)}(\varphi)=q^2\left(1+\sum_{n=0}^{\infty}q^{2n}\,\chi^{\mathfrak{su}(2)}_n(z)^2 \right)\,.
\end{align}
Applying the same idea to the whole character of the free vector multiplet we find\footnote{Here $\Phi$, $\Psi$ and $F$ represent the components of a free $\mathcal{N}=4$ vector multiplet: the scalars, fermions and field strength respectively.}
\begin{align}
\begin{split}
{\chi}(\mathbb{F})=(1+\chi^{\mathfrak{sp}(4)}_{[0,1]}(x,y))\,\chi(\Phi)+2\,\chi^{\mathfrak{sp}(4)}_{[1,0]}(x,y)\,\chi(\Psi)+2\,\chi(F)\,,
\end{split}
\end{align}
where
\begin{align}
\begin{split}
\chi(\Phi)&=q^2\,\left(1+\sum_{n=1}^{\infty}q^{2n}\,\chi^{\mathfrak{su}(2)}_n(z)^2\right)\,,\\
\chi(\Psi)&=q^3\,\left(\chi^{\mathfrak{su}(2)}_1(z)+\sum_{n=1}^{\infty}q^{2n}\,\chi^{\mathfrak{su}(2)}_n(z)\,\chi^{\mathfrak{su}(2)}_{n+1}(z)\right)\,,\\
\chi(F)&=q^4\,\left(\chi^{\mathfrak{su}(2)}_2(z)+\sum_{n=1}^{\infty}q^{2n}\,\chi^{\mathfrak{su}(2)}_n(z)\,\chi^{\mathfrak{su}(2)}_{n+2}(z)\right)\,,
\end{split}
\end{align}
The partition function at weak coupling is then simply obtained by taking all words made by these letters, where the order matters. Hence, we find
\begin{align}
\mathcal{Z}_{\text{weak}}=\sum_{k=0}^{\infty}\len^k\,{\chi}(\mathbb{F})^k\,,
\end{align}
where we have once again used a fugacity $\len$ to account for the length of operators. Note that each power of ${\chi}(\mathbb{F})$ can be expanded in characters of $\mathfrak{osp}(4^*|4)$ independently, for instance we have
\begin{align}
{\chi}(\mathbb{F})=\chi(B_1)^1_{0,[0,1]}+\sum_{n=0}^{\infty}\chi(A_1)^{a+b+n+1}_{2n,[0,0]}\,,
\end{align}
which contains only short and semi-short representations. On the other hand, for the next power we find 
\begin{align}
\begin{split}
({\chi}(\mathbb{F}))^2=&\chi(B_1)^2_{0,[0,2]}+\chi(B_1)^2_{0,[2,0]}+\sum_{n=1}^{\infty}(n+1)\,\chi(A_1)^{n+1}_{2(n-1),[0,1]}+\sum_{n=1}^{\infty}(2n+1)\,\chi(A_1)^{n+3/2}_{2n-1,[1,0]}\\
&+\sum_{n=0}^{\infty}\sum_{\D=n+2,\infty}(1+n-n^2+\lfloor(\tfrac{1}{2}+n)\D-\tfrac{1}{2}n\rfloor)\,\chi(L)^{\D}_{2n,[0,0]}\,,
\end{split}
\end{align}
while the formulas become more complicated for higher powers. We collect in table \ref{tab:longdegeneraciesweak} some results for the degeneracies for spin-less supermultiplets.

One can also organize the states by length, simply by expanding individual powers of $\chi(\mathbb{F})$. The result, for singlets of $\mathfrak{su}(2)\oplus \mathfrak{sp}(4)$, is given in table \ref{tab:longdegeneraciesfixedlength_weak}.

As opposed to the strongly-coupled limit, at weak coupling one witnesses the presence of semi-short multiplet, that can recombine with each other to form long multiplets at finite values of the coupling. An interesting questions is whether all semi-short multiplets are allowed, at least in principle, to recombine according to the rules \eqref{recombination}. To answer this question we consider a specialization of the fugacities that suppresses the contribution of long multiplets: this can be obtained by setting the character of one of the supercharges to zero. For example, we set $\chi^{\mathfrak{Q}}_{11}=0$ in \eqref{Qchar}, solving for $z$. This indeed sets 
\begin{align}
\left.\chi(L)_{s,[a,b]}^{\h}\right|_{z=-(q\,x)^{-1}}=0\,.
\end{align}
All pairs of semi-short multiplets that can (at least in principle) recombine with each other therefore sum to zero in this limit, and the remaining multiplets in the partition function are only those that cannot recombine. Since we find
\begin{align}
\left.\mathcal{Z}_{\text{Weak}}\right|_{z=-(q\,x)^{-1}}=1+\sum_{k=1}^{\infty}\chi_{\mathcal{D}_k}\,,
\end{align}
we conclude that all semi-short multiplets are allowed to recombine and we expect them to do so. One can then re-arrange the spectrum at weak coupling writing the partition function in terms of protected $\mathcal{D}_k$ multiplets and long multiplets only:
\begin{align}
\mathcal{Z}_{\text{Weak}}=1+\sum_{k=1}^{\infty}\chi_{\mathcal{D}_k}+\mathcal{Z}_{\text{Long}}\,,
\end{align}
where now notice that there is no longer a well-defined concept of length in $\mathcal{Z}_{\text{Long}}$, since recombination mixes multiplets of different length. We collect in Table \ref{tab:longdegeneraciesweak_recombined} the counting of states in terms of long multiplets only, after all semi-short multiplets have undergone recombination. Notice that the only differences with respect to Table \ref{tab:longdegeneraciesweak} arise, for each representation of $\mathfrak{su}(2)\oplus \mathfrak{sp}(4)$, for multiplets of the lowest possible dimension. Moreover, there are no differences for states in the representation $\{0,[0,0]\}$.
\begin{table}
\centering
 \begin{tabular}{| c || c|c|c|c|c|c|c|c|c|c|c|c|c|c|c|} 
 \hline
$\Delta$&
 $1$ & 
$2$  &
$3$ &
$4$  & 
$5$ &
$6$ &
$7$ &
$8$   \\
   \hline
    \hline
      $[0,0]$ & $1 $   & $ 2$    & $6 $  & $25 $   & $128 $  &  $758 $   & $4986$ &  $35550$ \\
      \hline
        $[0,1]$ & $-$   & $2$    & $6$  & $28$   & $167$  &  $1134$   & $8386$ &  $65995$ \\
      \hline
        $[0,2]$ & $-$   & $-$    & $3$  & $12$   & $76$  &  $588$   & $4972$  & $44260$ \\
      \hline
        $[0,3]$ & $-$   & $-$    & $-$  & $4$   & $20$  &  $160$   & $1520$ & $15388$ \\
      \hline
        $[0,4]$ & $-$   & $-$    & $-$  & $-$   & $5$  &  $30$   & $290$ & $3265$ \\
      \hline
        $[0,5]$ & $-$   & $-$    & $-$  & $-$   & $-$  &  $6$   & $42$ & $476$ \\
      \hline
        $[0,6]$ & $-$   & $-$    & $-$  & $-$   & $-$  &  $-$   & $7$ & $56$\\
      \hline
        $[0,7]$ & $-$   & $-$    & $-$  & $-$   & $-$  &  $-$   & $-$ & $8$\\
        \hline
      \hline
        $[2,0]$ & $-$   & $-$    & $2$  & $16$   & $128$  &  $1038$   & $8638$ &  $74000$ \\
      \hline
        $[2,1]$ & $-$   & $-$    & $-$  & $5$   & $55$  &  $570$   & $5750$ &  $57595$ \\
      \hline
        $[2,2]$ & $-$   & $-$    & $-$  & $-$   & $9$  &  $126$   & $1602$ &  $19341$ \\
      \hline
      $[2,3]$ & $-$   & $-$    & $-$  & $-$   & $-$  &  $14$   & $238$ &  $3584$ \\
      \hline
      $[2,4]$ & $-$   & $-$    & $-$  & $-$   & $-$  &  $-$   & $20$ &  $400$ \\
      \hline
       $[2,5]$ & $-$   & $-$    & $-$  & $-$   & $-$  &  $-$   & $-$ &  $27$ \\
      \hline
      \hline
       $[4,0]$ & $-$   & $-$    & $-$  & $-$   & $5$  &  $90$   & $1270$ &  $15947$ \\
      \hline
       $[4,1]$ & $-$   & $-$    & $-$  & $-$   & $-$  &  $14$   & $322$ &  $5516$ \\
      \hline
       $[4,2]$ & $-$   & $-$    & $-$  & $-$   & $-$  &  $-$   & $28$ &  $784$ \\
      \hline
\end{tabular}
\caption{Spectrum of spin-less, long multiplets at weak coupling after recombination of all semi-short multiplets is accounted for.}
\label{tab:longdegeneraciesweak_recombined}
\end{table}
To conclude, let us comment on what happens to short and semi-short multiplets after the character variables are specialized solving $\chi^{\mathfrak{Q}}_{11}=0$. The resulting characters turn out to depend only on the combinations $q^2\,x$ and $q^2\,y$, rather than on $(q,x,y)$ independently. Introducing then two new fugacities
\begin{align}
\tilde{q}=q\,\sqrt{x}\,, \qquad 
\tilde{z}=y/x\,,
\end{align}
one obtains, from the characters of short and semi-short multiplets, characters of the superalgebra $\mathfrak{osp}(2|2)$. This can be explained from the fact that setting $\chi^{\mathfrak{Q}}_{11}=0$ corresponds to considering the cohomology of the supercharge $\mathfrak{Q}_{11}$: only generators of $\mathfrak{osp}(4^*|4)$ that are in such cohomology survive and the resulting algebra is $\mathfrak{osp}(2|2)$.

\section{More details for constructing composite operators}\label{app:forOperatorConstruction}
\label{app:forOperatorConstuction}

\subsection{R-symmetry projectors and related structures}
\label{app:Rsymmprojectors}

In Section \ref{sec:toyexamples} we discussed how to construct an $\mathfrak{sl}(2)$ primary operator as a composite of two primaries as in \eqref{L2OusingD}, where the differential operator $\mathsf{D}$ given in \eqref{DoperatorsTomakeSinglets} is fixed, up to normalization, by  $\mathfrak{sl}(2)$  symmetry.
We will now extend that construction to the case of finite dimensional representations of $\mathfrak{sp}(4)$, see also Appendix of \cite{Goncalves:2023oyx}. In this case the generators of $\mathfrak{sp}(4)$ act on a space of five variables $(y^{ab},\ww_a)$ with $y^{ab}=y^{ba}$ and are given in  \eqref{SP4generatorsGeneric}.
We will denote the projectors as
\begin{equation}
\mathsf{P}^{\mathcal{R}_1,\mathcal{R}_2}_{\mathcal{R}}\,,
\end{equation}
where $\mathcal{R}_1=[a_1,b_1]$,  $\mathcal{R}_2=[a_2,b_2]$ and  $\mathcal{R}=[a,b]$ are Dynkin labels.
These projectors
 can be found by solving the condition
\begin{equation}
\label{equationforprojectors}
(\mathsf{P}_{12})^{\mathcal{R}_1,\mathcal{R}_2}_{\mathcal{R}}\,
\langle \mathsf{O}_{\mathcal{R}_1}(1)
\mathsf{O}_{\mathcal{R}_2}(2)
\mathsf{O}_{\mathcal{R}'}(3)
\rangle
\,=\,
\mathcal{N}^{\mathcal{R}_1,\mathcal{R}_2}_{\mathcal{R}}
\langle \mathsf{O}_{\mathcal{R}}(2)
\mathsf{O}_{\mathcal{R}'}(3)
\rangle\,.
\end{equation}
Where the factors $\mathcal{N}^{\mathcal{R}_1,\mathcal{R}_2}_{\mathcal{R}}$ are left unfixed at the moment.
The two-point functions entering the formula above are  the $\mathfrak{sp}(4)$ part of the two-point functions \eqref{twopointLong} namely, in some fixed normalization,
\begin{equation}
\langle \mathsf{O}_{[a,b]}(1)
\mathsf{O}_{[a',b']}(2)
\rangle=\delta_{a,a'}\delta_{b,b'}
(\ww_1 y_{12}\ww_2)^a (y_{12}^2)^b\,.
\end{equation}
The three-point functions are more complicated and in general present multiple structure. Due to the fact that the R-symmetry algebra $\mathfrak{sp}(4)$ coincides with the conformal algebra in three dimensions, the general form of three-point functions is known, see {\it e.g.} \cite{Giombi:2009wh}.
 For our purposes, namely to determine the necessary projectors via \eqref{equationforprojectors},  we will need only some specific three-point functions. They are collected in Tables \ref{tab:threepointfunctions0101}, \ref{tab:threepointfunctions1010}, \ref{tab:threepointfunctions2020}, \ref{tab:threepointfunctions2002} where the basic building blocks are defined as follows
 \begin{subequations}
\begin{align}
Y_{12}&= y_{12}^2\,,
\\
W_{12}&= \ww_1 y_{12}\ww_2\,,
\\
V_{12,3}&= y_{23}^2 ( \ww_3 y_{13}\ww_3) - y_{13}^2 ( \ww_3 y_{23}\ww_3) \,,
\\
U_{12,3}&=\ww_1 y_{13}\epsilon \, y_{32}\ww_2\,. 
\end{align}
\end{subequations}

\begin{table}[!ht]
\centering
 \begin{tabular}{| c || c | c|c| c|c|c|c|} 
\hline
$\mathcal{R}$
 & $[0,0]$ &  $[2,0]$    &  $[0,2]$    \\
   \hline
   \hline
    $\langle [0,1][0,1]\mathcal{R}]\rangle $
     &  $Y_{12}$   & $V_{12,3}$ &  $Y_{13}Y_{23}$
      \\
   \hline
\end{tabular}
\caption{Three-point functions of  type $\langle [0,1][0,1]\mathcal{R}]\rangle$.}
\label{tab:threepointfunctions0101}
\end{table}
%---------------------------------------
\begin{table}[!ht]
\centering
 \begin{tabular}{| c || c | c|c| c|c|c|c|} 
\hline
$\mathcal{R}$
 & $[0,0]$ &  $[0,1]$    &  $[2,0]$    \\
   \hline
   \hline
    $\langle [1,0][1,0]\mathcal{R}]\rangle $
     &  $W_{12}$   & $U_{12,3}$ &  $W_{13}W_{23}$
      \\
   \hline
\end{tabular}
\caption{Three-point functions of  type $\langle [1,0][1,0]\mathcal{R}]\rangle$.}
\label{tab:threepointfunctions1010}
\end{table}
%---------------------------------------
\begin{table}[!ht]
\centering
 \begin{tabular}{| c || c | c|c| c|c|c|c|c|c|c|} 
\hline
$\mathcal{R}$
 & $[0,0]$ &  $[0,1]$    &  $[2,0]$  & $[0,2]$  & $[2,1]$  & $[4,0]$ \\
   \hline
   \hline
    $\langle [2,0][2,0]\mathcal{R}]\rangle $
     &  $W_{12}^2$   & $W_{12}U_{12,3}$ &  $W_{12}W_{13}W_{23}$
      &  $U_{12,3}^2$
       &  $U_{12,3}W_{13}W_{23}$
       & $W_{13}^2W_{23}^2$
      \\
   \hline
\end{tabular}
\caption{Three-point functions of  type $\langle [2,0][2,0]\mathcal{R}]\rangle$.}
\label{tab:threepointfunctions2020}
\end{table}

%---------------------------------------
\begin{table}[!ht]
\centering
 \begin{tabular}{| c || c | c|c| c|c|c|c|c|} 
\hline
$\mathcal{R}$
 & $[0,2]$ &  $[2,1]$    &  $[2,0]$  & $[2,2]$  \\
   \hline
   \hline
    $\langle [2,0][0,2]\mathcal{R}]\rangle $
     &  $V_{32,1} Y_{23}$   & $U_{13,2}Y_{23}W_{13}$ &  $U_{13,2}^2$
      &  $W_{13}^2Y_{23}^2$     
      \\
   \hline
\end{tabular}
\caption{Three-point functions of  type $\langle [2,0][0,2]\mathcal{R}]\rangle$.}
\label{tab:threepointfunctions2002}
\end{table}

%Branching ratios reminder
%\begin{equation}
%[a,b]\mapsto [a]_{2b+a}\, \oplus \dots
%\end{equation}

%--------------------------------------------------------------------------------------------------------------------------------------------------------------------------
\paragraph{Explicit expressions for the projectors.}
We will now determine the projector by solving equation \eqref{equationforprojectors}. The projectors will have the schematic form, 
which follows from a $U(1)$ subgroup of the R-symmetry which scales $y$,
\begin{equation}
\mathsf{P}^{\mathcal{R}_1,\mathcal{R}_2}_{\mathcal{R}}\,\sim \left(\partial_y\right)^{\#}\,,
\quad 
\text{where}
\quad
\#=\tfrac{1}{2}
\left(
(2b_1+a_1)+(2b_2+a_2)-(2b+a)\right)\,.
\end{equation}
It should be remarked that the projectors are not uniquely fixed by  \eqref{equationforprojectors} because certain differential operators annihilate the operators. In fact the kernel of these operators singles out finite dimensional representations of $\mathfrak{sp}(4)$ in the space of polynomials in the $y$ variables. Up to these ambiguities, we determine the following projectors:

\noindent
Projectors for $\varphi\varphi$ bilinears:
 \begin{subequations}
\begin{align}
(\mathsf{P}_{12})^{[0,1],[0,1]}_{[0,0]}\,
&= 
\tfrac{1}{3}(\partial_{y_1}\cdot \partial_{y_1}+\partial_{y_2}\cdot \partial_{y_2})-\partial_{y_1}\cdot \partial_{y_2}\,,
\qquad  \qquad \
\mathcal{N}^{[0,1],[0,1]}_{[0,0]}=5\,,
\\
\label{P12010120}
(\mathsf{P}_{12})^{[0,1],[0,1]}_{[2,0]}\,
 &=
 (\epsilon \ww_1)\cdot
 \left( \partial_{y_2}- \partial_{y_1}\right)\cdot
 (\epsilon \ww_1)\,,
 \qquad 
  \qquad \qquad \qquad \!\!
\mathcal{N}^{[0,1],[0,1]}_{[2,0]}=1\,,
\\
(\mathsf{P}_{12})^{[0,1],[0,1]}_{[0,2]}\,
 &=
1\,,
 \qquad\qquad  \qquad\qquad 
  \qquad\qquad  \qquad\qquad\qquad
\mathcal{N}^{[0,1],[0,1]}_{[0,2]}=1\,.
\end{align}
\end{subequations}
Projectors for $\Psi\Psi$ bilinears:
 \begin{subequations}
 \label{AllprojectorsforPsiPsi}
\begin{align}
\label{P12101000}
(\mathsf{P}_{12})^{[1,0],[1,0]}_{[0,0]}\,
&= 
\partial_{ \ww_{1}}\cdot
\left(\partial_{y_1}-\partial_{ y_2}\right)
\cdot \partial_{\ww_{2}}\,,
\qquad  \qquad\qquad\qquad
\mathcal{N}^{[1,0],[1,0]}_{[0,0]}=6\,,
\\
(\mathsf{P}_{12})^{[1,0],[1,0]}_{[0,1]}\,
 &=
\partial_{ \ww_{1}} \epsilon\, \partial_{ \ww_{2}}
\,,
 \qquad  \qquad\qquad  \qquad\qquad \qquad\,\,\,\,\,\,\,\,\,\,\,
\mathcal{N}^{[1,0],[1,0]}_{[0,1]}=-2\,,
\\
(\mathsf{P}_{12})^{[1,0],[1,0]}_{[2,0]}\,
 &=
1\,,
 \qquad\qquad  \qquad\qquad 
  \qquad\qquad  \qquad\qquad \,\,\,\,\,
\mathcal{N}^{[1,0],[1,0]}_{[2,0]}=1\,.
\end{align}
\end{subequations}
Additional projectors for $\mathcal{O}$:
\begin{equation}
\left(\mathsf{P}_{12}\right)^{[2,0],[2,0]}_{[0,0]}
=(\partial_{y_1}\cdot \partial_{y_1}+\partial_{y_2}\cdot \partial_{y_2})\partial_{ \ww_{1}} \epsilon\, \partial_{\ww_{2}}
+\tfrac{1}{9}
(\partial_{ \ww_{1}} \partial_{y_1}\partial_{\ww_{1}})
(\partial_{ \ww_{2}} \partial_{y_2}\partial_{\ww_{2}})
+\tfrac{1}{2} (d_1)^b_a (d_2)^a_b\,,
\end{equation}
with $\mathcal{N}^{[2,0],[2,0]}_{[0,0]}=-40$.

\noindent
Additional projectors for $\widetilde{\mathcal{O}}$:
 \begin{subequations}
\begin{align}
(\mathsf{P}_{12})^{[2,0],[0,2]}_{[0,2]}\,
&= 
(\partial_{ \ww_{1}} \partial_{y_1}\partial_{\ww_{1}})
-\tfrac{45}{2}
(\partial_{ \ww_{1}} \partial_{y_2}\partial_{\ww_{1}})
\,,
\qquad  \qquad\quad\!
\mathcal{N}^{[2,0],[0,2]}_{[0,2]}=15\,,
\\
(\mathsf{P}_{12})^{[2,0],[2,0]}_{[0,2]}\,
 &=
\left(\partial_{ \ww_{1}} \epsilon\, \partial_{ \ww_{2}}\right)^2
\,,
 \qquad  \qquad\qquad  \qquad\qquad \qquad\,\,\,
\mathcal{N}^{[2,0],[2,0]}_{[0,2]}=12\,.
\end{align}
\end{subequations}
Notice that after the action of the differential operators above the coordinate associated to the second set of variables $(y_2,\ww_2)$ should be set equal to the first $(y_1,\ww_1)$ .
Above, we used the shorthand notation and definition
\begin{equation}
\partial_{y_i}\cdot \partial_{y_j}=\epsilon^{ac}\epsilon^{bd}\frac{\partial}{\partial y_i^{ab}}\frac{\partial}{\partial y_j^{cd}}\,,
\qquad
d^b_a=
\frac{\partial}{\partial w_c}
\frac{\partial}{\partial y^{ca}}
\frac{\partial}{\partial w_b}
-\tfrac{1}{2}\delta_a^b
\frac{\partial}{\partial w_c}
\frac{\partial}{\partial y^{ce}}
\frac{\partial}{\partial w_e}
\end{equation}
and the contraction of indices in \eqref{P12010120}, \eqref{P12101000} and so on  is understood.

%------------------------------------------------------------------------------------------------------------------
\paragraph{Recurrent bilinears.}
Let us start from the bilinears made of $\varphi$. Using the projectors introduced above they are
 \begin{subequations}
\begin{align}
\label{phi2singlet}
(\varphi^2)^{[0,0]}(t_1,t_2)&=(\mathsf{P}_{12})^{[0,1],[0,1]}_{[0,0]}\,:\varphi(t_1,y_1)\varphi(t_2,y_2):\big{|}_{y_2=y_1}\,,
\\
(\varphi^2)^{[2,0]}(t_1,t_2,y,\ww)&=(\mathsf{P}_{12})^{[0,1],[0,1]}_{[2,0]}\,:\varphi(t_1,y_1)\varphi(t_2,y_2):\big{|}_{y_2=y_1}\,,
\\
(\varphi^2)^{[0,2]}(t_1,t_2,y)&=:\varphi(t_1,y)\varphi(t_2,y):\,.
\end{align}
\end{subequations}
Concerning the bilinears in the fermionic operators $\Psi$, it is convenient to fist project in irreducible representations of the transverse spin as follows
\begin{subequations}
\label{transversespinProjectionsPsiPsi}
\begin{align}
\left(\Psi^2\right)^{s=2}_{\alpha \beta}(\mathsf{1},\mathsf{2})= &\,
\frac{1}{2} (\epsilon w_1)^a(\epsilon w_2)^b
\left(
\frac{\partial}{\partial \theta_1^{\alpha a}}
\frac{\partial}{\partial \theta_2^{\beta b}}
+
\frac{\partial}{\partial \theta_1^{\beta a}}
\frac{\partial}{\partial \theta_2^{\alpha b}}
\right)
:\theta\Psi(1)\theta\Psi(2):\,,
\\
\left(\Psi^2\right)^{s=0}(\mathsf{1},\mathsf{2})= &\,\,\,
(\epsilon w_1)^a(\epsilon w_2)^b
\,\epsilon^{\alpha \beta}
\frac{\partial}{\partial \theta_1^{\alpha a}}
\frac{\partial}{\partial \theta_2^{\beta b}}
:\theta\Psi(1)\theta\Psi(2):\,,
\end{align}
\end{subequations}
where $(\mathsf{1},\mathsf{2})$ is a shorthand for $(t_1,y_1,w_1,t_2,y_2,w_2)$.
From these objects we further project into $\mathfrak{sp}(4)$ R-symmetry irreducible components as
\begin{equation}
\label{ALLfermionicbilinearcompact}
(\Psi^2)^{[a,b],s}(t_1,t_2,y,\ww)=
(\mathsf{P}_{12})^{[1,0],[1,0]}_{[a,b]}
\left(\Psi^2\right)^{s}\big{|}_{y_i=y,\ww_i=\ww}\,,
\end{equation}
where $s=0,2$ and $[a,b]$ takes the value such that the corresponding projector exists, see \eqref{AllprojectorsforPsiPsi}. 
Finally, the only bilinear in $f$ that we will use is
%-------------------------------------------------------------------
\begin{align}
\label{f2bilinear}
(f^2)^{[0,0]}(t_1,t_2)=\epsilon^{\alpha\gamma}\epsilon^{\beta\delta}
\,:f_{\alpha\beta}(t_1)f_{\gamma\delta}(t_2):\,.
\end{align}
%-------------------------------------------------------------------

%------------------------------------------------------------------------------------------------------------------
\paragraph{Correlators involving the recurrent bilinears.}
%Some are in the main text, see \eqref{PhiPhiOL2prGENERATING}.
Let us start from the generating correlators involving one $\varphi$ bilinear
\begin{subequations}
\begin{align}
\langle \varphi(1) \varphi(2) \,
(\varphi^2)^{[0,0]}(t_3,t_4)
\rangle 
&=
\left(\tfrac{1}{t^2_{13} t^2_{24}}
+\tfrac{1}{t^2_{14} t^2_{23}}
\right)
Y_{12}\,,
\\
\label{phiphiphisqaredin20}
\langle \varphi(1) \varphi(2) \,
(\varphi^2)^{[2,0]}(t_3,t_4;y_3,\ww_3)
\rangle 
&=\left(\tfrac{1}{t_{13}^2t_{24}^2}-
\tfrac{1}{t_{23}^2t_{14}^2}
\right)
\,
V_{12,3}\,,
\\
\langle \varphi(1) \varphi(2) \,
(\varphi^2)^{[0,2]}(t_3,t_4,y_3)
\rangle 
&=
\left(\tfrac{1}{t^2_{13} t^2_{24}}
+\tfrac{1}{t^2_{14} t^2_{23}}
\right)
Y_{13}Y_{23}
\,,
\end{align}
\end{subequations}
where correlators are computed using Wick contraction (we dropped the suffix $(0)$ as no confusion should arise here),
compare to Table \ref{tab:threepointfunctions0101} for the R-symmetry structures.

Next we consider the case of bilinears in $\Psi$. First we compute the generating correlators involving the bilinears defined in \eqref{transversespinProjectionsPsiPsi} as
 \begin{subequations}
\begin{align}
\label{PsiPsiPspspinzero}
\langle \theta \Psi(1) \theta \Psi(2) \,
\left(\Psi^2\right)^{s=0}(\mathsf{3},\mathsf{4})
\rangle 
&=- \frac{4}{t_{13}^3t_{24}^3}\,
(\theta_1\cdot \theta_2)^{ab}
(\epsilon\, y_{13}\ww_3)_{a}
(\epsilon\, y_{24}\ww_4)_{b}
+(1\leftrightarrow 2)\,,
\\
\label{PsiPsiPspspin1}
\langle \theta \Psi(1) \theta \Psi(2) \,
\left(\Psi^2\right)^{s=2}_{\alpha \beta}(\mathsf{3},\mathsf{4})
\rangle 
&=\,\,\,\,\frac{4}{t_{13}^3t_{24}^3}\,
(\epsilon\, \theta_1\epsilon\, y_{13}\ww_3)_{(\alpha }
(\epsilon\, \theta_2\epsilon\, y_{24}\ww_4)_{\beta)}\,\,
+(1\leftrightarrow 2)\,,
\end{align}
\end{subequations}
where symmetrization is unit normalized ($v_{(\alpha\beta)}=\tfrac{1}{2}(v_{\alpha \beta}+v_{\beta \alpha})$) and $(\theta_1\cdot \theta_2)^{ab}=\epsilon_{\alpha\beta}
\theta_1^{\alpha a}
\theta_2^{\beta b}$.
From \eqref{PsiPsiPspspinzero} and  \eqref{PsiPsiPspspin1} we obtain
\begin{subequations}
\begin{align}
\langle \theta\Psi(1)\,\theta \Psi(2)(\Psi^2)^{[0,0],0}(t_3,t_4;3)\rangle\,&=
\, -6\,
\left(\tfrac{1}{t^3_{13} t^3_{24}}
-\tfrac{1}{t^3_{14} t^3_{23}}
\right)
\theta_{1}y_{12}\theta_{2}\,,
\\
\label{PsiPsiPsisqaredin01}
\langle \theta \Psi(1) \theta \Psi(2) \,
(\Psi^2)^{[0,1],0}(t_3,t_4;3)
\rangle 
&=
\, \,\,\,\,\left(\tfrac{1}{t_{13}^3t_{24}^3}-
\tfrac{1}{t_{14}^3t_{23}^3}
\right)
\,
\mathcal{J}_{12,3}\,,
\\
\label{PsiPsiPsisqaredin20}
\langle \theta \Psi(1) \theta \Psi(2) \,
(\Psi^2)^{[2,0],0}(t_3,t_4;3)
\rangle 
&=-4 \left(\tfrac{1}{t_{13}^3t_{24}^3}+
\tfrac{1}{t_{23}^3t_{14}^3}
\right)
\,
(\theta_1\cdot \theta_2)^{ab}
(\epsilon\, y_{13}\ww_3)_{a}
(\epsilon\, y_{23}\ww_3)_{b}\,,
\\
\label{PsiPsiPsisqaredin00spin1}
\langle \theta\Psi(1)\,\theta \Psi(2)(\Psi^2)^{[0,0],2}_{\alpha\beta}(t_3,t_4;3)\rangle\,&=
\, 6\,
\left(\tfrac{1}{t^3_{13} t^3_{24}}
+\tfrac{1}{t^3_{14} t^3_{23}}
\right)
(\theta_{1}y_{12}\theta_{2})_{(\alpha \beta)}\,,
\\
\label{PsiPsiPsisqaredin01SPIN1}
\langle \theta \Psi(1) \theta \Psi(2) \,
(\Psi^2)^{[0,1],2}_{\alpha \beta}(t_3,t_4;3)
\rangle 
&=-
\, \left(\tfrac{1}{t_{13}^3t_{24}^3}+
\tfrac{1}{t_{14}^3t_{23}^3}
\right)
\,
(\mathcal{J}_{\alpha \beta})_{12,3}\,,
\\
\label{PsiPsiPsisqaredin20SPIN1}
\langle \theta \Psi(1) \theta \Psi(2) \,
(\Psi^2)^{[2,0],2}_{\alpha \beta}(t_3,t_4;3)
\rangle 
&=4
\, \left(\tfrac{1}{t_{13}^3t_{24}^3}-
\tfrac{1}{t_{14}^3t_{23}^3}
\right)
(\epsilon\, \theta_1\epsilon\, y_{13}\ww_3)_{(\alpha }
(\epsilon\, \theta_2\epsilon\, y_{23}\ww_3)_{\beta)}
\,,
\end{align}
\end{subequations}
where $\mathcal{J}_{12,3}$ is defined in \eqref{J123def} and
\begin{equation}
\label{J123defwithspin}
(\mathcal{J}_{\alpha \beta})_{12,3}=
\epsilon_{(\alpha \alpha'}
\epsilon_{\beta) \beta'}
\theta_1^{a\alpha'}\theta_2^{b\beta'}
(\epsilon y_{13}\epsilon y_{23 }\epsilon)_{ab}\,,
\end{equation}
is its analogue with spin and satisfy 
$(\mathcal{J}_{\alpha \beta})_{12,3}=(\mathcal{J}_{\alpha \beta})_{21,3}$.
%---------------------------------------------------------------------

%------------------------------------------------------------------------------------------------------------------------------------------------------------------------------------------------------
\subsection{Acting with derivatives}
\label{app:actingwithderivatives}

The procedure described in Section \ref{sec:L4operatorsConstraint} to determine certain components of long supermultiplets in the free theory requires the explicit computation of a multitude of five-point functions. This is in principle straightforward since we just need to act on the generating correlators, {\it e.g.} \eqref{phiphiphiphiS}, with various differential operators of the form \eqref{DoperatorsTomakeSinglets}. In practice, we find it more convenient to compute them in closed form as functions of the three labels of the length four conformal primaries. The basic quantity that needs to be computed is given in equation \eqref{Ageneral}.
We also apply this procedure to compute the three- and four-point functions $\langle \mathcal{D}_2\mathcal{D}_2 \mathcal{O}^{(4)}\rangle$ and $\langle \mathcal{D}_2\mathcal{D}_1\mathcal{D}_1 \mathcal{O}^{(4)}\rangle$.

\paragraph{For $L=2$ and $L=3$ operators.}
For $L=2$ operators the relevant identity is
\begin{equation}
\label{acton2}
\mathsf{D}_{K}^{(\Delta_1,\Delta_2)}(\partial_1,\partial_2)
\,
\frac{1}{t_{1a}^{2\Delta_1}t_{2b}^{2\Delta_2}}
\Bigg{|}_{t_i=t_1}
=
\frac{1}{\Gamma(2\Delta_1)\Gamma(2\Delta_2)}
\frac{1}{t_{1a}^{2\Delta_1}t_{2b}^{2\Delta_2}}
\left(\frac{t_{ab}}{t_{1a}t_{1b}}\right)^K\,.
\end{equation}
This identity was already employed to derive the three-point function in \eqref{2and3GFT}.
The analogue identity for $L=3$ operators is\footnote{
When computing quantities of this type, it should be kept in mind that $(\partial_1+\partial_2)^{\ell}F(t_1,t_2)|_{t_2=t_1}=\partial_1^{\ell} F(t_1,t_1)$.
}
\begin{align}
\label{acton3general}
\begin{split}
\mathsf{D}_{K_2}^{(\Delta_1,\Delta_{2}+\Delta_3+K_1)}(\partial_1,\partial_2+\partial_3)
\mathsf{D}_{K_1}^{(\Delta_2,\Delta_3)}(\partial_2,\partial_3)
\,
\frac{1}{t_{1a}^{2\Delta_1}t_{2b}^{2\Delta_2}t_{3c}^{2\Delta_3}}
\Bigg{|}_{t_i=t_1}&
=
\\
=\frac{(-1)^{K_2}}{t_{1a}^{2\Delta_1}t_{1b}^{2\Delta_2}t_{1c}^{2\Delta_3}}
\left(\frac{t_{bc}}{t_{1b}t_{1c}}\right)^{K_1+K_2}
\frac{1}{\Gamma(2\Delta_1)\Gamma(2\Delta_2)\Gamma(2\Delta_3)}\,
%\frac{1}{t_{1a}^2t_{2b}^2t_{3c}^2}
&\mathcal{A}^{(K_1,K_2)}_{(\Delta_1,\Delta_2,\Delta_3)}(\tfrac{t_{1b}t_{ac}}{t_{1a}t_{bc}})
\,,
\end{split}
\end{align}
where
\begin{align}
\label{AforL3}
\begin{split}
\mathcal{A}^{(K_1,K_2)}_{\Delta_1,\Delta_2,\Delta_3}(z)=&
\frac{\Gamma(K_1+K_2+2\Delta_2)}{\Gamma(K_1+2\Delta_2)\Gamma(K_2+2K_1+2\Delta_2+2\Delta_3)}\times\\
&
\,\,\,\,\,\,\,\,\,\,\,\,\,\,\,\,\,\,\,\,\,\,{}_2F_1(-K_2,1-2K_1-K_2-2\Delta_2-2\Delta_3,1-K_1-K_2-2\Delta_2,z)\,.
\end{split}
\end{align}
Notice that $\mathcal{A}^{(K_1,K_2)}_{\Delta_1,\Delta_2,\Delta_3}(z)$ is independent of $\Delta_1$.
%-----
%\begin{equation}
%\label{acton3}
%\mathsf{D}_K^{(1,2+2n)}(\partial_1,\partial_2+\partial_3)
%\mathsf{D}_{2n}^{(1,1)}(\partial_2,\partial_3)
%\,
%\frac{1}{t_{1a}^2t_{2b}^2t_{3c}^2}
%\Big{|}_{t_i=t_1}
%=\frac{1}{t_{1a}^2t_{2b}^2t_{3c}^2}
%\left(-
%\frac{t_{bc}}{t_{1b}t_{1c}}
%\right)^{2n+K}\!\!
%F^{\{K,n\}}(\tfrac{t_{1b}t_{ac}}{t_{1a}t_{bc}})\,.
%\end{equation}

%\begin{equation}
%\label{Fspecial}
%F^{\{K,n\}}(z):=
%\frac{\Gamma(2+K+2n)}{\Gamma(2+2n)\Gamma(4+K+4n)}
%{}_2F_1(-K,-3-K-4n,-1-K-2n,z)\,.
%\end{equation}

%----------------------------------------------------------------------------------------------------------------------------------------------------------------------------------
\paragraph{Five-point functions involving a $L=4$ operator.}
To deal with $L=4$ operators we will need  to compute the following quantity 
\begin{align}
\label{Ageneral}
\begin{split}
&\mathsf{D}_K^{(k_1,k_2)}(\partial_1+\partial_2,\partial_3+\partial_4)
\mathsf{D}_{K_1}^{(\Delta_1,\Delta_1')}(\partial_1,\partial_2)\mathsf{D}_{K_2}^{(\Delta_2,\Delta_2')}(\partial_3,\partial_4) 
\frac{1}{t_{1a}^{2\Delta_1}t_{2b}^{2\Delta_1'}t_{3c}^{2\Delta_2}t_{4d}^{2\Delta_{2}'}}\Bigg{|}_{t_i=t_5}=
\\
&
\qquad
\qquad
\qquad
\qquad
=
\frac{(-1)^K}{
t_{5a}^{2\Delta_1}t_{5b}^{2\Delta_1'}t_{5c}^{2\Delta_2}t_{5d}^{2\Delta_{2}'}
}
W_2^{K_2}
\left(\frac{t_{ab}}{t_{5a}t_{5b}}\right)^{K+K_1+K_2}
\,\,
\mathcal{A}^{(K;\,K_1,K_2)}_{(\Delta_1,\Delta_1'|\Delta_2,\Delta_2')}\,,
\end{split}
\end{align}
where $k_1:=K_1+\Delta_1+\Delta_1'$, $k_2:=K_2+\Delta_2+\Delta_2'$.
The prefactor in  \eqref{Ageneral} is chosen so that 
$\mathcal{A}$ is a function of the two bosonic cross ratios for which we use the following definitions
\begin{equation}
\label{chi1andchi2forW1andW2}
\chi_1=\frac{t_{ab}t_{c5}}{t_{ac} t_{b5}}\,,
\quad
\chi_2=\frac{t_{ab}t_{d5}}{t_{ad} t_{b5}}\,,
\end{equation}
\begin{equation}
\label{W1andW2}
W_1=\frac{1}{2}\left(
1-\chi_1^{-1}-\chi_2^{-1}\right)\,,
\qquad
W_2=\chi_2^{-1}-\chi_1^{-1}=
\frac{t_{a5}t_{b5}t_{cd}}{t_{c5} t_{d5}t_{ab} }\,.
\end{equation}
%
%------------------------------
%The first step to compute \eqref{Ageneral} is to use the identity \eqref{acton2} twice. 
After some work, the expression above can be massaged to the form
\begin{align}
\label{AintermsofPhi}
\begin{split}
\mathcal{A}^{(K;\,K_1,K_2)}_{(\Delta_1,\Delta_1'|\Delta_2,\Delta_2')}=&
\frac{1}{\Gamma(2\Delta_1)\Gamma(2\Delta_1')\Gamma(2\Delta_2)\Gamma(2\Delta_2')\Gamma(2h_1)\Gamma(2h_2)}
\frac{(h_1-\delta_1)_K}{(2h_1)_K}\times
\,\,\,\,\,\,
\,\,\,\,\,\,
\\
&\,\,\,\,\,\,\,\,\,\,\,\,
\Phi(h_2+\delta_2,h_2-\delta_2;-K,1-2h_1-K;1-h_1-K+\delta_1|\chi_1^{-1},\chi_2^{-1})\,,
\end{split}
\end{align}
where we introduced the hypergeometric function\footnote{Technically, $\Phi$ is a Kampé de Fériet hypergeometric function. The general definition of this class of functions is 
\begin{align}
{ }^{p+q} F_{r+s}\left(\begin{array}{l}
a_1, \cdots, a_p: b_1, b_1{ }^{\prime} ; \cdots ; b_q, b_q{ }^{\prime} ; \\
c_1, \cdots, c_r: d_1, d_1{ }^{\prime} ; \cdots ; d_s, d_s{ }^{\prime} ;
\end{array} x, y\right)=\sum_{m=0}^{\infty} \sum_{n=0}^{\infty} \frac{\left(a_1\right)_{m+n} \cdots\left(a_p\right)_{m+n}}{\left(c_1\right)_{m+n} \cdots\left(c_r\right)_{m+n}} \frac{\left(b_1\right)_m\left(b_1{ }^{\prime}\right)_n \cdots\left(b_q\right)_m\left(b_q{ }^{\prime}\right)_n}{\left(d_1\right)_m\left(d_1{ }^{\prime}\right)_n \cdots\left(d_s\right)_m\left(d_s{ }^{\prime}\right)_n} \cdot \frac{x^m y^n}{m ! n !}
\end{align}
and one can observe that
\begin{align}
\Phi(a,b;c,d;e|x,y)={ }^{2+1} F_{2+0}\left(\begin{array}{l}
c,d: a,b; \\
a+b, e ; ;
\end{array} x, y\right)\,.
\end{align}
}
\begin{equation}
\label{Phiseries}
%\begin{split}
\Phi(a,b;c,d;e|x,y):=%&
\sum_{m,n=0}^{\infty}
\frac{1}{m!n!}\frac{\!(a)_m(b)_n}{\,\,\,(a+b)_{n+m}}\,
\frac{(c)_{n+m}(d)_{n+m}}{(e)_{n+m}}\,
x^m y^n\,.
\end{equation}
Above, we employed the standard notation for the Pochhammer symbol 
\begin{equation}
(a)_n=\frac{\Gamma(a+n)}{\Gamma(a)}\,.
\end{equation}
It should be notices that the function $\Phi$ entering \eqref{AintermsofPhi} is a polynomial in $\chi_1^{-1}$ and $\chi_2^{-1}$ since the series \eqref{Phiseries} terminates when the parameter $c=-K$ with $K\in \mathbb{Z}_{\geq 0}$.
Alternatively, one can write \eqref{Ageneral} in terms of the variables $W_1$ and $W_2$ as follows
\begin{equation}
\mathcal{A}^{(K;\,K_1,K_2)}_{(\Delta_1,\Delta_1'|\Delta_2,\Delta_2')}=
\frac{1}{\Gamma(2\Delta_1)\Gamma(2\Delta_1')\Gamma(2\Delta_2)\Gamma(2\Delta_2')\Gamma(2h_1)\Gamma(2h_2)}
\sum_{\ell=0}^K\frac{\Gamma(K+1)}{\Gamma(K+1-\ell)}\,W_1^{K-\ell} \mathcal{G}_{\ell}^{(h_1,h_2|\delta_1,\delta_2)}
\end{equation}
where $\delta_1=\Delta_1-\Delta_1'$ and $\delta_2=\Delta_2-\Delta_2'$ and
\begin{equation}
\label{mathcalFassum}
\mathcal{G}_{\ell}^{(h_1,h_2|\delta_1,\delta_2)}=\sum_{j=0}^{\ell}\,(-1)^j\,\mathsf{g}_{\ell-j}(h_1,\delta_1)\mathsf{g}_{j}(h_2,\delta_2)(W_2)^j\,.
\end{equation}
For the $L=4$ operators discussed in Sections \ref{sec:operatorsforthesuperprimary}, \ref{sec:operatorsforthesuperdescendant} we will use this result with $(\Delta_1,\Delta_1'|\Delta_2,\Delta_2')$ chosen among these five possibilities\footnote{Up to reordering of the points.}
\begin{equation}
\{(1,1|1,1),\,
(\tfrac{3}{2},\tfrac{3}{2}|1,1),\,
(2,2|1,1),\,
(\tfrac{3}{2},\tfrac{3}{2}|\tfrac{3}{2},\tfrac{3}{2}),\,
(1,2|\tfrac{3}{2},\tfrac{3}{2})
\}\,,
\end{equation}
where the first four correspond to $(\delta_1,\delta_2)=(0,0)$ while the last to  $(\delta_1,\delta_2)=(-1,0)$. We thus report the functions $\mathsf{g}_{j}(h,\delta)$ only for $\delta=0$ and $\delta=-1$:
\begin{equation}
\mathsf{g}_{j}(h,0)=
%\frac{m_{j/2}(h)}{8^{j/2} \Gamma(1+\tfrac{j}{2})}\,
%\delta_{j,\text{even}}=
%\begin{dcases*}
\begin{cases}
\dfrac{\Gamma(h+\tfrac{1}{2})}{4^j\, \Gamma(1+\tfrac{j}{2} ) \Gamma(h+\tfrac{j+1}{2} )} & \text{$j$ even}\\
0 & \text{$j$ odd}
\end{cases}
%\end{dcases*}
\end{equation}
\begin{equation}
\mathsf{g}_{j}(h,-1)=\begin{cases}
\dfrac{\Gamma(2h)}{ \Gamma(2h+j)}
\dfrac{(h+j)\Gamma(h+\tfrac{j}{2})}{2^j\, \Gamma(1+\tfrac{j}{2} ) \Gamma(h+1 )} & \text{$j$ even}\\
\dfrac{\Gamma(2h)}{ \Gamma(2h+j)}
\dfrac{\Gamma(h+\tfrac{j+1}{2})}{2^{j-1}\, \Gamma(1+\tfrac{j-1}{2} ) \Gamma(h+1 )}  & \text{$j$ odd}\end{cases}
\end{equation}
From these expression one can easily sum the expression  \eqref{mathcalFassum} for $\mathcal{G}_{\ell}^{(h_1,h_2|0,0)}$ and $\mathcal{G}_{\ell}^{(h_1,h_2|-1,0)}$ that we need and obtain (from now on $z:=W_2^{-2}$)
\begin{equation}
\mathcal{G}_{\ell}^{(h_1,h_2|0,0)}=
\frac{\delta_{\ell,\text{even}}\Gamma(h_2+\tfrac{1}{2})}{4^{\ell} \Gamma(1+\tfrac{\ell}{2})\Gamma(h_2+\tfrac{\ell+1}{2})}\,
z^{-\frac{\ell}{2}}\,
{}_2F_1(-\tfrac{\ell}{2},-\tfrac{\ell}{2}-h_2+\tfrac{1}{2},h_1+\tfrac{1}{2},z)\,,
\end{equation}
for $\ell$ even 
\begin{align}
\begin{split}
&\mathcal{G}_{\ell}^{(h_1,h_2|-1,0)}=
\frac{\Gamma(h_2+\tfrac{1}{2})}{4^{\ell} \Gamma(1+\tfrac{\ell}{2})\Gamma(h_2+\tfrac{\ell+1}{2})}
\,\times
\\
&
\left(
z^{-\frac{\ell}{2}}{}_2F_1(-\tfrac{\ell}{2},-\tfrac{\ell}{2}-h_2+\tfrac{1}{2},h_1+\tfrac{1}{2},z)
+
\frac{\ell (\ell+2h_2+1)}{h_1 (2h_1+1)}
z^{-\frac{\ell-2}{2}}{}_2F_1(-\tfrac{\ell-2}{2},-\tfrac{\ell-3}{2}-h_2,h_1+\tfrac{3}{2},z)
\right)\,,
\end{split}
\end{align}
and for $\ell$ odd
\begin{align}
\begin{split}
&\mathcal{G}_{\ell}^{(h_1,h_2|-1,0)}=
\frac{2\Gamma(h_2+\tfrac{1}{2})}{4^{\ell} h_1 \Gamma(\tfrac{\ell+1}{2})\Gamma(h_2+\tfrac{\ell}{2})}
\,\times
\\
&
\left(
z^{-\frac{\ell-1}{2}}{}_2F_1(\tfrac{1-\ell}{2},-\tfrac{\ell}{2}-h_2,h_1+\tfrac{1}{2},z)
-
\frac{\ell-1}{2h_1+1}
z^{-\frac{\ell-3}{2}}{}_2F_1(\tfrac{3-\ell}{2},-\tfrac{\ell-2}{2}-h_2,h_1+\tfrac{3}{2},z)
\right)\,.
\end{split}
\end{align}
%
%
%------------------------------------------------------------------------------------------------------------------
\paragraph{The example of $\langle \varphi\varphi\varphi\varphi\mathcal{O}\rangle$ .}
Let us employ the results  obtained above to obtain the five-point function 
\begin{equation}
\label{5pointCOMPUTEvarphiOKn1n2}
\langle \varphi(1) \varphi(2)  \varphi(3)  \varphi(4)
\mathcal{O}_{\varphi^4}^{\{K,n_1,n_2\}}(t_5)\rangle^{(0)}
\end{equation}
where $\mathcal{O}_{\varphi^4}$ is defined in  \eqref{OL4phi4operators}.
To compute this quantity, we will act on the correlator \eqref{phiphiphiphiS} with the differential operators used to define $\mathcal{O}_{\varphi^4}^{\{K,n_1,n_2\}}$.

The result is a sum of terms of the form \eqref{Ageneral}.
To organize these terms it is useful to collect the transformation properties of the quantities above under the permutation of points $t_a,t_b,t_c,t_d$ given by 
$(a\leftrightarrow b)$, $(c\leftrightarrow d)$ and $(a,b\leftrightarrow c,d)$:
\begin{subequations}
\begin{align}
 (W_1,W_2)\stackrel{(a\leftrightarrow b)}{\mapsto} & (-W_1,-W_2) \,, \,\,\,\,\,\,\qquad
\,\,\,\,\,\,\,\,\,\,\,\,\,\,\,\,\mathcal{A}^{(K;K_1,K_2)}_{(1,1|1,1)}\big{|}_{(a\leftrightarrow b)} =(-1)^K
 \mathcal{A}^{(K;K_1,K_2)}_{(1,1|1,1)}\,,
\\
 (W_1,W_2) \stackrel{(c\leftrightarrow d)}{\mapsto}&   (+W_1,-W_2)  \,, \,\,\,\,\,\,\qquad
\,\,\,\,\,\,\,\,\,\,\,\,\,\,\,\,\mathcal{A}^{(K;K_1,K_2)}_{(1,1|1,1)}\big{|}_{(c\leftrightarrow d)}=\,
  \mathcal{A}^{(K;K_1,K_2)}_{(1,1|1,1)}\,,
  \\
 (W_1,W_2) \stackrel{(a,b\leftrightarrow c,d)}{\mapsto} &  (-W_1W_2^{-1},W_2^{-1})  \,, \,\,\,\,\,\,\qquad
 \mathcal{A}^{(K;K_1,K_2)}_{(1,1|1,1)}\big{|}_{(a,b\leftrightarrow c,d)}=
 (-W_2)^{-K}  \mathcal{A}^{(K;K_1,K_2)}_{(1,1|1,1)}\,.
\end{align}
\end{subequations}
The finite group generated by these three transformations is of course eight dimensional.

Using these identities we see that the correlators \eqref{5pointCOMPUTEvarphiOKn1n2} can be written in the form \eqref{phi4terminSuperPR5pointfun} with the function $F^{(12)}_{\varphi\varphi\varphi \varphi}$ taking the form
\begin{equation}
\label{F12fromAA}
F^{(12)}_{\varphi\varphi\varphi \varphi}
\rightarrow 4 \left(
(-W_2)^{-K-h_1}
\mathcal{A}^{(K;\,2n_1,2n_2)}_{(1,1|1,1)}
  +
(+W_2)^{-K-h_2}
\mathcal{A}^{(K;\,2n_2,2n_1)}_{(1,1|1,1)}
  \right)\,,
\end{equation}
where $W_1$ and $W_2$ are as in \eqref{W1andW2} and \eqref{chi1andchi2forW1andW2} with $(t_a,t_b,t_c,t_d) \mapsto (t_1,t_2,t_3,t_4)$.
Notice that in this case we have $K_1=2n_1$,  $K_2=2n_2$, $h_1=2n_1+2$ and $h_2=2n_2+2$.
The remaining functions $F^{(13)}_{\varphi\varphi\varphi \varphi}$ and $F^{(14)}_{\varphi\varphi\varphi \varphi}$ are obtained by applying a suitable permutation of the points.
All the other five-point functions that we need are computed in a similar way.
%

%---------------------------
\paragraph{Some three- and four-point functions.}
Let us show how to compute the free theory correlator $\langle \mathcal{D}_2\mathcal{D}_1 \mathcal{D}_1 \mathcal{O}_{\varphi^4} \rangle$. This object will be used in the companion paper   \cite{Ferrero:2023gnu} to address the problem of operator mixing.
It can be obtained by taking the limit of \eqref{phi4terminSuperPR5pointfun} for $y_{12}^2\rightarrow 0$ first and $t_{12} \rightarrow 0 $ next. In this limit only $F^{(13)}_{\varphi\varphi\varphi \varphi}$ and  $F^{(14)}_{\varphi\varphi\varphi \varphi}$ contribute and are given by permuting the points of  \eqref{F12fromAA}, more  concretely
\begin{subequations}
\begin{align}
(W_1,W_2)\Big{|}_{(t_a,t_b,t_c,t_d,t_5)\mapsto (t_1,t_2,t_1,t_3,t_4)}&=(\tfrac{1}{2}(1-x),x)\,,\\
(W_1,W_2)\Big{|}_{(t_a,t_b,t_c,t_d,t_5)\mapsto (t_1,t_3,t_2,t_1,t_4)}&=(\tfrac{1}{2}(1-x^{-1}),-x^{-1})\,,
\end{align}
\end{subequations}
where $x=\tfrac{t_{13}t_{24}}{t_{12}t_{34}}$.
We thus need to consider the quantity 
\begin{equation}
\mathcal{Z}^{(K;\,K_1,K_2)}_{(\Delta_1,\Delta_1'|\Delta_2,\Delta_2')}:=
\mathcal{A}^{(K;\,K_1,K_2)}_{(\Delta_1,\Delta_1'|\Delta_2,\Delta_2')}\Big{|}_{W_1=\tfrac{1}{2}(1-z),W_2=z}
\,.
\end{equation}
From \eqref{AintermsofPhi} and \eqref{Phiseries} it is easy to recognize that it is given by
\begin{align}
\label{Zappendix3F2}
\begin{split}
\mathcal{Z}^{(K;\,K_1,K_2)}_{(\Delta_1,\Delta_1'|\Delta_2,\Delta_2')}=&
\frac{1}{\Gamma(2\Delta_1)\Gamma(2\Delta_1')\Gamma(2\Delta_2)\Gamma(2\Delta_2')\Gamma(2h_1)\Gamma(2h_2)}
\frac{(h_1-\delta_1)_K}{(2h_1)_K}\times
\,\,\,\,\,\,
\,\,\,\,\,\,
\\
&\,\,\,\,\,\,\,\,\,\,\,\,\,\,\,\,\,\,\,\,\,\,\,\,
{}_3F_{2}(-K,1-K-2h_1,h_2-\delta_2; 2h_2, 1-K-h_1+\delta_1,z)\,,
\end{split}
\end{align}
where, as above, $h_1:=K_1+\Delta_1+\Delta_1'$, $h_2:=K_2+\Delta_2+\Delta_2'$ and $\delta_1=\Delta_1-\Delta_1'$, $\delta_2=\Delta_2-\Delta_2'$.
Assembling the various pieces we arrive at
\begin{equation}
\langle \mathcal{D}_2\mathcal{D}_1 \mathcal{D}_1 \mathcal{O}_{\varphi^4}^{\{K,n_1,n_2\}} \rangle=
\delta_{K,\text{even}}
\frac{y_{12}^2}{t_{12}^2}\,\frac{y_{13}^2}{t_{13}^2}\,
\left(\frac{t_{12}}{t_{14}t_{24}}\right)^{\h}
8\left(x^{h_2}\,
\mathcal{Z}^{(K;\,2n_1,2n_2)}_{(1,1|1,1)}(x)+
x^{h_1}\,
\mathcal{Z}^{(K;\,2n_2,2n_1)}_{(1,1|1,1)}(x)\right)\,,
\end{equation}
where $h_i=2n_i+2$ and  $\h=K+h_1+h_2$.
With a simple further OPE, corresponding to setting $x=1$,  we can recover the three-point function 
\begin{equation}
\langle \mathcal{D}_2(1)\mathcal{D}_2(2) \mathcal{O}_{\varphi^4}^{\{K,n_1,n_2\}}(t_3)\rangle^{(0)}\,=\,
16 \,\,\mathcal{Z}^{(K;\,2n_1,2n_2)}_{(1,1|1,1)}(1)\,
\delta_{K,\text{even}}
\left(\frac{y_{12}^2}{t_{12}^2}\right)^2\,
\left(\frac{t_{12}}{t_{13}t_{23}}\right)^{K+\Delta_1+\Delta_2}\,,
\end{equation}
where, from \eqref{Zappendix3F2},
\begin{equation}
\label{sharppinD2D2O}
\mathcal{Z}^{(K;\,2n_1,2n_2)}_{(1,1|1,1)}(1)=
\frac{1}{\Gamma(2\h_1)\Gamma(2\h_2)}
\frac{\Gamma(K+1)}{4^{K} \Gamma(1+\tfrac{K}{2})}\,
\frac{\Gamma(\h_1+\tfrac{1}{2})}{\Gamma(\h_1+\tfrac{K+1}{2})}\,
\frac{\Gamma(\h_2+\tfrac{1}{2})}{\Gamma(\h_2+\tfrac{K+1}{2})}\,
\frac{\Gamma(\h_1+\h_2+K)}{\Gamma(\h_1+\h_2+\tfrac{K}{2})} \,.
\end{equation}

\section{Blockology}\label{app:blocks}

In this appendix we collect some conventions and definitions related to (super)conformal blocks, and give expressions for the superconformal blocks relevant for $\langle \mathcal{D}_{k_1}\,\mathcal{D}_{k_2}\,\mathcal{D}_{k_3}\,\mathcal{D}_{k_4}\rangle$ correlators. 

\subsection{Bosonic conformal blocks}

Let us start by stating our conventions for $\mathfrak{sl}(2)$ conformal blocks: given a four-point function of $\mathfrak{sl}(2)$ primaries $\mathcal{O}$
\begin{align}\label{sl24pt}
\langle \mathcal{O}_{\h_1}(t_1)\, \mathcal{O}_{\h_2}(t_2)\, \mathcal{O}_{\h_3}(t_3)\, \mathcal{O}_{\h_4}(t_4)\rangle=\frac{1}{t_{12}^{\h_1+\h_2}\,t_{34}^{\h_3+\h_4}}\,\left(\frac{t_{24}}{t_{14}}\right)^{\h_{12}}\,\left(\frac{t_{14}}{t_{13}}\right)^{\h_{34}}G(\chi)\,,\quad
\chi=\frac{t_{12}\,t_{34}}{t_{13}\,t_{24}}\,,
\end{align}
with $\h_{ij}=\h_i-\h_j$, the conformal blocks in the ``direct'' channel $t_2\to t_1$ ($\chi\to 0$) are eigenfunctions of the quadratic $\mathfrak{sl}(2)$ Casimir operator acting on the first two points, with eigenvalue
\begin{align}
\mathfrak{c}_{\mathfrak{sl}(2)}(\h)=\h(\h-1)\,,
\end{align}
namely they are functions $g_h(\chi)$ satisfying the differential equation
\begin{align}
\widehat{\mathfrak{C}}_{\mathfrak{sl}(2)}\,g_h(\chi)=\mathfrak{c}_{\mathfrak{sl}(2)}\,g_h(\chi)\,,
\end{align}
where
\begin{align}
\widehat{\mathfrak{C}}_{\mathfrak{sl}(2)}=(1-\chi)\,\chi^2\,\frac{\partial^2}{\partial\chi^2}-(1-\h_{12}+\h_{34})\,\chi^2\,\frac{\partial}{\partial\chi}+\h_{12}\,h_{34}\,\chi\,.
\end{align}
The solution with suitable boundary conditions is
\begin{align}
g_{\h}(\chi)=\chi^{\h}\,{}_2F_1(\h-\h_{12},\h+\h_{34},2\h;\chi)\,.
\end{align}

Imagine now to suppress the spacetime dependence and to have four operators $\mathcal{O}$ transforming in $[0,k]$ representations of $\mathfrak{sp}(4)$ and depending on coordinates $y^{ab}$. This is not a scenario that we actually consider, but it is useful to separate the spacetime and R-symmetry dependence for a moment in order to introduce some definitions. We then write
\begin{align}\label{sp44pt}
\begin{split}
\langle \mathcal{O}_{k_1}(y_1)\,\mathcal{O}_{k_2}(y_2)\,\mathcal{O}_{k_3}(y_3)\,\mathcal{O}_{k_4}(y_4)\rangle&=(y^2_{12})^{\tfrac{1}{2}(k_1+k_2)}\,(y^2_{34})^{\tfrac{1}{2}(k_3+k_4)}\,\left(\frac{y^2_{14}}{y^2_{24}}\right)^{\tfrac{1}{2}k_{12}}\,\left(\frac{y^2_{13}}{y^2_{14}}\right)^{\tfrac{1}{2}k_{34}}\,G(\zeta_1,\zeta_2)\,,\\
\zeta_1\,\zeta_2&=\frac{y^2_{12}\,y^2_{34}}{y^2_{13}\,y^2_{24}}\,, \qquad
(1-\zeta_1)(1-\zeta_2)=\frac{y^2_{14}\,y^2_{23}}{y^2_{13}\,y^2_{24}}\,, 
\end{split}
\end{align}
where $y^2_{ij}=\det y^{ab}_{12}$ as usual. The R-symmetry conformal blocks in the ``direct'' channel $y_2\to y_1$ ($\zeta_i\to 0$) are eigenfunctions of the quadratic $\mathfrak{sp}(4)$ Casimir operator acting on the first two points, with eigenvalue
\begin{align}
\mathfrak{c}_{\mathfrak{sp}(4)}(a,b)=\tfrac{1}{4}a(a+4)+\tfrac{1}{2}b(b+3)+\tfrac{1}{2}ab\,,
\end{align}
namely they are functions $\mathbb{B}_{[a,b]}(\zeta_1,\zeta_2)$ satisfying the differential equation
\begin{align}
\widehat{\mathfrak{C}}_{\mathfrak{sp}(4)}\,\mathbb{B}_{[a,b]}(\zeta_1,\zeta_2)=\mathfrak{c}_{\mathfrak{sp}(4)}(a,b)\,\mathbb{B}_{[a,b]}(\zeta_1,\zeta_2)\,,
\end{align}
where
\begin{align}
\begin{split}
\widehat{\mathfrak{C}}_{\mathfrak{sp}(4)}=&(\zeta_1+\zeta_2)\,\frac{k_{12}\,k_{34}}{4}+(1-\zeta_1)\,\zeta_1^2\,\frac{\partial^2}{\partial \zeta_1^2}+(1-\zeta_2)\,\zeta_2^2\,\frac{\partial^2}{\partial \zeta_2^2}\\
&+\frac{1}{\zeta_1-\zeta_2}\left[\frac{k_{12}-k_{34}}{2}\,(\zeta_1-\zeta_2)\,\zeta_1+(\zeta_2-\zeta_1^2)\right]\,\zeta_1\,\frac{\partial}{\partial \zeta_1}\\
&+\frac{1}{\zeta_2-\zeta_1}\left[\frac{k_{12}-k_{34}}{2}\,(\zeta_2-\zeta_1)\,\zeta_2+(\zeta_1-\zeta_2^2)\right]\,\zeta_2\,\frac{\partial}{\partial \zeta_2}\,.
\end{split}
\end{align}
Imposing suitable boundary conditions, $\mathbb{B}_{[a,b]}(\zeta_1,\zeta_2)$ are symmetric polynomials of degree $\tfrac{a+b}{2}$ in $\zeta_{1}^{-1}$ and $\zeta_{2}^{-1}$, normalized in such a way that the leading power of $\zeta_{1,2}$ in the small $\zeta_{1,2}$ expansion appears with unit coefficient.  Being related by analytic continuation to three-dimensional conformal blocks,  they do not admit a closed-form expression: we collect some results in a $\mathtt{Mathematica}$ notebook included with the submission.  

\subsection{$\langle \mathcal{D}_{k_1}\,\mathcal{D}_{k_2}\,\mathcal{D}_{k_3}\,\mathcal{D}_{k_4}\rangle$}

For $\langle \mathcal{D}_{k_1}\,\mathcal{D}_{k_2}\,\mathcal{D}_{k_3}\,\mathcal{D}_{k_4}\rangle$ correlators, we use the prefactor \eqref{K_1234}, which is just a superposition of \eqref{sl24pt} and \eqref{sp44pt}. The superconformal blocks are then functions $\mathfrak{G}_{\h,s,[a,b]}(\chi,\zeta_1,\zeta_2)$ that are eigenfunctions of the quadratic Casimir of $\mathfrak{osp}(4^*|4)$ acting on the first two operators: they satisfy for supermultiplets in a representation $\mathcal{R}$ we have
\begin{align}
\widehat{\mathfrak{C}}_{\mathfrak{osp}(4^*|4)}\,\mathfrak{G}_{\mathcal{R}}(\chi,\zeta_1,\zeta_2)=\mathfrak{c}_{\mathfrak{osp}(4^*|4)}(\mathcal{R})\,\mathfrak{G}_{\mathcal{R}}(\chi,\zeta_1,\zeta_2)\,,
\end{align}
where the superconformal Casimir acting on \eqref{eq: correlation function in K and A} gives
\begin{align}
\begin{split}
\widehat{\mathfrak{C}}_{\mathfrak{osp}(4^*|4)}=&\widehat{\mathfrak{C}}_{\mathfrak{sl}(2)}-2\,\widehat{\mathfrak{C}}_{\mathfrak{sp}(4)}-\frac{2\,\chi\,(1-\chi)\,(\chi\,(\zeta_1+\zeta_2)-2\,\zeta_1\,\zeta_2)}{(\chi-\zeta_1)\,(\chi-\zeta_2)}\,\frac{\partial}{\partial \chi}-\sum_{i=1}^2\frac{4\,\chi\,\zeta_i\,(1-\zeta_i)}{\chi-\zeta_i}\,\frac{\partial}{\partial \zeta_i}\,,
\end{split}
\end{align}
with eigenvalue
\begin{align}\label{supercasimireigenvalue}
\mathfrak{c}_{\mathfrak{osp}(4^*|4)}(\mathcal{R})=\h\,(\h+3)+\frac{1}{4}\,s\,(s+2)-\frac{1}{2}\,a^2-a\,(b+2)-b\,(b+3)\,,
\end{align}
where the Dynkin labels of $\mathcal{R}$ are $\omega(\mathcal{R})=\{h,s,[a,b]\}$. In practice, a convenient way to find superconformal blocks for external half-BPS operators is to make an ansatz in terms of sums of products of suitable blocks for the bosonic subalgebra (reflecting the structure of the exchanged supermultiplets) and to demand that this satisfies the superconformal Ward identities. Here we follow this method and generalize the results of \cite{Liendo:2018ukf}.

Let us now specialize to the case $k_1=k_4=p$ and $k_2=k_3=q$. For the exchange of short $\mathfrak{D}_k$ supermultiplets with $\omega(\mathcal{D}_k)=\{k,0,[0,k]\}$ (note that $\mathfrak{c}_{\mathfrak{osp}(4^*|4)}(\mathfrak{D}_k)=0$) we find the blocks\footnote{We suppress the dependence on the cross-ratios, but it should be understood that $g_{\h}$ is a functions of $\chi$ and $\mathbb{B}_{[a,b]}$ is a function of $\zeta_1,\zeta_2$.}
\begin{align}\label{shortsuperblock}
\begin{split}
\mathfrak{G}_{\mathcal{D}_k}(\chi,\zeta_1,\zeta_2)=&\mathbb{B}_{[0,k]}\,g_k+\frac{(k+p-q)^2\,(k-p+q)^2}{8\,(1-k)\,(1+2k)\,k^2}\,\mathbb{B}_{[2,k-2]}\,g_{k+1}\\
&+\frac{(k+p-q)^2(1+k+p-q)^2(k-p+q)^2(1+k-p+q)^2}{16\,k\,(1-k)\,(1+k)^2\,(1-2k)\,(1+2k)^2\,(3+2k)}\,\mathbb{B}_{[0,k-2]}\,g_{k+2}\,.
\end{split}
\end{align}
The other type of exchanged representation we are interested in is spin-less long multiplets $\mathcal{L}^{\h}_{0,[a,b]}$\footnote{Conformal blocks for semi-short representations can be obtained by taking a suitable limit where the conformal dimension approaches the unitarity bound. }. For these representations we find
\begin{align}\label{longsuperblock}
\begin{split}
\mathfrak{G}_{\h,[a,b]}(\chi,\zeta_1,\zeta_2)&=\mathbb{B}_{[a,b]}\,g_{\h}\\
&+\left(
c_1\,\mathbb{B}_{[a+2,b]}
+c_2\,\mathbb{B}_{[a+2,b-2]}
+c_3\,\mathbb{B}_{[a,b]}
+c_4\,\mathbb{B}_{[a-2,b+2]}
+c_5\,\mathbb{B}_{[a-2,b]}
\right)\,g_{\h+1}\\
&+\left(
c_6\,\mathbb{B}_{[a+4,b-2]}
+c_7\,\mathbb{B}_{[a+2,b]}
+c_8\,\mathbb{B}_{[a+2,b-2]}
+c_9\,\mathbb{B}_{[a,b+2]}
+c_{10}\,\mathbb{B}_{[a,b]}
\right.\\
&\left.+\,\,\,
c_{11}\,\mathbb{B}_{[a,b-2]}
+c_{12}\,\mathbb{B}_{[a-2,b+2]}
+c_{13}\,\mathbb{B}_{[a-2,b]}
+c_{14}\,\mathbb{B}_{[a-4,b+2]}
\right)\,g_{\h+2}\\
&+\left(
c_{15}\,\mathbb{B}_{[a+2,b]}
+c_{16}\,\mathbb{B}_{[a+2,b-2]}
+c_{17}\,\mathbb{B}_{[a,b]}
+c_{18}\,\mathbb{B}_{[a-2,b+2]}
+c_{19}\,\mathbb{B}_{[a-2,b]}
\right)\,g_{\h+3}\\
&+c_{20}\,\mathbb{B}_{[a,b]}\,g_{\h+4}\,,
\end{split}
\end{align}
where the coefficients $c_i$ ($i=1,...,20$) are rational functions of $\h$, $a$, $b$ and $p-q$,  given by 
\begingroup\makeatletter\def\f@size{9}\check@mathfonts
\begin{align}\label{longsuperblockcoeff}
\begin{split}
c_1=&-\frac{a+b-\h}{1+a+b-\h}\,,\quad
c_2=\frac{(2+a+2b)^2(b+p-q)^2(b-p+q)^2(3+b+\h)}{16(1-b)b^2(1+b)(1+a+2b)(3+a+2b)(2+b+\h)}\,,\\
c_3=&-\frac{(p-q)^2}{b(2+b)(1+a+b)(3+a+b)\h(2+\h)}\,P_2\,,\quad
c_4=-\frac{(1-b+\h)a^2}{(1-a)(1+a)(b-\h)}\,,\\
c_5=&\frac{a^2(2+a+2b)^2(1+a+b+p-q)^2(1+a+b-p+q)^2(4+a+b+\h)}{16(1-a)(1+a)(a+b)(1+a+b)^2(2+a+b)(1+a+2b)(3+a+2b)(3+a+b+\h)}\,,\\
c_6=&-\frac{(b+p-q)^2(b-p+q)^2(a+b-\h)(3+b+\h)}{16(1-b)b^2(1+b)(1+a+b-\h)(2+b+\h)}\,,\\
c_7=&-\frac{(p-q)^2(1-b+\h)(a+b-\h)(3+b+\h)}{2b(2+b)(1+a+b-\h)(1+\h)(2+\h)}\,,\\
c_8=&-\frac{(2+a+2b)^2(p-q)^2(b+p-q)^2(b-p+q)^2(a+b-\h)(3+b\h)(4+a+b\h)}{32(1-b)b^2(1+b)(1+a+b)(3+a+b)(1+a+2b)(3+a+2b)(1+\h)(2+\h)(2+b+\h)}\,,\\
c_9=&-\frac{(1-b+\h)(a+b-\h)}{(b-\h)(1+a+b-\h)}\,, \\
c_{10}=&\frac{\left[2(1-a)(3+a)b(2+b)(1+a+b)(3+a+b)(1+a+2b)(5+a+2b)\right]^{-1}}{(b-\h)(1+a+b-\h)(1+\h)(2+\h)(2+b+\h)(3+a+b+\h)(1+2\h)(5+2\h)}\,P_8\,,\\
c_{11}=&-\frac{(a+2b)^2(2+a+2b)^2(b+p-q)^2(b-p+q)^2(1+a+b+p-q)^2(1+a+b-p+q)^2}{256(1-b)b^2(1+b)(a+b)(1+a+b)^2(2+a+b)(-1+a+2b)(1+a+2b)^2}\\
&\times \frac{(3+b+\h)(4+a+b+\h)}{(3+a+2b)(2+b+\h)(2+a+b+\h)}\,,\\
c_{12}=&\frac{a^2(p-q)^2(1-b+\h)(a+b-\h)(4+a+b+\h)}{2(1-a)(1+a)(1+a+b)(3+a+b)(b-\h)(1+\h)(2+\h)}\,,\\
c_{13}=&\frac{a^2(2+a+2b)^2(p-q)^2(1+a+b+p-q)^2(q+a+b-p+q)^2(1-b+\h)}{32(1-a)(1+a)b(2+b)(a+b)(1+a+b)^2(2+a+b)(1+a+2b)(3+a+2b)}\\
&\times \frac{(3+b+\h)(4+a+b+\h)}{(1+\h)(2+\h)(3+a+b+\h)}\,,\\
c_{14}=&\frac{(2-a)^2a^2(1+a+b+p-q)^2(1+a+b-p+q)^2(1-b+\h)(4+a+b+\h)}{16(3-a)(1-a)^2(1+a)(a+b)(1+a+b)^2(2+a+b)(b-\h)(3+a+b+\h)}\,,\\
c_{15}=&-\frac{(1-b+\h)(3+b+\h)(2+p-q+\h)^2(2-p+q+\h)^2}{4(b-\h)(2+\h)^2(2+b+\h)(3+2\h)(5+2\h)}\,c_1\,,\\
c_{16}=&\frac{(a+b-\h)(2-p+q+\h)^2(2+p-q+\h)^2(4+a+b+\h)}{4(1+a+b-\h)(2+\h)^2(3+a+b+\h)(3+2\h)(5+2\h)}\,c_2\,,\\
c_{17}=&-\frac{(1-b+\h)(a+b-\h)\h(2+b+\h)(4+a+b+\h)(2+p-q+\h)^2(2-p+q+\h)^2}{4(b-\h)(1+a+b-\h)(1+\h)(2+\h)(3+\h)(2+b+\h)(3+a+b+\h)(3+2\h)(5+2\h)}\,c_3\,,\\
c_{18}=&\frac{(a+b-\h)(4+a+b+\h)(2+p-q+\h)^2(2-p+q+\h)^2}{4(1+a+b-\h)(2+\h)^2(3+a+b+\h)(3+2\h)(5+2\h)}\,c_4\,,\\
c_{19}=&-\frac{(1-b+\h)(2-p+q+\h)^2(2+p-q+\h)^2(3+b+\h)}{4(b-\h)(2+\h)^2(2+b+\h)(3+2\h)(5+2\h)}\,c_5\,,\\
c_{20}=&-\frac{(1-b+\h)(a+b-\h)(3+b+\h)(4+a+b+\h)(2+p-q+\h)^2)(2-p+q+\h)^2}{16(b-\h)(1+a+b-\h)(2+\h)^2(3+\h)^2(2+b+\h)(3+a+b+\h)}\\
&\times \frac{(3+p-q+\h)^2(3-p+q+\h)^2}{(3+2\h)(5+2\h)^2(7+2\h)}\,,
\end{split}
\end{align}
\endgroup
where $P_2$ and $P_8$ are polynomials in $\h$ of degree two and eight,  respectively,  whose coefficients are polynomials in $a$,  $b$ and $p-q$. For $P_2$ we have found
\begin{align}
P_2&=b(b+2)(1+a+b)(3+a+b)-\tfrac{1}{2}\h(\h+3)(a^2+2a(b+2)+(b+1)(b+2))\,,
\end{align}
while a closed form expression for $P_8$ is harder to obtain, but can easily be found on a case-by-case basis by demanding that the blocks \eqref{longsuperblock} vanish when evaluated for $\zeta_1=\zeta_2=\chi$.

Finally, if one is interested in the case $k_1=k_2=p$ and $k_3=k_4=p$, the blocks are exactly the same after setting $p=q$ in the coefficients appearing in \eqref{shortsuperblock} and \eqref{longsuperblockcoeff}.

\section{OPE rules}\label{app:operules}

In this appendix we collect some selection rules for three-point functions involving short $\mathcal{D}_k$ multiplets or long $\mathcal{L}^{\h}_{s,[a,b]}$ multiplets, in all possible combinations. We emphasize certain point that are relevant in the body of the paper or for \cite{Ferrero:2023gnu}.
\vspace{0.2cm}
\paragraph{Three short operators.}
In a $\langle\mathcal{D}_{k_1}\,\mathcal{D}_{k_2}\,\mathcal{D}_{k_3}\rangle$ three-point function, all fermionic coordinates can be set to zero using the fermionic generators of $\mathfrak{osp}(4^*|4)$. For given $k_1$ and $k_2$, the values of $k_3$ for which the three-point function is non-vanishing are fixed by the requirement that the $\mathfrak{sp}(4)$ representation $[0,k_3]$ appears in $[0,k_1]\otimes [0,k_2]$. We have
\begin{align}\label{sp4[0,b][0,d]}
[0,b]\otimes [0,d]=\sum_{n=0}^{\text{min}(b,d)}\sum_{m=0}^n[2\,\text{min}(b,d)-2n,|b-d|+2m]\,,
\end{align}
and we are interested only in the case $n=\text{min}(b,d)$, so one of the two sums collapses and we find the allowed values
\begin{align}\label{D1D2D3}
k_3=|k_1-k_2|\,,\,|k_1-k_2|+2\,,\,\ldots\,,\,k_1+k_2\,,
\end{align}
from which one should exclude $k_3=0$ (if present) as the $\mathcal{D}_k$ multiplets only exist for $k>0$. $k_3=0$ can be interpreted as a contribution of the identity operator.
\vspace{0.2cm}
\paragraph{Two short and one long operator.} In this case one has exactly as many fermionic coordinates in the supermultiplets as the number of fermionic generators in the superalgebra, so that there is a superconformal transformation that sets all fermionic coordinates to zero. For a $\langle \mathcal{D}_{k_1}\,\mathcal{D}_{k_2}\,\mathcal{L}^{\h}_{s,[a,b]}\rangle$ three-point function, the relevant question is then whether $\{s,[a,b]\}$ appears in the tensor product $\{0,[0,k_1]\}\otimes \{0,[0,k_2]\}$. Clearly, $\mathfrak{su}(2)$ selection rules imply that necessarily $s=0$. Moreover, in principle one should be able to read off the allowed values of $a$ and $b$ directly from \eqref{sp4[0,b][0,d]}. However, there is a subtlety related to harmonic singularities in superspace: the requirement that these are absent effectively lowers the upper bound of the second sum in \eqref{sp4[0,b][0,d]} from $n$ to $n-1$. In other words, while \eqref{sp4[0,b][0,d]} would allow values of $a+b$ in the same range as for $k_3$ in \eqref{D1D2D3}, the maximum value for $a+b$ is actually restricted to be $k_1+k_2-2$, see the discussion around \eqref{SSLrestriction}. Summarizing, 
\begin{align}
\mathcal{D}_{k_1}\times \mathcal{D}_{k_2}=\mathcal{I}\,\delta_{k_1,k_2}+\sum_{k_3=|k_1-k_2|}^{k_1+k_2}\mathcal{D}_{k_3}+\sum_{\h} \sum_{n=0}^{\text{min}(k_1,k_2)}\sum_{m=0}^{n-1}\mathcal{L}^{\h}_{0,[2\,\text{min}(k_1,k_2)-2n,|k_1-k_2|+2m]}\,.
\end{align}
This is in agreement with the results of \cite{Liendo:2018ukf}.
\vspace{0.2cm}
\paragraph{One short and two long operators.} This is the first case where the number of fermionic coordinates in the relevant supermultiplets exceeds the number of fermionic generators in the superalgebra. A convenient strategy is the following: imagine to set to zero all fermionic coordinates of the two long supermultiplets. In this frame, the three-point function $\langle \mathcal{D}_k\,\mathcal{L}^{\h_1}_{s_1,[a_1,b_1]}\,\mathcal{L}^{\h_2}_{s_2,[a_2,b_2]}\rangle$ (where now all operators are thought of as superfields) is given by a sum of three-point functions where the three operators are the two superconformal primaries of the long multiplets and any component of the short one, where the coefficients are fermionic coordinates of the $\mathcal{D}_k$ superfield. Thus, given certain Dynkin labels $\{s_1,[a_1,b_1]\}$ and $\{s_2,[a_2,b_2]\}$ one should check if any of the representations listed in \eqref{B1[0,b]} (or, equivalently, \eqref{DkStructure}) appears in their tensor product. Let us consider the special case $\langle \mathcal{D}_k\,\mathcal{L}^{\h_1}_{0,[0,0]}\,\mathcal{L}^{\h_2}_{s_2,[a_2,b_2]}\rangle$: in this case since one of the two long operators is a singlet of $\mathfrak{su}(2)\oplus \mathfrak{sp}(4)$ the allowed $\{s_2,[a_2,b_2]\}$ are those that match the Dynkin labels of states in $\mathcal{D}_k$ multiplets. In particular, note that for $k=2$ we have non-trivial three-point functions with $\{s_2,[a_2,b_2]\}=\{0,[0,0]\}$, due to the level-four state in \eqref{DkStructure}. A useful alternative point of view on this can be obtained by setting the fermionic coordinates of $\mathcal{D}_2$ and $\mathcal{L}^{\h_1}_{0,[0,0]}$ to zero and exploring the components of $\mathcal{L}^{\h_2}_{s_2,[a_2,b_2]}$. From this perspective, the three-point function with $\{s_2,[a_2,b_2]\}=\{0,[0,0]\}$ is non-vanishing due to the presence of a (unique!) $[0,2]_0$ state at level four in \eqref{L00Structure}. Two results that are relevant in the body of the paper and in \cite{Ferrero:2023gnu} are
\begin{align}
\mathcal{D}_1\times \mathcal{L}^{\h_1}_{0,[0,0]}=\mathcal{D}_1\oplus \sum_{\h_2}\left(\mathcal{L}^{\h_2}_{0,[0,1]}\oplus\mathcal{L}^{\h_2}_{1,[1,0]}\oplus\mathcal{L}^{\h_2}_{2,[0,0]}\right)\,,
\end{align}
and
\begin{align}
\mathcal{D}_2\times \mathcal{L}^{\h_1}_{0,[0,0]}=\mathcal{D}_2\oplus \sum_{\h_2}\left(\mathcal{L}^{\h_2}_{0,[0,2]}\oplus\mathcal{L}^{\h_2}_{1,[1,1]}\oplus\mathcal{L}^{\h_2}_{0,[2,0]}\oplus\mathcal{L}^{\h_2}_{2,[0,1]}\oplus\mathcal{L}^{\h_2}_{1,[1,0]}\oplus \mathcal{L}^{\h_2}_{0,[0,0]}\right)\,.
\end{align}
Also notice that $\mathcal{L}^{\h_1}_{0,[0,0]}\times \mathcal{L}^{\h_2}_{0,[0,0]}\supset \mathcal{D}_2$, due to the presence of a singlet representation at level for in the $\mathcal{D}_2$ multiplet.
\vspace{0.2cm}
\paragraph{Three long operators.} As in the previous case the easiest way to carry out the analysis is to switch off the fermionic coordinates for the superfields corresponding to two of the long multiplets. It is then immediate to realize that a three-point function of the type $\langle \mathcal{L}^{\h_1}_{s_1,[a_1,b_1]}\,\mathcal{L}^{\h_2}_{s_2,[a_2,b_2]}\,\mathcal{L}^{\h_3}_{s_3,[a_3,b_3]} \rangle$ is non-vanishing if (at least) one of the states in $\mathcal{L}^{\h_1}_{s_1,[a_1,b_1]}$ (listed in \eqref{L[a,b]_s}) can appear in the tensor product of the representations of the two superconformal primaries in the other multiplets. The simplest case arises when the superconformal primaries of two of the long multiplets are scalars of $\mathfrak{su}(2)\oplus \mathfrak{sp}(4)$, say the last two. In this case the three-point function is non-zero for the values of $\{s_1,[a_1,b_1]\}$ such that the multiplet $\mathcal{L}^{\h_1}_{s_1,[a_1,b_1]}$ contains at least one singlet. This happens for the values $\{s_1,[a_1,b_1]\}$ collected in Table \ref{tab:L00xL00}, where we have also included how many singlet states are contained in each supermultiplet that can contribute.
\begin{table}[!ht]
\centering
 \begin{tabular}{| c || c|c|c|c|c|} 
 \hline
$s$ & 0 & 1 & 2 & 3 & 4\\
\hline
\hline
$[0,0]$ & 3 & $-$ & 2 & $-$ & 1\\
\hline
$[0,1]$ & $1$ & $-$ & $3$ & $-$ & $-$\\
\hline
$[1,0]$ & $-$ & $4$ & $-$ & $2$ & $-$\\
\hline
$[0,2]$ & $1$ & $-$ & $-$ & $-$ & $-$\\
\hline
$[1,1]$ & $-$ & $2$ & $-$ & $-$ & $-$\\
\hline
$[2,0]$ & $2$ & $-$ & $1$ & $-$ & $-$\\
\hline
\end{tabular}
\caption{Values of $\{s,[a,b]\}$ for which the three-point function $\langle \mathcal{L}^{\h_1}_{s,[a,b]}\,\mathcal{L}^{\h_2}_{0,[0,0]}\,\mathcal{L}^{\h_3}_{0,[0,0]} \rangle$ is non-vanishing. We have used the symbol ``$-$'' for cases where the correlator is forbidden by selection rules, while in allowed cases we have written the number of independent three-point structures, due to the presence of multiple singlet representations in the multiplet $\mathcal{L}^{\h_1}_{s,[a,b]}$.}
\label{tab:L00xL00}
\end{table}
Let us stress in particular that $\langle \mathcal{L}^{\h_1}_{0,[0,0]}\,\mathcal{L}^{\h_2}_{0,[0,0]}\,\mathcal{L}^{\h_3}_{0,[0,0]} \rangle$ is non-vanishing due to the presence of three singlets in the $\mathcal{L}^{\h_1}_{0,[0,0]}$: one at level zero, one at level four and one at level eight, as one can easily argue from \eqref{L00Structure}.

\providecommand{\href}[2]{#2}\begingroup\raggedright\endgroup


\begin{thebibliography}{100}

\bibitem{Ferrero:2023gnu}
P.~Ferrero and C.~Meneghelli, {\it {Unmixing the Wilson line defect CFT. Part
  II: analytic bootstrap}},  \href{http://arxiv.org/abs/2312.12551}{{\tt
  arXiv:2312.12551}}.

\bibitem{Ferrero:2021bsb}
P.~Ferrero and C.~Meneghelli, {\it {Bootstrapping the half-BPS line defect CFT
  in $\mathcal{N}=4$ SYM at strong coupling}},
  \href{http://arxiv.org/abs/2103.10440}{{\tt arXiv:2103.10440}}.

\bibitem{Rattazzi:2008pe}
R.~Rattazzi, V.~S. Rychkov, E.~Tonni, and A.~Vichi, {\it {Bounding scalar
  operator dimensions in 4D CFT}},  {\em JHEP} {\bf 0812} (2008) 031,
  [\href{http://arxiv.org/abs/0807.0004}{{\tt arXiv:0807.0004}}].

\bibitem{Cardy:1984bb}
J.~L. Cardy, {\it {Conformal Invariance and Surface Critical Behavior}},  {\em
  Nucl. Phys. B} {\bf 240} (1984) 514--532.

\bibitem{Cardy:1989ir}
J.~L. Cardy, {\it {Boundary Conditions, Fusion Rules and the Verlinde
  Formula}},  {\em Nucl. Phys. B} {\bf 324} (1989) 581--596.

\bibitem{Liendo:2012hy}
P.~Liendo, L.~Rastelli, and B.~C. van Rees, {\it {The Bootstrap Program for
  Boundary CFT}},  {\em JHEP} {\bf 1307} (2013) 113,
  [\href{http://arxiv.org/abs/1210.4258}{{\tt arXiv:1210.4258}}].

\bibitem{Gaiotto:2013nva}
D.~Gaiotto, D.~Mazac, and M.~F. Paulos, {\it {Bootstrapping the 3d Ising twist
  defect}},  {\em JHEP} {\bf 03} (2014) 100,
  [\href{http://arxiv.org/abs/1310.5078}{{\tt arXiv:1310.5078}}].

\bibitem{Gliozzi:2015qsa}
F.~Gliozzi, P.~Liendo, M.~Meineri, and A.~Rago, {\it {Boundary and Interface
  CFTs from the Conformal Bootstrap}},  {\em JHEP} {\bf 05} (2015) 036,
  [\href{http://arxiv.org/abs/1502.07217}{{\tt arXiv:1502.07217}}].

\bibitem{Gliozzi:2016cmg}
F.~Gliozzi, {\it {Truncatable bootstrap equations in algebraic form and
  critical surface exponents}},  {\em JHEP} {\bf 10} (2016) 037,
  [\href{http://arxiv.org/abs/1605.04175}{{\tt arXiv:1605.04175}}].

\bibitem{Billo:2016cpy}
M.~Billò, V.~Gonçalves, E.~Lauria, and M.~Meineri, {\it {Defects in conformal
  field theory}},  {\em JHEP} {\bf 04} (2016) 091,
  [\href{http://arxiv.org/abs/1601.02883}{{\tt arXiv:1601.02883}}].

\bibitem{Agmon:2020pde}
N.~B. Agmon and Y.~Wang, {\it {Classifying Superconformal Defects in Diverse
  Dimensions Part I: Superconformal Lines}},
  \href{http://arxiv.org/abs/2009.06650}{{\tt arXiv:2009.06650}}.

\bibitem{Barrat:2020vch}
J.~Barrat, P.~Liendo, and J.~Plefka, {\it {Two-point correlator of chiral
  primary operators with a Wilson line defect in $ \mathcal{N} $ = 4 SYM}},
  {\em JHEP} {\bf 05} (2021) 195, [\href{http://arxiv.org/abs/2011.04678}{{\tt
  arXiv:2011.04678}}].

\bibitem{Barrat:2021yvp}
J.~Barrat, A.~Gimenez-Grau, and P.~Liendo, {\it {Bootstrapping holographic
  defect correlators in $ \mathcal{N} $ = 4 super Yang-Mills}},  {\em JHEP}
  {\bf 04} (2022) 093, [\href{http://arxiv.org/abs/2108.13432}{{\tt
  arXiv:2108.13432}}].

\bibitem{Gimenez-Grau:2023fcy}
A.~Gimenez-Grau, {\it {The Witten Diagram Bootstrap for Holographic Defects}},
  \href{http://arxiv.org/abs/2306.11896}{{\tt arXiv:2306.11896}}.

\bibitem{Meneghelli:2022gps}
C.~Meneghelli and M.~Tr\'epanier, {\it {Bootstrapping string dynamics in the 6d
  \ensuremath{\mathscr{N}} = (2, 0) theories}},  {\em JHEP} {\bf 07} (2023)
  165, [\href{http://arxiv.org/abs/2212.05020}{{\tt arXiv:2212.05020}}].

\bibitem{Chen:2023yvw}
J.~Chen, A.~Gimenez-Grau, and X.~Zhou, {\it {Defect two-point functions in 6d
  (2,0) theories}},  \href{http://arxiv.org/abs/2310.19230}{{\tt
  arXiv:2310.19230}}.

\bibitem{Maldacena:1998im}
J.~M. Maldacena, {\it {Wilson loops in large N field theories}},  {\em Phys.
  Rev. Lett.} {\bf 80} (1998) 4859--4862,
  [\href{http://arxiv.org/abs/hep-th/9803002}{{\tt hep-th/9803002}}].

\bibitem{Drukker:2006xg}
N.~Drukker and S.~Kawamoto, {\it {Small deformations of supersymmetric Wilson
  loops and open spin-chains}},  {\em JHEP} {\bf 07} (2006) 024,
  [\href{http://arxiv.org/abs/hep-th/0604124}{{\tt hep-th/0604124}}].

\bibitem{Drukker:2005kx}
N.~Drukker and B.~Fiol, {\it {All-genus calculation of Wilson loops using
  D-branes}},  {\em JHEP} {\bf 02} (2005) 010,
  [\href{http://arxiv.org/abs/hep-th/0501109}{{\tt hep-th/0501109}}].

\bibitem{Gomis:2006sb}
J.~Gomis and F.~Passerini, {\it {Holographic Wilson Loops}},  {\em JHEP} {\bf
  08} (2006) 074, [\href{http://arxiv.org/abs/hep-th/0604007}{{\tt
  hep-th/0604007}}].

\bibitem{Gomis:2006im}
J.~Gomis and F.~Passerini, {\it {Wilson Loops as D3-Branes}},  {\em JHEP} {\bf
  01} (2007) 097, [\href{http://arxiv.org/abs/hep-th/0612022}{{\tt
  hep-th/0612022}}].

\bibitem{Giombi:2017cqn}
S.~Giombi, R.~Roiban, and A.~A. Tseytlin, {\it {Half-BPS Wilson loop and
  AdS$_2$/CFT$_1$}},  {\em Nucl. Phys.} {\bf B922} (2017) 499--527,
  [\href{http://arxiv.org/abs/1706.00756}{{\tt arXiv:1706.00756}}].

\bibitem{Giombi:2020kvo}
S.~Giombi and B.~Offertaler, {\it {Wilson loops in $ \mathcal{N} $ = 4 SO(N)
  SYM and D-branes in AdS$_{5}$ \texttimes{}
  \ensuremath{\mathbb{R}}\ensuremath{\mathbb{P}}$^{5}$}},  {\em JHEP} {\bf 10}
  (2021) 016, [\href{http://arxiv.org/abs/2006.10852}{{\tt arXiv:2006.10852}}].

\bibitem{Kiryu:2018phb}
N.~Kiryu and S.~Komatsu, {\it {Correlation Functions on the Half-BPS Wilson
  Loop: Perturbation and Hexagonalization}},  {\em JHEP} {\bf 02} (2019) 090,
  [\href{http://arxiv.org/abs/1812.04593}{{\tt arXiv:1812.04593}}].

\bibitem{Grabner:2020nis}
D.~Grabner, N.~Gromov, and J.~Julius, {\it {Excited States of One-Dimensional
  Defect CFTs from the Quantum Spectral Curve}},  {\em JHEP} {\bf 07} (2020)
  042, [\href{http://arxiv.org/abs/2001.11039}{{\tt arXiv:2001.11039}}].

\bibitem{Cavaglia:2021bnz}
A.~Cavagli\`a, N.~Gromov, J.~Julius, and M.~Preti, {\it {Integrability and
  conformal bootstrap: One dimensional defect conformal field theory}},  {\em
  Phys. Rev. D} {\bf 105} (2022), no.~2 L021902,
  [\href{http://arxiv.org/abs/2107.08510}{{\tt arXiv:2107.08510}}].

\bibitem{Cavaglia:2022qpg}
A.~Cavagli\`a, N.~Gromov, J.~Julius, and M.~Preti, {\it {Bootstrability in
  defect CFT: integrated correlators and sharper bounds}},  {\em JHEP} {\bf 05}
  (2022) 164, [\href{http://arxiv.org/abs/2203.09556}{{\tt arXiv:2203.09556}}].

\bibitem{Giombi:2018qox}
S.~Giombi and S.~Komatsu, {\it {Exact Correlators on the Wilson Loop in
  $\mathcal{N}=4$ SYM: Localization, Defect CFT, and Integrability}},  {\em
  JHEP} {\bf 05} (2018) 109, [\href{http://arxiv.org/abs/1802.05201}{{\tt
  arXiv:1802.05201}}]. [Erratum: JHEP 11, 123 (2018)].

\bibitem{Giombi:2018hsx}
S.~Giombi and S.~Komatsu, {\it {More Exact Results in the Wilson Loop Defect
  CFT: Bulk-Defect OPE, Nonplanar Corrections and Quantum Spectral Curve}},
  {\em J. Phys. A} {\bf 52} (2019), no.~12 125401,
  [\href{http://arxiv.org/abs/1811.02369}{{\tt arXiv:1811.02369}}].

\bibitem{Giombi:2020amn}
S.~Giombi, J.~Jiang, and S.~Komatsu, {\it {Giant Wilson loops and
  AdS$_{2}$/dCFT$_{1}$}},  {\em JHEP} {\bf 11} (2020) 064,
  [\href{http://arxiv.org/abs/2005.08890}{{\tt arXiv:2005.08890}}].

\bibitem{Giombi:2021zfb}
S.~Giombi, S.~Komatsu, and B.~Offertaler, {\it {Large charges on the Wilson
  loop in $ \mathcal{N} $ = 4 SYM: matrix model and classical string}},  {\em
  JHEP} {\bf 03} (2022) 020, [\href{http://arxiv.org/abs/2110.13126}{{\tt
  arXiv:2110.13126}}].

\bibitem{Giombi:2022anm}
S.~Giombi, S.~Komatsu, and B.~Offertaler, {\it {Large Charges on the Wilson
  Loop in $\mathcal{N}=4$ SYM: II. Quantum Fluctuations, OPE, and Spectral
  Curve}},  \href{http://arxiv.org/abs/2202.07627}{{\tt arXiv:2202.07627}}.

\bibitem{Liendo:2016ymz}
P.~Liendo and C.~Meneghelli, {\it {Bootstrap equations for $ \mathcal{N} $ = 4
  SYM with defects}},  {\em JHEP} {\bf 01} (2017) 122,
  [\href{http://arxiv.org/abs/1608.05126}{{\tt arXiv:1608.05126}}].

\bibitem{Liendo:2018ukf}
P.~Liendo, C.~Meneghelli, and V.~Mitev, {\it {Bootstrapping the half-BPS line
  defect}},  {\em JHEP} {\bf 10} (2018) 077,
  [\href{http://arxiv.org/abs/1806.01862}{{\tt arXiv:1806.01862}}].

\bibitem{Polchinski:2011im}
J.~Polchinski and J.~Sully, {\it {Wilson Loop Renormalization Group Flows}},
  {\em JHEP} {\bf 10} (2011) 059, [\href{http://arxiv.org/abs/1104.5077}{{\tt
  arXiv:1104.5077}}].

\bibitem{Beccaria:2017rbe}
M.~Beccaria, S.~Giombi, and A.~Tseytlin, {\it {Non-supersymmetric Wilson loop
  in $ \mathcal{N} $ = 4 SYM and defect 1d CFT}},  {\em JHEP} {\bf 03} (2018)
  131, [\href{http://arxiv.org/abs/1712.06874}{{\tt arXiv:1712.06874}}].

\bibitem{Beccaria:2019dws}
M.~Beccaria, S.~Giombi, and A.~A. Tseytlin, {\it {Correlators on
  non-supersymmetric Wilson line in $ \mathcal{N}=4 $ SYM and
  AdS$_{2}$/CFT$_{1}$}},  {\em JHEP} {\bf 05} (2019) 122,
  [\href{http://arxiv.org/abs/1903.04365}{{\tt arXiv:1903.04365}}].

\bibitem{Beccaria:2021rmj}
M.~Beccaria, S.~Giombi, and A.~A. Tseytlin, {\it {Higher order RG flow on the
  Wilson line in $ \mathcal{N} $ = 4 SYM}},  {\em JHEP} {\bf 01} (2022) 056,
  [\href{http://arxiv.org/abs/2110.04212}{{\tt arXiv:2110.04212}}].

\bibitem{Beccaria:2022bcr}
M.~Beccaria, S.~Giombi, and A.~A. Tseytlin, {\it {Wilson loop in general
  representation and RG flow in 1D defect QFT}},  {\em J. Phys. A} {\bf 55}
  (2022), no.~25 255401, [\href{http://arxiv.org/abs/2202.00028}{{\tt
  arXiv:2202.00028}}].

\bibitem{Cuomo:2021rkm}
G.~Cuomo, Z.~Komargodski, and A.~Raviv-Moshe, {\it {Renormalization Group Flows
  on Line Defects}},  {\em Phys. Rev. Lett.} {\bf 128} (2022), no.~2 021603,
  [\href{http://arxiv.org/abs/2108.01117}{{\tt arXiv:2108.01117}}].

\bibitem{Giombi:2022pas}
S.~Giombi, S.~Komatsu, and B.~Offertaler, {\it {Chaos and the reparametrization
  mode on the AdS$_2$ string}},  \href{http://arxiv.org/abs/2212.14842}{{\tt
  arXiv:2212.14842}}.

\bibitem{Cornagliotto:2017dup}
M.~Cornagliotto, M.~Lemos, and V.~Schomerus, {\it {Long Multiplet Bootstrap}},
  \href{http://arxiv.org/abs/1702.05101}{{\tt arXiv:1702.05101}}.

\bibitem{Kos:2018glc}
F.~Kos and J.~Oh, {\it {2d small N=4 Long-multiplet superconformal block}},
  {\em JHEP} {\bf 02} (2019) 001, [\href{http://arxiv.org/abs/1810.10029}{{\tt
  arXiv:1810.10029}}].

\bibitem{Buric:2020buk}
I.~Buri\'c, V.~Schomerus, and E.~Sobko, {\it {The superconformal $X$-ing
  equation}},  {\em JHEP} {\bf 10} (2020) 147,
  [\href{http://arxiv.org/abs/2005.13547}{{\tt arXiv:2005.13547}}].

\bibitem{Buric:2020qzp}
I.~Buri\'c, V.~Schomerus, and E.~Sobko, {\it {Crossing symmetry for long
  multiplets in 4D $ \mathcal{N} $ = 1 SCFTs}},  {\em JHEP} {\bf 04} (2021)
  130, [\href{http://arxiv.org/abs/2011.14116}{{\tt arXiv:2011.14116}}].

\bibitem{Gunaydin:1990ag}
M.~Gunaydin and R.~J. Scalise, {\it {Unitary Lowest Weight Representations of
  the Noncompact Supergroup Osp(2m*/2n)}},  {\em J. Math. Phys.} {\bf 32}
  (1991) 599--606.

\bibitem{Dorey:2018klg}
N.~Dorey and A.~Singleton, {\it {An Index for Superconformal Quantum
  Mechanics}},  \href{http://arxiv.org/abs/1812.11816}{{\tt arXiv:1812.11816}}.

\bibitem{phdthesis}
A.~Singleton, {\it The geometry and representation theory of superconformal
  quantum mechanics},  {\em University of Cambridge PhD thesis} (2016).

\bibitem{Gimenez-Grau:2019hez}
A.~Gimenez-Grau and P.~Liendo, {\it {Bootstrapping line defects in
  $\mathcal{N}=2$ theories}},  {\em JHEP} {\bf 03} (2020) 121,
  [\href{http://arxiv.org/abs/1907.04345}{{\tt arXiv:1907.04345}}].

\bibitem{Liendo:2015cgi}
P.~Liendo, C.~Meneghelli, and V.~Mitev, {\it {On correlation functions of BPS
  operators in $3d$ $\mathcal{N}=6$ superconformal theories}},  {\em Commun.
  Math. Phys.} (2016) 1--33, [\href{http://arxiv.org/abs/1512.06072}{{\tt
  arXiv:1512.06072}}].

\bibitem{Zhou:2017zaw}
X.~Zhou, {\it {On Superconformal Four-Point Mellin Amplitudes in Dimension
  $d>2$}},  {\em JHEP} {\bf 08} (2018) 187,
  [\href{http://arxiv.org/abs/1712.02800}{{\tt arXiv:1712.02800}}].

\bibitem{Alday:2021odx}
L.~F. Alday, C.~Behan, P.~Ferrero, and X.~Zhou, {\it {Gluon Scattering in AdS
  from CFT}},  {\em JHEP} {\bf 06} (2021) 020,
  [\href{http://arxiv.org/abs/2103.15830}{{\tt arXiv:2103.15830}}].

\bibitem{Dolan:2001tt}
F.~Dolan and H.~Osborn, {\it {Superconformal symmetry, correlation functions
  and the operator product expansion}},  {\em Nucl.Phys.} {\bf B629} (2002)
  3--73, [\href{http://arxiv.org/abs/hep-th/0112251}{{\tt hep-th/0112251}}].

\bibitem{Eden:2001ec}
B.~Eden and E.~Sokatchev, {\it {On the OPE of 1/2 BPS short operators in N=4
  SCFT(4)}},  {\em Nucl.Phys.} {\bf B618} (2001) 259--276,
  [\href{http://arxiv.org/abs/hep-th/0106249}{{\tt hep-th/0106249}}].

\bibitem{Arutyunov:2001qw}
G.~Arutyunov, B.~Eden, and E.~Sokatchev, {\it {On nonrenormalization and OPE in
  superconformal field theories}},  {\em Nucl.Phys.} {\bf B619} (2001)
  359--372, [\href{http://arxiv.org/abs/hep-th/0105254}{{\tt hep-th/0105254}}].

\bibitem{Eden:2001wg}
B.~Eden, S.~Ferrara, and E.~Sokatchev, {\it {(2,0) superconformal OPEs in D =
  6, selection rules and nonrenormalization theorems}},  {\em JHEP} {\bf 0111}
  (2001) 020, [\href{http://arxiv.org/abs/hep-th/0107084}{{\tt
  hep-th/0107084}}].

\bibitem{Ferrara:2001uj}
S.~Ferrara and E.~Sokatchev, {\it {Universal properties of superconformal OPEs
  for 1/2 BPS operators in 3 <= D <= 6}},  {\em New J.Phys.} {\bf 4} (2002) 2,
  [\href{http://arxiv.org/abs/hep-th/0110174}{{\tt hep-th/0110174}}].

\bibitem{Fitzpatrick:2011dm}
A.~L. Fitzpatrick and J.~Kaplan, {\it {Unitarity and the Holographic
  S-Matrix}},  {\em JHEP} {\bf 10} (2012) 032,
  [\href{http://arxiv.org/abs/1112.4845}{{\tt arXiv:1112.4845}}].

\bibitem{Braun:2003rp}
V.~M. Braun, G.~P. Korchemsky, and D.~M\"uller, {\it {The Uses of conformal
  symmetry in QCD}},  {\em Prog. Part. Nucl. Phys.} {\bf 51} (2003) 311--398,
  [\href{http://arxiv.org/abs/hep-ph/0306057}{{\tt hep-ph/0306057}}].

\bibitem{Henn:2005mw}
J.~Henn, C.~Jarczak, and E.~Sokatchev, {\it {On twist-two operators in N=4
  SYM}},  {\em Nucl. Phys. B} {\bf 730} (2005) 191--209,
  [\href{http://arxiv.org/abs/hep-th/0507241}{{\tt hep-th/0507241}}].

\bibitem{LiE1992}
M.~A.~A. van Leeuwen, A.~M. Cohen, and B.~Lisser, {\em {LiE}, A package for Lie
  group computations}.
\newblock Computer Algebra Nederland, Amsterdam, 1992.

\bibitem{Alday:2017zzv}
L.~F. Alday, J.~Henriksson, and M.~van Loon, {\it {Taming the
  $\epsilon$-expansion with large spin perturbation theory}},  {\em JHEP} {\bf
  07} (2018) 131, [\href{http://arxiv.org/abs/1712.02314}{{\tt
  arXiv:1712.02314}}].

\bibitem{Henriksson:2022rnm}
J.~Henriksson, {\it {The critical O(N) CFT: Methods and conformal data}},
  \href{http://arxiv.org/abs/2201.09520}{{\tt arXiv:2201.09520}}.

\bibitem{Aprile:2017bgs}
F.~Aprile, J.~M. Drummond, P.~Heslop, and H.~Paul, {\it {Quantum Gravity from
  Conformal Field Theory}},  {\em JHEP} {\bf 01} (2018) 035,
  [\href{http://arxiv.org/abs/1706.02822}{{\tt arXiv:1706.02822}}].

\bibitem{Alday:2017xua}
L.~F. Alday and A.~Bissi, {\it {Loop Corrections to Supergravity on $AdS_5
  \times S^5$}},  {\em Phys. Rev. Lett.} {\bf 119} (2017), no.~17 171601,
  [\href{http://arxiv.org/abs/1706.02388}{{\tt arXiv:1706.02388}}].

\bibitem{Alday:2020tgi}
L.~F. Alday, S.~M. Chester, and H.~Raj, {\it {6d (2,0) and M-theory at
  1-loop}},  {\em JHEP} {\bf 01} (2021) 133,
  [\href{http://arxiv.org/abs/2005.07175}{{\tt arXiv:2005.07175}}].

\bibitem{Alday:2021ajh}
L.~F. Alday, A.~Bissi, and X.~Zhou, {\it {One-loop gluon amplitudes in AdS}},
  {\em JHEP} {\bf 02} (2022) 105, [\href{http://arxiv.org/abs/2110.09861}{{\tt
  arXiv:2110.09861}}].

\bibitem{Behan:2022uqr}
C.~Behan, {\it {Holographic S-fold theories at one loop}},
  \href{http://arxiv.org/abs/2202.05261}{{\tt arXiv:2202.05261}}.

\bibitem{Alday:2022rly}
L.~F. Alday, S.~M. Chester, and H.~Raj, {\it {M-theory on AdS$_{4}$\texttimes{}
  S$^{7}$ at 1-loop and beyond}},  {\em JHEP} {\bf 11} (2022) 091,
  [\href{http://arxiv.org/abs/2207.11138}{{\tt arXiv:2207.11138}}].

\bibitem{Aprile:2017xsp}
F.~Aprile, J.~M. Drummond, P.~Heslop, and H.~Paul, {\it {Unmixing
  Supergravity}},  {\em JHEP} {\bf 02} (2018) 133,
  [\href{http://arxiv.org/abs/1706.08456}{{\tt arXiv:1706.08456}}].

\bibitem{Aprile:2018efk}
F.~Aprile, J.~Drummond, P.~Heslop, and H.~Paul, {\it {Double-trace spectrum of
  $N=4$ supersymmetric Yang-Mills theory at strong coupling}},  {\em Phys. Rev.
  D} {\bf 98} (2018), no.~12 126008,
  [\href{http://arxiv.org/abs/1802.06889}{{\tt arXiv:1802.06889}}].

\bibitem{Aprile:2019rep}
F.~Aprile, J.~Drummond, P.~Heslop, and H.~Paul, {\it {One-loop amplitudes in
  AdS$_{5}$\texttimes{}S$^{5}$ supergravity from $ \mathcal{N} $ = 4 SYM at
  strong coupling}},  {\em JHEP} {\bf 03} (2020) 190,
  [\href{http://arxiv.org/abs/1912.01047}{{\tt arXiv:1912.01047}}].

\bibitem{Beisert:2003tq}
N.~Beisert, C.~Kristjansen, and M.~Staudacher, {\it {The Dilatation operator of
  conformal N=4 superYang-Mills theory}},  {\em Nucl. Phys. B} {\bf 664} (2003)
  131--184, [\href{http://arxiv.org/abs/hep-th/0303060}{{\tt hep-th/0303060}}].

\bibitem{Beisert:2003yb}
N.~Beisert and M.~Staudacher, {\it {The N=4 SYM integrable super spin chain}},
  {\em Nucl. Phys. B} {\bf 670} (2003) 439--463,
  [\href{http://arxiv.org/abs/hep-th/0307042}{{\tt hep-th/0307042}}].

\bibitem{Beisert:2004ry}
N.~Beisert, {\it {The Dilatation operator of N=4 super Yang-Mills theory and
  integrability}},  {\em Phys. Rept.} {\bf 405} (2004) 1--202,
  [\href{http://arxiv.org/abs/hep-th/0407277}{{\tt hep-th/0407277}}].

\bibitem{Barrat:2021tpn}
J.~Barrat, P.~Liendo, G.~Peveri, and J.~Plefka, {\it {Multipoint correlators on
  the supersymmetric Wilson line defect CFT}},
  \href{http://arxiv.org/abs/2112.10780}{{\tt arXiv:2112.10780}}.

\bibitem{Barrat:2022eim}
J.~Barrat, P.~Liendo, and G.~Peveri, {\it {Multipoint correlators on the
  supersymmetric Wilson line defect CFT II: Unprotected operators}},
  \href{http://arxiv.org/abs/2210.14916}{{\tt arXiv:2210.14916}}.

\bibitem{Bliard:2023zpe}
G.~J.~S. Bliard, {\em {Perturbative and non-perturbative analysis of defect
  correlators in AdS/CFT}}.
\newblock PhD thesis, Humboldt U., Berlin, 2023.
\newblock \href{http://arxiv.org/abs/2310.18137}{{\tt arXiv:2310.18137}}.

\bibitem{Rosenhaus:2018zqn}
V.~Rosenhaus, {\it {Multipoint Conformal Blocks in the Comb Channel}},  {\em
  JHEP} {\bf 02} (2019) 142, [\href{http://arxiv.org/abs/1810.03244}{{\tt
  arXiv:1810.03244}}].

\bibitem{Parikh:2019dvm}
S.~Parikh, {\it {A multipoint conformal block chain in $d$ dimensions}},  {\em
  JHEP} {\bf 05} (2020) 120, [\href{http://arxiv.org/abs/1911.09190}{{\tt
  arXiv:1911.09190}}].

\bibitem{Fortin:2019zkm}
J.-F. Fortin, W.~Ma, and W.~Skiba, {\it {Higher-Point Conformal Blocks in the
  Comb Channel}},  {\em JHEP} {\bf 07} (2020) 213,
  [\href{http://arxiv.org/abs/1911.11046}{{\tt arXiv:1911.11046}}].

\bibitem{Fortin:2020yjz}
J.-F. Fortin, W.-J. Ma, and W.~Skiba, {\it {Six-point conformal blocks in the
  snowflake channel}},  {\em JHEP} {\bf 11} (2020) 147,
  [\href{http://arxiv.org/abs/2004.02824}{{\tt arXiv:2004.02824}}].

\bibitem{Fortin:2020bfq}
J.-F. Fortin, W.-J. Ma, and W.~Skiba, {\it {Seven-point conformal blocks in the
  extended snowflake channel and beyond}},  {\em Phys. Rev. D} {\bf 102}
  (2020), no.~12 125007, [\href{http://arxiv.org/abs/2006.13964}{{\tt
  arXiv:2006.13964}}].

\bibitem{Fortin:2020zxw}
J.-F. Fortin, W.-J. Ma, and W.~Skiba, {\it {All Global One- and Two-Dimensional
  Higher-Point Conformal Blocks}},  \href{http://arxiv.org/abs/2009.07674}{{\tt
  arXiv:2009.07674}}.

\bibitem{Hoback:2020pgj}
S.~Hoback and S.~Parikh, {\it {Towards Feynman rules for conformal blocks}},
  {\em JHEP} {\bf 01} (2021) 005, [\href{http://arxiv.org/abs/2006.14736}{{\tt
  arXiv:2006.14736}}].

\bibitem{Hoback:2020syd}
S.~Hoback and S.~Parikh, {\it {Dimensional reduction of higher-point conformal
  blocks}},  {\em JHEP} {\bf 03} (2021) 187,
  [\href{http://arxiv.org/abs/2009.12904}{{\tt arXiv:2009.12904}}].

\bibitem{Poland:2021xjs}
D.~Poland and V.~Prilepina, {\it {Recursion relations for 5-point conformal
  blocks}},  {\em JHEP} {\bf 10} (2021) 160,
  [\href{http://arxiv.org/abs/2103.12092}{{\tt arXiv:2103.12092}}].

\bibitem{Buric:2021ywo}
I.~Buric, S.~Lacroix, J.~A. Mann, L.~Quintavalle, and V.~Schomerus, {\it
  {Gaudin models and multipoint conformal blocks: general theory}},  {\em JHEP}
  {\bf 10} (2021) 139, [\href{http://arxiv.org/abs/2105.00021}{{\tt
  arXiv:2105.00021}}].

\bibitem{Buric:2021ttm}
I.~Buric, S.~Lacroix, J.~A. Mann, L.~Quintavalle, and V.~Schomerus, {\it
  {Gaudin models and multipoint conformal blocks. Part II. Comb channel
  vertices in 3D and 4D}},  {\em JHEP} {\bf 11} (2021) 182,
  [\href{http://arxiv.org/abs/2108.00023}{{\tt arXiv:2108.00023}}].

\bibitem{Buric:2021kgy}
I.~Buric, S.~Lacroix, J.~A. Mann, L.~Quintavalle, and V.~Schomerus, {\it
  {Gaudin models and multipoint conformal blocks III: comb channel coordinates
  and OPE factorisation}},  {\em JHEP} {\bf 06} (2022) 144,
  [\href{http://arxiv.org/abs/2112.10827}{{\tt arXiv:2112.10827}}].

\bibitem{Goncalves:2019znr}
V.~Gon\c{c}alves, R.~Pereira, and X.~Zhou, {\it {$20'$ Five-Point Function from
  $AdS_5\times S^5$ Supergravity}},  {\em JHEP} {\bf 10} (2019) 247,
  [\href{http://arxiv.org/abs/1906.05305}{{\tt arXiv:1906.05305}}].

\bibitem{Alday:2022lkk}
L.~F. Alday, V.~Gon\c{c}alves, and X.~Zhou, {\it {Supersymmetric Five-Point
  Gluon Amplitudes in AdS Space}},  {\em Phys. Rev. Lett.} {\bf 128} (2022),
  no.~16 161601, [\href{http://arxiv.org/abs/2201.04422}{{\tt
  arXiv:2201.04422}}].

\bibitem{Goncalves:2023oyx}
V.~Gon\c{c}alves, C.~Meneghelli, R.~Pereira, J.~Vilas~Boas, and X.~Zhou, {\it
  {Kaluza-Klein Five-Point Functions from $\textrm{AdS}_5\times S_5$
  Supergravity}},  \href{http://arxiv.org/abs/2302.01896}{{\tt
  arXiv:2302.01896}}.

\bibitem{Alday:2023kfm}
L.~F. Alday, V.~Gon\c{c}alves, M.~Nocchi, and X.~Zhou, {\it {Six-Point AdS
  Gluon Amplitudes from Flat Space and Factorization}},
  \href{http://arxiv.org/abs/2307.06884}{{\tt arXiv:2307.06884}}.

\bibitem{Aharony:2008ug}
O.~Aharony, O.~Bergman, D.~L. Jafferis, and J.~Maldacena, {\it {N=6
  superconformal Chern-Simons-matter theories, M2-branes and their gravity
  duals}},  {\em JHEP} {\bf 10} (2008) 091,
  [\href{http://arxiv.org/abs/0806.1218}{{\tt arXiv:0806.1218}}].

\bibitem{Drukker:2009hy}
N.~Drukker and D.~Trancanelli, {\it {A Supermatrix model for N=6 super
  Chern-Simons-matter theory}},  {\em JHEP} {\bf 02} (2010) 058,
  [\href{http://arxiv.org/abs/0912.3006}{{\tt arXiv:0912.3006}}].

\bibitem{Cardinali:2012ru}
V.~Cardinali, L.~Griguolo, G.~Martelloni, and D.~Seminara, {\it {New
  supersymmetric Wilson loops in ABJ(M) theories}},  {\em Phys. Lett. B} {\bf
  718} (2012) 615--619, [\href{http://arxiv.org/abs/1209.4032}{{\tt
  arXiv:1209.4032}}].

\bibitem{Bianchi:2017ozk}
L.~Bianchi, L.~Griguolo, M.~Preti, and D.~Seminara, {\it {Wilson lines as
  superconformal defects in ABJM theory: a formula for the emitted radiation}},
   {\em JHEP} {\bf 10} (2017) 050, [\href{http://arxiv.org/abs/1706.06590}{{\tt
  arXiv:1706.06590}}].

\bibitem{Drukker:2019bev}
N.~Drukker et~al., {\it {Roadmap on Wilson loops in 3d
  Chern\textendash{}Simons-matter theories}},  {\em J. Phys. A} {\bf 53}
  (2020), no.~17 173001, [\href{http://arxiv.org/abs/1910.00588}{{\tt
  arXiv:1910.00588}}].

\bibitem{Bianchi:2020hsz}
L.~Bianchi, G.~Bliard, V.~Forini, L.~Griguolo, and D.~Seminara, {\it {Analytic
  bootstrap and Witten diagrams for the ABJM Wilson line as defect CFT$_{1}$}},
   {\em JHEP} {\bf 08} (2020) 143, [\href{http://arxiv.org/abs/2004.07849}{{\tt
  arXiv:2004.07849}}].

\bibitem{Alday:2020dtb}
L.~F. Alday and X.~Zhou, {\it {All Holographic Four-Point Functions in All
  Maximally Supersymmetric CFTs}},  {\em Phys. Rev. X} {\bf 11} (2021), no.~1
  011056, [\href{http://arxiv.org/abs/2006.12505}{{\tt arXiv:2006.12505}}].

\bibitem{Bissi:2021hjk}
A.~Bissi, G.~Fardelli, and A.~Manenti, {\it {Rebooting quarter-BPS operators in
  $ \mathcal{N} $ = 4 super Yang-Mills}},  {\em JHEP} {\bf 04} (2022) 016,
  [\href{http://arxiv.org/abs/2111.06857}{{\tt arXiv:2111.06857}}].

\bibitem{Cordova:2016emh}
C.~Cordova, T.~T. Dumitrescu, and K.~Intriligator, {\it {Multiplets of
  Superconformal Symmetry in Diverse Dimensions}},
  \href{http://arxiv.org/abs/1612.00809}{{\tt arXiv:1612.00809}}.

\bibitem{Fuchs:1997jv}
J.~Fuchs and C.~Schweigert, {\em {Symmetries, Lie algebras and representations:
  a graduate course for physicists}}.
\newblock Cambridge Univ. Press, Cambridge, 1997.

\bibitem{Bianchi:2006ti}
M.~Bianchi, F.~A. Dolan, P.~J. Heslop, and H.~Osborn, {\it {N=4 superconformal
  characters and partition functions}},  {\em Nucl. Phys. B} {\bf 767} (2007)
  163--226, [\href{http://arxiv.org/abs/hep-th/0609179}{{\tt hep-th/0609179}}].

\bibitem{Hanany:2014dia}
A.~Hanany and R.~Kalveks, {\it {Highest Weight Generating Functions for Hilbert
  Series}},  {\em JHEP} {\bf 10} (2014) 152,
  [\href{http://arxiv.org/abs/1408.4690}{{\tt arXiv:1408.4690}}].

\bibitem{Giombi:2009wh}
S.~Giombi and X.~Yin, {\it {Higher Spin Gauge Theory and Holography: The
  Three-Point Functions}},  {\em JHEP} {\bf 09} (2010) 115,
  [\href{http://arxiv.org/abs/0912.3462}{{\tt arXiv:0912.3462}}].

\end{thebibliography}
\end{document}